\documentclass[10pt]{article}

\usepackage{graphicx, epstopdf, caption}
\usepackage{amsmath, amssymb, mathtools}
\usepackage{color, paralist, longtable}
\usepackage[text={6.5in,9in},centering]{geometry}
\usepackage[numbers, comma, square, sort&compress]{natbib}
\usepackage[hypertexnames=false, colorlinks=true, linkcolor=blue, citecolor=blue, urlcolor=blue]{hyperref}
\usepackage[nottoc]{tocbibind}

\newcommand{\rev}[1]{{#1}}
\setcaptionmargin{0.25in}

\newtheorem{Theorem}{Theorem}[section]
\newtheorem{Hypothesis}[Theorem]{Hypothesis}
\newtheorem{Corollary}[Theorem]{Corollary}
\newtheorem{Definition}[Theorem]{Definition}
\newtheorem{Lemma}[Theorem]{Lemma}
\newtheorem{Proposition}[Theorem]{Proposition}
\newtheorem{Remark}[Theorem]{Remark}

\newenvironment{Acknowledgment}%
{\begin{trivlist}\item[]\textbf{Acknowledgments }}{\end{trivlist}}

\newcommand{\qed}{\hspace*{\fill}$\rule{0.3\baselineskip}{0.35\baselineskip}$}

\newenvironment{Proof}[1][.]%
{\begin{trivlist}\item[]\textbf{Proof#1 }}%
{\qed\end{trivlist}}

\makeatletter\@addtoreset{equation}{section}\makeatother

\newcommand{\C}{\mathbb{C}}

\newcommand{\R}{\mathbb{R}}
\newcommand{\Z}{\mathbb{Z}}

\def\Re{\mathop\mathrm{Re}\nolimits}    % real part
\def\Im{\mathop\mathrm{Im}\nolimits}    % imaginary part

\newcommand{\codim}{\mathop\mathrm{codim}\nolimits}
\newcommand{\id}{\mathop\mathrm{id}\nolimits}
\newcommand{\ind}{\mathop\mathrm{ind}\nolimits}
\newcommand{\Ns}{\mathrm{ker}}
\newcommand{\Rg}{\mathrm{Rg}}

\newcommand{\rmO}{\mathrm{O}}
\newcommand{\rmo}{\mathrm{o}}
\newcommand{\rmd}{\mathrm{d}}
\newcommand{\rme}{\mathrm{e}}
\newcommand{\rmi}{\mathrm{i}}

\begin{document}
\pagenumbering{roman}
\title{Spiral waves: linear and nonlinear theory}

\author{%
Bj\"orn Sandstede\\
Division of Applied Mathematics\\
Brown University\\
Providence, USA
\and
Arnd Scheel\\
School of Mathematics\\
University of Minnesota\\
Minneapolis, USA
}

\date{\today}
\maketitle

\begin{abstract}
Spiral waves are striking self-organized coherent structures that organize spatio-temporal dynamics in dissipative, spatially extended systems. In this paper, we provide a conceptual approach to various properties of spiral waves. Rather than studying existence in a specific equation, we study properties of spiral waves in general reaction-diffusion systems. We show that many features of spiral waves are robust and to some extent independent of the specific model analyzed. To accomplish this, we present a suitable analytic framework, spatial radial dynamics, that allows us to  rigorously characterize features such as the shape of spiral waves and their eigenfunctions, properties of the linearization, and finite-size effects. We believe that our framework can also be used to study spiral waves further and help analyze bifurcations, as well as provide guidance and predictions for experiments and numerical simulations. From a technical point of view, we introduce non-standard function spaces for the well-posedness of the existence problem which allow us to understand properties of spiral waves using dynamical systems techniques, in particular exponential dichotomies. Using these pointwise methods, we are able to bring tools from the analysis of one-dimensional coherent structures such as fronts and pulses to bear on these inherently two-dimensional defects. 
\end{abstract}

\vspace*{10mm}
\centerline{\textbf{2010 Mathematics Subject Classification}: 35B36, 35B40, 37L15}
\clearpage

\tableofcontents
\clearpage
\pagenumbering{arabic}
\setcounter{page}{1}

%%%%%%%%%%%%%%%%%%%%%%%%%%%%%%%%%%%%%%%%%%%%%%%%%%%%%%%%%%%%%%%%%%%%%%%%%

\section{Introduction}

\begin{figure}[b]
\centering
\includegraphics[scale=0.7]{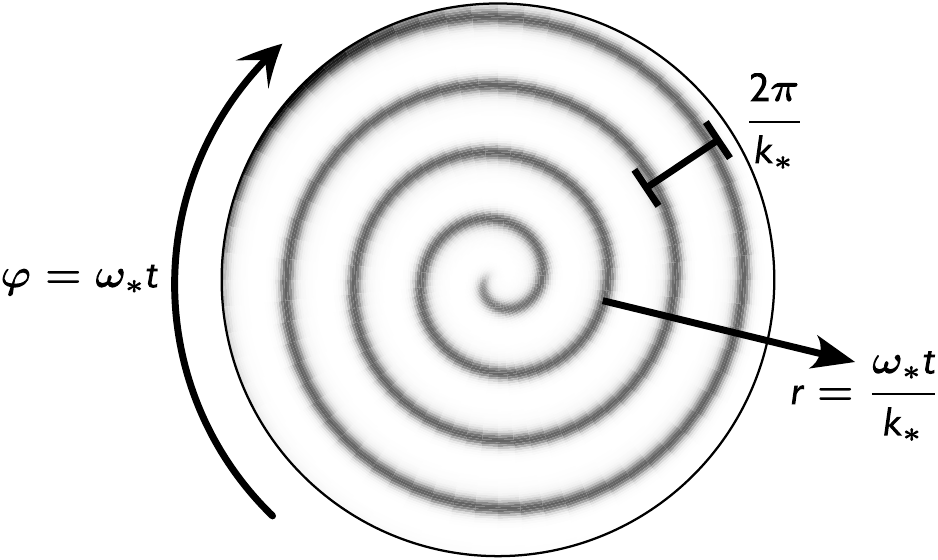}
\caption{Shown is a contour plot of an Archimedean planar spiral wave for a fixed value of time. The spiral wave rotates rigidly with temporal frequency $\omega_*$ \rev{around its center or core}, and consecutive spiral arms are approximately equidistant in the radial direction with period $2\pi/k_*$, where $k_*$ is the spatial wavenumber. \rev{Each spiral arm moves with speed approximately equal to $\omega_*/k_*$ in the radial direction.}}\label{fi:1}
\end{figure}

Spiral waves have been observed in numerous experiments, for instance in the Belousov--Zhabotinsky reaction, during the oxidation of carbon-monoxide on platinum surfaces, during arrhythmias in cardiac tissue, and as transient states during the aggregation of the slime mold \emph{Dictyostelium discoideum}. Archimedean spiral waves, \rev{which are illustrated in Figure~\ref{fi:1},} have also been found in numerical simulations of many different reaction-diffusion systems. Their importance is owed to  their prominent role in organizing the collective spatio-temporal dynamics, but possibly also to their aesthetic appeal. Observing spiral wave dynamics, one immediately notices both the topological nature of these defects, where a constant phase line terminates at the center, as well as the active nature of the center which emits waves that propagate away from the spiral center.

\paragraph{Excitable media.} \rev{Excitable media are characterized by the presence of a stable homogeneous rest state so that small perturbations return to this rest state, while large perturbations above a certain threshold lead to excitation waves.} Early interest in spiral waves was motivated by self-organized excitation waves in muscle tissue; see \cite{wiener} for an early reference and \cite{winfree} for a comprehensive exposition and review of this earlier literature. Early works focused on the organization of excitation waves into spiral structures, ignoring or postulating dynamics in the center of the spiral. However, in the context of excitable media, the core of the spiral is thought of as the key organizing element, creating sequences of excitation waves that emanate from the center in a medium that might otherwise simply return to a uniform rest state. More mathematical approaches, many in the context of the FitzHugh--Nagumo equation and mean-curvature description of excitation waves, resolved the core structure in spiral waves to a much more refined degree; see, for instance, \cite{keener1,keener2,bernoff} and \cite{belmonte,kapral,karma,barkleyrecipe} for more recent perspectives. Among the outcomes are accurate predictions of the frequencies of spiral waves in the singular fast-reaction limit.

\paragraph{Oscillatory media.} \rev{Spiral waves were studied also in oscillatory media, which are characterized by the presence of stable time-periodic oscillations. In this scenario,} one would try to describe a spatially extended system that exhibits temporal oscillations at every point in physical space through a  scalar variable that monitors the phase of the oscillation. One thereby obtains a map from the two-dimensional spatial domain into the circle. Spiral waves now correspond to the states where this phase variable has a non-trivial winding number away from the center of rotation. The core can then be thought of as merely a necessary phase singularity. Similar to the difficulty in excitable media, the core region is not easily resolved within the context of simple approximations. The first consistent results on existence of spiral waves focused on reaction-diffusion systems that coupled phase and amplitude of oscillations. In the simplest form, the kinetics possess a gauge symmetry, and the resulting equations are referred to as $\lambda$-$\omega$-systems or complex Ginzburg--Landau equations in the literature. In a peculiar limit where dispersion of oscillations can be eliminated, these systems reduce to the classical Ginzburg--Landau model of superconductivity, where spiral waves correspond to stationary vortices. Existence of spiral waves in these systems with gauge symmetry was established in a series of papers \cite{gre,greenberg2,hagan,kopell1,kopell2} and later extended to systems without gauge symmetry, but near Hopf bifurcation \cite{s-siam}. We refer to \cite{kramerworld} for an overview of dynamics in oscillatory media as captured by the complex Ginzburg--Landau equation and to \cite{pismen} for a broader discussion including both oscillatory and excitable media. 

Both these perspectives can be explored in the FitzHugh--Nagumo system, which, depending on reaction rates and levels of input currents, can be excitable or oscillatory. In the excitable regime, without further stimulus, the kinetics return to a stable rest state. Changing the input current as a homotopy parameter, stable periodic oscillations arise through a Hopf bifurcation and develop quickly into relaxation oscillations. Clearly, properties of the medium change quite dramatically during this homotopy. Nevertheless, spirals typically exist throughout and appear almost oblivious to these changes in the medium. Only in the regime of weak excitability, when a short temporal stimulus of small size is not sufficient to trigger an excitation in the kinetics, does one see spiral waves disappear.

\paragraph{Instabilities \rev{of spiral waves.}} Much of more recent theoretical and experimental work has focused on the phenomenology of instabilities of spiral waves. The interest was stimulated to a large extent by observations of spiral instabilities leading to breakup and spatio-temporal turbulence in reaction-diffusion systems, but also in cardiac tissue, where spiral waves and their instabilities are thought to be responsible for cardiac arrhythmias, tachycardia, and  ventricular fibrillation (see \cite{exc} for a collection of more recent contributions to the role of spiral waves in cardiac tissue). We refer to Figure~\ref{cf:1} for simulations that illustrate some of the instabilities we will describe in the next paragraphs.

The \emph{meander instability}  is an apparent instability of the spiral tip motion. It is  often supercritical and leads to two-frequency dynamics, where the spiral tip evolves on epicycloids. At parameter values when the relative  direction in which the two super-imposed circular motions occur changes sign one observes a drifting trajectory of the spiral tip. Frequency locking is not observed. The effect of the meandering motion of the tip are waves of compression and expansion in the far-field, which organize along super-spirals that rotate in the same or in the opposite direction of the primary spiral,  with the transition happening at the drifting transition; see \cite{swinneymeander,swinneymeander2} for examples of experimental analysis of transitions, and \cite{barkleyeuclid,fssw,ssw1,ssw2,golubitskymeander} for theoretical explanations based on effective tip motion on the Euclidean group. More complicated tip dynamics have also been observed; see \cite{hyperroessler,winfreehyper} for (numerical) experiments and \cite{ashwindrift,fiedlerturaevhyper} for theory. 

More dramatic instabilities cause \emph{spiral breakup}, where the compression and expansion of the waves emitted by the spiral wave grow in time and space, leading to filamentation and complex dynamics; see for instance  \cite{Ouyang1996} for experiments and  \cite{hagmer,PhysRevE.48.R1635,PhysRevLett.82.1160,PhysRevA.46.R2992} for analysis.  The compression and expansion can be modulated in the lateral direction of wave trains, leading to different fragmentation phenomenologies; see \cite{ogawa,panfilovlateral}. Spatio-temporal growth of perturbations has been described in terms of properties of dispersion relations at wave trains \cite{ss-spst} and the resulting subcritical instabilities are often very sensitive to noise and domain size.

A related instability results in \emph{alternans}, which are characterized by the property that the spiral arms are elongated and shortened periodically in time. Alternans have approximately twice the temporal period of the spiral waves from which they bifurcate. They have been implicated in the transition from tachycardia to fibrillation  \cite{Rosenbaum:1994bt,Pastore:1999dq}, and we refer to the review article \cite{Alonso:2016et} and the special issue \cite{Cherry:2017je} for analysis, modeling, and computations of alternans, and to \cite{dodson2019} for recent spectral computations.

A different type of \emph{period-doubling} instabilities can be associated with a period-doubling instability of the oscillations in the medium, which leads to line defects and slow drifting of the spiral core; see \cite{Yoneyama1995,224501PhysRevLett.88.} for experimental observations, \cite{kapralpd} for numerical explorations and analysis, \cite{sspd} for analysis, and \cite{dodson2019} for recent spectral computations.

During the creation of spirals from initial conditions and in the evolution of disturbances near instability thresholds, characteristic transport of disturbances can be observed. Spirals are formed when the core sends out waves so that the part of the domain occupied by the rigidly-rotating Archimedean structure grows in time. This outward transport is crucial even when the spiral is apparently rotating inwards and the  apparent phase of wave trains propagates towards the center of rotation \cite{kapralpd}. Spirals are notably insensitive to perturbations far away from the core and easily regenerate even after large perturbations in the far field. The super-spiral patterns that appear at meandering instabilities grow temporally outward from the core, yet with a weakly decaying amplitude;  disturbances that lead to far-field breakup grow outward both temporally and spatially; disturbances in core breakup appear to grow at first in the core region only; spirals near the period-doubling regime rotate inwards, yet disturbances are transported away from the core. 

It is this phenomenology of robustness and instabilities that motivates the analysis presented here, hopefully putting both analysis and numerical simulations on a more precise footing. Before delving into our setup, we caution the reader that the transport properties described and exploited here may be different for spiral waves observed in other circumstances, such as the often multi-armed slowly rotating waves in B\'enard convection \cite{Bestehorn1993} or the spiral arms of galaxies \cite{bertin1996spiral}.

\paragraph{Setup and conceptual assumptions.}
Our approach to the analysis of spiral waves is largely model-independent and provides a framework in which the phenomena mentioned above can be analyzed systematically. Rather than making assumptions directly on the system that guarantee, for instance, excitability, gauge invariants, or closeness to a Hopf bifurcation, we make conceptual assumptions that require the existence of particular solutions.

We consider general reaction-diffusion systems
\begin{equation}
\label{e:rdgen}
u_t=D \Delta u + f(u),\qquad u\in \R^N, \ x\in \Omega,\ t\geq0, 
\end{equation}
where either $\Omega=\R^2$ or $\Omega=\{|x|<R\}$ with $R\gg 1$ supplemented with appropriate boundary conditions at $|x|=R$. We assume that $D>0$ is a diagonal diffusion matrix with strictly positive entries on the diagonal and that the nonlinearity $f:\R^N\to\R^N$ describing the kinetics is of class $C^p$ with $p$ sufficiently large.

\rev{We are interested in spiral waves that exhibit an asymptotic spatially-periodic structure as indicated in Figure~\ref{fi:1} and formalize this characterization through the following assumptions. First, we consider (\ref{e:rdgen}) with  $x\in\R$ in one space dimension and assume that the resulting system admits a spatio-temporally periodic wave-train solution of the form $u(x,t)=u_\infty(kx-\omega t)$, where the profile $u_\infty(\xi)$ is $2\pi$-periodic (so that $u_\infty(\xi)=u_\infty(\xi+2\pi)$ for all $\xi\in\mathbb{R}$) for an appropriate temporal frequency $\omega\neq0$ and spatial wavenumber $k\neq0$. The wavelength or spatial period of the wave train is therefore $2\pi/k$. Next, we consider (\ref{e:rdgen}) on the unbounded plane $\Omega=\R^2$, since this allows us to characterize the shape of spiral waves in an asymptotic sense far away from the center of rotation. In polar coordinates $(r,\varphi)$, which are related via $x=r(\cos\varphi,\sin\varphi)$ to the Cartesian coordinate $x\in\R^2$,} this characterization \rev{(which we will make more precise in \S\ref{s:mr.asw})} roughly reads
\begin{equation}\label{e:rdspsh}
u(x,t)=u_*(r,\varphi-\omega t)\sim u_\infty(k r+\varphi-\omega t+\theta(r)),\qquad \text{with }\theta'(r)\to 0 \text{ as } r\to\infty,
\end{equation}
\rev{where $u(x,t)$ is the solution written in Cartesian coordinates, and $u_*(r,\varphi)$ is the spiral-wave profile written in polar coordinates. Note that the spiral wave} rotates around the origin with constant angular velocity and resembles a periodic wave train along any fixed ray emanating from the origin. Some of our results study the effect of finite domain size by truncation to large bounded disks $\Omega=\{|x|<R\}$. A key message of these results is that the effect of this restriction is very weak, and in fact exponentially small in $R$.  Besides this characterization of spiral waves through their limiting shape far away from the center of rotation, we note that the choice of an unbounded domain also introduces spatial translation in addition to rotation as a symmetry of the equation, a property that has been recognized as crucial to understanding the behavior of spiral waves especially near meandering transitions \cite{barkleyeuclid}.

We note that our results all require $D>0$, thus excluding some prominent prototypical models. We do not claim that our results readily extend to the case of vanishing diffusivities in one or more species. It appears that most phenomena observed for systems where diffusivity vanishes in one or more components are quite robust in regards to introducing small diffusion into these components. On the other hand, the vanishing of diffusivities appears to introduce  structure that might be helpful in understanding some of the instabilities listed above and an adaptation and extension of the results presented here could well shed light on these phenomena. 

\paragraph{Scope of results.}
Our main results can be roughly grouped into three categories. 

The first set of results is concerned with the characterization of spiral waves as special equilibria of \eqref{e:rdgen} in a corotating frame:
\begin{compactenum}
\item \emph{Conceptual characterization:} we give a precise far-field description of spiral waves refining \eqref{e:rdspsh};
\item \emph{Asymptotics:} we derive universal expansions of $\theta(r)$ in terms of properties of the wave train $u_\infty$;
\item \emph{Group velocity and multiplicity:} we clarify the role of the group velocity of the asymptotic wave trains for properties of spiral waves, especially local multiplicity and uniqueness;
\item \emph{Robustness:} we show that spiral waves exist for open classes of reaction diffusion systems, that is, they persist and vary continuously in an appropriate sense upon variations of system parameters.
\end{compactenum}

\begin{figure}
\centering\includegraphics[width=0.9\textwidth]{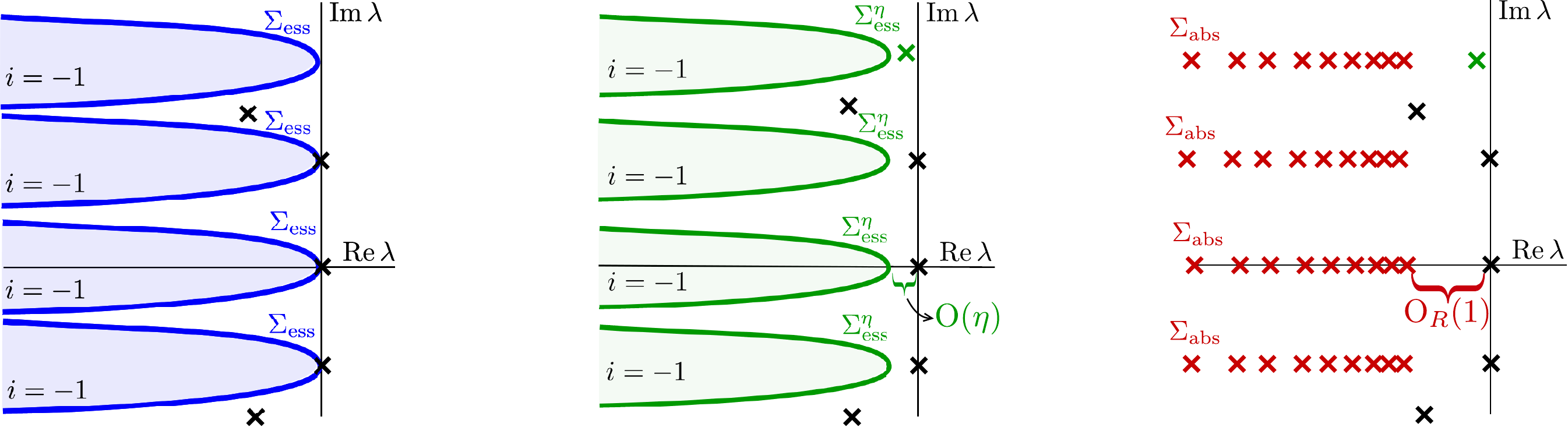}
\caption{Spectra of spiral waves in $L^2(\R^2)$ (left), $L^2_\eta(\R^2)$ (center), and $L^2(|x|\leq R)$ (right). The essential spectrum is periodic with vertical period $\rmi\omega_*$, and the borders of regions with constant Fredholm index are given by the Floquet spectra of wave trains. For positive group velocities, the Fredholm index to the left of the Fredholm border is $i=-1$. Exponential weights push spectral borders associated with positive group velocities to the left, and the resulting spectra generically move smoothly with the weight $\eta$. Eigenvalues do not depend on the exponential weight but may emerge from essential spectra; examples for the latter are translation and rotation eigenvalues at $\pm\rmi\omega_*$ and $0$, respectively, and the green eigenvalue near $2\rmi\omega_*$. On large bounded disks of radius $R\gg1$, eigenvalues cluster along curves given by the absolute spectrum of wave trains that do not depend on the radius $R$. We refer to Figures~\ref{cf:6} and~\ref{cf:5} for numerically computed examples.}\label{f:lin}
\end{figure}

The second set of results is concerned with properties of the linearization $\mathcal{L}_*$ about a spiral wave. In a corotating frame $\psi=\varphi-\omega t$, spiral waves are equilibria, and the goal is then to relate the phenomenology of instabilities described above to properties of the linearization. Our results characterize the spectral properties of this linear operator:
\begin{compactenum}
\item \emph{Essential spectra:} we characterize the essential spectrum of $\mathcal{L}_*$  and Fredholm indices of $\mathcal{L}_*-\lambda$ in terms of spectra and (generalized) group velocities of the asymptotic wave train;
\item \emph{Exponential weights:} we describe the change of essential spectra when $\mathcal{L}_*$ is considered in spaces of exponentially weighted functions in terms of group velocities of the asymptotic wave train; we show in particular that, in a typical stable scenario, the essential spectrum has strictly negative real part in spaces of functions with small exponential radial growth, reflecting outward transport of the oscillatory phase;
\item \emph{Point spectra:} we describe the shape of eigenfunctions and resonance poles in the far field, giving predictions for the phenomenology of instabilities caused by point spectrum;
\item \emph{Adjoints and response to perturbations:} we characterize properties of adjoint eigenfunctions and prove in particular that adjoint eigenfunctions associated with translation and rotation modes are typically exponentially localized, thus explaining on a linear level the robustness of tip motion of spiral waves with respect to perturbations in the far field.
\end{compactenum}
Figure~\ref{f:lin} illustrates spectra of the linearization, Fredholm indices, group velocities, and point spectra schematically.

\begin{figure}
\centering\includegraphics[width=0.9\textwidth]{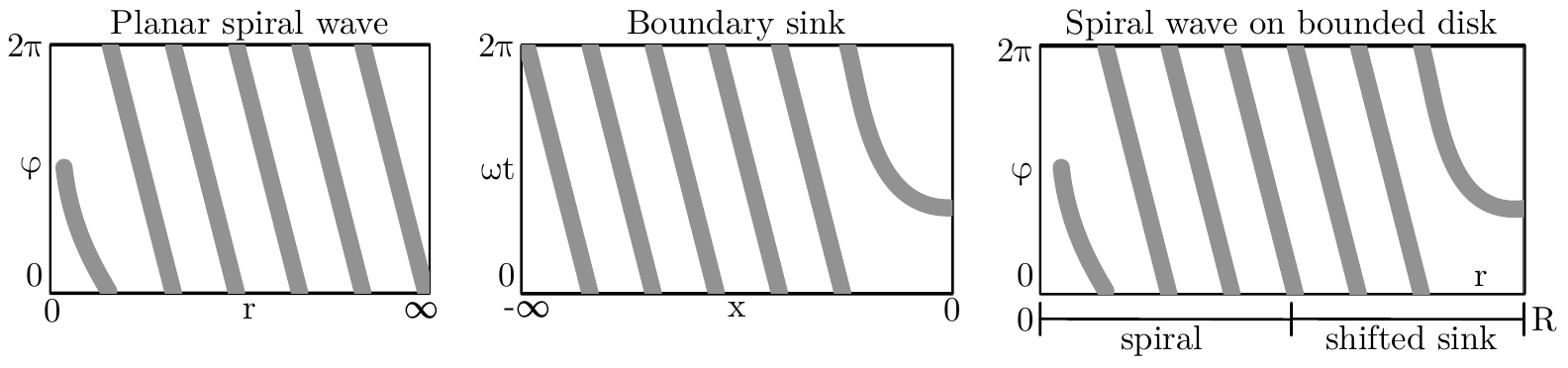}
\caption{\rev{From left to right, we show contour plots of a planar spiral wave $u_*(r,\varphi)$ in polar coordinates $(r,\varphi)$ connecting its core at $r=0$ with a wave train in the far field at $r=\infty$, a one-dimensional boundary sink $u_\mathrm{bs}(x,\omega t)$ in $(x,\omega t)$ coordinates for $x<0$ connecting the same wave train at $x=-\infty$ to Neumann boundary conditions at $x=0$, and a truncated spiral wave in polar coordinates $(r,\varphi)$ on a disk of radius $R\gg1$ with Neumann boundary conditions at $r=R$ that consists of the planar spiral wave glued together with the boundary sink (shifted by $R$ to the right) to accommodate the boundary conditions at $r=R$.}}
\label{f:gss}
\end{figure}

The last set of results is concerned with finite-size effects. We add a conceptual assumption on the interaction of the wave trains with boundary conditions: typically, wave trains are not compatible with a boundary condition, that is, $u_\infty(kx-\omega t)$ is not a solution to the reaction-diffusion system in $x<0$ when, say, Neumann boundary conditions are imposed at $x=0$. Since wave trains are time periodic, we therefore assume the existence of a time-periodic solution $u_\mathrm{bs}(x,\omega t)$ on $x<0$ that satisfies the boundary condition at $x=0$ and converges to the wave train $u_\mathrm{bs}(x,\omega t)\sim u_\infty(kx-\omega t)$ as $x\to-\infty$. For these boundary layers, the wave trains transport small disturbances \rev{from $x=-\infty$ towards the boundary at $x=0$}, and we therefore refer to these solutions as boundary sinks. We can now envision patching the spiral wave with such a boundary sink to obtain a solution on a large but finite disk \rev{as illustrated in Figure~\ref{f:gss}. Our results show the existence of truncated spiral waves and characterize their spectra:}
\begin{compactenum}
\item \emph{Truncation by gluing:} we prove the existence of rotating waves on disks of radius $R$ for sufficiently large $R$ whose profiles consist of the spiral wave glued together with a boundary sink;
\item \emph{Spectra of truncated spirals:} we show that spectra of the linearizations around truncated spiral waves converge as $R\to\infty$; the limit consists of a continuous part and a discrete part;
\item \emph{Extended point spectrum:} the discrete part of the limiting spectrum consists of the union of the spectra of $\mathcal{L}_*$ considered on the plane in suitable exponentially weighted spaces and the boundary sink considered on $\mathbb{R}^-$;
\item \emph{Absolute spectra:} the continuous part of the limiting spectrum is \emph{not} given by the essential spectrum but by semi-algebraic curves, which we refer to as the absolute spectrum, belonging to the wave trains.
\end{compactenum}
\rev{See again Figure~\ref{f:lin} for a schematic representation of the results on spectra.}

\paragraph{Techniques.} 
Our approach to the analysis of spiral waves is based on the method of spatial dynamics, casting existence and eigenvalue problems as evolution problems in the radial direction and using pointwise matching and gluing constructions in determining existence, bifurcation, and spectral properties. This method has been used extensively in the study of existence and bifurcation problems for elliptic equations starting with the pioneering work of Kirchg\"assner \cite{kirch82} \rev{and continued later for instance in \cite{Mielke1, Mielke2, Mielke3} to capture small-amplitude solutions. Most relevant for our perspective} here are the adaptation to radial dynamics \cite{scheelradial} and to bifurcation to spiral waves \cite{s-siam}. While in all of those examples, solutions are constructed as small perturbations of a spatially constant trivial solution, our approach is global in nature and can be compared with \cite{ssdefect} where properties  of time-periodic solutions asymptotic to wave trains in the far field are classified based on conceptual assumptions, not necessarily assuming that solutions are close to a trivial state. In such a global context, spatial dynamics are based on a pointwise description of the linear operator as an evolution problem via exponential dichotomies. \rev{In the context of elliptic equations on multi-dimensional domains, exponential dichotomies were first constructed in \cite{pss} and later used in \cite{ss-fred} to clarify the relation to Fredholm properties of the related elliptic operator, building on earlier work \cite{Coppel, Palmer1, Palmer2} for ordinary differential equations. Later work on exponential dichotomies for multi-dimensional domains includes, for instance, \cite{Beck1, Beck2, Latushkin1, Latushkin2, Latushkin3}.}

The approach via spatial dynamics allows us to utilize dynamical systems methods which provide powerful tools to study fine asymptotics of solutions to differential equations, in particular characterizing exponential asymptotics and the analysis of neutral, non-exponential modes via center-manifold reduction and geometric blowup. These fine asymptotics are essential here in many places, in particular when characterizing the asymptotic behavior of the phase function $\theta(r)$ of spiral waves in \eqref{e:rdspsh} in the far field, the shape of eigenfunctions in the point spectrum representing super spirals of compression and expansion, or the clustering of eigenfunctions near the absolute spectrum in large bounded disks. 

Many of the constructions here have been used in related but simpler situations \cite{ssdefect,ssgluing}. A major complication for spiral waves stems from the fact that there is no simple way to compactify at infinity: treating the Laplacian in radial coordinates $\partial_{rr}+\frac{1}{r}\partial_r + \frac{1}{r^2}\partial_{\varphi\varphi}$ as a non-autonomous dynamical system in $r$, we notice that the derivative operator in $\varphi$ disappears at $r=\infty$ due to the factor $\frac{1}{r^2}$. On the other hand, we see that the derivative $\omega\partial_\varphi$, introduced by passing to a corotating frame, is unbounded relative to the Laplacian so that the operator $\Delta + \omega \partial_\varphi$ is not sectorial. We overcome these difficulties by choosing appropriate anisotropic function spaces with norms based on $r^{-1}|\partial_\varphi u|+|\partial_\varphi^{\frac12} u|$ and using compactifications at infinity only on finite-dimensional reduced center manifolds. 

The challenges arising here are somewhat unique and not readily comparable to other work on defects in the literature. We remark however that a similar question of truncation of defects has been analyzed in \cite{pacardriviere} for Ginzburg--Landau vortices. The problem there is quite different as the relevant linear operators are mostly self-adjoint, and much more information is accessible explicitly. On the other hand, the absence of convective transport necessitates the use of algebraically weighted spaces, and the techniques are generally quite different from our approach here. 

\paragraph{Outline.} We present background material on wave trains in \S\ref{s:wts} before stating our main results in \S\ref{s:mr}. Section~\ref{s:wt} presents proofs of the main properties of wave trains collected in \S\ref{s:wts}. In \S\ref{s:ed}, we develop the framework of exponential dichotomies in the context of spiral waves, laying the basis for all later technical analyses. Using these exponential dichotomies, we study Fredholm properties of the linearization in \S\ref{s:fred}. \rev{We establish robustness of spiral waves and derive far-field expansions in \S\ref{s:rob} and analyze point spectra in \S\ref{s:ps}}. The \rev{next} three sections are concerned with the truncation of spiral waves to large disks: we cover the gluing construction with boundary sinks in \S\ref{s:trunc}, analyze the accumulation points of spectra for operators in large disks in \S\ref{s:abs}, and finally describe the limits of spectra including the effect of boundary sinks in \S\ref{s:absglue}. We conclude with a discussion, focusing in particular on the implications of our results to observations in experiments and simulations, in \S\ref{s:appl}.

%%%%%%%%%%%%%%%%%%%%%%%%%%%%%%%%%%%%%%%%%%%%%%%%%%%%%%%%%%%%%%%%%%%%%%%%%

\section{Background material on wave trains}\label{s:wts}

We consider the reaction-diffusion system 
\begin{equation}\label{e:rds0}
u_t = D u_{xx} + f(u), \qquad x\in\R, \quad u\in\R^N,
\end{equation}
where we may think of $u\in\R^N$ as a vector of chemical
concentrations. Furthermore, $D=\mathrm{diag}(d_j)>0$ is a positive,
diagonal diffusion matrix and $f$ is a smooth nonlinearity.
\rev{We refer to the coordinate system $(x,t)$ as the laboratory frame to distinguish it from coordinate frames that move with a travelling wave. Note that velocities of movement and transport depend on the underlying reference system.} 

We assume that (\ref{e:rds0}) has a wave-train solution
$u(x,t)=u_\infty(kx-\omega t)$ for a certain non-zero wavenumber $k$,
non-zero temporal frequency $\omega$, and wave speed $c=\omega/k$,
where the function $u_\infty$ is $2\pi$-periodic in its argument
$\xi=kx-\omega t$. Note that any such wave train $u_\infty(\xi)$ is a
$2\pi$-periodic solution of the ordinary differential equation (ODE)
\begin{equation}\label{e:wtode}
-\omega u^\prime = k^2 D u^{\prime\prime} + f(u).
\end{equation}
We are interested in the linearization of (\ref{e:rds0}) at the wave train \rev{and specifically in spectral information in the laboratory wave as this is the frame in which we will later view spiral waves. It is easier to compute the spectrum of the relevant linear operator in the frame that moves with the wave train, and we will therefore do this first in \S\ref{s:wts.co}, before we transfer these results in \S\ref{s:wts.lab} to the laboratory frame in which we will need the spectral information. In \S\ref{s:wts.ndr}, we will show that wave trains typically come in families where the profile $u_\infty$ and the temporal frequency $\omega$ are parametrized by the spatial wavenumber $k$. In \S\ref{s:wts.morse}, we will explore a spatial-dynamics formulation of the linear eigenvalue problem associated with a wave train, introduce and calculate relative Morse indices that can be thought of as the difference of the dimensions of generalized unstable and stable eigenspaces of a spatial-dynamics operator, and link the relative Morse index to group velocities -- these concepts and calculations will be used throughout the remainder of the paper. Finally, in \S\ref{s:wts.2d}, we consider instabilities of plane waves that are transverse to the direction of propagation.}

In passing, we remark that much of the discussion in this section can be presented in a simpler way by exploiting Floquet theory for parabolic equations (as developed, for instance, in \cite{kuc}). We prefer the slightly more complicated approach below since it naturally generalizes to travelling waves which are not necessarily spatially periodic \cite{ss-fred} and, in particular, provides us with a framework that we will encounter again when we study spiral waves.

\subsection{Spectra of wave trains in the co-moving frame}\label{s:wts.co}

In \rev{the} scaled co-moving coordinates $\xi=kx-\omega t$, the
reaction-diffusion system (\ref{e:rds0}) becomes
\begin{equation}\label{e:rdsco}
u_t = k^2 D u_{\xi\xi} + \omega u_\xi + f(u), \qquad \xi\in\R, \quad u\in\R^N,
\end{equation}
where $u(\xi,t)=u_\infty(\xi)$ is an equilibrium solution. Linearizing
(\ref{e:rdsco}) at this equilibrium $u_\infty$, we obtain the
differential operator
\begin{equation}\label{e:rdscol}
\mathcal{L}_\mathrm{co} :=
k^2 D \partial_{\xi\xi} + \omega\partial_\xi + f^\prime(u_\infty(\xi)),
\end{equation}
\rev{which we consider as an unbounded operator on $L^2(\R,\C^N)$ with domain $H^2(\R,\C^N)$}. The spectrum of $\mathcal{L}_\mathrm{co}$ \rev{on $L^2(\R,\C^N)$, given by the set of $\lambda\in\C$ for which $\mathcal{L}_\mathrm{co}-\lambda$ does not have a bounded inverse,} can be computed using the Bloch-wave ansatz
\[
u(\xi) = \rme^{\nu\xi/k} u_\mathrm{p}(\xi),
\]
where $\nu\in\rmi\R$ and $u_\mathrm{p}$ is $2\pi$-periodic in $\xi$. \rev{Denoting by $c=\omega/k$ the phase velocity of the wave train in the laboratory frame, we arrive} at the family of operators $\hat{\mathcal{L}}_\mathrm{co}(\nu)$ defined by
\begin{equation}\label{e:fam}
\hat{\mathcal{L}}_\mathrm{co}(\nu) u_\mathrm{p} = D (k\partial_\xi+\nu)^2 u_\mathrm{p}
+ c (k\partial_\xi+\nu) u_\mathrm{p} + f^\prime(u_\infty(\xi)) u_\mathrm{p},
\end{equation}
\rev{which we consider as unbounded operators on $L^2(S^1,\C^N)$ with domain $H^2(S^1,\C^N)$, where $S^1:=\R/2\pi\Z$. For each $\nu\in\rmi\R$, the spectrum of $\hat{\mathcal{L}}_\mathrm{co}(\nu)$ on $L^2(S^1,\C^N)$ is a discrete set in $\C$, and the union over $\nu\in\rmi\R$ of the spectra} of $\hat{\mathcal{L}}_\mathrm{co}(\nu)$ on $L^2(S^1,\C^N)$ gives the spectrum of
$\mathcal{L}_\mathrm{co}$ on $L^2(\R,\C^N)$; see, for instance, \cite{gar}. Thus,
the spectrum of $\mathcal{L}_\mathrm{co}$ is given by curves of the form
$\lambda=\lambda_\mathrm{co}(\nu)$ where $\nu\in\rmi\R$. These curves are
referred to as the (linear) dispersion curves.  Alternatively, we can
rewrite the eigenvalue problem
\[
\mathcal{L}_\mathrm{co} u = \lambda u
\]
as the ordinary differential equation
\begin{align}\label{e:rdscol0}
k u_\xi = &  v  \\ \nonumber
k v_\xi = &  -D^{-1}[c v + f^\prime(u_\infty(\xi)) u - \lambda u]
\end{align}
with $2\pi$-periodic coefficients. We denote by $\Phi(\lambda)$ the
associated period map which maps an initial value to the solution of (\ref{e:rdscol0}) evaluated at $\xi=2\pi$. In particular, the ODE (\ref{e:rdscol0}) has \rev{a solution that is bounded uniformly in $\xi\in\R$} if and only if the Evans function\footnote{\rev{Notation: we will never include a symbol for the identity operator when writing down scalar multiples of the identity.}} \cite{gar}
\begin{equation}\label{e:drper}
E(\lambda,\nu) := \det\left[\Phi(\lambda)-\rme^{2\pi\nu/k}\right] = 0
\end{equation}
vanishes for some $\nu\in\rmi\R$. The set of all $\lambda$ for which
$E(\lambda,\nu)=0$ has a purely imaginary solution $\nu$ is the
spectrum of $\mathcal{L}_\mathrm{co}$ on $L^2(\R,\C^N)$; see again \cite{gar}.
Since (\ref{e:drper}) defines an analytic function of $\lambda$ and
$\nu$, we can solve (\ref{e:drper}) for $\lambda$ as functions of
$\nu$ and find again an at most countable set of solution curves of
the form $\lambda=\lambda_\mathrm{co}(\nu)$ with $\nu\in\rmi\R$.
For any element $\lambda$ in the spectrum with
$\partial_\lambda E(\lambda,\nu)\neq0$ for some $\nu\in\rmi\R$, we can
solve $E(\lambda,\nu)=0$ locally for $\lambda=\lambda_\mathrm{co}(\nu)$ as
a function of $\nu$. For such elements, the \emph{linear group velocity}
\[
c_\mathrm{g,l} := -\frac{\rmd\lambda_\mathrm{co}}{\rmd\nu} + \frac{\omega}{k}
= -\frac{\rmd\lambda_\mathrm{co}}{\rmd\nu} + c
\]
in the original laboratory frame is well defined.
If $\lambda\in\rmi\R$, then the first term
$-\rmd\lambda_\mathrm{co}/\rmd\nu$ is the derivative of the temporal
frequency $\lambda$ of solutions of the linearized PDE with respect to
the spatial wavenumber $\nu$: this term gives the group velocity in
the co-moving frame, i.e. the velocity with which wave packets with
wavenumbers close to $\nu$ would propagate. The second term $\omega/k$
compensates for the moving frame in which we computed the group
velocity. Note that a dispersion curve $\lambda_\mathrm{co}(\nu)$ has a
vertical tangent precisely at points where $c_\mathrm{g,l}$ is real.
Note also that $E(0,0)=0$ since
$(u_\infty^\prime(\xi),ku_\infty^{\prime\prime}(\xi))$ is a bounded
solution of (\ref{e:rdscol0}) with $\lambda=0$ and $\nu=0$.

\subsection{The nonlinear dispersion relation}\label{s:wts.ndr}

\rev{The next result shows that, under an appropriate nondegeneracy assumption, wave trains} come in one-parameter families, where \rev{the profile and the} temporal frequency $\omega=\omega(k)$ \rev{depend smoothly} on the wavenumber $k$.

\begin{Proposition}[Families of wave trains and nonlinear group velocities]\label{p:nldr}
Assume that $u_\infty(\xi)$ is a $2\pi$-periodic solution of (\ref{e:wtode}) for $(k,\omega)=(k_*,\omega_*)$ with $k_*,\omega_*\neq0$. We also assume that the associated Evans function satisfies $\partial_\lambda E(0,0)\neq0$.  
\begin{compactenum}[(i)]
\item\rev{There are then smooth functions $u_\infty(\xi;k)$ that are $2\pi$-periodic in $\xi$ and a smooth function $\omega(k)$ both defined for each $k$ near $k_*$ with $u_\infty(\xi;k_*)=u_\infty(\xi)$ and $\omega(k_*)=\omega_*$ so that $(u_\infty(\cdot;k),\omega(k),k)$ satisfies (\ref{e:wtode}) for each $k$ near $k_*$. We refer to the function $\omega(k)$ as the \emph{nonlinear dispersion relation} and call its derivative $c_\mathrm{g,nl}(k):=\omega^\prime(k)$ the \emph{nonlinear group velocity}.}
\item The linear group velocity at $\lambda=\nu=0$ and the nonlinear group velocity coincide so that
\begin{equation}\label{e:ndr}
c_\mathrm{g} := c_\mathrm{g,l}\Big|_{\lambda=0,\nu=0} = c_\mathrm{g,nl}(k_*),
\end{equation}
and we refer to the common value $c_\mathrm{g}$ as ``the'' group velocity of the wave train in the laboratory frame.
\item Moreover, $\partial_\lambda E(0,0)\neq0$ implies that the kernel of $\hat{\mathcal{L}}_\mathrm{co}(0)$ on $L^2(S^1,\C^N)$ is one-dimensional and the kernel of the $L^2$-adjoint $\hat{\mathcal{L}}_\mathrm{co}(0)^*$ on $L^2(S^1,\C^N)$ is spanned by a single function $u_\mathrm{ad}(\xi)$. We find
\[
c_\mathrm{g,nl}(k_*)=\omega^\prime(k_*) =
- \frac{2k_* \langle u_\mathrm{ad},D u_\infty^{\prime\prime}\rangle}{\langle u_\mathrm{ad},u_\infty^\prime\rangle},
\]
where $\langle\cdot,\cdot\rangle$ denotes the standard inner product in $L^2(S^1,\C^N)$. 
\end{compactenum}
\end{Proposition}

Proposition~\ref{p:nldr} is proved in \S\ref{s:wt.1}. 

\subsection{\rev{Floquet spectra} of wave trains in the laboratory frame}\label{s:wts.lab}

In \rev{\S\ref{s:wts.co}}, we computed the spectra of wave trains in the co-moving frame. \rev{Here, we will demonstrate how we can compute the spectrum of the linearization in the laboratory frame $x$}. The linearization in the laboratory frame is the linear, non-autonomous parabolic equation
\begin{equation}\label{e:rdsl}
u_t = D u_{xx} + f^\prime(u_\infty(kx-\omega t)) u.
\end{equation}
Stability information is encoded in the associated linear period map
$\Psi_\mathrm{st}:L^2(\R,\C^N)\to L^2(\R,\C^N)$ that maps an initial function
$u(\cdot,0)$ at $t=0$ to the solution $u(\cdot,2\pi/\omega)$ of (\ref{e:rdsl}) evaluated at $t=2\pi/\omega$.

\begin{Definition}[Floquet spectrum]\label{d:fsp}
 We define the \emph{Floquet spectrum} of the wave
train as the set $\Sigma_\mathrm{st}$ of \rev{those $\lambda\in\C$} for which
$[\Psi_\mathrm{st}-\rme^{2\pi\lambda/\omega}]$ does not have a bounded inverse on \rev{$L^2(\R,\C^N)$}. 
\end{Definition}

The following Lemma~\ref{l:floquet} is proved in \S\ref{s:equiv}.

\begin{Lemma}[Floquet spectrum vs spectrum in the co-moving frame]\label{l:floquet}
The Floquet spectrum $\Sigma_\mathrm{st}$ of the linearization $\Psi_\mathrm{st}$ in the
laboratory frame can be computed from the dispersion curves
$\lambda_\mathrm{co}(\nu)$ with $\nu\in\rmi\R$ of the linearization $\mathcal{L}_\mathrm{\rev{co}}$ posed in the co-moving frame (\ref{e:rdscol}) on $L^2(\R,\C^N)$ by adding the speed of the co-moving
frame to the group velocity:
\begin{align}
\lambda\in\mathrm{spec}\,\mathcal{L}_\mathrm{co}
& \qquad \Longleftrightarrow \qquad
\lambda = \lambda_\mathrm{co}(\nu) \mbox{ for some } \nu\in\rmi\R, \nonumber\\ 
\lambda\in\mathrm{spec}\,\Psi_\mathrm{st}
& \qquad\Longleftrightarrow \qquad 
\lambda = \lambda_\mathrm{st}(\nu) := \lambda_\mathrm{co}(\nu)-c\nu+\rmi\omega\ell \mbox{ for some } \nu\in\rmi\R, \ell\in\Z.\label{e:trans}
\end{align}
\end{Lemma}

\rev{The relation (\ref{e:trans}) implies in particular that the group
velocity transforms according to simple Galilean addition of velocities: 
\rev{$-\rmd\lambda_\mathrm{st}/\rmd\nu$} in the laboratory frame is obtained
from the group velocity \rev{$-\rmd\lambda_\mathrm{co}/\rmd\nu$} in the co-moving
frame by adding the speed of the coordinate frame $c=\omega/k$.}
We refer to the curves $\lambda_\mathrm{st}(\nu)$ as the dispersion curves in the
laboratory frame. Typically, an element $\lambda$ of the Floquet spectrum lies on precisely one dispersion curve.

\begin{Remark}[Floquet periodicity]\label{r:perspec}
Note that the eigenvalue problem in the \rev{laboratory} frame is invariant under the transformation 
$u\mapsto\rme^{\rmi\ell\omega t}u$, $\nu\mapsto\nu-\rmi\ell k$ and
$\lambda\mapsto\lambda+\rmi\omega\ell$ for any $\ell\in\Z$, where we
satisfy the requirement that $u$ needs to be $2\pi$-periodic. Hence,
the Floquet spectrum is invariant under translations by integer
multiples of $\rmi\omega$. This periodicity represents
precisely the ambiguity in the definition of the temporal Floquet
exponent $\lambda_\mathrm{st}$ as the logarithm of the Floquet multiplier.
\end{Remark}

\begin{Definition}[Spectrally stable wave trains]\label{d:seval}
We say that a wave train is spectrally stable if its Floquet spectrum is contained in $\Re\lambda<0$ with the exception of a simple dispersion curve at $\lambda=0$ (and, by Floquet periodicity, at $\lambda\in\omega\rmi\Z$). Here, we say that a dispersion curve at $\lambda$ is simple if $E(\lambda+c\nu,\nu)$ has precisely one purely imaginary
root $\nu$ and $\partial_\lambda E(\lambda+c\nu,\nu)\neq0$ where
$c=\omega/k$. Simple dispersion curves are given as analytic curves $\lambda(\nu)$ parametrized by $\nu\in\rmi\R$ that we shall orient with
decreasing(!) $\Im\nu$ so that curves point upward in the complex
plane at points of positive group velocity.
\end{Definition}

\begin{Remark}[Bloch waves]\label{r:eigfun}
To each spectral value $\lambda_\mathrm{co}(\nu)$ for a given
$\nu\in\rmi\R$, there corresponds an almost-eigenfunction
$u(\xi)=\rme^{\nu\xi/k}u_\mathrm{p}(\xi;\lambda,\nu)$ of
$\mathcal{L}_\mathrm{co}$, where the Bloch-wave function
$u_\mathrm{p}(\cdot;\lambda,\nu)$ is $2\pi$-periodic. An almost
eigenfunction of $\lambda_\mathrm{st}(\nu)$ in the laboratory frame is obtained
by substituting $\xi=kx-\omega t$ such that
\[
u(x,t) = \rme^{\lambda_\mathrm{co} t} \rme^{\nu(kx-\omega t)/k} u_\mathrm{p}(kx-\omega t;\lambda_\mathrm{co},\nu) = \rme^{\lambda_\mathrm{st} t} \rme^{\nu x} u_\mathrm{p}(kx-\omega t;\lambda_\mathrm{co},\nu).
\]
\end{Remark}

\begin{Remark}[Exponential weights]\label{r:expweight}
If we consider (\ref{e:rdsco}) or (\ref{e:rdsl}) in  $L^2$-spaces
with exponential weights
\[
L^2_\eta(\R,\C^N) := \{ u\in L^2_\mathrm{loc};\; |u|_{L^2_\eta}<\infty \},
\qquad
|u|_{L^2_\eta}^2 := \int_\R |u(x)\rme^{\eta x}|^2 \,\rmd x,
\]
all the above results apply if we fix $\Re\nu=-\eta$. In particular,
consider a point $\lambda_\mathrm{st}(\nu)$ on a dispersion curve with real
group velocity $c_\mathrm{g,l}$. The real part of the dispersion curve
$\lambda_\mathrm{st}(\nu;\eta)$ in the exponentially weighted space moves
according to
\[
\frac{\partial\lambda_\mathrm{st}(\nu;\eta)}{\partial\eta} =
\frac{\partial\lambda_\mathrm{st}(\nu-\eta;0)}{\partial\eta} =
-\frac{\partial\lambda_\mathrm{st}(\nu)}{\partial\nu} = c_\mathrm{g,l}.
\]
In particular, exponential weights with negative exponents stabilize
elements in the spectrum with positive group velocities. This is in
accordance with the intuition that transport towards $x\to\infty$ is
stabilized by a weight function $\rme^{\eta x}$ with $\eta<0$.
\end{Remark}

Note that we used the Cauchy--Riemann equation in the above remark,
since the (real) exponential weight shifts the real part of the
eigenvalue $\lambda$ with $-\rmd\Re\lambda/\rmd\Re\nu$,
whereas the group velocity is traditionally defined via the imaginary
parts $\rmd\Im\lambda/\rmd\Im\nu$. Since the eigenvalue
problems are analytic in $\lambda$, both derivatives coincide.

\subsection{Relative Morse indices and spatial eigenvalues}\label{s:wts.morse}

If we substitute the Floquet ansatz $u(x,t)=\rme^{\lambda t}\tilde{u}(x,\omega t)$ into (\ref{e:rdsl}),
change coordinates\footnote{\rev{Notation: The variables $\xi$ and $\sigma$ are both equal to $kx-\omega t$. We will use $\xi$ for the reaction-diffusion operators and $\sigma$ for spatial-dynamics formulations.}} by replacing the temporal time-variable $t$ by $\sigma=kx-\omega t$, and write $u$ for $\tilde{u}$, we obtain the autonomous equation
\begin{align*}
u_x = &  -k \partial_\sigma u + v \\
v_x = &  -k \partial_\sigma v - D^{-1}[\omega\partial_\sigma u
+ f^\prime(u_\infty(\sigma)) u - \lambda u],
\end{align*}
which we also write as $\mathbf{u}_x=\mathcal{A}_\infty(\lambda)\mathbf{u}$. 

\begin{Lemma}[Spectra from spatial dynamics in the steady frame]\label{l:equiv}
A complex number $\lambda$ is in the Floquet spectrum if and only if the spectrum of 
$\mathcal{A}_\infty(\lambda)$, considered as a closed operator on
$H^{\frac12}(S^1,\C^N)\times L^2(S^1,\C^N)$ with domain
$H^{\frac32}(S^1,\C^N)\times H^1(S^1,\C^N)$, intersects the imaginary axis. 
Furthermore, the spectrum of $\mathcal{A}_\infty(\lambda)$ is a
countable set $\{\nu_j(\lambda)\}_{j\in\Z}$ of isolated eigenvalues $\nu_j(\lambda)$ with finite multiplicity. If ordered by increasing real part, the spatial eigenvalues $\nu_j$ satisfy $\Re\nu_j\to\pm\infty$ as $j\to\pm\infty$.
\end{Lemma}

Lemma~\ref{l:equiv}, which is proved in \S\ref{s:equiv},
therefore leads us to consider the spatial eigenvalue problem
\begin{align}\label{e:spmtw}
\nu u = &  -k \partial_\sigma u + v \\ \nonumber
\nu v = &  -k \partial_\sigma v - D^{-1}[\omega\partial_\sigma u
+ f^\prime(u_\infty(\sigma)) u - \lambda u]
\end{align}
with $2\pi$-periodic boundary conditions for $(u,v)$. As in the
preceding lemma, we denote the eigenvalues of $\mathcal{A}_\infty(\lambda)$
by $\nu_j(\lambda)$, repeat them by multiplicity, and order them by increasing real part so that
\[
\ldots\leq \Re\nu_{-(j+1)} \leq \Re\nu_{-j} \leq\ldots\leq
\Re\nu_{-1} \leq \Re\nu_0 \leq \Re\nu_1 \leq\ldots\leq
\Re\nu_{j-1} \leq \Re\nu_{j} \leq\ldots.
\]
The spatial eigenvalues $\nu_j(\lambda)$ are precisely the solutions of
$E(\lambda+c\nu,\nu)=0$ for fixed $\lambda$. Since the
$\nu_j=\nu_j(\lambda)$ are eigenvalues of an analytic family of
operators, we can follow each individual eigenvalue in the parameter
$\lambda$ although the labelling might jump for certain values of
$\lambda$.

We will next normalize the labeling with respect to the relabeling transformation $\nu_j\mapsto \nu_{j+1}$, $j\in\Z$. We therefore start with a value $\lambda=\lambda_\mathrm{inv}\gg1$ such that
$\mathcal{L}_\mathrm{\rev{co}}-\lambda_\mathrm{inv}$ has a bounded inverse. We fix the
labelling of the spatial eigenvalues belonging to
$\lambda_\mathrm{inv}$ by requiring that $\Re\nu_{-1}<0<\Re\nu_0$,
where we use that none of the $\nu_j$ is purely imaginary since we are
in the resolvent set.

\begin{figure}
\centering\includegraphics[width=0.7\textwidth]{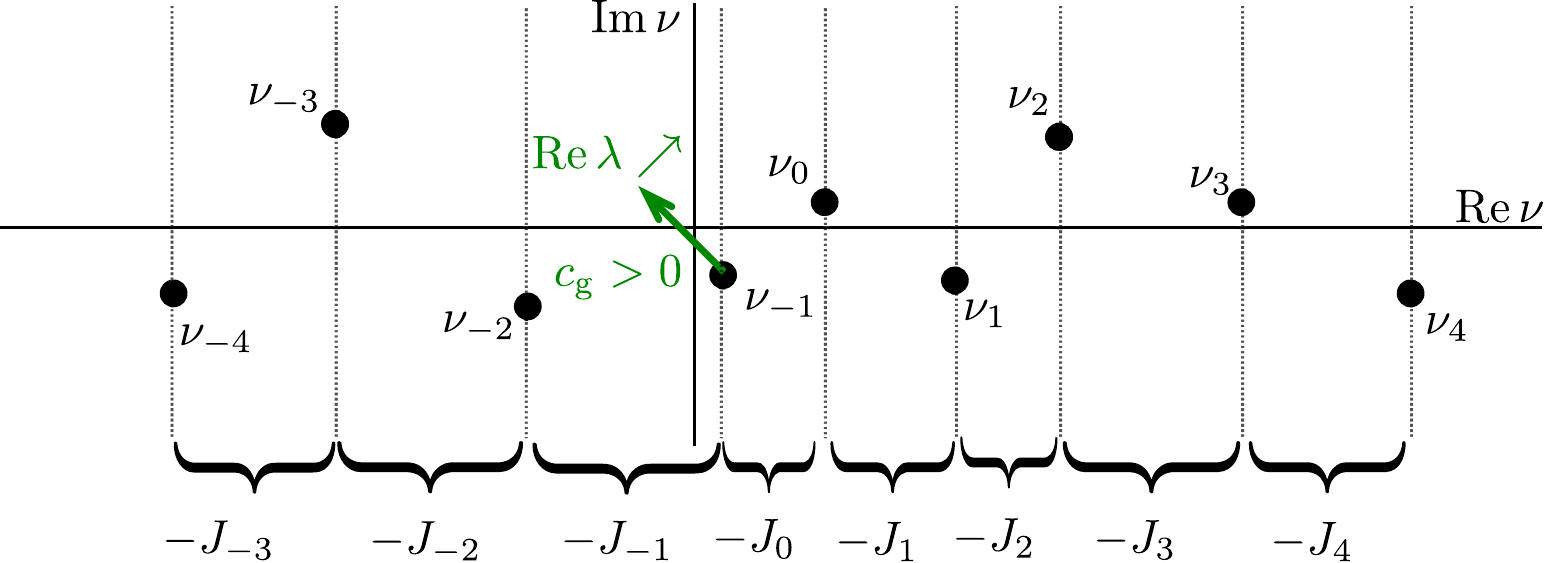}
\caption{Schematic representation of spatial Floquet exponents $\nu_j$ ordered by real part for a fixed $\lambda\in\C$, with corresponding (negative!) spectral gap intervals $-J_j$. Eigenvalues move left (or right) as $\lambda$ is varied depending on the sign of the group velocity. The relative Morse index $i_\mathrm{M}$ changes from $i_\mathrm{M}=1$ in the picture shown to  $i_\mathrm{M}=0$ as $\Re\lambda\nearrow$ increases through zero and $\nu_{-1}(\lambda)$ follows the green arrow.}\label{f:spatial_eigenvalues}
\end{figure}

Following the spatial eigenvalues $\nu_j(\lambda)$ from this region in the complex $\lambda$-plane defines a unique labelling of the eigenvalues except at points where some of the spatial eigenvalues have equal real part. In each of those cases, however, only a finite number of spatial eigenvalues share the same fixed real part since $\Re\nu_j\to\pm\infty$ as $j\to\pm\infty$ by Lemma~\ref{l:equiv}. In other words, if two spatial eigenvalues have the same real part for some value of $\lambda$, then there are $j\in\Z$ and $m\geq1$ so that $\Re\nu_{j-1}<\Re\nu_j=\Re\nu_{j+\ell}<\Re\nu_{j+m+1}$ for $\ell=1,\ldots,m$ (we note that there could be many possible real-part resonances occurring simultaneously for different real parts: each of these real-part resonances involves only finitely many spatial eigenvalues though). We can therefore continue labelling the spatial eigenvalues in a consistent fashion through any such real-part resonance by changing the indices of only those finitely many eigenvalues that are involved in a real-part resonance at a specific real part, within the set of indices associated with these same finitely many eigenvalues. 

\begin{Definition}[Relative Morse index]\label{d:rM}
For each $\lambda$ that does not belong to the Floquet spectrum of the
wave trains, we define the \emph{relative Morse index}
$i_\mathrm{M}(\lambda)$ as the negative index of the first spatial eigenvalue
with positive real part. In other words, $i_\mathrm{M}(\lambda)$ is
the unique index for which
\[
\ldots\leq \Re\nu_{-i_\mathrm{M}(\lambda)-1}(\lambda) < 0 <
\Re\nu_{-i_\mathrm{M}(\lambda)}(\lambda) \leq\ldots
\]
\end{Definition}

The following definition will allow us to relate spatial eigenvalues and exponential weights.

\begin{Definition}[Spatial spectral gaps]\label{d:j}
For each $\ell\in\Z$, we define $J_{\ell}(\lambda):=(-\Re\nu_{\ell}(\lambda),-\Re\nu_{\ell-1}(\lambda))$, assuming the ordering in Definition~\ref{d:rM}. Note that $J_\ell(\lambda)$ will be empty if $\,\Re\nu_{\ell}(\lambda)=\Re\nu_{\ell+1}(\lambda)$. Also note that the intervals are defined with the negative signs of the $\Re\nu_j$ such that for all $\ell\in\Z$ and all $\eta\in J_\ell$ we have $\Re\nu_j+\eta>0$ for $j\geq\ell$ and $\Re\nu_j+\eta<0$ for $j\leq\ell-1$. See Figure~\ref{f:spatial_eigenvalues} for a schematic representation of Floquet exponents and spectral gap intervals and Figure~\ref{cf:5} for numerically computed spatial Floquet exponents $\nu_j$. 
\end{Definition}

The next remark, which follows directly from our definitions, relates the relative Morse index at $\lambda=0$ and the nonlinear group velocity.

\begin{Remark}[Relative Morse indices and group velocities]\label{r:morse}
Assume the wave train is spectrally stable (see Definition~\ref{d:seval}), then we have $i_\mathrm{M}(\lambda)=0$ for all $\lambda>0$. If, in addition, its nonlinear group velocity $c_\mathrm{g}$ is positive, then a single spatial eigenvalue $\nu$ of $\mathcal{A}_\infty(\lambda)$ crosses through the origin from left to right when $\lambda$ decreases through zero, and this spatial eigenvalue is therefore given by $\nu_{-1}(\lambda)$. In particular, the unstable dimension increases by one as $\lambda$ decreases through zero, and we therefore have $i_\mathrm{M}(\lambda)=+1$ for $\lambda$ to the left of the critical dispersion curve and $J_0(0)=(-\Re\nu_0(0),0)\subset\R^-$. Similarly, the relative Morse index to the left of the critical spectral curve is $-1$ if the group velocity is negative.
\end{Remark}

\subsection{Transverse stability of wave trains}\label{s:wts.2d}

We conclude this section by collecting some properties of wave trains in two
space dimensions. We consider (\ref{e:rds0}) on $\R^2$,
\[
u_t = D (\partial_{xx}+\partial_{yy}) u + f(u), \qquad (x,y)\in\R^2,
\]
and notice that wave trains appear as plane waves $u(x,y,t)=u_\infty(kx-\omega t)$ that are independent of $y$. \rev{We say that the plane wave admits a transverse instability if it is stable with respect to perturbations that depend only on $x$ but becomes unstable when we allow perturbations to depend on $x$ and $y$. The stability of a plane wave} with respect to two-dimensional perturbations in the co-moving frame $\xi=kx-\omega t$ is determined by the linearized eigenvalue problem
\[
D (k^2\partial_{\xi\xi}+\partial_{yy}) u + \omega \partial_\xi u + f^\prime(u_\infty(\xi)) u = \lambda u,
\]
and the Fourier--Bloch ansatz \rev{$u(\xi,y)=\rme^{\nu_\perp y} v(\xi)$ with $\nu_\perp\in\rmi\R$} then leads to the spectral problem
\begin{equation}\label{e:rdscoly}
\mathcal{L}_{\rev{\perp}}(\nu_\perp) v := D (k^2\partial_{\xi\xi}+\nu_\perp^2) v
+ \omega \partial_\xi v + f^\prime(u_\infty(\xi)) v = \lambda v
\end{equation}
with $v\in L^2(S^1,\C^N)$. We focus on the long-wavelength stability $\nu_\perp\sim0$ of the translational eigenfunction $v=u_\infty^\prime$ with $\nu_\perp=0$ and denote by $u_\mathrm{ad}$ the generator of the kernel of the $L^2$-adjoint of \rev{$\mathcal{L}_\perp(0)=\hat{\mathcal{L}}_\mathrm{co}(0)$ posed on $L^2(S^1,\C^N)$}.

\begin{Lemma}[Transverse long-wavelength stability]\label{l:2d}
Assume that $u_\infty$ is a wave train whose eigenvalue at $\lambda=0$ is algebraically simple in the co-moving frame so that $\partial_\lambda E(0,0)\neq0$, then for each $\nu_\perp\sim0$ \rev{the operator $\mathcal{L}_\perp(\nu_\perp)$ has} a unique eigenvalue $\lambda_\perp(\nu_\perp)$ close to zero, and we
have the expansion $\lambda_\perp(\nu_\perp)=d_\perp\nu_\perp^2+\rev{\rmO(\nu_\perp^4)}$ where
\begin{equation}\label{e:dperp}
d_\perp = \frac{\langle u_\mathrm{ad},Du_\infty^\prime\rangle_{L^2(S^1)}}%
{\langle u_\mathrm{ad},u_\infty^\prime\rangle_{L^2(S^1)}}.
\end{equation}
In particular, the wave trains are spectrally unstable with respect to
long-wavelength transverse perturbations if $d_\perp<0$ (note $\nu_\perp\in\rmi\R$).
\end{Lemma}

Lemma~\ref{l:2d} is proved in \S\ref{s:wt.1}.
For later use, we remark that the eigenfunctions \rev{$u(\xi;\nu_\perp)$ to
\[
\mathcal{L}_\perp(\nu_\perp) u = \lambda_\perp(\nu_\perp) u
\]
can be chosen to be differentiable with respect to $\nu_\perp$ after a suitable normalization and that the second derivative $u_{\nu_\perp\nu_\perp}\rev{(\xi;0)}$ satisfies the equation
\begin{equation}\label{e:nunu}
\mathcal{L}_\perp(0) u_{\nu_\perp\nu_\perp} =
\hat{\mathcal{L}}_\mathrm{co}(0) u_{\nu_\perp\nu_\perp} =
2 (D u_\infty^\prime - d_\perp u_\infty^\prime)
\end{equation}
independent of the normalization.}

%%%%%%%%%%%%%%%%%%%%%%%%%%%%%%%%%%%%%%%%%%%%%%%%%%%%%%%%%%%%%%%%%%%%%%%%%

\section{Main results}\label{s:mr}

\rev{We present our main definitions and results. We define planar Archimedean spiral waves formally in \S\ref{s:mr.asw}, characterize the spectra of their PDE linearization in \S\ref{s:mr.fredholm}, provide asymptotic expansions and robustness results of planar spiral waves in \S\ref{s:mr.asym}, establish far-field expansions of eigenfunctions in \S\ref{s:mr.exp}, discuss persistence results for planar spiral waves to large bounded disks in \S\ref{s:mr.psw}, characterize the spectra of spiral waves under restriction and truncation to bounded disks in \S\ref{s:spectratrunc} and \S\ref{s:mr.tsp}, respectively, and describe scenarios in \S\ref{ss:spiralperp} for which the spectral mapping theorem fails for planar spiral waves. The proofs of these results are provided in subsequent sections.}

\subsection{Archimedean spiral waves}\label{s:mr.asw}

We are interested in Archimedean spiral waves of planar reaction-diffusion systems,
\begin{equation}\label{e:rds-cc}
u_t = D \Delta u  + f(u), \qquad x\in\R^2, \ u\in\R^N,
\end{equation}
that we shall characterize as solutions with particular spatio-temporal behavior. To do so, we view (\ref{e:rds-cc}) in polar coordinates \rev{$(r,\varphi)\in\R^+\times(\R/2\pi\Z)$ with $x=r(\cos\varphi,\sin\varphi)\in\R^2$ for which (\ref{e:rds-cc}) becomes
\begin{equation}\label{e:rds-}
u_t = D \Delta_{r,\varphi} u  + f(u), \qquad u(r,\varphi,t)\in\R^N,
\end{equation}
where
\[
\Delta_{r,\varphi} := \partial_{rr}+\frac{1}{r}\partial_r + \frac{1}{r^2}\partial_{\varphi\varphi}
\]
is the Laplacian expressed in polar coordinates.
}

\begin{Definition}[Spiral waves]
We say that a rigidly rotating solution $u(r,\varphi,t)=u_*(r,\varphi-\omega_* t)$ of (\ref{e:rds-}) with $\omega_*> 0$ is an  \emph{(Archimedean) spiral wave} if there exists a smooth $2\pi$-periodic non-constant function $u_\infty(\vartheta)$, a smooth function $\theta(r)$ with $\theta^\prime(r)\to0$ as
$r\to\infty$, and a non-zero constant $k_*$ such that
\[
|u_*(r,\cdot-\omega_*t) -
u_\infty(k_*r+\theta(r)+\cdot-\omega_*t)|_{C^1(\R/2\pi\Z)} \to 0
\mbox{ as } r\to\infty,
\]
where the profile $u_\infty(\cdot)$ is a wave-train solution of the one-dimensional reaction-diffusion system (\ref{e:rds0}). 
In other words, Archimedean spiral waves are asymptotic to wave trains $u_\infty$ far from the center of rotation and therefore approximately constant along arcs $k_*r+\varphi\equiv \omega_* t$, that rotate rigidly in time around the origin.
\end{Definition}

In the corotating frame $\psi=\varphi-\omega_* t$, rotating waves are equilibria and satisfy 
\begin{equation}\label{e:rds}
0 = D \Delta_{r,\psi} u + \omega_* \partial_\psi u + f(u), \qquad u=u(r,\varphi)\in\R^N.
\end{equation}
Note that the condition $\theta^\prime(r)\to0$ as $r\to\infty$ implies that $\theta(r)/r\to0$ as $r\to\infty$.

\subsection{Fredholm properties of the linearization at spiral waves}\label{s:mr.fredholm}

\rev{Upon linearizing the reaction-diffusion system (\ref{e:rds}) in the corotating frame at the spiral wave $u_*$, we obtain a system of the form $u_t=\mathcal{L}_*u$. We will always consider the resulting linear operator $\mathcal{L}_*$ in Cartesian coordinates so that it is given by}
\begin{equation}\label{e:lrds}
\mathcal{L}_* = D \Delta + \omega_* \partial_\psi + f^\prime(u_*(r,\psi)),
\end{equation}
\rev{where $\Delta$ is the Laplacian in Cartesian coordinates, $\partial_\psi$ is given in Cartesian coordinates $x=(x_1,x_2)$ by $\partial_\psi=x_1\partial_{x_2}-x_2\partial_{x_1}$, and we consider the profile $u_*(r,\psi)=u_*(r(x),\psi(x))$ also in Cartesian coordinates $x\in\R^2$. We are interested in spectral properties of the operator $\mathcal{L}_*$ on $L^2(\R^2,\C^N)$. Note that $\mathcal{L}_*$ is closed and densely defined on $L^2(\R^2,\C^N)$ as a bounded perturbation of the commuting operators $\Delta$ and $\partial_\psi=x_1\partial_{x_2}-x_2\partial_{x_1}$, and its domain contains the intersection of the domains $H^2(\R^2,\R^N)$ and $\{u\in L^2(\R^2,\R^N):\; \partial_\psi u\in L^2(\R^2,\R^N)\}$ of these two operators. Furthermore, $\mathcal{L}_*$ generates a strongly continuous semigroup on $L^2$ since $D\Delta$ and $\omega_*\partial_\psi$ generate commuting contraction semigroups on $L^2(\R^2,\C^N)$.}

%We are interested in spectral properties of the operator
%$\mathcal{L}_*$ on $L^2(\R^2,\C^N)$. Since the operators $\Delta$ and
%$\partial_\psi=x_1\partial_{x_2}-x_2\partial_{x_1}$ commute,
%$\mathcal{L}_*$ is densely defined as its domain includes the
%intersection of the domains $H^2(\R^2,\R^N)$ and $\{u\in
%L^2(\R^2,\R^N):\; \partial_\psi u\in L^2(\R^2,\R^N)\}$ of these two
%operators. More precisely, note that $D\Delta$ with domain
%$H^2(\R^2,\R^N)$ generates a contraction semigroup on $L^2(\R^2,\C^N)$.
%We also know that the family of operators
%$\mathcal{R}_\varphi:u(x)\mapsto u(R_{-\varphi}x)$, where $R_\varphi$
%is the rotation by the angle $\varphi$ in the plane, is a contraction
%semigroup on $L^2(\R^2,\C^N)$ whose generator $\omega_*\partial_\psi$
%has domain $\{u\in L^2(\R^2,\R^N):\; \partial_\psi u\in
%L^2(\R^2,\R^N)\}$. Since the semigroups generated by $D\Delta$ and
%$\omega_*\partial_\psi$ commute, it follows for instance from
%\cite[Theorem~II.2.7]{engelnagel} that their product semigroup is also
%a contraction semigroup which is generated by
%$D\Delta+\omega_*\partial_\psi$ defined on a dense subset of
%$L^2(\R^2,\C^N)$ that includes the intersection of the domains of
%$D\Delta$ and $\omega_*\partial_\psi$. Finally, $\mathcal{L}_*$ is
%closed and densely defined as a bounded perturbation of the operator
%$D\Delta+\omega_*\partial_\psi$ and generates a strongly continuous
%semigroup on $L^2(\R^2,\C^N)$ with the same domain by
%\cite[Theorem~III.1.3]{engelnagel}.

\rev{We say that a closed, densely defined, linear operator $\mathcal{T}$ defined on a Hilbert space $H$ is Fredholm if its range $\Rg(\mathcal{T})$ is closed in $H$ and both its null space $\Ns(\mathcal{T})$ and the complement of its range $\Rg(\mathcal{T})$ are finite-dimensional. The index of a Fredholm operator is $\ind(T):=\dim\Ns(\mathcal{T})-\codim\Rg(\mathcal{T})$.}

\begin{Definition}[Spectrum]\label{d:spec}
We call the set
\[
\Sigma_*:=\{\lambda\in\mathbb{C}: \mathcal{L}_*-\lambda \mbox{ does not have a bounded inverse on } L^2(\R^2,\C^N)\}
\]
the spectrum of $\mathcal{L}_*$. We write $\Sigma_*=\Sigma_\mathrm{pt}\overset{\cdot}{\cup}\Sigma_\mathrm{fb}\overset{\cdot}{\cup}\Sigma_{i\neq0}$ where
\begin{compactitem}
\item Point spectrum: $\Sigma_\mathrm{pt}=\{\lambda\in\mathbb{C}: \mathcal{L}_*-\lambda \mbox{ is Fredholm with index 0 and the kernel of } \mathcal{L}_*-\lambda \mbox{ is nontrivial}\}$,
\item Fredholm boundary: $\Sigma_\mathrm{fb}=\{\lambda\in\mathbb{C}: \mathcal{L}_*-\lambda \mbox{ is not Fredholm}\}$,
\item Fredholm spectrum: $\Sigma_{i\neq0}=\{\lambda\in\mathbb{C}: \mathcal{L}_*-\lambda \mbox{ is Fredholm with nonzero index}\}$,
\end{compactitem}
and call the set $\Sigma_\mathrm{ess}:=\Sigma_\mathrm{fb}\cup\Sigma_{i\neq0}$ the essential spectrum.
\end{Definition}

The following result characterizes the essential and Fredholm spectra of spiral waves in terms of the spectra of their asymptotic wave trains.

\begin{Theorem}[\rev{Fredholm} properties of linearization]\label{t:fm}
The linear operator $\mathcal{L}_*-\lambda$ posed on $L^2(\R^2,\C^N)$ is Fredholm if and only if $\lambda$ does not belong to the Floquet spectrum $\Sigma_\mathrm{st}$ (see Definition~\ref{d:fsp}) of the asymptotic wave train: \rev{in other words, we have} $\Sigma_\mathrm{fb}=\Sigma_\mathrm{st}$. Furthermore, if $\lambda$ does not belong to the Floquet spectrum of the asymptotic wave train, then the Fredholm index of $\mathcal{L}_*-\lambda$ is given by
\begin{equation}\label{e:fred=morse}
\ind(\mathcal{L}_*-\lambda) = -i_\mathrm{M}(\lambda),
\end{equation}
where $i_\mathrm{M}(\lambda)$ is the relative Morse index associated with the linearization at the asymptotic wave train from Definition~\ref{d:rM}.
\end{Theorem}

\begin{figure}
\centering\includegraphics[width=.95\textwidth]{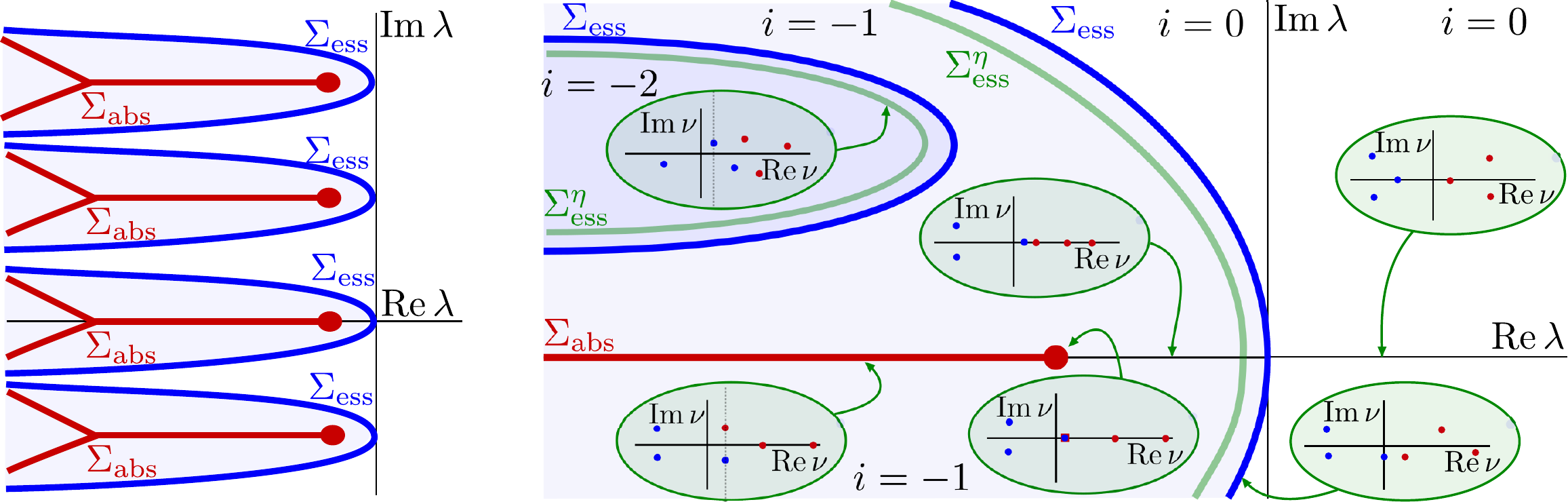}
\caption{Left panel: schematic diagram of essential spectra of spiral waves showing periodicity of essential and absolute spectra with period $\rmi\omega_*$. Branch points and triple junctions are the generic singularities of absolute spectra \cite{ssr}. Right panel: zoom into spectra, showing shaded regions that correspond, from left to right, to the Fredholm indices $i=\ind(\mathcal{L}_*-\lambda)=-2,-1,0$. Blue curves indicate the Floquet spectra of wave trains (corresponding to the Fredholm boundaries of spiral waves), green curves are the Floquet spectra of wave trains in exponentially weighted spaces with weight $\eta<0$, and the red curve corresponds to part of the absolute spectrum. Oval green insets show the spatial Floquet exponents of wave trains at the indicated locations $\lambda\in\C$, illustrating in particular the crossing of Floquet exponents on the imaginary axis at Floquet spectra, the direction of crossing relating to the Fredholm index, and the roots with equal real part at the absolute spectrum. We refer to Figure~\ref{cf:5} for numerically computed spatial and temporal spectra.}\label{f:spectra_schematic}
\end{figure}

Theorem~\ref{t:fm} is proved in \S\ref{s:fred}. We illustrate this and the following results in the schematic representation of spiral spectra in Figure~\ref{f:spectra_schematic}. Note that Remark~\ref{r:perspec} implies the following result. 

\begin{Corollary}[Floquet periodicity of Fredholm properties]\label{c:spectruminvariant}
The operator $\mathcal{L}_*-\lambda$ is Fredholm of index $i$ if and only if  $\mathcal{L}_*-(\lambda+\rmi\omega_*)$ is Fredholm of index $i$. In other words, the property of being Fredholm and the Fredholm index are periodic with period $\rmi\omega_*$ in the complex plane. 
\end{Corollary}

Note that this periodicity demonstrates quite graphically that the linearization at a spiral wave is not a sectorial operator: vertical periodicity precludes the possibility that the spectrum is contained in a sector $\{\lambda;\,|\Im\lambda|\leq C_1-C_2\Re\lambda\}$ for some $C_1,C_2>0$. From a different perspective, although the Laplacian  $\Delta$ is sectorial, $\partial_\psi$ is neither sectorial nor bounded relative to $\Delta$ on $L^2(\R^2,\C^N)$, and $\mathcal{L}_*$ therefore need not be and is in fact not sectorial. 

Recall that we oriented the dispersion curves $\lambda_\mathrm{st}(\nu)$ of a wave train in the laboratory frame so that curves point upward at
points of positive group velocity and downward at points of negative
group velocity; see Definition~\ref{d:seval}.

\begin{Corollary}\label{c:simplefloquet}
If $\lambda$ is a simple element of the Floquet spectrum of the
asymptotic wave train (see Definition~\ref{d:seval}) that lies on the dispersion curve
$\lambda_\mathrm{st}(\nu)$, then the Fredholm index of $\mathcal{L}_*-\lambda$
increases by one upon crossing the dispersion curve $\lambda_\mathrm{st}(\nu)$
from left to right (left and right are, of course, relative to the
curve's orientation).
\end{Corollary}

\begin{Definition}[Spiral waves as wave sources]
We say that the spiral wave $u_*(r,\varphi)$ \emph{emits a spectrally stable wave train} if the asymptotic wave train $u_\infty$ (i) is spectrally stable according to Definition~\ref{d:seval} and (ii) has positive group velocity, that is, the group velocity is directed away from the origin in polar coordinates.
\end{Definition}

We discuss properties of spiral sinks, whose asymptotic wave trains have negative group velocity, briefly in Remark~\ref{r:spiralsinks}.

\begin{Corollary}\label{c:emits}
Assume that a spiral wave emits a spectrally stable wave train, then
the linearization $\mathcal{L}_*-\lambda$ has Fredholm index $-1$ in
the connected component of the Fredholm region to the left of the
dispersion curve that contains $\lambda=0$.
\end{Corollary}

Corollaries~\ref{c:simplefloquet} and~\ref{c:emits} follow from Remark~\ref{r:morse} and Theorem~\ref{t:fm}.

We may also consider the linearization $\mathcal{L}_*$ on a space of
functions equipped with an exponential weight
\[
L^2_\eta(\R^2,\C^N) :=
\{ u\in L^2_\mathrm{loc} ;\; |u|_{L^2_\eta}<\infty \}, \qquad
|u|_{L^2_\eta}^2 := \int_{x\in\R^2} |u(x)\rme^{\eta|x|}|^2 \,\rmd x.
\]
For any $\eta\in\R$ and $i\in\Z$, we define
\begin{equation}\label{calf}
\mathcal{F}_i^\eta(\mathcal{L}_*) := \{ \lambda\in\C ;\;
\mathcal{L}_*-\lambda \mbox{ is Fredholm in } L^2_\eta(\R^2,\C^N)
\mbox{ with index } i \}.
\end{equation}
Recall the definition of the spatial eigenvalues $\nu_j(\lambda)$ from
\S\ref{s:wts.morse}.

\begin{Proposition}\label{p:extpt}
For each fixed $\lambda\in\C$, the operator $\mathcal{L}_*-\lambda$ is Fredholm with index zero in the space $L^2_\eta(\R^2,\C^N)$ for all $\eta\in J_0(\lambda)=(-\Re\nu_0(\lambda),-\Re\nu_{-1}(\lambda))$. Fix any such rate $\eta$ and consider the connected component $\mathcal{S}$ of $\mathcal{F}_0^\eta(\mathcal{L}_*)$ that contains $\lambda$, then either the entire connected component $\mathcal{S}$ lies in the spectrum of $\mathcal{L}_*$ posed on $L^2_\eta$ or else $\mathcal{S}$ contains only isolated eigenvalues with finite algebraic multiplicity of $\mathcal{L_*}$ on $L^2_\eta$. In either case, the spectrum, and in the latter case also the geometric and algebraic multiplicities of eigenvalues, do not depend on the choice of the rate $\eta\in J_0(\lambda)$.
\end{Proposition}

Proposition~\ref{p:extpt} is proved in \S\ref{s:fred}. As we shall see later, if $\lambda$ is an eigenvalue of $\mathcal{L}_*$ on the space $L^2_\eta$ for some $\eta\in J_0(\lambda)$, then $\lambda$ is close to an eigenvalue of the spiral wave considered on a large but finite disk. Thus, we are led to the following two definitions which adapt the terminology from \cite{ss-trunc} to the infinite-dimensional setup.
 
\begin{Definition}[Absolute spectrum]\label{d:absspec}
We call the set of $\lambda\in\mathbb{C}$ for which $J_0(\lambda)$ is empty, that is, where $\Re\nu_0(\lambda)=\Re\nu_{-1}(\lambda)$, the \emph{absolute spectrum} $\Sigma_\mathrm{abs}$ of $\mathcal{L}_*$.
\end{Definition}
 
\begin{Definition}[Extended point spectrum]\label{d:extspec}
We say that $\lambda\in\mathbb{C}$ is in the extended point spectrum of $\mathcal{L}_*$ if (i) $\lambda\notin\Sigma_\mathrm{abs}$ and (ii) the kernel of $\mathcal{L}_*-\lambda$ is nontrivial in $L^2_\eta$ for some $\eta\in J_0(\lambda)$.
\end{Definition}
 
The next corollary provides estimates for eigenfunctions and adjoint eigenfunctions associated with elements in the extended point spectrum. The result for eigenfunctions follows directly from the definition of the extended point spectrum, while the estimates for the adjoint eigenfunctions follow from the fact that the dual of $L^2_\eta$, computed with respect to the usual $L^2$ scalar product, is given by $L^2_{-\eta}$.
 
\begin{Corollary}[Localization of eigenfunctions and adjoint eigenfunctions]\label{e:eigloc}
Suppose that $\lambda$ belongs to the extended point spectrum and let $u$ be an associated eigenfunction or generalized eigenfunction of $\mathcal{L}_*$, and $u_\mathrm{ad}$ be the associated eigenfunction, or generalized eigenfunction, of the adjoint operator $\mathcal{L}_*^*$, then for each $\eta\in J_0(\lambda)$ there exists $C(\eta)>0$ such that
\begin{align*}
\|u(r,\cdot)\|_{H^1(S^1)}+\|(\nabla_x u)(r,\cdot)\|_{H^1(S^1)} & \leq
C(\eta)\,\rme^{\eta r} \\
\|u_\mathrm{ad}(r,\cdot)\|_{H^1(S^1)}+\|(\nabla_x u_\mathrm{ad})(r,\cdot)\|_{H^1(S^1)} & \leq
C(\eta)\,\rme^{-\eta r}
\end{align*}
\rev{for $r\geq1$, where $\nabla_x u$ is the gradient of $u$ in Cartesian coordinates so that $|(\nabla_x u)(r,\cdot)|^2=|\partial_r u(r,\cdot)|^2+|\frac{1}{r}\partial_\psi u(r,\cdot)|^2$.}
\end{Corollary}

An application of Remark~\ref{r:morse}, Proposition~\ref{p:extpt}, and Corollary~\ref{e:eigloc} to $\lambda=0,\pm \rmi\omega_*$ gives the following result.

\begin{Corollary}[Stabilization of spectrum and symmetries]\label{c:stexp}
Assume that a spiral wave emits a spectrally stable wave train; then there is an $\eta_*<0$ such that the essential spectrum of the linearization $\mathcal{L}_*$ considered as a closed operator on $L^2_\eta$ is strictly contained in the open left half-plane for all
$\eta_*<\eta<0$. For these values of $\eta$, the spectrum contains the eigenvalues $\{0,\pm\rmi\omega_*\}$ with associated eigenfunctions $\partial_\psi u_*$ and $(\partial_x\pm\rmi\partial_y)u_*$, respectively. The adjoint eigenfunctions associated with the eigenvalues $\{0,\pm\rmi\omega_*\}$ are exponentially localized with rate $\eta$.
\end{Corollary}

In other words, $\lambda=0,\pm\rmi\omega_*$ belong to the extended point spectrum, and the adjoint eigenfunctions belonging to the elements $\lambda=0,\pm\rmi\omega_*$ of the extended point spectrum are exponentially localized. We will give refined asymptotics rather than upper bounds for eigenfunctions in \S\ref{s:mr.exp} below.

In the next section, we shall consider robustness of spiral waves. We therefore introduce the following characterization of spiral waves with ``minimal kernel''.

\begin{Definition}[Transverse spirals]\label{d:transverse}
We say that a spiral is transverse if (i) it emits a spectrally stable wave train and (ii) for all  $\eta<0$ sufficiently small the eigenvalue $\lambda=0$ of  $\mathcal{L}_*$ considered as a closed operator on $L^2_\eta$ is algebraically simple.
\end{Definition}

For the sake of simplicity, we shall state most of our results for transverse spiral waves, although we can significantly relax the assumption of spectral stability of wave trains.

\subsection{Asymptotics and robustness of spiral waves}\label{s:mr.asym}

We have the following far-field expansion of Archimedean spiral waves that emit stable wave trains.

\begin{Proposition}[Far-field expansion]\label{p:exppha}
Assume that the reaction-diffusion system (\ref{e:rds-}) admits a transverse spiral wave as characterized in Definition~\ref{d:transverse}. 
For each $K<\infty$, we then have the following expansion:
\begin{align}
u_*(r,\psi) = &  u_\infty(k_*r+\theta_*(r)+\psi)
+ \sum_{j=1}^{K} u_j(k_*r+\theta_*(r)+\psi)\frac{1}{r^j} + \rmO\left(\frac{1}{r^{K+1}}\right),
\nonumber \\ \label{e:expand}
\theta_*(r) = &  \frac{k_*d_\perp}{c_\mathrm{g}} \log r
+ \sum_{j=1}^{K} \theta_j \frac{1}{r^j} +  \rmO\left(\frac{1}{r^{K+1}}\right), 
\end{align}
for $r\gg1$, with coefficients $\theta_j$ and smooth $2\pi$-periodic functions $u_j$ that can be calculated recursively, and with error terms that are bounded uniformly in $\psi$.
In the expansions for $\theta$, the factor $c_\mathrm{g}$ denotes the group velocity (\ref{e:ndr}) of the asymptotic wave trains, and $d_\perp$ is the transverse diffusion
coefficient of the wave trains defined in (\ref{e:dperp}).
The first term in the expansion for the spiral wave is given explicitly through
\[
u_1(\vartheta) =  k_* \left( \frac{d_\perp}{c_\mathrm{g}}
\partial_k u_\infty(\vartheta) - \frac{1}{2} u_{\nu_\perp\nu_\perp}(\vartheta\rev{;0}) \right),
\]
where 
$\partial_k u_\infty$ denotes  the derivative of the family of wave trains with respect to the
wavenumber $k$, and the transverse correction $u_{\nu_\perp\nu_\perp}$ is defined
in (\ref{e:nunu}).
\end{Proposition}

When $d_\perp>0$, we have $0<\theta_*^\prime\sim \frac{1}{r}$ for large $r$, and the wavenumber therefore decreases towards the asymptotic value at the wave train. This corresponds to waves emitted by the spiral appearing to ``decompress'' as waves travel away from the center; see Figure~\ref{cf:3} for a numerical illustration of this phenomenon. For spirals that emit spectrally stable wave trains that are transversely unstable in two dimensions, so that $d_\perp<0$, we have $\theta_*^\prime<0$; see Figure~\ref{mf:fhn} for a numerical example.

Note that Proposition~\ref{p:exppha} justifies the use of the term Archimedean for spiral waves despite the logarithmic phase correction given by $\theta_*(r)$. Indeed, the local wavelength $L(r)$, i.e.\ the distance between consecutive spiral arms, converges to a constant as $r\to\infty$ since (\ref{e:expand}) implies that 
\[
u_\infty(k_*r+\theta_*(r)) = u_\infty(k_*(r+L(r))+\theta_*(r+L(r))) \quad\mbox{gives}\quad
L(r) = \frac{2\pi}{k_*} \left(1- \frac{k_*d_\perp}{c_\mathrm{g}r} + \rmO(1/r^2) \right).
\]

Transverse spiral waves are robust in that they persist upon changing parameters in the nonlinearity. To make this more precise, we consider a reaction-diffusion system
\begin{equation}\label{e:rdspa}
u_t = D \Delta u + f(u;\mu)
\end{equation}
whose kinetics $f(u;\mu)$ depends smoothly on a parameter $\mu$ and look for rotating waves  $u(r,\psi)$ as solutions to 
\begin{equation}\label{e:spwp}
D \Delta_{r,\psi} u + \omega \partial_\psi u + f(u;\mu) = 0,
\end{equation}
for a certain frequency $\omega(\mu)$.

\begin{Theorem}[Robustness of transverse spirals]\label{t:rob}
If the \rev{steady-state equation (\ref{e:spwp})} with $\mu=0$ admits a
transverse  spiral wave $u_*(r,\psi)$, then the spiral is
robust. More precisely, there exists a family of spiral
waves $u(r,\psi;\mu)$ with frequencies $\omega=\omega_*(\mu)$ and
asymptotic phases $\theta_*(r;\mu)$ so that $u(r,\psi;0)=u_*(r,\psi)$,
$\omega(0)=\omega_*$, and
\[
|u(r,\cdot;\mu) - u_\infty(k_*(\mu)r+\theta_*(r;\mu)+\cdot;\mu)|_{C^1}
\to 0 \mbox{ as } r\to\infty.
\]
Here, $u_\infty(\xi;\mu)$ is the (unique) wave train for the problem
(\ref{e:rdspa}) in one space dimension with frequency $\omega_*(\mu)$
and wavenumber $k_*(\mu)$. The frequency $\omega_*(\mu)$, the
wavenumber $k_*(\mu)$, and the phase $\theta_*(r;\mu)$ depend smoothly
on the parameter $\mu$, and the derivative of the phase
$\theta^\prime(r;\mu)$ converges to zero uniformly in $\mu$. The
asymptotic wavenumber is selected according to the $\mu$-dependent
nonlinear dispersion relation $\omega_*(k)$ of the wave trains
via $\omega_*(k_*(\mu);\mu)=\omega_*(\mu)$. For each $\mu$, the
far-field expansion of Proposition~\ref{p:exppha} holds.
\end{Theorem}

Proposition~\ref{p:exppha} and Theorem~\ref{t:rob} are proved in
\S\ref{s:rob}.

\subsection{Far-field expansions of eigenfunctions}\label{s:mr.exp}

When spiral waves undergo bifurcations that involve isolated
eigenvalues, the shape of the associated eigenfunctions gives useful
clues as to the spatial structure of patterns that bifurcate from the
spiral wave.

\begin{Proposition}[Lower bounds on eigenfunction decay]\label{p:evalgrowth}
Take an element $\lambda$ of the extended point spectrum: by definition, there is then an $\eta_0\in J_0(\lambda)$ such that the kernel of $\mathcal{L}_*-\lambda$ in $L^2_{\eta_0}$ is nontrivial. Let \rev{$u\neq0$} be a nontrivial element of this kernel in $L^2_{\eta_0}$ and assume that there is $\eta_1\in J_{-1}(\lambda)$ (which is defined in Definition~\ref{d:j}) so that the kernel of $\mathcal{L}_*-\lambda$ in $L^2_{\eta_1}$ is trivial. For each $\eta\in J_{-1}(\lambda)$, there is then a $C(\eta)>0$ such that
\[
|u(r,\cdot)|_{H^1(S^1)} \geq C(\eta) \rme^{-\eta r}.
\]
\end{Proposition}

The next proposition gives an expansion of eigenfunctions in the far
field.

\begin{Proposition}[Far-field expansions of eigenfunctions]\label{p:asyeig}
Assume that $\lambda$ lies on a simple dispersion curve
$\lambda_\mathrm{st}(\nu)$ with $\nu\in\rmi\R$ that separates the set
$\mathcal{F}_0^0(\mathcal{L}_*)$ defined in (\ref{calf}) from
$\mathcal{F}_{-1}^0(\mathcal{L}_*)$. In addition, assume that
$\lambda$ lies in the extended point spectrum and has geometric
multiplicity one. Lastly, we assume that the kernel of
$\mathcal{L}_*-\lambda$ in $L^2_\eta$ with $\eta\in J_{-1}(\lambda)$
is trivial. Denote by $u(r,\psi;\lambda)$ the eigenfunction. We then
have the expansion
\begin{align*}
u(r,\psi;\lambda) = & 
a(r) \left[ u_\mathrm{wt}(k_*r+\theta_*^\prime(r)+\psi)
+ \rmO\left(\frac{1}{r}\right) \right], \\
a(r) = &  r^{\alpha} \rme^{\nu r}
\left[1+\rmO\left(\frac{1}{r}\right)\right], \\
\alpha = &  \frac{\langle u_\mathrm{ad},
[(2k_*d_\perp/c_\mathrm{g})\partial_\vartheta+1] D v_\mathrm{wt}
+ f^{\prime\prime}(u_\mathrm{wt})[u_1,u_\mathrm{wt}]\rangle}{c_\mathrm{g,l}\,\langle u_\mathrm{ad},u_\mathrm{wt}\rangle},
\end{align*}
where the scalar products are taken in $L^2(S^1,\C^N)$, $u_\mathrm{wt}$ and
$u_\mathrm{ad}$ are the eigenfunctions of $\hat{\mathcal{L}}_\mathrm{\rev{co}}(\nu)$ and
$\hat{\mathcal{L}}_\mathrm{\rev{co}}^\mathrm{ad}(\nu)$, respectively, corresponding to the
eigenvalue $\lambda_\mathrm{co}=\lambda_\mathrm{st}(\nu)+c_*\nu$, and
$v_\mathrm{wt}=(k_*\partial_\vartheta+\nu)u_\mathrm{wt}$. The terms $u_1$ and
$\theta_*(r)$ appear in Proposition~\ref{p:exppha}, and
$c_\mathrm{g,l}$ is the linear group velocity of $\lambda_\mathrm{st}(\nu)$
given by
\[
c_\mathrm{g,l} = - \frac{2\langle u_\mathrm{ad},D v_\mathrm{wt}\rangle}%
{\langle u_\mathrm{ad},u_\mathrm{wt}\rangle}.
\]
\end{Proposition}

We will prove Propositions~\ref{p:evalgrowth} and~\ref{p:asyeig} in \S\ref{s:ps}. We refer to Proposition~\ref{p:absspecdiag} for a generalization of Proposition~\ref{p:asyeig} and remark that results analogous to Propositions~\ref{p:evalgrowth} and~\ref{p:asyeig} hold for the adjoint linearization.

\subsection{Persistence of spiral waves on large disks}\label{s:mr.psw}

Assume that the reaction-diffusion system (\ref{e:rds-}) admits a
transverse planar Archimedean spiral wave $u_*(r,\psi)$ with temporal
frequency $\omega_*$. The issue discussed here is whether this spiral
wave persists on large disks. In other words, is there a spiral wave
to the equation
\begin{align*}
u_t = &  D \Delta u  + f(u), \qquad |x|<R, \\
0 = &  a u + b \frac{\partial u}{\partial\vec{n}}, \quad\qquad |x|=R,
\end{align*}
for all large $R\gg1$, where $\vec{n}$ denotes the outer unit normal
of the disk of radius $R$ centered at zero, and where $a^2+b^2=1$. 
We
show that this is indeed true under the following natural hypothesis. It will be clear from our analysis that the results carry over to much more general types of boundary conditions, for instance nonlinear Robin boundary conditions $\frac{\partial u}{\partial\vec{n}}=g(u)$. 

\begin{Hypothesis}[Boundary sink]\label{d:bsi}
Given a spectrally stable  wave train $u_\infty$ with \rev{fixed} wavenumber \rev{$k_*$}, frequency \rev{$\omega_*>0$}, and \rev{positive} group velocity $c_\mathrm{g}>0$, we say that the one-dimensional equation
\begin{eqnarray}\label{e:bs}
u_t & = &  D u_{xx} + f(u), \qquad x\in(-\infty,0) \\ \nonumber
0 & = &  a u(0,t) + b u_x(0,t)
\end{eqnarray}
has a boundary sink if \rev{(\ref{e:bs}) admits a time-periodic solution $u(x,t)=u_\mathrm{bs}(x,\omega_*t)$ with $u_\mathrm{bs}(x,\tau)=u_\mathrm{bs}(x,\tau+2\pi)$ for all $(x,\tau)$} such that
\[
|u_\mathrm{bs}(x,\cdot)-u_\infty(k_* x-\cdot)|_{C^1(S^1)}\to0
\mbox{ as } x\to -\infty.
\]
We say that the boundary sink is non-degenerate if the linearized equation 
\begin{align}\label{e:rds1dtrunclin}
u_t = &  D u_{xx} + f'(u_\mathrm{bs}(x,\omega_*t))u, \qquad x\in(-\infty,0) \\ \nonumber
0 = &  a u(0,t) + b u_x(0,t),
\end{align}
does not possess an exponentially localized, time-periodic solution, that is, for any smooth     solution $u$ to \eqref{e:rds1dtrunclin} with $u(x,t+\frac{2\pi}{\omega_*})=u(x,t)$, we have 
\[
 \int_{0}^{2\pi/\omega_*} \int_{-\infty}^0 \rme^{-\eta x} (|u(x,t)^2+|u_x(x,t)|^2)\,\rmd x\,\rmd t=\infty
\]
for any $\eta>0$.
\end{Hypothesis}

\rev{Similar to our assumptions on the existence of wave trains or spiral waves, Hypothesis~\ref{d:bsi} is, in general, difficult to verify. However, we will prove in Lemma~\ref{l:tr} that boundary sinks arise typically as one-parameter families that are parametrized by the wavenumber $k$ of the asymptotic wave train. Furthermore, their existence near homogeneous oscillations with wavenumbers $k$ close to zero was shown in \cite[\S6.8 on p46]{ssdefect}.}

The term boundary sink is intuitive as the group velocity \rev{of the wave train at $x=-\infty$ is positive so that perturbations near $x=-\infty$ are transported towards the boundary at $x=0$, where they are annihilated by the boundary.} Non-degeneracy can be interpreted as absence of the Floquet exponent $\lambda=0$ in the extended point spectrum. In fact, the discussion in \S\ref{s:wts.morse} shows that $J_0(\lambda)\supset (-\delta,0)$ for some $\delta>0$ since $c_\mathrm{g}>0$. Choosing the exponential weight $\eta>0$, we then conclude that the linearization at the wave trains is hyperbolic with relative Morse index zero, which implies that the linearization is Fredholm of index zero when the operator is equipped with boundary conditions at $x=0$ \cite{ssmorse}. The absence of a periodic solution to the linearization then implies that the linearized operator does not have a Floquet exponent $\lambda=0$. We emphasize that non-degeneracy is a meaningful assumption since the ``trivial'' time-periodic \rev{solution $\partial_t u_\mathrm{bs}$} to the linearization is not exponentially localized as $\omega_*\neq0$.

\rev{We emphasize that the boundary sink connects the wave train at $x=-\infty$ with the Neumann boundary conditions at $x=0$. This feature will allow us to glue together the spiral wave $u_*(r,\psi)$ and the shifted boundary sink $u_\mathrm{bs}(r-R,\psi)$ at $r=R-\log R$, where both patterns are close to the asymptotic wave train, to obtain a truncated spiral wave on the disk $0\leq r\leq R$ that satisfies Neumann boundary conditions at $r=R$; see Figure~\ref{f:gss} for an illustration. The next theorem formalizes this expectation.}

\begin{Theorem}[Gluing spirals with boundary sinks]\label{t:trunc}
Assume the existence of (i) a transverse spiral wave $u_*$ (see Definition~\ref{d:transverse}) and (ii) a non-degenerate boundary sink (see Definition~\ref{d:bsi}) with the same asymptotic wave train $u_\infty$, frequency $\omega_*$, and wavenumber $k_*$. Then there are positive numbers $\delta$, $C$, $\kappa$ and $R_*$ with $0<\delta<1$ so that the following is true. For each $R>R_*$, there are a unique frequency $\omega=\omega(R)$ with $|\omega-\omega_*|<\delta$ and a unique smooth function $u_R(r,\rev{\psi})$ with
\[
|\omega(R)-\omega_*| + 
\sup_{0\leq r\leq R-\kappa^{-1}\log R} |u_R(r,\rev{\psi})-u_*(r,\rev{\psi})| +
\sup_{R-\kappa^{-1}\log R\leq r\leq R} |u_R(r,\rev{\psi})-u_\mathrm{bs}(r-R,\rev{\psi})| \leq \delta
\]
such that the pair $(u,\omega)=(u_R(r,\psi),\omega(R))$ satisfies the system
\begin{align*}
0 = &  D \Delta_{r,\psi} u + \omega u_{\rev{\psi}} + f(u), \qquad 0\leq r<R \\
0 = &  a u + b \rev{u_r}, \qquad\qquad\qquad r=R.
\end{align*}
Furthermore, we have the estimates
\begin{eqnarray}
|\omega(R)-\omega_*| & \leq & C \rme^{-\kappa R} \nonumber \\ \label{est:trunc}
|u_R(r,\rev{\psi})-u_*(r,\rev{\psi})| & \leq & \frac{C}{R^{1-\delta}}\rme^{-\kappa(R-\kappa^{-1}\log R - r)}, \qquad 0\leq r\leq R-\kappa^{-1}\log R \\ \nonumber
|u_R(r,\rev{\psi})-u_\mathrm{bs}(r-R,\rev{\psi})| & \leq & \frac{C}{R^{1-\delta}}, \qquad R-\kappa^{-1}\log R\leq r\leq R
\end{eqnarray}
uniformly in $R>R_*$.
\end{Theorem}

Theorem~\ref{t:trunc} is proved in \S\ref{s:trunc}.

\subsection{Spectra of spiral waves restricted to large disks}\label{s:spectratrunc}

In the previous section, we provided conceptual assumptions guaranteeing that the existence of a spiral wave on $x\in\R^2$ implies the existence of spiral waves in large disks. The results establish in particular the convergence of profiles as the size of the disk increases. We now pair these results with analogous convergence results for properties of the linearization. 

It turns out that in addition to contributions from the spectrum of the spiral wave, there is a contribution from the boundary condition that we shall specify first. Consider therefore the linearization at the asymptotic wave train in the steady frame, restricted to $x<0$ and equipped with boundary conditions,
\begin{align}\label{e:wtl}
u_t-\lambda u & = Du_{xx}+f'(u_\infty(kx-\omega t))u,\qquad x<0,
\\ \notag
au+bu_x & =0,\qquad x=0.
\end{align}

\begin{Definition}[Boundary spectrum]\label{d:bs}
We define the boundary spectrum $\Sigma_\mathrm{bdy}$ of wave trains as the set of $\lambda\not\in\Sigma_\mathrm{abs}$ for which there exists a $2\pi/\omega$-periodic solution to \eqref{e:wtl} with $u(x,0)\in L^2_\eta$ for some $\eta\in J_0(\lambda)$.
\end{Definition}

We shall need some mild non-degeneracy assumptions on the absolute spectrum, which we defined in Definition~\ref{d:absspec}. The first non-degeneracy condition is concerned with the dispersion relation, asserting roughly that the absolute spectrum consists of algebraically simple curves; compare for instance \cite{ssr}.

\begin{Definition}[Simple absolute spectrum]\label{d:absgen}
We say that the absolute spectrum is simple at a point $\lambda_*\in\C$ if (i) $J_{\pm 1}(\lambda)$ are non-trivial and (ii) the two critical spatial eigenvalues with  equal real part split non-trivially upon varying $\lambda$, that is, 
\[
\Re\nu_{-2}(\lambda_*)<\Re\nu_{-1}(\lambda_*)=\Re\nu_{0}(\lambda_*)<\Re\nu_{1}(\lambda_*),\qquad \nu_{-1}(\lambda_*)\neq\nu_{0}(\lambda_*),\quad \text{and} \quad\frac{\rmd \nu_0}{\rmd\lambda}\neq \frac{\rmd \nu_{-1}}{\rmd\lambda},
\]
at $\lambda=\lambda_*$. 
\end{Definition}

Many results on the absolute spectrum can be extended without this simplicity assumption  \cite{ps} but we shall not attempt such a generalization in this context. 

\begin{Definition}[Resonances in the absolute spectrum --- informal]\label{d:absgen2}
We say that a point $\lambda_*$ in the simple part of the absolute spectrum is resonant if each nontrivial element $u(r,\psi)$ of the kernel of $\mathcal{L}_*-\lambda_*$ in $L^2_\eta$ with $\eta\in J_1(\lambda_*)$ converges to $u_0(\psi)\rme^{\nu_0(\lambda_*)r}$ or $u_{-1}(\psi)\rme^{\nu_{-1}(\lambda_*)r}$ (but not to a linear combination of both) as $r\to\infty$, where the functions $u_0$ or $u_{-1}$ may vanish. We refer to Definitions~\ref{d:resonance} and~\ref{d:resonance2} for a precise definition of resonance.
\end{Definition}

\rev{We define the linear operator}
\begin{eqnarray}\label{e:sstarR}
\mathcal{L}_{*,R} u & = & D\Delta u + \omega_* \partial_\psi u + f^\prime(u_*\rev{(r,\psi)}) u \quad\mbox{ for } |x|<R \\ \nonumber
au+b\frac{\partial u}{\partial n} & = & 0 \quad\mbox{ at } |x|=R
\end{eqnarray}
\rev{in Cartesian coordinates on $L^2(\{|x|<R\})$ with dense domain $\{ u\in H^2(\{|x|<R\}):\; (au+b\frac{\partial u}{\partial n})|_{|x|=R}=0 \}$, where the domain is well defined due to standard trace theorems. Note that $\mathcal{L}_{*,R}$ has compact resolvent as a relatively compact perturbation of $D\Delta$, and its spectrum on $L^2(\{|x|<R\})$ consists therefore entirely of discrete point spectrum for each fixed $R$.}

\begin{Theorem}[Spectra of truncated linearization]\label{t:tl}
Assume the existence of a transverse spiral wave $u_*(r,\rev{\psi})$ (see Definition~\ref{d:transverse}) with frequency $\omega_*>0$. Recall the Definitions~\ref{d:absspec} and~\ref{d:extspec} of the absolute spectrum $\Sigma_\mathrm{abs}$ and the extended point spectrum $\Sigma_\mathrm{ext}$, respectively, and Definition~\ref{d:bs} of the boundary spectrum $\Sigma_\mathrm{bdy}$. Assume that there exists a dense subset in the absolute spectrum where the absolute spectrum is (i) simple (see Definition~\ref{d:absgen}) and (ii) not resonant (Definition~\ref{d:absgen2}) for both the spiral wave linearization $\mathcal{L}_*$  and the boundary linearization \eqref{e:wtl}. Moreover, we assume that $\Sigma_\mathrm{ext}$ and $\Sigma_\mathrm{bdy}$ do not intersect. \rev{For the spectrum of the operator $\mathcal{L}_{*,R}$ defined in (\ref{e:sstarR}),} we then have
\[
\mathrm{spec}\,\mathcal{L}_{*,R}\longrightarrow \Sigma_\mathrm{abs}\cup \Sigma_\mathrm{ext}\cup \Sigma_\mathrm{bdy}
\]
as $R\to\infty$ in the Hausdorff distance on each fixed compact subset of $\C$. Note that \rev{$\Sigma_\mathrm{abs}$} consists of semi-algebraic curves whereas \rev{$\Sigma_\mathrm{ext}\cup\Sigma_\mathrm{bdy}$} is discrete. Convergence to the discrete part preserves multiplicity and is exponential in $R$ for $\Sigma_\mathrm{ext}$ and algebraic in $R$ for $\Sigma_\mathrm{bdy}$. Convergence to the continuous part \rev{$\Sigma_\mathrm{abs}$} is understood in the sense that for each fixed $\lambda_*\in\Sigma_\mathrm{abs}$ \rev{the number of eigenvalues of $\mathcal{L}_{*,R}$ in $\mathcal{U}(\lambda_*)$ converges to infinity as $R\to\infty$ for each fixed neighborhood $\mathcal{U}(\lambda_*)$ of $\lambda_*$}.
\end{Theorem}

Theorem~\ref{t:tl} is proved in \S\ref{s:abs}.

\begin{Remark}[Absolute spectra versus pseudo-spectra]
We emphasize that eigenvalues accumulate along curves that are not given by the Fredholm boundaries or the essential spectrum and instead lie strictly to the left of the Fredholm boundaries. The limiting curves \rev{lie in the absolute spectrum and} are, just as the Fredholm boundaries, periodic in the complex plane with period $\rmi\omega_*$ and determined solely by the \rev{linear} dispersion relation of the wave trains. \rev{It is possible to prove that the norm of the resolvent of $\mathcal{L}_{*,R}$ grows exponentially in $R$ in regions where the Fredholm index of the linearization is not zero; see \cite{ss-trunc} for a precise statement in a context of travelling waves on the real line. Thus, the $\varepsilon$-pseudo spectra of $\mathcal{L}_{*,R}$, defined as the set of $\lambda$ so that the resolvent has norm $1/\varepsilon$, fill large regions between the absolute spectrum and the Fredholm boundary for $\varepsilon\geq\varepsilon(R)$ with $\varepsilon(R)\to0$ as $R\to\infty$.}
\end{Remark}

\subsection{Spectra of truncated spiral waves}\label{s:mr.tsp}

This section extends the results from \S\ref{s:spectratrunc} by including the corrections to the nonlinear spiral wave profile considered in \S\ref{s:trunc}. The solutions constructed can be thought of as spiral waves glued to a boundary sink that corrects for the influence of the boundary conditions.

\rev{We therefore define the linear operator}
\begin{eqnarray}\label{e:ltr}
\mathcal{L}_{\mathrm{s},R} u & = & D\Delta u + \omega(R) \partial_{\rev{\psi}} u + f^\prime(u_R\rev{(r,\psi)}) u \quad\mbox{ for } |x|<R, \\ \nonumber
au+b\frac{\partial u}{\partial n} & = & 0 \quad\mbox{ at } |x|=R
\end{eqnarray}
\rev{in Cartesian coordinates on $L^2(\{|x|<R\})$ with dense domain $\{ u\in H^2(\{|x|<R\}):\; (au+b\frac{\partial u}{\partial n})|_{|x|=R}=0 \}$, where $u_R$ and $\omega_R$ are profile and frequency of the truncated spiral wave from Theorem~\ref{t:trunc}. The spectrum of $\mathcal{L}_{*,R}$ consists of discrete point spectrum for each fixed $R$.}

We are interested in the convergence of the spectrum of  $\mathcal{L}_{\mathrm{s},R}$ as $R\to\infty$. The results are very similar to the results presented in  \S\ref{s:spectratrunc}. The main correction due to the gluing procedure accounts for the boundary sink by replacing the boundary spectrum $\Sigma_\mathrm{bdy}$ in the results of \S\ref{s:spectratrunc} with the extended point spectrum of the boundary sink. To be precise, consider the linearization at the boundary sink $u_\mathrm{bs}(x,t)$ in the Floquet form 
\begin{align}
u_t-\lambda u&=Du_{xx}+f'(u_\mathrm{bs}(x,t))u,\qquad &x<0,\notag\\
au+bu_x &=0,\qquad &x=0, \label{e:bsl} \\ \notag
u(x,t)&=u(x,t+2\pi/\omega), & \forall (x,t)\in \mathbb{R}^-\times\mathbb{R}^+.
\end{align}
Since boundary sinks converge to the asymptotic wave trains of the spiral waves, we can again use the absolute spectrum $\Sigma_\mathrm{abs}$ of the asymptotic wave trains. We can also define the space $L^2_\eta(\R^-)$ of functions with exponentially weighted norms given by
\[
 |u|_{L^2_\eta}^2=\int_{x=-\infty}^0 |u(x)\rme^{\eta x}|^2\rmd x,
\]
where the rates $\eta$ will be related to the exponential growth rates $\nu_j(\lambda)$ that we identified in Definition~\ref{d:j}.

\begin{Definition}[Extended point spectrum of boundary sinks]\label{d:sbs}
We define the extended point spectrum $\Sigma_\mathrm{extbs}$ of the boundary sink as the set of $\lambda\not\in\Sigma_\mathrm{abs}$ for which there exists a nontrivial solution $u(x,t)$ to \eqref{e:bsl} with $u(x,0)\in L^2_\eta(\R^-)$ for some $\eta\in J_0(\lambda)$ (where $J_0(\lambda)$ was defined in Definition~\ref{d:j}).
\end{Definition}

The following main result closely mimics Theorem~\ref{t:tl}. 

\begin{Theorem}[Spectra of  truncated spirals]\label{t:tl2}
Consider the linearization  $\mathcal{L}_{\mathrm{s},R}$ at the truncated spiral \eqref{e:ltr}.
Assume that there exists a dense subset in the absolute spectrum where the absolute spectrum is (i) simple (see Definition~\ref{d:absgen}) and (ii) not resonant (see Definition~\ref{d:absgen2}) for the linearizations $\mathcal{L}_*$ about the spiral wave and \eqref{e:bsl} about the boundary sink. Moreover, we assume that the extended point spectra of spiral wave and boundary sink do not intersect. We then have convergence 
\[
\mathrm{spec}\,\mathcal{L}_{\mathrm{s},R}\longrightarrow \Sigma_\mathrm{abs}\cup \Sigma_\mathrm{ext}\cup \Sigma_\mathrm{extbs}
\]
as $R\to\infty$ in the Hausdorff distance uniformly on each fixed compact subset of $\mathbb{C}$. Note that \rev{$\Sigma_\mathrm{abs}$} consists of semi-algebraic curves whereas \rev{$\Sigma_\mathrm{ext}\cup\Sigma_\mathrm{extbs}$} is discrete. Convergence to $\Sigma_\mathrm{ext}$ is exponential and convergence to $\Sigma_\mathrm{extbs}$ algebraic in $R$ (and both preserve multiplicity), while convergence to \rev{$\Sigma_\mathrm{abs}$} is understood in the sense that for each fixed $\lambda_*\in\Sigma_\mathrm{abs}$ \rev{the number of eigenvalues of $\mathcal{L}_{\mathrm{s},R}$ in $\mathcal{U}(\lambda_*)$ converges to infinity as $R\to\infty$ for each fixed neighborhood $\mathcal{U}(\lambda_*)$ of $\lambda_*$}.
\end{Theorem}

Theorem~\ref{t:tl2} is proved in \S\ref{s:absglue}. 

\subsection{Transverse instability of spiral waves}\label{ss:spiralperp}

We note that none of our results about spectra or Fredholm properties of the linearization $\mathcal{L}_*$ at a spiral wave requires assumptions on the transverse stability of the asymptotic wave train belonging to the spiral wave. We show here that transverse instabilities \rev{of the wave trains} become important when considering decay or growth properties of the $C^0$-semigroup $\rme^{\mathcal{L}_*t}$ generated by $\mathcal{L}_*$ on $L^2(\R^2,\C^N)$. \rev{In particular, we will show that a transverse instability of the asymptotic wave train implies linear instability of the planar spiral wave --- we refer to this instability mechanism as a transverse instability of a planar spiral wave.}

\begin{Lemma}\label{l:spiralperp}
Assume that $u_*(r,\varphi)$ is a transverse spiral wave and that its asymptotic wave train $u_\infty(kx-\omega t)$ is unstable with respect to transverse perturbations so that there are constants $\gamma>0$ and $\lambda_*\in\C$ with $\Re\lambda_*>0$ as well as a nontrivial $2\pi$-periodic function $v_\infty(\xi)$ with
\begin{equation}\label{e:tr}
D (k^2\partial_{\xi\xi}-\gamma^2) v_\infty + \omega_* \partial_\xi v_\infty + f^\prime(u_\infty(\xi)) v_\infty = \lambda_* v_\infty.
\end{equation}
Under these assumptions, we have
\[
\inf\left\{ a\in\R:\; \exists M_a\geq1: \|\rme^{\mathcal{L}_*t}\|\leq M_a \rme^{at} \;\forall t\geq0 \right\} \geq \Re\lambda_* > 0,
\]
where $\rme^{\mathcal{L}_*t}$ denotes the $C^0$-semigroup generated by the linearization $\mathcal{L}_*$ at the spiral wave $u_*$ on $L^2(\R^2,\C^N)$.
\end{Lemma}

We will prove Lemma~\ref{l:spiralperp} in \S\ref{s:ps}. We briefly discuss a few \rev{implications} of the preceding lemma.

Recall that spectrally stable wave trains can be unstable with respect to transverse perturbations as our definition of spectral stability of wave train pertains only to perturbations in the direction of propagation. Assume that $u_*$ is a transverse spiral wave whose extended point spectrum lies on or to the left of the imaginary axis. It then follows from the results in \S\ref{s:mr.fredholm} that the entire spectrum of the linearization $\mathcal{L}_*$ at $u_*$ lies on or to the left of the imaginary axis. If the spectral mapping theorem held for $\mathcal{L}_*$, we could conclude that the semigroup generated by $\mathcal{L}_*$ could grow at most weakly exponentially. However, Lemma~\ref{l:spiralperp} shows that if the asymptotic wave train is transversely unstable, then the spectral mapping theorem cannot hold for the linearization. The reason for the exponential growth of the semigroup is the fact that the resolvent of $\mathcal{L}_*$ in $L^2(\R^2,\C^N)$ cannot be bounded uniformly along the vertical line $\Re\lambda=\Re\lambda_*$ (we will prove this in \S\ref{s:ps}).

On large bounded domains, Theorems~\ref{t:tl} and~\ref{t:tl2} imply that the spectrum of the linearization at the truncated spiral wave will, inside each fixed bounded region in the complex plane, lie on or to the left of the imaginary axis for all sufficiently large radii $R$. \rev{Hence, if the transverse instability of the asymptotic wave train generates unstable eigenvalues in the spectrum of the truncated spiral wave that are bounded away from the imaginary axis, then these eigenvalues $\lambda$ must diverge with $|\Im\lambda|\to\infty$ as $R\to\infty$. We note that we have not proved that transverse instabilities of the asymptotic wave train create unstable point spectrum of the truncated spiral wave, though we expect that they do.}

%%%%%%%%%%%%%%%%%%%%%%%%%%%%%%%%%%%%%%%%%%%%%%%%%%%%%%%%%%%%%%%%%%%%%%%%%

\section{Wave trains}\label{s:wt}

\rev{We give the proofs of the results stated in \S\ref{s:wts}. Specifically, we consider one-parameter families of wave trains and transverse instabilities of wave trains in \S\ref{s:wt.1}, characterize the spectra in the laboratory frame in \S\ref{s:equiv}, and compare properties of the PDE linearization and the spatial dynamical system in \S\ref{s:AL}. In \S\ref{s:rmi}, we give a different but equivalent definition of the relative Morse index of the wave trains that will be useful later.}

\subsection{Proofs of Proposition~\ref{p:nldr} and Lemma~\ref{l:2d}}\label{s:wt.1}

We begin with the proof of Proposition~\ref{p:nldr}. We want to solve
the equation
\begin{equation}\label{e:perode}
\mathcal{F}(u,k,\omega) := k^2 D u_{\xi\xi} + \omega u_\xi + f(u) = 0
\end{equation}
\rev{in $H^2(S^1,\R^N)\times\R^2$ near the solution $(u,k,\omega)=(u_\infty,k_*,\omega_*)$}. The linearization of this equation at $u_\infty$ gives the linear operator $\hat{\mathcal{L}}_\mathrm{co}(0)$
\[
\hat{\mathcal{L}}_\mathrm{co}(0) = k_*^2 D \partial_{\xi\xi} + \omega_*\partial_\xi + f^\prime(u_\infty(\xi))
\]
in $L^2(S^1,\C^N)$. The condition $\partial_\lambda E(0,0)\neq0$ of simplicity of the linear dispersion relation, where the Evans function $E$ was defined in \eqref{e:drper}, guarantees that \rev{the eigenvalue $\lambda=0$ of $\hat{\mathcal{L}}_\mathrm{co}(0)$ has algebraic multiplicity one; see \cite{gar}. In particular, the kernel of $\hat{\mathcal{L}}_\mathrm{co}(0)$ is one-dimensional and spanned by $u_\infty^\prime$, the kernel of the adjoint operator $\hat{\mathcal{L}}_\mathrm{co}^\mathrm{ad}(0)$ is spanned by a nonzero function $u_\mathrm{ad}$, and the $L^2(S^1,\C^N)$-scalar product $\langle u_\mathrm{ad},u_\infty^\prime\rangle\neq0$ of $u_\infty^\prime$ and $u_\mathrm{ad}$ does not vanish. We can now apply Lyapunov--Schmidt reduction: First, we solve (\ref{e:perode}) projected spectrally onto the range $\Rg(\hat{\mathcal{L}}_\mathrm{co}(0))$ of $\hat{\mathcal{L}}_\mathrm{co}(0)$ near $(u,k,\omega)=(u_\infty,k_*,\omega_*)$ for $v=u-u_\infty\in(\R u_\infty^\prime)^\perp\in H^2(S^1,\R^N)$ using the implicit function theorem. It remains to project (\ref{e:perode}), evaluated at the solution $u=u_\infty+v(k,\omega)$ of the previous step, spectrally onto the one-dimensional null space of $\hat{\mathcal{L}}_\mathrm{co}(0)$, which gives the equation $h(k,\omega):=\langle u_\mathrm{ad},\mathcal{F}(u_\infty+v(k,\omega),k,\omega)\rangle=0$, where $h(k_*,\omega_*)=0$ and $\partial_\omega h(k_*,\omega_*)=\langle u_\mathrm{ad},u_\infty^\prime\rangle\neq0$. We can therefore solve the reduced equation for $\omega$ as a function of $k$ using the implicit function theorem.}

The derivative $\omega^\prime(k)$ of the nonlinear dispersion relation
can be computed as follows. Evaluating (\ref{e:perode}) along the
family $u(\xi;k)$ of periodic solutions that we found in the preceding
paragraph, taking the derivative with respect to $k$, and evaluating
at $k=k_*$ gives
\begin{equation}\label{e:uk}
\hat{\mathcal{L}}_\mathrm{co}(0) \frac{\partial u}{\partial k}(\xi;k) =
- \left[ 2k_* D \partial_{\xi\xi} u_\infty
+ \frac{\rmd\omega}{\rmd k}(k_*) \partial_\xi u_\infty \right];
\end{equation}
see \cite[\S4]{dsss} for details. 
Projecting with $u_\mathrm{ad}$ onto the kernel of $\hat{\mathcal{L}}_\mathrm{co}(0)$ gives
the expression
\[
c_\mathrm{g,nl} = \frac{\rmd\omega}{\rmd k}(k_*) =
- \frac{2k_* \langle u_\mathrm{ad},D u_\infty^{\prime\prime}\rangle}%
{\langle u_\mathrm{ad},u_\infty^\prime\rangle}.
\]
To see that the linear and nonlinear group velocity coincide, we take
the derivative of the eigenvalue problem
\[
\hat{\mathcal{L}}_\mathrm{co}(\nu) u(\nu) = \lambda_\mathrm{co}(\nu) u(\nu),
\]
with respect to $\nu$, evaluate at
$\lambda=\nu=0$, and project onto the kernel of $\hat{\mathcal{L}}_\mathrm{co}(0)$ using
$u_\mathrm{ad}$. The resulting expression for
$[-\rmd\lambda_\mathrm{co}/\rmd\nu(0)+c]$ coincides with (\ref{e:ndr}).
This completes the proof of Proposition~\ref{p:nldr}.

To prove Lemma~\ref{l:2d}, we solve the eigenvalue problem
(\ref{e:rdscoly})
\[
\lambda v = D (k^2\partial_{\xi\xi}+\nu_\perp^2) v
+ \omega \partial_\xi v + f^\prime(u_\infty(\xi)) v
\]
near $(v,\lambda,\nu_\perp)=(u_\infty^\prime,0,0)$ using Lyapunov--Schmidt reduction on $L^2(S^1,\C^N)$. The reduced equation on the kernel is
\[
\lambda \langle u_\mathrm{ad},u_\infty^\prime\rangle =
\langle u_\mathrm{ad},D u_\infty^\prime\rangle \nu_\perp^2 + \rmO(\nu_\perp^4),
\]
which proves the lemma.

\subsection{Proofs of Lemmas~\ref{l:floquet} and~\ref{l:equiv}}\label{s:equiv}

We consider the linear non-autonomous parabolic equation
\begin{equation}\label{e:rdsln}
u_t = D u_{xx} + f^\prime(u_\infty(kx-\omega t)) u
\end{equation}
and are interested in the set of $\lambda$ for which
$\Psi_\mathrm{st}-\rme^{2\pi\lambda/\omega}$ does not have a bounded inverse,
where
\[
\Psi_\mathrm{st}: \quad L^2(\R,\C^N) \longrightarrow L^2(\R,\C^N), \quad
u(\cdot,0) \longmapsto u(\cdot,2\pi/\omega)
\]
is the period map associated with (\ref{e:rdsln}). If we substitute
the Floquet ansatz $u(x,t)=\rme^{\lambda t}\tilde{u}(x,\omega t)$ into
(\ref{e:rdsln}), and use $\tau=\omega t$, we can rewrite
(\ref{e:rdsln}) as the differential equation
\begin{align}\label{e:mtwl}
u_x = &  v \\ \nonumber
v_x = &  -D^{-1}[-\omega\partial_\tau u
+ f^\prime(u_\infty(kx-\tau)) u - \lambda u],
\end{align}
where we replaced $\tilde{u}$ by $u$. Imposing $2\pi$-periodic boundary conditions in $\tau$, we can write this equation in
the abstract form
\begin{equation}\label{e:mtwabs}
\mathbf{u}_x = \tilde{\mathcal{A}}(x;\lambda) \mathbf{u},
\end{equation}
where $\tilde{\mathcal{A}}(x;\lambda)$ is a closed operator on
$Y:=H^{\frac12}(S^1,\C^N)\times L^2(S^1,\C^N)$ with domain
$Y^1=H^1(S^1,\C^N)\times H^{\frac12}(S^1,\C^N)$; see \cite{ss-fred}.

\begin{Lemma}[{\cite[Theorems~2.6 and~2.8(i)]{ss-fred}}]
The closed operator $\tilde{\mathcal{T}}_\lambda$,
\[
\tilde{\mathcal{T}}_\lambda =
\frac{\rmd}{\rmd x} - \tilde{\mathcal{A}}(\cdot;\lambda): \quad
L^2(\R,Y) \longrightarrow L^2(\R,Y),
\]
with domain $L^2(\R,Y^1)\cap H^1(\R,Y)$ has a bounded inverse if and
only if $\lambda$ does not belong to the Floquet spectrum of $\Psi_\mathrm{st}$.
\end{Lemma}

The differential equation (\ref{e:mtwl}) is non-autonomous in the
spatial evolution variable $x$. However, if we change coordinates by
replacing the time variable $\tau$ by $\sigma=kx-\tau$, we
obtain the autonomous equation
\begin{align}\label{e:mtwl2}
u_x = &  - k\partial_\sigma u + v\\ \nonumber
v_x = &  - k\partial_\sigma v - D^{-1}[\omega\partial_\sigma u
+ f^\prime(u_\infty(\sigma)) u - \lambda u],
\end{align}
which we also write as
\begin{equation}\label{e:mtwabs2}
\mathbf{u}_x = \mathcal{A}_\infty(\lambda) \mathbf{u},
\end{equation}
where $\mathcal{A}_\infty(\lambda)$ is a closed operator on $Y$ with domain
$H^{\frac32}(S^1,\C^N)\times H^1(S^1,\C^N)$.

\begin{Lemma}\label{l:eqinv0}
The operator $\mathcal{T}_\lambda$
\[
\mathcal{T}_\lambda =
\frac{\rmd}{\rmd x} - \mathcal{A}_\infty(\lambda): \quad
L^2(\R,Y) \longrightarrow L^2(\R,Y)
\]
with domain
\[
\mathcal{D}(\mathcal{T}_\lambda) =
\{ (u,v) \in L^2(\R,Y^1);\;
(\partial_x+k\partial_\sigma)(u,v) \in L^2(\R,Y) \}
\]
is closed. It has a bounded inverse if and only if
$\tilde{\mathcal{T}}_\lambda$ does.
\end{Lemma}
\begin{Proof}
We refer to \cite[\S2.2]{hss} for the proof that
$\mathcal{T}_\lambda$ is closed on $L^2(\R,Y)$. The statement about
invertibility is obvious as both operators are conjugated by a
transformation of the independent variables.
\end{Proof}

The key is now that it is far easier to check invertibility of
$\mathcal{T}_\lambda$ as this involves only the $x$-independent
operator $\mathcal{A}_\infty(\lambda)$. Particular solutions to
(\ref{e:mtwl2}) with exponential growth $\rme^{\nu x}$ can be readily
constructed provided $\nu$ is an eigenvalue of $\mathcal{A}_\infty(\lambda)$.
Note that $\mathcal{A}_\infty(\lambda)$ has compact resolvent so that its
spectrum is discrete.

\begin{Lemma}\label{l:eqinv}
The operator $\mathcal{T}_\lambda$ has a bounded inverse if and only
if $\mathcal{A}_\infty(\lambda)$ is hyperbolic, i.e., if none of its
eigenvalues is purely imaginary. In particular, $\lambda$ is in the
Floquet spectrum if and only if $\mathcal{A}_\infty(\lambda)$ has a purely
imaginary eigenvalue $\nu$.
\end{Lemma}
\begin{Proof}
If there is an eigenvalue $\nu$ of $\mathcal{A}_\infty(\lambda)$ with
$\Re\nu=0$, then we can construct an almost eigenfunction as in
\cite{hen,ssw2}, and $\mathrm{T}_\lambda$ does not have a bounded
inverse. On the other hand, suppose that all eigenvalues of
$\mathcal{A}_\infty(\lambda)$ have non-zero real part. Transforming back to
the $\tau=kx-\sigma$ variable, this excludes the existence of bounded,
purely imaginary Floquet exponents of (\ref{e:mtwl}) with Floquet
eigenfunctions $(u,v)(x+2\pi/k,\tau)=\rme^{\rmi\gamma}(u,v)(x,\tau)$
for some $\gamma\in\R$. Floquet theory \cite{mie,ss-fred} for
(\ref{e:mtwl}) shows that $\tilde{\mathcal{T}}_\lambda$ is then
invertible, and therefore $\mathcal{T}_\lambda$ is invertible as well
on account of Lemma~\ref{l:eqinv0}.
\end{Proof}

It remains to study the eigenvalue problem (\ref{e:spmtw}),
\begin{align*}
\nu u = &  -k \partial_\sigma u + v \\
\nu v = &  -k \partial_\sigma v - D^{-1}[\omega\partial_\sigma u
+ f^\prime(u_\infty(\sigma)) u - \lambda u]
\end{align*}
with $2\pi$-periodic boundary conditions for $(u,v)$. This is
equivalent to the generalized eigenvalue problem
\begin{equation}\label{e:cgcomp}
D (k\partial_\sigma+\nu)^2 u + \omega\partial_\sigma u
+ f^\prime(u_\infty(\sigma)) u = \lambda u
\end{equation}
for $\nu$, again with $2\pi$-periodic boundary conditions for $u$ and
its derivative. Adding the term $c\nu u$ with $c=\omega/k$ on both sides, we find that
$u$ needs to be a $2\pi$-periodic solution of
\begin{equation}\label{e:flspd}
D (k\partial_\sigma+\nu)^2 u + c (k\partial_\sigma+\nu) u
+ f^\prime(u_\infty(\sigma)) u = (\lambda+\omega\nu/k) u.
\end{equation}
Comparing (\ref{e:fam}) and (\ref{e:flspd}), we have found a way to
compute the spectrum in the co-moving frame: for $\nu\in\rmi\R$, we have
\[
\lambda_\mathrm{co}=\lambda+c\nu\in\mathrm{spec}\,\mathcal{L}
\quad\Longleftrightarrow\quad
\lambda_\mathrm{st}=\lambda=\lambda_\mathrm{co}-c\nu\in\mathrm{spec}\,\Psi_\mathrm{st}.
\]
Note that (\ref{e:cgcomp}) implies that $\lambda_\mathrm{st}(\nu+\rmi k\ell)=\lambda_\mathrm{st}(\nu)-\rmi\omega\ell$ for all $\ell\in\mathbb{Z}$ and $\nu\in\mathbb{C}$.
This completes the proof of Lemma~\ref{l:floquet} and of the first
part of Lemma~\ref{l:equiv}. The remaining statements in Lemma~\ref{l:equiv} regarding the spatial
eigenvalues $\nu$ of the operator $\mathcal{A}_\infty(\lambda)$ can be proved
easily using Fourier series; see \cite{ss-fred} for similar
arguments.

\subsection{\rev{Comparison of PDE and spatial-dynamics linearizations}}\label{s:AL}

We remark that elements of the null space of
\[
\hat{\mathcal{L}}_\mathrm{\rev{co}}(\nu) - \lambda_\mathrm{\rev{co}} = D (k\partial_\sigma+\nu)^2
+ \omega\partial_\sigma + f^\prime(u_\infty(\sigma)) - \lambda, \qquad
\rev{\lambda_\mathrm{co} = \lambda + c\nu}
\]
and
\[
\mathcal{A}_\infty(\lambda) - \nu =
\left(\begin{array}{cc} -(k\partial_\sigma+\nu) & \id \\
- D^{-1}[\omega\partial_\sigma + f^\prime(u_\infty(\sigma)) - \lambda]
& -(k\partial_\sigma+\nu) \end{array}\right)
\]
are related. If $u$ is an eigenfunction of
$\hat{\mathcal{L}}_\mathrm{\rev{co}}(\nu)$ associated with the temporal eigenvalue
$\lambda_\mathrm{\rev{co}}$, then $\mathbf{u}=(u,(k\partial_\sigma+\nu)u)$ is an
eigenfunction of $\mathcal{A}_\infty(\lambda)$ associated with the spatial
eigenvalue $\nu$, and vice versa. Furthermore, $u_\mathrm{ad}$ is an
eigenfunction of the $L^2$-adjoint
\[
\hat{\mathcal{L}}_\mathrm{\rev{co}}^\mathrm{ad}(\nu) = D (-k\partial_\sigma+\nu)^2
+ c (-k\partial_\sigma+\nu) + f^\prime(u_\infty(\rev{\sigma}))^*
\]
associated with the eigenvalue $\lambda_\mathrm{\rev{co}}$ if and only if
$\mathbf{u}_\mathrm{ad}=(\rev{D}(-k\partial_\sigma+\nu)u_\mathrm{ad},\rev{D}u_\mathrm{ad})$
is an eigenfunction of the formal adjoint
\[
\left(\begin{array}{cc} k\partial_\sigma &
[\omega\partial_\sigma - f^\prime(u_\infty(\sigma))^* + \lambda] D^{-1}
\\ \id & k\partial_\sigma \end{array}\right)
\]
of $\mathcal{A}_\infty(\lambda)$ to the eigenvalue $\nu$.

\subsection{The relative Morse index}\label{s:rmi}

We give an equivalent definition of the relative Morse index $i_\mathrm{M}(\lambda)$ that we defined in \S\ref{s:wts.morse}. Recall that $\lambda$ belongs to the Floquet spectrum of the wave trains if \rev{and only if} there exists a purely imaginary eigenvalue $\nu\in\rmi\R$ of the operator $\mathcal{A}_\infty(\lambda)$ defined in (\ref{e:mtwabs2}).

We define the reference operator
\begin{equation}\label{e:Aref}
\mathcal{A}_\mathrm{ref} := \left(\begin{array}{cc} - k\partial_\sigma & \id \\
- D^{-1} \omega \partial_\sigma + 1 & - k\partial_\sigma \end{array}\right)
\end{equation}
on $Y=H^{\frac12}\times L^2$ with domain $H^{\frac32}\times H^1$. Using explicit Fourier-series calculations, we see that the operator $\mathcal{A}_\mathrm{ref}$ is invertible on $Y$ and that there are bounded stable and unstable projections $P_\mathrm{ref}^\mathrm{s}$ and $P_\mathrm{ref}^\mathrm{u}=\id-P_\mathrm{ref}^\mathrm{s}$ on $Y$ that commute with $\mathcal{A}_\mathrm{ref}$ on its domain such that $\Re\mathrm{spec}\,\mathcal{A}_\mathrm{ref}|_{\Rg(P_\mathrm{ref}^\mathrm{s})}<0$ and $\Re\mathrm{spec}\,\mathcal{A}_\mathrm{ref}(\lambda)|_{\Rg(P_\mathrm{ref}^\mathrm{u})}>0$.

\begin{Proposition}\label{p:mor}
If $\lambda$ is not in the Floquet spectrum of the wave train $u_\infty$, then the following is true.
\begin{compactenum}[(i)]
\item There exist bounded stable and unstable projections $P_\mathrm{wt}^\mathrm{s}(\lambda)$ and $P_\mathrm{wt}^\mathrm{u}(\lambda)=\id-P_\mathrm{wt}^\mathrm{s}(\lambda)$ on $Y$ that commute with $\mathcal{A}_\infty(\lambda)$ on its domain such that
\[
\Re\mathrm{spec}\,\mathcal{A}_\infty(\lambda)|_{\Rg(P_\mathrm{wt}^\mathrm{s}(\lambda))}<0 \quad\mbox{and}\quad
\Re\mathrm{spec}\,\mathcal{A}_\infty(\lambda)|_{\Rg(P_\mathrm{wt}^\mathrm{u}(\lambda))}>0.
\]
\item The operator $P_\mathrm{ref}^\mathrm{u}-P_\mathrm{wt}^\mathrm{u}(\lambda):Y\to Y$ is compact.
\item If $\tilde{\lambda}$ is also not in the Floquet spectrum of the wave train $u_\infty$, then the operators
\begin{equation}\label{e:pfred}
P_\mathrm{wt}^\mathrm{u}(\tilde{\lambda}):\Rg(P_\mathrm{wt}^\mathrm{u}(\lambda))\to\Rg(P_\mathrm{wt}^\mathrm{u}(\tilde{\lambda}))
\end{equation}
and
\[
\iota(\lambda,\tilde{\lambda}): \quad
\Rg(P_\mathrm{wt}^\mathrm{u}(\lambda)) \times \Rg(P_\mathrm{wt}^\mathrm{s}(\tilde{\lambda})) \longrightarrow Y, \quad
(\mathbf{u}^\mathrm{u},\mathbf{u}^\mathrm{s}) \longmapsto \mathbf{u}^\mathrm{u}+\mathbf{u}^\mathrm{s}.
\]
are Fredholm operators with the same Fredholm index, which we denote by $i_{P_\mathrm{wt}}(\lambda,\tilde{\lambda})$.
\item Choose $\lambda_\mathrm{inv}\gg1$ so large that $[\lambda_\mathrm{inv},\infty)$ belongs to the resolvent set of $\Psi_\mathrm{st}$, then $i_\mathrm{M}(\lambda)=i_{P_\mathrm{wt}}(\lambda,\lambda_\mathrm{inv})$, where $i_\mathrm{M}(\lambda)$ was defined in \S\ref{s:wts.morse}.
\end{compactenum}
\end{Proposition}

\begin{Proof}
Statement~(i) and the claims for the operator in (\ref{e:pfred}) were proved in \cite[Theorems~5.1 and~5.2]{ssmorse}. Alternatively, these statements were proved in \cite{ss-fred} for (\ref{e:mtwabs}), and since the evolution operators and projections of the exponential dichotomies of (\ref{e:mtwabs}) and (\ref{e:mtwabs2}) are conjugated by the strongly continuous shift generated by $\mathbf{u}_x=-k\partial_\sigma\mathbf{u}$, the results also hold for (\ref{e:mtwabs2}).

Statement~(ii) was proved in \cite[Remark~4.1]{ss-fred}. Alternatively, we can use the results in \S\ref{s:ed} below: Using the notation introduced there, the stable projections $\mathcal{P}_\mathrm{ref}^\mathrm{s}$ of the operator $\tilde{\mathcal{A}}_\infty$ defined in (\ref{e:at}) and $\mathcal{P}_m^\mathrm{s}$ of the right-hand side of (\ref{e:ainfd1}) differ only in the finite-dimensional space $\Rg(Q_m)$, and we conclude that the difference of these projections is compact. Since the projections $\mathcal{P}_m^\mathrm{s}$ of (\ref{e:ainfd1}) converge in norm to the stable projections $\mathcal{P}_\infty^\mathrm{s}$ of the right-hand side of (\ref{e:ainfd}), we see that the difference $\mathcal{P}_\infty^\mathrm{s}-\mathcal{P}_\mathrm{ref}^\mathrm{s}$ is also compact. Finally, we note that the operator on the right-hand side of (\ref{e:ainfd}) corresponds to the operator $\mathcal{A}_\infty(\lambda)$.

To prove the statement about the operator $\iota(\lambda,\tilde{\lambda})$, we note that the map (\ref{e:pfred}) and the operator
\[
\tilde{\iota}(\lambda,\tilde{\lambda}): \quad
\Rg(P_\mathrm{wt}^\mathrm{u}(\lambda)) \times \Rg(P_\mathrm{wt}^\mathrm{s}(\tilde{\lambda})) \longrightarrow Y, \quad
(\mathbf{u}^\mathrm{u},\mathbf{u}^\mathrm{s}) \longmapsto
P_\mathrm{wt}^\mathrm{u}(\tilde{\lambda}) \mathbf{u}^\mathrm{u} + \mathbf{u}^\mathrm{s}
\]
share the same Fredholm and Fredholm index properties. The difference of $\iota(\lambda,\tilde{\lambda})$ and $\tilde{\iota}(\lambda,\tilde{\lambda})$ is given by $P_\mathrm{wt}^\mathrm{u}(\lambda)-P_\mathrm{wt}^\mathrm{u}(\tilde{\lambda})$, which is compact by (ii) (add and subtract $P_\mathrm{ref}^\mathrm{u}$). This completes of the proof of (iii). Finally, statement~(iv) can be proved using Fourier series as in \cite{ss-fred} or \cite[\S5]{ssmorse}.
\end{Proof}

%%%%%%%%%%%%%%%%%%%%%%%%%%%%%%%%%%%%%%%%%%%%%%%%%%%%%%%%%%%%%%%%%%%%%%%%%

\section{Exponential dichotomies}\label{s:ed}

Throughout this section, we assume that  $u_*(r,\psi)$,
$u_\infty(\psi)$, and  $\theta(r)$ are smooth functions, written in
polar coordinates \rev{$(r,\psi)$}, such that for some non-zero number
$k_*$
\[
|u_*(r,\cdot) - u_\infty(k_*r+\theta(r)+\cdot)|_{C^1(S^1)} \to 0
\mbox{ as } r\to\infty,
\]
and $\theta^\prime(r)\to0$ as $r\to\infty$. Note that we do not assume
that $u_*$ is an Archimedean spiral wave or even a solution to the
reaction-diffusion equation (\ref{e:rds-}). We are interested in the
linear eigenvalue problem $\mathcal{L}_*u=\lambda u$ where
\[
\mathcal{L}_* =
D \left[ \partial_{rr} + \frac{1}{r} \partial_r
+ \frac{1}{r^2} \partial_{\psi\psi} \right]
+ \omega_* \partial_\psi + f^\prime(u_*(r,\psi)).
\]
We assume that $\omega_*$ is non-zero. We rewrite this eigenvalue
problem as a first-order differential equation
\begin{align}\label{e:lspdyn}
u_r = &  v \\ \nonumber
v_r = &  - \frac{1}{r} v - \frac{1}{r^2} \partial_{\psi\psi} u
- D^{-1}[\omega_*\partial_{\psi} u + f^\prime(u_*(r,\psi)) u - \lambda u]
\end{align}
in the spatial ``time''-variable $r$. The system
(\ref{e:lspdyn}) can be viewed as an abstract linear ordinary
differential equation
\begin{equation}\label{e:sdabs}
\mathbf{u}_r = \mathcal{A}(r;\lambda) \mathbf{u}, \qquad \mathbf{u}=(u,v)
\end{equation}
on the Banach space $X:=H^1(S^1,\C^N)\times L^2(S^1,\C^N)$. For each fixed $r$, the operator $\mathcal{A}(r;\lambda)$ is closed on $X$ with domain $X^1:=H^2(S^1,\C^N)\times H^1(S^1,\C^N)$.

We say that a function $\mathbf{u}\in C^0(J,X)$ is a solution of
(\ref{e:sdabs}) on an interval $J\in\R^+$ if for each $r$ in the
interior of $J$ the function $\mathbf{u}(r)$ is continuous with values in $X^1$ and
differentiable in $r$ as a function into $X$, and satisfies
(\ref{e:sdabs}) in $X$.

In \S\ref{s:edc} and \S\ref{s:edff}, we construct exponential dichotomies in the core region $0<r<R$ and the far-field $r>R$, respectively. \rev{In \S\ref{s:ced}, we describe different ways in which the resulting projections can be compared to each other. Section~\ref{s:eda} deals with exponential dichotomies for the adjoint differential equations. Finally, we consider exponential dichotomies in weighted spaces in \S\ref{s:edweights} and discuss exponential trichotomies, where we allow for neutral center directions, in \S\ref{s:edcenter}.} For background, we refer to \cite{pss} for basic results on exponential dichotomies in this infinite-dimensional, ill-posed setting, in particular \rev{for} a result on robustness of dichotomies, and to \cite{ss-fred} for a slightly different approach based on Galerkin approximations.

\subsection{Exponential dichotomies in the core region}\label{s:edc}

Upon introducing the logarithmic radial time $s=\log r$ as in
\cite{s-siam}, the eigenvalue problem (\ref{e:lspdyn}) becomes
\begin{align}\label{e:coreprelim}
u_s = & \rme^s v \\ \nonumber
v_s = & -v - \rme^{-s} \partial_{\psi\psi} u - \rme^{s} D^{-1}[
\omega_*\partial_{\psi} u + f^\prime(u_*(\rme^s,\psi)) u - \lambda u].
\end{align}
We introduce the new variable $w=\rme^s v$ so that (\ref{e:coreprelim}) becomes
\begin{align}\label{e:lspdyn0}
u_s = & w \\ \nonumber
w_s = & -\partial_{\psi\psi} u - \rme^{2s} D^{-1}[
\omega_*\partial_{\psi} u + f^\prime(u_*(\rme^s,\psi)) u - \lambda u],
\end{align}
which corresponds to the PDE
\begin{equation}\label{e:mz}
D (u_{ss} + u_{\psi\psi}) + \rme^{2s}
(\omega_* u_\psi + f^\prime(u_*(\rme^s,\psi)) u - \lambda u) = 0.
\end{equation}
We consider (\ref{e:lspdyn0}) on the Hilbert space
$X=H^1(S^1,\C^N)\times L^2(S^1,\C^N)$ and write it as the abstract
differential equation
\begin{equation}\label{e:sdabs0}
\mathbf{u}_s = \mathcal{A}_\mathrm{core}(s;\lambda) \mathbf{u}.
\end{equation}
Our goal is to show that (\ref{e:sdabs0}) has a dichotomy on
$(-\infty,s_*]$ for each fixed $s_*\in\R$. We begin with the limiting
equation
\begin{align}\label{e:lspdyn0inf}
u_s = &  w \\ \nonumber
w_s = &  -\partial_{\psi\psi} u,
\end{align}
which is obtained by formally taking $s=-\infty$ in (\ref{e:lspdyn0}).
Equation (\ref{e:lspdyn0inf}) can be readily solved using Fourier
series. The resulting solutions can be distinguished by their growth
or decay properties which give a decomposition of $X$ into the
following spaces:
\begin{align*}
\rev{E_{-\infty}^\mathrm{ss}} = & \mathrm{span} \left\{
u_0 \rme^{\rmi\ell\psi} \begin{pmatrix} 1 \\ -\ell \end{pmatrix};\; \ell\in\mathbb{Z}\setminus\{0\}, u_0\in\C^N \right\} \\
\rev{E_{-\infty}^\mathrm{uu}} = & \mathrm{span} \left\{
u_0 \rme^{\rmi\ell\psi} \begin{pmatrix} 1 \\ \ell \end{pmatrix};\; \ell\in\mathbb{Z}\setminus\{0\}, u_0\in\C^N \right\} \\
\rev{E_{-\infty}^\mathrm{c}} = & \mathrm{span} \left\{
\begin{pmatrix} u_0 \\ w_0 \end{pmatrix};\; u_0,w_0\in\C^N \right\}.
\end{align*}
Note that solutions to initial data in $\rev{E_{-\infty}^\mathrm{ss}}$ exist for
$s>0$ and decay exponentially with rate $1$, while solutions to
initial data in $\rev{E_{-\infty}^\mathrm{uu}}$ exist for $s<0$ and decay again
exponentially with rate $1$. Solutions with initial data
$(u,v)\in \rev{E_{-\infty}^\mathrm{c}}=\C^N\times\C^N$ are given by $(u+sv,v)$. Associated with these subspaces are projections $P^\mathrm{ss/uu/c}_{-\infty}$ which project onto $E^\mathrm{ss/uu/c}_{-\infty}$, respectively, along the other spectral subspaces.  For
elements $(u,v)\in \rev{E_{-\infty}^\mathrm{c}}$, we define the complementary projections $\tilde{P}^\mathrm{ker}_{-\infty}(s)=\id - \tilde{P}^\mathrm{gker}_{-\infty}(s)$ and the asymptotic linear generator of the evolution via
\[
\tilde{P}^\mathrm{ker}_{-\infty}(s) {u \choose w} := {u - sw\choose 0}, \qquad
\tilde{P}^\mathrm{gker}_{-\infty}(s){u \choose w} := {s w \choose w}, \qquad
A^\mathrm{c}_{-\infty} :=
\left(\begin{array}{cc} 0 & 1 \\ 0 & 0 \end{array}\right).
\]
We can extend these projections to $X$ through $P^\mathrm{ker/gker}_{-\infty}(s):=\tilde{P}^\mathrm{ker/gker}_{-\infty}(s)P^\mathrm{c}_{-\infty}$.
The following proposition states that solutions to the full equation
(\ref{e:lspdyn0}) behave in the same fashion as those of the limiting
equation (\ref{e:lspdyn0inf}).

\begin{Proposition}\label{p:dc}
For any fixed $s_*\in\R$, the following is true. There exists a
constant $C>0$ and strongly continuous families $P_-^\mathrm{uu}(s)$,
$P_-^\mathrm{ss}(s)$ and $P_-^\mathrm{c}(s)$ of complementary
projections on $X$, all defined for $-\infty<s<s_*$, as well as linear
evolution operators $\Phi_-^\mathrm{ss}(s;\sigma)$,
$\Phi_-^\mathrm{uu}(\sigma;s)$ and $\Phi_-^\mathrm{c}(s;\sigma)$
on $X$ which are strongly continuous in $(s,\sigma)$ for
$-\infty<\sigma\leq s\leq s_*$ and differentiable in $(s,\sigma)$ for
$-\infty<\sigma<s< s_*$, such that the following is true on $X$:
\begin{compactitem}
\item\emph{Compatibility}.
We have $\Phi_-^\mathrm{ss}(\sigma;\sigma)=P_-^\mathrm{ss}(\sigma)$,
$\Phi_-^\mathrm{uu}(\sigma;\sigma)=P_-^\mathrm{uu}(\sigma)$, and
\[
\id =
P_-^\mathrm{c}(s) + P_-^\mathrm{uu}(s) + P_-^\mathrm{ss}(s)
\]
for all $s<s_*$. The projections are bounded in norm uniformly in $s$.
\item\emph{Instability}.
For any $\mathbf{u}_0\in X$, $\Phi_-^\mathrm{uu}(s,\sigma)\mathbf{u}_0$
is a solution of (\ref{e:sdabs0}) with
\[
|\Phi_-^\mathrm{uu}(s;\sigma)\mathbf{u}_0|_{X}
\leq C \rme^{-|s-\sigma|}\, |\mathbf{u}_0|_{X},
\]
where $s\leq\sigma\leq s_*$.
\item\emph{Stability}.
For any $\mathbf{u}_0\in X$, $\Phi_-^\mathrm{ss}(s,\sigma)\mathbf{u}_0$
is a solution of (\ref{e:sdabs0}) with
\[
|\Phi_-^\mathrm{ss}(s;\sigma)\mathbf{u}_0|_{X}
\leq C \rme^{-|s-\sigma|}\, |\mathbf{u}_0|_{X},
\]
where $\sigma\leq s\leq s_*$.
\item\emph{Neutral directions}.
We have $\dim\Rg(P_-^\mathrm{c}(s))=2N$. For any $\mathbf{u}_0\in X$,
$\Phi_-^\mathrm{c}(s,\sigma)\mathbf{u}_0$ is a solution of
(\ref{e:sdabs0}) with
\[
|\Phi_-^\mathrm{c}(s;\sigma)\mathbf{u}_0|_{X}
\leq C \rev{(1+|s-\sigma|)}\, |\mathbf{u}_0|_{X},\qquad P_-^\mathrm{c}(s)=\Phi_-^\mathrm{c}(s;s),\qquad |P_-^\mathrm{c}(s)-P_{-\infty}^\mathrm{c}|_{L(X)}=\rmO(\rme^{2s})\]
where $\sigma,s\leq s_*$. We can decompose even further
\[
P_-^\mathrm{c}(s) = P_-^\mathrm{ker}(s) + P_-^\mathrm{gker}(s),
\qquad
\|P_-^\mathrm{ker}(s)\|_{L(X)}+\|P_-^\mathrm{gker}(s)\|_{L(X)} \leq C \rev{(1+|s|)}
\]
with 
\begin{align*}
 \Phi_-^\mathrm{ker}(s;\sigma)&:=\Phi_-^\mathrm{c}(s;\sigma)P_-^\mathrm{ker}(\sigma)=P_-^\mathrm{ker}(s)\Phi_-^\mathrm{c}(s;\sigma)\\
 \Phi_-^\mathrm{gker}(s;\sigma)&:=\Phi_-^\mathrm{c}(s;\sigma)P_-^\mathrm{gker}(\sigma)=P_-^\mathrm{gker}(s)\Phi_-^\mathrm{c}(s;\sigma).
\end{align*}
There exist bounded transformations $T(s):E^\mathrm{c}_{-\infty}\to E^\mathrm{c}_-(s)\subset X$ with 
\[
\|T(s)-\id\|=\rmO(\rme^{2s}),\quad
\Phi_-^\mathrm{c}(s,\sigma) = T(s) \rme^{A^\mathrm{c}_{-\infty}(s-\sigma)} T^{-1}(\sigma),\quad
T^{-1}(\sigma) P_-^\mathrm{ker/gker}(\sigma) = \tilde{P}_{-\infty}^\mathrm{ker/gker}(\sigma) T^{-1}(\sigma).
\]
In particular, for $u_\mathrm{ker}=P_-^\mathrm{ker}(\sigma)u_\mathrm{ker}$, \rev{the last identity implies that}
%\[
%\Phi_-^\mathrm{c}(s,\sigma) u_\mathrm{ker} = T(s) T^{-1}(\sigma) u_\mathrm{ker} = (\id+\rmO(\rme^{2s})+\rmO(\rme^{2\sigma})) u_\mathrm{ker}.
%\]
\begin{eqnarray*}
\Phi_-^\mathrm{c}(s,\sigma) u_\mathrm{ker} & = &
\rev{ T(s) \rme^{A^\mathrm{c}_{-\infty}(s-\sigma)} T^{-1}(\sigma) u_\mathrm{ker}
= T(s) \rme^{A^\mathrm{c}_{-\infty}(s-\sigma)} \tilde{P}_{-\infty}^\mathrm{ker}(\sigma) T^{-1}(\sigma) u_\mathrm{ker}} \\ & = & T(s) T^{-1}(\sigma) u_\mathrm{ker}
= (\id+\rmO(\rme^{2s})+\rmO(\rme^{2\sigma})) u_\mathrm{ker}.
\end{eqnarray*}
% For a given $\mathbf{u}_0\in X$, let
% $\mathbf{u}_\mathrm{ker}=P_-^\mathrm{ker}(\sigma)\mathbf{u}_0$ and
% $\mathbf{u}_\mathrm{gker}=P_-^\mathrm{gker}(\sigma)\mathbf{u}_0$, then
% \begin{align*}
% |\Phi_-^\mathrm{c}(s;\sigma)\mathbf{u}_\mathrm{ker}
% - P^\mathrm{ker}_{-\infty}\mathbf{u}_\mathrm{ker}|_{X} & = 
% \rmO(\rme^{-2\min(|s|,|\sigma|)}) |\mathbf{u}_\mathrm{ker}|_{X} \\
% |\Phi_-^\mathrm{c}(s;\sigma)\mathbf{u}_\mathrm{gker}
% - \rme^{A^\mathrm{c}_{-\infty}(s-\sigma)} P^\mathrm{gker}_{-\infty}
% \mathbf{u}_\mathrm{gker}|_{X} & = 
% \rmO(|s-\sigma|\rme^{-2\min(|s|,|\sigma|)}) |\mathbf{u}_\mathrm{gker}|_{X}
% \end{align*}
% for all $s,\sigma<s_*$.
\item\emph{Invariance}.
The solutions $\Phi_-^\mathrm{ss}(s;\sigma)\mathbf{u}_0$,
$\Phi_-^\mathrm{uu}(s;\sigma)\mathbf{u}_0$, and
$\Phi_-^\mathrm{c}(s;\sigma)\mathbf{u}_0$ satisfy
\begin{align*}
\Phi_-^\mathrm{ss}(s;\sigma)\mathbf{u}_0 \in \Rg(P_-^\mathrm{ss}(s))
& \mbox{ for all }  \sigma\leq s\leq s_* \\
\Phi_-^\mathrm{uu}(s;\sigma)\mathbf{u}_0 \in \Rg(P_-^\mathrm{uu}(s))
& \mbox{ for all }  s\leq\sigma\leq s_* \\
\Phi_-^\mathrm{c}(s;\sigma)\mathbf{u}_0 \in \Rg(P_-^\mathrm{c}(s))
& \mbox{ for all }  s,\sigma\leq s_*.
\end{align*}
\end{compactitem}
\end{Proposition}
\begin{Proof}
Since the perturbation
\[
\rme^{2s} D^{-1}[\omega_*\partial_{\psi} u + f^\prime(u_*(\rme^s,\psi)) u - \lambda u],
\]
is bounded in $X$ and converges to zero as $s\to-\infty$, we can
apply the results in \cite{pss} to (\ref{e:lspdyn0}). Note that the
technical hypothesis \cite[(H5)]{pss} is satisfied on account of
\cite[Theorem~2.5]{mz} which applies to (\ref{e:mz}). As a
consequence, for any $0<\delta<1$, there are projections
$P_-^\mathrm{ss}(s)$, $P_-^\mathrm{uu}(s)$ and
$P_-^\mathrm{c}(s)$ as well as evolution operators
$\Phi_-^\mathrm{ss}(s,\sigma)$, $\Phi_-^\mathrm{uu}(s,\sigma)$ and
$\Phi_-^\mathrm{c}(s,\sigma)$ of the full problem (\ref{e:lspdyn0})
such that
\begin{align*}
\|\Phi_-^\mathrm{ss}(s,\sigma)\| + \|\Phi_-^\mathrm{uu}(\sigma,s)\|
& \leq  C \rme^{-|s-\sigma|}
\qquad \mbox{ for all } \sigma<s<s_* \\
\|\Phi_-^\mathrm{c}(\sigma,s)\| & \leq  C \rme^{\delta|s-\sigma|}
\qquad \mbox{ for all } s,\sigma<s_*.
\end{align*}
In fact, since the perturbation converges to zero exponentially as
$s\to-\infty$, the projection $P_-^\mathrm{c}(s)$ converges with rate
$\rev{\rme^{2s}}$ to the orthogonal projection onto $\rev{E_{-\infty}^\mathrm{c}}$ as
$s\to-\infty$. The equation in the center subspace $\Rg(P_-^\mathrm{c}(s))$ can therefore be projected onto the fixed reference frame $\rev{E_{-\infty}^\mathrm{c}}$, where it is an $\rmO(\rme^{2s})$-perturbation of $u_c'=A^c_{-\infty}u_c$. As a consequence, following for instance \cite[Chapter~3.8]{cl}, the solutions are foliated over this asymptotic equation in the form stated. 
% 
% Using \cite[Chapter~3.8]{cl}, we conclude that the solutions for
% (\ref{e:lspdyn0}) with initial data in $P_-^\mathrm{c}(s)$ are an
% $\rmO(\rme^{2s})$-correction to the solutions of the asymptotic ODE
% on $E_-^\mathrm{c}$.
Lastly, differentiability with respect to the
initial time $\sigma$ can be shown as in
\cite[Lemma~5.5]{ss-fred}.
\end{Proof}

\rev{We can now define the stable and unstable projections and dichotomies in the core region that we will rely on in the remainder of this paper. We set}
\begin{eqnarray}\label{e:cdich}
& P_-^\mathrm{s}(s) := P_-^\mathrm{ss}(s) + P_-^\mathrm{gker}(s), \quad
P_-^\mathrm{u}(s) := P_-^\mathrm{uu}(s) + P_-^\mathrm{ker}(s)
& \\ \nonumber &
\Phi_-^\mathrm{s}(s,\sigma):= \Phi_-^\mathrm{ss}(s,\sigma)+
\underbrace{\Phi_-^\mathrm{c}(s,\sigma)P_-^\mathrm{gker}(\sigma)}_{=:\Phi_-^\mathrm{gker}(s,\sigma)},\quad
\Phi_-^\mathrm{u}(s,\sigma):= \Phi_-^\mathrm{uu}(s,\sigma)+
\underbrace{\Phi_-^\mathrm{c}(s,\sigma)P_-^\mathrm{ker}(\sigma)}_{=:\Phi_-^\mathrm{ker}(s,\sigma)} &
\end{eqnarray}
\rev{and refer to these operators from now on as the exponential dichotomies in the core region.}

\subsection{Exponential dichotomies in the far field}\label{s:edff}

Recall the eigenvalue problem (\ref{e:lspdyn})
\begin{align*}
u_r = &  v \\
v_r = &  - \frac{1}{r} v - \frac{1}{r^2} \partial_{\psi\psi} u
- D^{-1}[\omega_*\partial_{\psi} u + f^\prime(u_*(r,\psi)) u - \lambda u],
\end{align*}
which we write as the abstract system
\[
\mathbf{u}_r = \mathcal{A}(r;\lambda) \mathbf{u}
\]
on the Banach space $X=H^1(S^1,\C^N)\times L^2(S^1,\C^N)$.
Alternatively, we can consider this equation in Archimedean
coordinates $\vartheta=k_*r+\theta(r)+\psi$. In these coordinates,
(\ref{e:lspdyn}) becomes
\begin{align}\label{e:lspdyncomov}
u_r = &  -(k_*+\theta^\prime(r))\partial_\vartheta u + v \\ \nonumber
v_r = &  -(k_*+\theta^\prime(r))\partial_\vartheta v
- \frac{1}{r} v - \frac{1}{r^2} \partial_{\vartheta\vartheta} u
- D^{-1}[\omega_*\partial_\vartheta u
+ f^\prime(u_*(r,\vartheta-k_*r-\theta(r))) u - \lambda u],
\end{align}
which we write as
\begin{equation}\label{e:arch}
\mathbf{u}_r = \mathcal{A}_\mathrm{arch}(r;\lambda) \mathbf{u},
\end{equation}
again on the Banach space $X=H^1(S^1,\C^N)\times L^2(S^1,\C^N)$. We
equip $X$ with the $r$-dependent norm \cite{s-siam}
\begin{equation}\label{e:spdnorm}
|\mathbf{u}(r)|_{X_r}^2 :=
\frac{1}{r^2} |u|_{H^1}^2 + |u|_{H^{\frac12}}^2 + |v|_{L^2}^2
\end{equation}
and write $X_r$ whenever the $r$-dependence of the norm needs to be
emphasized. Similarly, we equip the common domain
$X^1=H^2(S^1,\C^N)\times H^1(S^1,\C^N)$ of $\mathcal{A}(r,\lambda)$
and $\mathcal{A}_\mathrm{arch}(r,\lambda)$ with the $r$-dependent norm
\begin{equation}\label{e:spdnorm1}
|\mathbf{u}(r)|_{X_r^1}^2 :=
\frac{1}{r^4} |u|_{H^2}^2 + |u|_{H^{\frac32}}^2 + |v|_{H^1}^2.
\end{equation}
We are interested in comparing solutions to (\ref{e:lspdyncomov}) with
solutions to the asymptotic equation
\begin{align}\label{e:lspdyninf}
u_r = &  v \\ \nonumber
v_r = & 
- D^{-1}[\omega_*\partial_\psi u + f^\prime(u_\infty(k_*r+\psi)) u - \lambda u]
\end{align}
for the wave trains. We therefore introduce the variable\footnote{\rev{Notation: We use the same letter $\vartheta$ for the spiral-wave coordinate $\vartheta=k_*r+\theta(r)+\psi$ and the wave-train coordinate $\vartheta=k_*r+\psi$ as this makes it easier to compare the formulations (\ref{e:lspdyncomov}) and (\ref{e:lspdyncomovinf}).}} $\vartheta=k_*r+\psi$, which transforms (\ref{e:lspdyninf}) into
\begin{align}\label{e:lspdyncomovinf}
u_r = &  -k_*\partial_\vartheta u + v \\ \nonumber
v_r = &  -k_*\partial_\vartheta v - D^{-1}[
\omega_*\partial_\vartheta u + f^\prime(u_\infty(\vartheta)) u - \lambda u]
\end{align}
posed on $Y=H^{\frac12}(S^1,\C^N)\times L^2(S^1,\C^N)$, where the right-hand side coincides with the operator $\mathcal{A}_\infty(\lambda)$ that we discussed in \S\ref{s:wts.morse}. Note that (\ref{e:lspdyncomovinf}) coincides with (\ref{e:spmtw}) which describes the eigenvalue problem of the wave train in the laboratory frame. Before we proceed, we recall that our assumptions on $u_*$ imply that
\begin{equation}\label{e:ca}
f^\prime(u_*(r,\psi)) \longrightarrow f^\prime(u_\infty(k_*r+\theta(r)+\psi))
\quad \mbox{as} \quad r \longrightarrow \infty.
\end{equation}
In the Archimedean coordinates $\vartheta=k_*r+\theta(r)+\psi$, (\ref{e:ca}) becomes
\[
f^\prime(u_*(r,\vartheta-k_*r-\theta(r))) \longrightarrow f^\prime(u_\infty(\vartheta))
\quad \mbox{as} \quad r \longrightarrow \infty.
\]
% since $\theta^\prime(r)\to0$ as $r\to\infty$ implies that $\theta(r)/r\to0$ as $r\to\infty$.
Neither (\ref{e:lspdyncomov}) nor (\ref{e:lspdyncomovinf}) admits a semiflow in the variable $r$. Instead, we are interested in exponential dichotomies.

\begin{Definition}[{\cite[\S2.1]{pss}}]\label{d:dich}
We say that (\ref{e:lspdyncomov}) has an exponential dichotomy on a
subinterval $J\subset\R^+$ if there exist constants $C$ and $\eta$, 
 strongly continuous family of projections $P^\mathrm{s/u}:J\to\mathrm{L}(X)$. $P^\mathrm{s}(r)+P^\mathrm{u}(r)=\id$,  and families of linear operators  $\Phi^\mathrm{s}(r;\rho)$ and
$\Phi^\mathrm{u}(r;\rho)$ such
that the following is true:
\begin{compactitem}
\item\emph{Stability}.
For any $\rho\in J$ and $\mathbf{u}_0\in X$, there exists a solution
$\Phi^\mathrm{s}(r;\rho)\mathbf{u}_0$ of (\ref{e:sdabs}) that is
defined for $r\geq\rho$ in $J$, is continuous in $(r,\rho)$ for
$r\geq\rho$ and differentiable in $(r,\rho)$ for $r>\rho$,
and we have $\Phi^\mathrm{s}(\rho;\rho)\mathbf{u}_0=P^\mathrm{s}(\rho)\mathbf{u}_0$
as well as
\[
|\Phi^\mathrm{s}(r;\rho)\mathbf{u}_0|_{X_r} \leq
C \rme^{-\eta|r-\rho|}\, |\mathbf{u}_0|_{X_\rho},
\]
for all $r\geq\rho$ such that $r,\rho\in J$.
\item\emph{Instability}.
For any $\rho\in J$ and $\mathbf{u}_0\in X$, there exists a solution
$\Phi^\mathrm{u}(r;\rho)\mathbf{u}_0$ of (\ref{e:sdabs}) that is
defined for $r\leq\rho$ in $J$, is continuous in $(r,\rho)$ for
$r\leq\rho$ and differentiable in $(r,\rho)$ for $r<\rho$,
and we have
$\Phi^\mathrm{u}(\rho;\rho)\mathbf{u}_0=P^\mathrm{u}(\rho)\mathbf{u}_0$
as well as
\[
|\Phi^\mathrm{u}(r;\rho)\mathbf{u}_0|_{X_r} \leq
C \rme^{-\eta|r-\rho|}\, |\mathbf{u}_0|_{X_\rho},
\]
for all $r\leq\rho$ such that $r,\rho\in J$.
\item\emph{Invariance}.
The solutions $\Phi^\mathrm{s}(r;\rho)\mathbf{u}_0$ and
$\Phi^\mathrm{u}(r;\rho)\mathbf{u}_0$ satisfy 
\begin{align*}
\Phi^\mathrm{s}(r;\rho)\mathbf{u}_0 \in \Rg(P^\mathrm{s}(r))
\quad\mbox{ for all }\quad r\geq\rho \mbox{ with } r,\rho\in J,\nonumber\\
\Phi^\mathrm{u}(r;\rho)\mathbf{u}_0 \in \Rg(P^\mathrm{u}(r))
\quad\mbox{ for all }\quad r\leq\rho \mbox{ with } r,\rho\in J.
\end{align*}
\item\emph{Regularity}.
The solution operators give strong solutions, that is, $\Phi^\mathrm{s}(r;\rho)\mathbf{u}_0$ and $\Phi^\mathrm{u}(\rho;r)\mathbf{u}_0$ are differentiable in $r$ and $\rho$ for all $r\geq\rho$ and all initial conditions $\mathbf{u}_0$ in a dense subset of $X$.
\end{compactitem}
\end{Definition}

\begin{Remark}[Strong solutions]\label{r:strong}
We note that our characterization of differentiability in Definition~\ref{d:dich} is slightly different from the one adopted in \cite{pss}. Differentiability on a dense subset is sufficient to guarantee uniqueness of the continuous evolution operators as a continuous extension of strong solutions and provides a convenient way to guarantee uniqueness in the context of strongly continuous semigroups, where the dense subset is usually chosen as the domain of the generator. Alternatively, one can require differentiability in $r$ for $r>\rho$ for all initial conditions, which was the approach taken in \cite{pss}: we shall recover this property in our situation as well. 
\end{Remark}

Proposition~\ref{p:mor} shows that the asymptotic equation
(\ref{e:lspdyncomovinf}) has an exponential dichotomy on $Y$ if and only
if $\lambda$ is not in the Floquet spectrum of the wave trains. Our goal is to prove that (\ref{e:lspdyncomov}) has an exponential dichotomy on $X_r=H^1\times L^2$ for large $r$ if the asymptotic equation (\ref{e:lspdyncomovinf}) has an exponential dichotomy on $Y=H^{\frac12}\times L^2$. In addition, it will be useful to understand the relation between the projections associated with these two dichotomies. The following lemma, which follows immediately from Lemma~\ref{l:iso} below, allows us to compare these projections.

\begin{Lemma}\label{l:xy}
There are constants $C>0$ and $R_*>0$ so that the operators
\[
\mathcal{I}_r: \quad
X_r \longrightarrow Y, \quad
(u,v) \longmapsto (A_\infty^{-\frac12}A(r)^{\frac12}u,v)
\]
with
\[
A_\infty := -D^{-1}\omega_*\partial_\vartheta+1, \quad
A(r) := -\frac{1}{r^2}\partial_{\vartheta\vartheta} -D^{-1}\omega_*\partial_\vartheta+1
\]
are isomorphisms with $\|\mathcal{I}_r\|_{\mathrm{L}(X_r,Y)}+\|\mathcal{I}_r^{-1}\|_{\mathrm{L}(Y,X_r)}\leq C$ for all $r\geq R_*$.
\end{Lemma}

We can now define the bounded projections
\begin{equation}\label{e:wtinf}
P^\mathrm{s/u}_\infty(r,\lambda) := \mathcal{I}_r^{-1} P^\mathrm{s/u}_\mathrm{wt}(\lambda) \mathcal{I}_r \in \mathrm{L}(X_r)
\end{equation}
whenever the asymptotic equation (\ref{e:lspdyncomovinf}) has an exponential dichotomy on $Y$ with projections $P^\mathrm{s/u}_\mathrm{wt}(\lambda)$. \rev{Note that, even though the projections $P^\mathrm{s}_\mathrm{wt}(\lambda)$ on $Y$ do not depend on $r$, the resulting projections $P^\mathrm{s}_\infty(r,\lambda)$ on $X_r$ will depend on $r$ since the isomorphisms $\mathcal{I}_r:X_r\to Y$ depend on $r$.} We can now state our main result.

\begin{Proposition}\label{p:edff}
Assume that the asymptotic equation (\ref{e:lspdyncomovinf}) has an
exponential dichotomy on $Y$. There are positive constants $C$, $\eta$, and $R_*$ such that \eqref{e:lspdyncomov} has an exponential dichotomy on $[R_*,\infty)$. Furthermore, we have $\|P^\mathrm{s}(r)-P_\infty^\mathrm{s}(r)\|_{\mathrm{L}(X_r)}\leq C r^{-\frac13}$, where $P^\mathrm{s}(r)$ and $P_\infty^\mathrm{s}(r)$ denote the projections associated with \rev{(\ref{e:lspdyncomov}) and (\ref{e:lspdyncomovinf})}, respectively, \rev{on $X_r$}.
\end{Proposition}

The remainder of this section is devoted to the proof of Proposition~\ref{p:edff}. The main idea is to use robustness of exponential dichotomies as established, for instance, in \cite{pss}. There, we constructed exponential dichotomies to perturbed equations using a mild integral formulation and a fixed point argument in exponentially weighted spaces. In our current case, perturbation terms are small and involve terms of order $\rmO(\frac{1}{r})$. In the proof, we will encounter two main difficulties. 

First, comparing (\ref{e:lspdyncomov}) and (\ref{e:lspdyncomovinf}), we see that the perturbation term $-\frac{1}{r^2}\partial_{\psi\psi}$ is not bounded (not even relative to the remaining principal part $D^{-1}\omega_*\partial_\psi$). Thus, the main technical point of the proof of Proposition~\ref{p:edff} is to show that this unbounded term does not matter as far as exponential dichotomies are concerned. 

An additional difficulty stems from the fact that the corotating frame, passing from \eqref{e:lspdyninf} to \eqref{e:lspdyncomovinf}, changes regularity properties of the equation. In particular, the equation in the corotating coordinates is not bisectorial, as one can readily verify by calculating the spectrum of the leafing order operator, setting terms involving $f^\prime$ to zero. On the other hand, the equation in the corotating frame still has smoothing properties, as one can see either directly using spectral computations or by noticing that solution of  \eqref{e:lspdyninf} transform back to solutions of \eqref{e:lspdyncomovinf} using the simple shear transformation.

The following lemma will be crucial for addressing the unbounded perturbation $-\frac{1}{r^2}\partial_{\psi\psi}$. 

\begin{Lemma}\label{l:iso}
There are constants $c_1,c_2,C>0$ such that the following is true for each $r\geq1$:
\begin{compactenum}[(i)]
\item The operator $A_\infty^{\frac12}:H^{\frac12}(S^1)\to L^2(S^1)$ is well defined and an isomorphism.
\item The operator $A(r)^{\frac12}:H^1(S^1)\to L^2(S^1)$ is well defined, bounded uniformly in $r$, and continuously differentiable in $r$ with
\[
\left\|\left[ \partial_r (A(r)^{\frac12}) \right] A(r)^{-\frac12} \right\|_{\mathrm{L}(L^2)} \leq \frac{C}{r}.
\]
\item The operator defined on $X_r\to L^2(S^1)\times L^2(S^1)$ through $(u,v)\mapsto (A(r)^{\frac12}u,v)$ is an isomorphism with
\[
c_1 \left( |A(r)^{\frac12}u|_{L^2}^2 + |v|_{L^2}^2 \right)
\leq |\mathbf{u}|_{X_r}^2 \leq
c_2 \left( |A(r)^{\frac12}u|_{L^2}^2 + |v|_{L^2}^2 \right)
\]
for all $\mathbf{u}=(u,v)\in X_r$.
\end{compactenum}
\end{Lemma}

\begin{Proof}
Using Fourier series, it is straightforward to prove (i) and (iii) and to see that $A(r)^{\frac12}:H^1\to L^2$ is well defined and bounded. Continuous differentiability of $A(r)^{\frac12}$ is also clear since the family of operators is analytic in $r$. We compute the derivative
\[
\partial_r (A(r)^{\frac12}) =
\partial_r \left[ -\frac{1}{r^2}\partial_{\vartheta\vartheta} 
-D^{-1}\omega_*\partial_\vartheta+1 \right]^{\frac12} =
\frac{1}{r^3} \partial_{\vartheta\vartheta} A(r)^{-\frac12} =
\frac{1}{r} \left[ \frac{1}{r^2} \partial_{\vartheta\vartheta} \right]
A(r)^{-\frac12}
\]
and conclude that
\[
\left\|\left[ \partial_r A(r)^{\frac12} \right]
\rev{A(r)^{-\frac12}} \right\|_{\mathrm{L}(L^2)}
\leq \frac{1}{r}
\left\| \frac{1}{r^2} \partial_{\vartheta\vartheta} A(r)^{-1}
\right\|_{\mathrm{L}(L^2)} \leq \frac{C}{r}
\]
for some constant $C>0$ that is independent of $r$.
\end{Proof}

Next, we introduce the notation $T(r) := -(k_*+\theta^\prime(r))\partial_\vartheta$ and
\[
B := -D^{-1}[f^\prime(u_\infty(\vartheta))-\lambda]-1, \qquad
S(r) := -D^{-1}[f^\prime(u_*(r,\vartheta-k_*r-\theta(r))) - f^\prime(u_\infty(\vartheta))]
\]
so that (\ref{e:lspdyncomov}) becomes
\begin{align}\label{e:lspdyncomovabs}
u_r = &  T(r) u + v \\ \nonumber
v_r = &  T(r) v + A(r) u + B u - \frac{v}{r} + S(r) u.
\end{align}
Note that the last two terms in the second equation in
(\ref{e:lspdyncomovabs}) converge to zero in norm as $r\to\infty$.
The operator $T(r)$ generates the shift in $\vartheta$. The principal
term in (\ref{e:lspdyncomovabs}) is the system
\[
\mathbf{u}_r = \left(\begin{array}{cc} 0 & 1 \\ A(r) & 0 \end{array}\right) \mathbf{u}.
\]
It is convenient to symmetrize the principal term by using the transformation
\[
\hat{u}(r) = A(r)^{\frac12} u(r).
\]
Lemma~\ref{l:iso}(iii) shows that the $|\cdot|_{X_r}$ norm for $(u,v)$
and the ordinary $L^2$-norm for $(\hat{u},v)$ are equivalent so that
we can simply take $(\hat{u},v)\in L^2(S^1,\C^N)\times L^2(S^1,\C^N)$.
Equation (\ref{e:lspdyncomovabs}) becomes
\begin{align*}
\hat{u}_r = &  T(r) \hat{u} + A(r)^{\frac12} v
+ [\partial_r A(r)^{\frac12}] A(r)^{-\frac12} \hat{u} \\
v_r = &  T(r) v + A(r)^{\frac12} \hat{u} + B A(r)^{-\frac12} \hat{u}
- \frac{v}{r} + S(r) A(r)^{-\frac12} \hat{u},
\end{align*}
where we used that $A(r)^{\frac12}T(r)\rev{A(r)^{-\frac12}}=T(r)$. If we now use that
$S(r)$ converges to zero as $r\to\infty$, we get
\begin{align}\label{e:lspdyncotrafo0}
\hat{u}_r = &  T(r) \hat{u} + A(r)^{\frac12} v + \rmo_r(1) u \\ \nonumber
v_r = &  T(r) v + A(r)^{\frac12} \hat{u} + B A(r)^{-\frac12} \hat{u}
+ \rmo_r(1) \hat{u} + \rmo_r(1) v.
\end{align}
Proposition~\ref{p:edrobust} states that we can safely neglect the
$\rmo_r(1)$ terms in (\ref{e:lspdyncotrafo0}) since we are only
interested in proving the existence of exponential dichotomies. We
then exploit the transformation $w^\pm=\hat{u}\pm v$ that
diagonalizes the principal part and leads to
\begin{align*}
w^+_r = &  A(r)^{\frac12} w^+ + T(r) w^+ + \frac12 B A(r)^{-\frac12} \rev{(w^+ + w^-)} \\
w^-_r = &  - A(r)^{\frac12} w^- + T(r) w^- \rev{\,-\, \frac12 B A(r)^{-\frac12}(w^+ + w^-)}
\end{align*}
or
\begin{equation}\label{e:ed4}
w_r = [\tilde{\mathcal{A}}(r) + \mathcal{K}(r)] w,
\end{equation}
where $w=(w^+,w^-)\in L^2(S^1,\C^{2N})$ for each $r$ and
\[
\tilde{\mathcal{A}}(r) := \left( \begin{array}{cc} A(r)^{\frac12}+T(r) & 0 \\
0 & -A(r)^{\frac12}+T(r) \end{array} \right), \qquad
\mathcal{K}(r) := \frac12 \left( \begin{array}{rr}
B A(r)^{-\frac12} & \rev{B A(r)^{-\frac12}} \\ \rev{-B A(r)^{-\frac12}} & -B A(r)^{-\frac12}
\end{array} \right).
\]
\rev{Next, we carry out the analogous analysis for the asymptotic equation (\ref{e:lspdyncomovinf}). Using Lemma~\ref{l:iso}(i), we see that the isomorphism
\begin{equation}\label{e:it}
Y \longrightarrow L^2\times L^2, \quad
(u,v) \longmapsto w=(A_\infty^{\frac12}u+v,A_\infty^{\frac12}u-v)
\end{equation}
transforms (\ref{e:lspdyncomovinf}) posed on $Y$ into}
\begin{equation}\label{e:ed1}
\mathbf{w}_r = [\tilde{\mathcal{A}}_\infty + \mathcal{K}_\infty] \mathbf{w}
\end{equation}
\rev{posed on $L^2\times L^2$,} where
\begin{equation}\label{e:at}
\tilde{\mathcal{A}}_\infty :=
\left( \begin{array}{cc} A_\infty^{\frac12}-k_*\partial_\vartheta & 0 \\
0 & -A_\infty^{\frac12}-k_*\partial_\vartheta \end{array} \right), \qquad
\mathcal{K}_\infty := \frac12
\left( \begin{array}{rr} B A_\infty^{-\frac12} & \rev{B A_\infty^{-\frac12}} \\
\rev{-B A_\infty^{-\frac12}} & -B A_\infty^{-\frac12} \end{array} \right).
\end{equation}
\rev{Since we assumed in Proposition~\ref{p:edff} that $\lambda$ is chosen so that the asymptotic equation (\ref{e:lspdyncomovinf}) has an exponential dichotomy with $r$-independent projections, we can use the $r$-independent isomorphism (\ref{e:it}) to construct an exponential dichotomy of (\ref{e:ed1}) on $L^2\times L^2$ with an $r$-independent stable projection denoted by $\mathcal{P}^\mathrm{s}_\infty$.} We have the following result.

\begin{Lemma}\label{l:ed1}
There are positive constants $C_0$, $\eta_0$, and $R_*$ so that the following is true. For each $R\geq R_*$, equation (\ref{e:ed4}) has an exponential dichotomy for $r\geq R$ with projections $\mathcal{P}^\mathrm{s}_R(r)$ that satisfy
\[
\|\mathcal{P}^\mathrm{s}_R(r)-\mathcal{P}^\mathrm{s}_\infty\|_{\mathrm{L}(L^2\times L^2)}\leq C R^{-\frac13}
\]
for $r\geq R$.
\end{Lemma}

\begin{Proof}
We denote by $Q_m$ the orthogonal projection onto the subspace of $L^2(S^1,\C^{2N})$ spanned by $\rme^{\pm\rmi\ell\vartheta}w_0$ for $0\leq|\ell|\leq m$ and $w_0\in\C^{2N}$. We write $w_m=Q_m w$ and $w_m^\perp=(\id-Q_m)w$ for any $w\in L^2$. Note that both
$\tilde{\mathcal{A}}_\infty$ and $\tilde{\mathcal{A}}(r)$ commute with
$Q_m$. Using this fact, it is not difficult to prove that there is a constant $C>0$ such that
\[
\|A_\infty^{-\frac12}(\id-Q_m)\|_{\mathrm{L}(L^2)}
+ \|A(r)^{-\frac12}(\id-Q_m)\|_{\mathrm{L}(L^2)}
\leq \frac{C}{\sqrt{m}}
\]
uniformly in $m\geq1$ and $r\geq1$, which implies that
\begin{equation}\label{e:regest}
\|\mathcal{K}_\infty (\id-Q_m)\|_{\mathrm{L}(L^2)}
+ \|(\id-Q_m)\mathcal{K}_\infty\|_{\mathrm{L}(L^2)}
+ \|\mathcal{K}(r)(\id-Q_m)\|_{\mathrm{L}(L^2)}
+ \|(\id-Q_m)\mathcal{K}(r)\|_{\mathrm{L}(L^2)}
\leq \frac{C}{\sqrt{m}}
\end{equation}
uniformly in $m\geq1$ and $r\geq1$. Writing (\ref{e:ed1}) in the components given by $(w_m,w_m^\perp)$, we obtain the equation
\begin{equation}\label{e:ainfd}
\left(\begin{array}{c} \mathbf{w}_m \\ \mathbf{w}_m^\perp \end{array}\right)_r
= \left(\begin{array}{ccc}
\tilde{\mathcal{A}}_\infty+Q_m\mathcal{K}_\infty Q_m&\ & Q_m\mathcal{K}_\infty (\id-Q_m)\\
(\id-Q_m)\mathcal{K}_\infty Q_m &\ & \tilde{\mathcal{A}}_\infty +(\id-Q_m)\mathcal{K}_\infty (\id-Q_m)\end{array}\right)
\left(\begin{array}{c} \mathbf{w}_m \\ \mathbf{w}_m^\perp \end{array}\right),
\end{equation}
which, by assumption, has an exponential dichotomy on $L^2\times L^2$. \rev{Note that differentiability of solutions as stated in Definition~\ref{d:dich} follows since derivatives in $(r,\varphi)$ exist for sufficiently regular initial conditions $\mathbf{u}_0$ and therefore so do the derivatives in the corotating coordinates $(r,\psi)$.} In a similar fashion, we see that the evolution operators gain regularity, that is, they map $X$ to $X^\alpha$ for each given $\alpha>0$. For small $\alpha$, this follows as in \cite{pss}, and we can bootstrap these arguments to larger values of $\alpha$ using uniqueness of the Cauchy problem. 

Using \eqref{e:regest} and robustness of dichotomies \cite{pss}, we see that there are constants $C>0$ and $m_*\geq1$ so that the system
\begin{equation}\label{e:ainfd1}
\left(\begin{array}{c} \mathbf{w}_m \\ \mathbf{w}_m^\perp \end{array}\right)_r
= \left(\begin{array}{ccc}
\tilde{\mathcal{A}}_\infty+Q_m\mathcal{K}_\infty Q_m&\ & 0\\
0 &\ & \tilde{\mathcal{A}}_\infty \end{array}\right)
\left(\begin{array}{c} \mathbf{w}_m \\ \mathbf{w}_m^\perp \end{array}\right)
\end{equation}
also has an exponential dichotomy for each $m\geq m_*$ and that the difference between the dichotomies of \eqref{e:ainfd} and \eqref{e:ainfd1} is bounded by $C/\sqrt{m}$ uniformly in $m\geq m_*$. The existence of these perturbed dichotomy operators and decay estimates follow from a contraction argument for the mild variation-of-constant formula in \cite{pss}. Differentiability is obtained by exploiting smoothing properties of the unperturbed solution operators and the fact that the perturbation $\rev{\mathcal{K}}$ preserves higher regularity so that $\rev{\mathcal{K}(r)}:X^\alpha\to X^\alpha$ for $\alpha>0$ since the coefficients that appear in $\mathcal{K}_\infty$ are smooth. Thus, derivatives in $(r,\varphi)$ exist for $r>\rho$ or for smooth initial data $\mathbf{u}_0$ for $r\geq\rho$. 

Our next goal is to show that
\begin{equation}\label{e:ard1}
\left(\begin{array}{c} \mathbf{w}_m \\ \mathbf{w}_m^\perp \end{array}\right)_r
= \left(\begin{array}{ccc}
\tilde{\mathcal{A}}(r)+Q_m\mathcal{K}(r) Q_m&\ & 0\\
0 &\ & \tilde{\mathcal{A}}(r) \end{array}\right)
\left(\begin{array}{c} \mathbf{w}_m \\ \mathbf{w}_m^\perp \end{array}\right)
\end{equation}
has an exponential dichotomy and that its projections are close to those of (\ref{e:ainfd1}). Since (\ref{e:ainfd1}) and (\ref{e:ard1}) are both diagonal, it suffices to analyze the range and kernel of $Q_m$ separately.
On the kernel of $Q_m$, we can use that both $\tilde{\mathcal{A}}_\infty$ and $\tilde{\mathcal{A}}(r)$ commute with $Q_m$ and that $T(r)$ generates the unitary shift flow in $\vartheta$ that commutes with the evolution of $w_r=A(r)^{\frac12}w$ to prove that each of the systems
\[
(\mathbf{w}_m^\perp)_r = \tilde{\mathcal{A}}_\infty \mathbf{w}_m^\perp, \qquad
(\mathbf{w}_m^\perp)_r = \tilde{\mathcal{A}}(r) \mathbf{w}_m^\perp
\]
have an exponential dichotomy on $L^2\times L^2$ with constant $C$ and rate larger than $\sqrt{m}$ uniformly in $m\geq1$ and that their stable projections coincide.
On the range of $Q_m$, we see that there is a constant $C>0$ so that
\[
\|Q_m(\mathcal{K}(r)-\mathcal{K}_\infty)Q_m\|_{\mathrm{L}(L^2)} \leq \frac{C}{r}
\]
uniformly in $m\geq m_*$. Similarly, an explicit Fourier-series computation shows that
\[
\|Q_m (\tilde{\mathcal{A}}_\infty-\tilde{\mathcal{A}}(r))Q_m\|_{\mathrm{L}(L^2)} \leq C\left(\frac{m}{r}+\frac{m^{3/2}}{r^2}\right)
\]
uniformly in $m\geq m_*$ and $r\geq1$.
Thus, there are constants $C$ and $R_*\geq1$ so that for each $R\geq R_*$ equation (\ref{e:ard1}) with $m:=R^{\frac23}$ has an exponential dichotomy with projections whose difference to those of (\ref{e:ainfd1}) can be bounded by $CR^{-\frac13}$ uniformly in $r\geq R$. One also explicitly verifies that solutions are differentiable in $r$ and the initial radial time $\rho$ for sufficiently smooth $\mathbf{u}_0$ and that the evolution operators map the space $L^2\times L^2$ into $H^M\times H^M$ for each $M$ and each $r\neq\rho$. 

Finally, using again (\ref{e:regest}), we can conclude as above that the system
\[
\left(\begin{array}{c} \mathbf{w}_m \\ \mathbf{w}_m^\perp \end{array}\right)_r
= \left(\begin{array}{ccc}
\tilde{\mathcal{A}}(r)+Q_m\mathcal{K}(r) Q_m&\ & Q_m\mathcal{K}(r) (\id-Q_m)\\
(\id-Q_m)\mathcal{K}(r) Q_m &\ & \tilde{\mathcal{A}}(r) +(\id-Q_m)\mathcal{K}(r) (\id-Q_m)\end{array}\right)
\left(\begin{array}{c} \mathbf{w}_m \\ \mathbf{w}_m^\perp \end{array}\right)
\]
with $m=R^{\frac23}$ also has an exponential dichotomy for $r\geq R$ with constants and rates that are independent of $R$ and that the difference of its projections to those of (\ref{e:ainfd}) is bounded by $CR^{-\frac13}$ uniformly in $r\geq R$, since $1/\sqrt{m}=R^{-\frac13}$ when $m=R^{\frac23}$. The existence of evolution operators for the perturbed equation is obtained from the mild variation-of-constant formula in \cite{pss}, and differentiability on a dense subset can be obtained using again the fact that the perturbation preserves smoothness.

This completes the proof of the lemma.
\end{Proof}

The preceding lemma shows that there are constants $C,\eta,R_*$ so that for each $R\geq R_*$ equation (\ref{e:ed4}) has an exponential dichotomy with constant $C$ and rate $\eta$ for $r\geq R$ with projections $\mathcal{P}^\mathrm{s}_R(r)$ that satisfy
\begin{equation}\label{e:closeed}
\|\mathcal{P}^\mathrm{s}_R(r)-\mathcal{P}^\mathrm{s}_\infty\|_{\mathrm{L}(L^2\times L^2)}\leq C R^{-\frac13}, \qquad r\geq R.
\end{equation}
Note that the ranges of $\mathcal{P}^\mathrm{s}_{R_1}(r)$ and $\mathcal{P}^\mathrm{s}_{R_2}(r)$ agree for all $r\geq\max\{R_1,R_2\}$. We can therefore use (\ref{e:closeed}) to conclude that for each $r\geq R_*$ we can write the kernel of $\mathcal{P}^\mathrm{s}_{R_*}(r)$ as the graph of an operator from the kernel of $\mathcal{P}^\mathrm{s}_{r}(r)$ into the range of $\mathcal{P}^\mathrm{s}_{r}(r)$ and that the norm of this operator is bounded by $C$ uniformly in $r$. Using \cite[(3.20)]{pss}, we see that
\[
\|\mathcal{P}^\mathrm{s}_{R_*}(2r) - \mathcal{P}^\mathrm{s}_{r}(2r)\|_{\mathrm{L}(L^2\times L^2)} \leq C\rme^{-\eta r}
\]
for $r\geq R_*$. Hence, we obtain
\[
\|\mathcal{P}^\mathrm{s}_{R_*}(2r) - \mathcal{P}^\mathrm{s}_\infty\|
\leq \|\mathcal{P}^\mathrm{s}_{R_*}(2r) - \mathcal{P}^\mathrm{s}_{r}(2r)\|
+ \|\mathcal{P}^\mathrm{s}_{r}(2r) - \mathcal{P}^\mathrm{s}_\infty\|
\leq C\rme^{-\eta r} + \frac{C}{r^{\frac13}}
\leq \frac{\tilde{C}(R_*)}{(2r)^{\frac13}}
\]
for $r\geq R_*$ where $\|\cdot\|:=\|\cdot\|_{\mathrm{L}(L^2\times L^2)}$. This completes the proof of Proposition~\ref{p:edff}.\qed

\begin{Remark}\label{r:halpha}
We chose to construct the exponential dichotomies in the space
$X=H^1(S^1,\C^N)\times L^2(S^1,\C^N)$. In fact, the same construction
works in the spaces
$X=H^{1+\alpha}(S^1,\C^N)\times H^\alpha(S^1,\C^N)$ for any
$\alpha\geq0$.
\end{Remark}

For later reference, we state here more formally a result on robustness of exponential dichotomies that we used repeatedly in the proof. 

\begin{Proposition}[\cite{pss}]\label{p:edrobust}
Exponential dichotomies are robust. More precisely, if an abstract equation of the form (\ref{e:sdabs}) has an exponential dichotomy on an interval $J\subset\R^+$ with constants $C$ and $\eta$, then, for any choice of $\varepsilon>0$ and $\tilde{\eta}$ with $0<\tilde{\eta}<\eta$, there are constants $\tilde{C}$ and $\tilde{\delta}>0$ such that the perturbed system
\begin{equation}\label{e:sdabspert}
\mathbf{u}_r = \mathcal{A}(r;\lambda) \mathbf{u} + \mathcal{B}(r) \mathbf{u}
\end{equation}
with
\[
\|\mathcal{B}(r)\|_{\mathrm{L}(X_r)} \leq \tilde{\delta}
\]
has an exponential dichotomy on $J$ with constants $\tilde{C}$ and $\tilde{\eta}$, and the projections of (\ref{e:sdabspert}) are $\varepsilon$-close to the projections of (\ref{e:sdabs}). If, in fact,
\[
\|\mathcal{B}(r)\|_{\mathrm{L}(X_r)} = \rmo_r(1),
\]
then we can choose $\tilde{\eta}=\eta$.
\end{Proposition}

% \begin{Proof}
%  The result is a consequence of the variation-of-constant mild formulation for exponential dichotomies in \cite{pss}, which gives solution operators for the perturbed equation.  Regularity is obtained by exploiting smoothing properties of the unperturbed solution operators and the fact that the perturbation $\mathcal{B}$ preserves higher regularity, $\mathcal{B}(r):X^\alpha\to X^\alpha$. One thereby obtains derivatives in $r$ and $\varphi$ for $r>\rho$ or for smooth initial data $\mathbf{u}_0$. The shear transformation to $r$ and $\psi$ inducecd by the coroating frame preserves this smoothing property. One can now establish differentiability properties of perturbations of the wave train dichotomies  by inspecting the fixed-point formulas. One starts by establishing continuity of solutions in $X^\alpha$ provided the initial condition belongs to $X^\alpha$. For $\alpha$ large enough, one can then differentiate the solution in $X$. Differentiability for $r>\rho$ can be obtained by first noticing that solutions to the mild formulation possess the cocycle property by uniqueness \cite{pss}, and using that the evolution operators map $X^\alpha$ into $X^{\alpha+\beta}$ for some $\beta>0$, which allows one to conclude differentiability for  $r>\rho$ due to regularity of the initial condition at this new initial radial time $r$. 
% \end{Proof}
% % 

\subsection{Comparing core and far-field dichotomies}\label{s:ced}

We now discuss briefly the relation between the core and far-field coordinates that we used to construct exponential dichotomies. We started with the system (\ref{e:sdabs})
\[
\mathbf{u}_r = \mathcal{A}(r;\lambda) \mathbf{u}, \qquad \mathbf{u}=(u,v).
\]
In the core region, we used the new coordinates $(u,w):=(u,rv)$ with \rev{$(u,w)\in X=H^1(S^1,\C^N)\times L^2(S^1,\C^N)$} equipped with the usual $H^1\times L^2$ norm and proved \rev{in \S\ref{s:edc}} the existence of exponential dichotomies with projections $P_-^\mathrm{u}(s)$ with $s=\log r$ for the core equation (\ref{e:sdabs0}) on $X$. \rev{For $r\leq1$, we have} $|(u,w)|_X=r|(u,v)|_{X_r}$ since
\[
|(u,w)|_X^2 = |u|_{H^1}^2 + |w|_{L^2}^2 = |u|_{H^1}^2 + |rv|_{L^2}^2 = 
r^2 \left( \frac{1}{r^2} |u|_{H^1}^2 + |v|_{L^2}^2\right) = r^2 |(u,v)|_{X_r}^2.
\]
We define the linear isomorphism
\begin{equation}\label{e:jr}
\mathbf{j}(r):\, X \longrightarrow X_r, \quad
(u,w) \longmapsto (u,v):=\left(u,\frac{w}{r}\right)
\end{equation}
which has norm $1/r$. The projections
\[
\widehat{P}^\mathrm{s/u}_-(r) := \mathbf{j}(r) P^\mathrm{s/u}_-(\log r) \mathbf{j}(r)^{-1} \in \mathrm{L}(X_r)
\]
then provide exponential dichotomies of (\ref{e:sdabs}) on $X_r$ in the core region. Using the isomorphism 
\begin{equation}\label{e:compare}
X_r\longrightarrow L^2\times L^2, \quad
(u,v) \longmapsto (\rev{\hat{u}},v)=(A(r)^{\frac12}u,v)
\end{equation}
considered in Lemma~\ref{l:iso}(iii), we can also define projections $\widehat{\mathcal{P}}^\mathrm{s/u}_-(r)$ of (\ref{e:sdabs}) on $L^2\times L^2$. In the far-field region, we used the variables $(u,v)\in X_r$ for (\ref{e:lspdyncomov}) and the variables $(\rev{\hat{u}},v)\in L^2\times L^2$ for the system (\ref{e:lspdyncotrafo0}) to construct exponential dichotomies with projections $\mathcal{P}_+^\mathrm{s}(r)$ for (\ref{e:lspdyncotrafo0}) on $L^2\times L^2$ and exponential dichotomies with projections $P_+^\mathrm{s}(r)$ for (\ref{e:lspdyncomov}) on $X_r$. Lemma~\ref{l:iso}(iii) shows that the constants and rates of the exponential dichotomies on $X_r$ and $L^2\times L^2$ agree. To compare the far-field projections of the spiral wave with those of the asymptotic wave trains, we can use either of the following equivalent approaches:
\begin{compactenum}[(i)]
\item Relate the spiral-wave projections $\mathcal{P}_+^\mathrm{s}(r)$ with the $r$-independent wave-train projections $\mathcal{P}_\infty^\mathrm{s}$ of (\ref{e:ed1}) in the $(\rev{\hat{u}},v)$ variables on the space $L^2\times L^2$.
\item Relate the spiral-wave projections $P_+^\mathrm{s}(r)$ with the $r$-dependent wave-train projections $P_\infty^\mathrm{s}(r)$ defined in (\ref{e:wtinf}) in the $(u,v)$ variables on $X_r$.
\end{compactenum}

\subsection{Exponential dichotomies for the adjoint equation}\label{s:eda}

In the preceding sections, we proved the existence of exponential dichotomies for the linearizations (\ref{e:lspdyn0}) and (\ref{e:lspdyncomov}) in the core and the far field, respectively. In this section, we relate appropriate adjoint equations on the PDE and spatial dynamics level.

We focus first on the adjoint systems in the far field. Using the notation
\[
A(r) = -\frac{1}{r^2}\partial_{\vartheta\vartheta} -D^{-1}\omega_*\partial_\vartheta+1, \qquad
\rev{\tilde{B}(r)} := -D^{-1} \rev{(f^\prime(u_*(r,\vartheta-k_*r-\theta(r))) - \lambda)} -1,
\]
we can write \rev{the eigenvalue problem for the} PDE linearization of the spiral wave as
\begin{equation}\label{e:a1}
u_{rr} = (A(r)+\rev{\tilde{B}(r)}) u - \frac{1}{r} u_r.
\end{equation}
Written as spatial dynamical system in the Archimedean coordinates $\vartheta=k_*r+\theta(r)+\psi$, we obtain
\begin{equation}\label{e:a0}
\begin{pmatrix} u \\ v \end{pmatrix}_r =
\begin{pmatrix} T(r) & 1 \\ A(r) + \rev{\tilde{B}(r)} & T(r) - \frac{1}{r} \end{pmatrix}
\begin{pmatrix} u \\ v \end{pmatrix}
\end{equation}
posed on $X_r$. Using again $\hat{u}:=A^{\frac12}(r)u$ and defining $C(r):=[\partial_r A^{\frac12}(r)] A^{-\frac12}(r)$, we arrive at the system
\begin{equation}\label{e:a2}
\begin{pmatrix} \hat{u} \\ v \end{pmatrix}_r =
\begin{pmatrix} T(r) + C(r) & A^{\frac12}(r) \\
A^{\frac12}(r) + \rev{\tilde{B}(r)} A^{-\frac12}(r) & T(r) - \frac{1}{r} \end{pmatrix}
\begin{pmatrix} \hat{u} \\ v \end{pmatrix},
\end{equation}
which is posed on $L^2\times L^2$. Taking the $L^2\times L^2$ adjoint of (\ref{e:a2}), we obtain the system
\begin{equation}\label{e:a3}
\begin{pmatrix} \tilde{z} \\ \tilde{w} \end{pmatrix}_r =
- \begin{pmatrix} T(r)^* + C(r)^* & A^{\frac12}(r)^* + A^{-\frac12}(r)^* \rev{\tilde{B}(r)}^* \\
A^{\frac12}(r)^* & T(r)^* - \frac{1}{r} \end{pmatrix}
\begin{pmatrix} \tilde{z} \\ \tilde{w} \end{pmatrix}
\end{equation}
posed also on $L^2\times L^2$. Writing
\begin{equation}\label{e:a8}
\begin{pmatrix} \tilde{z} \\ \tilde{w} \end{pmatrix} = 
r \begin{pmatrix} -\hat{z} \\ w \end{pmatrix},
\end{equation}
equation (\ref{e:a3}) becomes
\begin{equation}\label{e:a4}
\begin{pmatrix} \hat{z} \\ w \end{pmatrix}_r =
\begin{pmatrix} T(r) - C(r)^* - \frac{1}{r} & A^{\frac12}(r)^* + A^{-\frac12}(r)^* \rev{\tilde{B}(r)}^* \\
A^{\frac12}(r)^* & T(r) \end{pmatrix}
\begin{pmatrix} \hat{z} \\ w \end{pmatrix}.
\end{equation}
Next, we let $\hat{z}=A^{-\frac12}(r)^*z$ and obtain
\begin{equation}\label{e:a5}
\begin{pmatrix} z \\ w \end{pmatrix}_r =
\begin{pmatrix} T(r) - \frac{1}{r} & A(r)^* + \rev{\tilde{B}(r)}^* \\ 1 & T(r) \end{pmatrix}
\begin{pmatrix} z \\ w \end{pmatrix}
\end{equation}
posed on $X_r^*$ so that $w$ is a solution to the $L^2$-adjoint
\begin{equation}\label{e:a6}
w_{rr} = (A(r)^* + \rev{\tilde{B}(r)}^*) w - \frac{1}{r} w_r
\end{equation}
of the PDE linearization (\ref{e:a1}) of the spiral wave in the coordinates $(r,\psi)$.

We now summarize the conclusions we can draw from the computations carried out above. First, we can apply the approach developed in \S\ref{s:edff} also to the adjoint system (\ref{e:a3}) to conclude that (\ref{e:a3}) and (\ref{e:a5}) have exponential dichotomies on $L^2\times L^2$ and $X_r^*$, respectively. Alternatively, the arguments in the proof of \cite[Lemma~5.1]{ss-fred} show that if $\Phi^j(r;\rho)$ with $j=\mathrm{s,u}$ denote the exponential dichotomies of (\ref{e:a2}) on $L^2\times L^2$, then the exponential dichotomies $\Phi^\mathrm{s}_\mathrm{adj}(r;\rho)$ of (\ref{e:a3}) on $L^2\times L^2$ are given by $\Phi^\mathrm{s}_\mathrm{adj}(r;\rho)=\Phi^\mathrm{u}(\rho;r)^*$. Second, it follows from the form of the right-hand sides of (\ref{e:a2}) and (\ref{e:a3}) that
\begin{equation}\label{e:a7}
\frac{\rmd}{\rmd r} \left\langle \begin{pmatrix} \hat{u}(r) \\ v(r) \end{pmatrix},
\begin{pmatrix} \tilde{z}(r) \\ \tilde{w}(r) \end{pmatrix} \right\rangle_{L^2\times L^2}
= 0
\end{equation}
for all $r$ for any two solutions of (\ref{e:a2}) and (\ref{e:a3}).

\begin{Lemma}\label{l:abddff}
Suppose that (\ref{e:a2}) posed on $L^2\times L^2$ has an exponential dichotomy on \rev{$[R_*,\infty)$} with stable projection $\mathcal{P}^\mathrm{s}(r)$, \rev{then there is an $R\geq R_*$ such that the following is true}. If $\mathbf{u}(r)$ is a solution of (\ref{e:a2}) on $[r_*,\infty)$ \rev{for some $r_*\geq R$} that is uniformly bounded, then $\mathbf{u}(r)\in\Rg(\mathcal{P}^\mathrm{s}(r))$ for $r\geq r_*$. In particular, not only do solutions with initial data in the range $\Rg(P^\mathrm{s}(r_*))$ exist and are uniformly bounded in $r>r_*$, but any solution with these properties must lie in this range.
\end{Lemma}

\begin{Proof}
Step~1: We apply the results from \S\ref{s:edff} and the preceding arguments to the asymptotic wave-train system and the associated adjoint system and denote the resulting stable/unstable projections by $\mathcal{P}^\mathrm{s/u}_\infty$ and $\mathcal{P}^\mathrm{s/u}_{\mathrm{adj},\infty}$, respectively. We claim that
\begin{equation}\label{e:a10}
\Rg(\mathcal{P}^\mathrm{s}_{\mathrm{adj},\infty}) \oplus \Rg(\mathcal{P}^\mathrm{s}_\infty) = L^2 \times L^2.
\end{equation}
To prove this, take any two initial conditions $\mathbf{u}_0\in\Rg(\mathcal{P}^\mathrm{s}_\infty)$ and $\mathbf{w}_0\in\Rg(\mathcal{P}^\mathrm{s}_{\mathrm{adj},\infty})$ and denote the corresponding solutions of the asymptotic systems belonging to (\ref{e:a2}) and (\ref{e:a3}) by $\mathbf{u}(r)$ and $\mathbf{w}(r)$, respectively. Equation (\ref{e:a7}) then implies that
\[
\langle\mathbf{u}_0,\mathbf{w}_0\rangle = \langle\mathbf{u}(r),\mathbf{w}(r)\rangle = 0
\]
for $r\geq0$, since $\mathbf{u}(r)$ and $\mathbf{w}(r)$ both decay to zero as $r\to\infty$. We conclude that
$\Rg(\mathcal{P}^\mathrm{s}_{\mathrm{adj},\infty}) \perp \Rg(\mathcal{P}^\mathrm{s}_\infty)$. We can apply the same argument to the unstable projections and arrive at
\[
\Rg(\mathcal{P}^\mathrm{s}_{\mathrm{adj},\infty}) \perp \Rg(\mathcal{P}^\mathrm{s}_\infty), \qquad
\Rg(\mathcal{P}^\mathrm{u}_{\mathrm{adj},\infty}) \perp \Rg(\mathcal{P}^\mathrm{u}_\infty).
\]
Since we have
\[
\Rg(\mathcal{P}^\mathrm{s}_{\mathrm{adj},\infty})^\perp \oplus \Rg(\mathcal{P}^\mathrm{u}_{\mathrm{adj},\infty})^\perp = L^2 \times L^2, \qquad
\Rg(\mathcal{P}^\mathrm{s}_\infty) \oplus \Rg(\mathcal{P}^\mathrm{u}_\infty) = L^2 \times L^2,
\]
we conclude that (\ref{e:a10}) is true as claimed.

Step~2: We turn to the $r$-dependent equations (\ref{e:a2}) and (\ref{e:a3}), We have shown above that the adjoint system (\ref{e:a3}) associated with (\ref{e:a2}) has an exponential dichotomy on $[R,\infty)$ with stable projection $\mathcal{P}^\mathrm{s}_\mathrm{adj}(r)$. The results proved in \S\ref{s:edff} show that $\mathcal{P}^\mathrm{s}_\mathrm{adj}(r)\to\mathcal{P}^\mathrm{s}_{\mathrm{adj},\infty}$ and $\mathcal{P}^\mathrm{s}(r)\to\mathcal{P}^\mathrm{s}_\infty$ as $r\to\infty$. In particular, we conclude from (\ref{e:a10}) that $\Rg(\mathcal{P}^\mathrm{s}_\mathrm{adj}(r))\oplus\Rg(\mathcal{P}^\mathrm{s}(r))=L^2\times L^2$ for all $r\geq R$, possibly after making $R$ larger. The same argument involving (\ref{e:a7}) as in the first step of our proof then implies that $\Rg(\mathcal{P}^\mathrm{s}(r))=[\Rg(\mathcal{P}^\mathrm{s}_\mathrm{adj}(r))]^\perp$ for all $r\geq R$.

Step~3: Suppose now that $\mathbf{u}(r)$ is a solution of (\ref{e:a2}) on $[r_*,\infty)$ so that $\sup_{r\geq r_*}|\mathbf{u}(r)|_{L^2\times L^2}\leq M$. We need to show that $\mathbf{u}(r_*)\in\Rg(P^\mathrm{s}(r_*))$. For each initial condition $\mathbf{w}_0\in\Rg(\mathcal{P}^\mathrm{s}_\mathrm{adj}(r_*))$, the associated solution $\mathbf{w}(r)$ of (\ref{e:a3}) exists and decays exponentially as $r$ increases. Equation (\ref{e:a7}) therefore implies that
\[
\langle\mathbf{u}(r_*),\mathbf{w}_0\rangle = \langle\mathbf{u}(r),\mathbf{w}(r)\rangle = 0
\]
for all $\mathbf{w}_0\in\Rg(\mathcal{P}^\mathrm{s}_\mathrm{adj}(r_*))$, and we conclude that $\mathbf{u}(r)\perp\Rg(\mathcal{P}^\mathrm{s}_\mathrm{adj}(r))$ for all $r\geq r_*$. Step~2 implies $\mathbf{u}(r)\in\Rg(\mathcal{P}^\mathrm{s}(r))$ for $r\geq r_*$ as claimed.
\end{Proof}

We state the following lemma, which is the analogue of Lemma~\ref{l:abddff} for the core region.

\begin{Lemma}\label{l:abddcore}
Suppose that (\ref{e:lspdyn0}) posed on $H^1\times L^2$ has an exponential dichotomy on $(-\infty,\log R]$ with unstable projection $P^\mathrm{u}_-(s)$. If $\mathbf{u}(s)$ is a solution of (\ref{e:lspdyn0}) on $(-\infty,\log r_*]$ that is uniformly bounded, then $\mathbf{u}(s)\in\Rg(P^\mathrm{u}_-(s))$ for \rev{$s\leq\log r_*$}.
\end{Lemma}

The proof of Lemma~\ref{l:abddcore} follows from taking the adjoint of (\ref{e:lspdyn0}) with respect to the $L^2\times L^2$ inner product and relating the resulting equation to $L^2$-adjoint of the PDE linearization about the spiral wave using (\ref{e:a8}) without the factor $r$. The details are much simpler than those for the far-field region, and we therefore omit them.

\subsection{Exponential dichotomies in exponentially weighted spaces}\label{s:edweights}

Next, we summarize the changes needed to obtain the existence of exponential dichotomies of the wave-train and spiral-wave systems we investigated in \S\ref{s:wt} and \S\ref{s:edc}-\ref{s:eda}, respectively, in exponentially weighted spaces. We begin with the spatial eigenvalue problem
\begin{equation}\label{eta:nwt}
\mathbf{u}_x = \mathcal{A}_\infty(\lambda) \mathbf{u}
\end{equation}
of the wave train that we defined in (\ref{e:mtwl2}) and (\ref{e:mtwabs2}). Using the rate $\eta\in J_0(\lambda)$, we introduce the new variable
\begin{equation}\label{eta:var}
\mathbf{v} := \rme^{\eta r} \mathbf{u}
\end{equation}
and obtain the new system
\begin{equation}\label{eta:wt}
\mathbf{v}_x = [\mathcal{A}_\infty(\lambda)+\eta] \mathbf{v}
=: \mathcal{A}_\infty^\eta(\lambda) \mathbf{v}.
\end{equation}
Since the spatial eigenvalues of $\mathcal{A}_\infty^\eta(\lambda)$ are given by $\nu_j(\lambda)+\eta$, it follows from Definition~\ref{d:j} and $\eta\in J_0(\lambda)$ that $\mathcal{A}_\infty^\eta(\lambda)$ is invertible and has relative Morse index zero. Next, we consider the spatial dynamical-systems formulation (\ref{e:lspdyn})
\begin{align}\label{eta:nowt}
u_r = & v \\ \nonumber
v_r = & - \frac{1}{r} v - \frac{1}{r^2} \partial_{\psi\psi} u
- D^{-1}[\omega_*\partial_{\psi} u + f^\prime(u_*(r,\psi)) u - \lambda u]
\end{align}
associated with the operator $\mathcal{L}_*-\lambda$, which we write as before as
\begin{equation}\label{eta:nff}
\mathbf{u}_r = \mathcal{A}(r;\lambda) \mathbf{u}.
\end{equation}
Using the transformation (\ref{eta:var}), this system becomes
\begin{equation}\label{eta:ff}
\mathbf{v}_x = [\mathcal{A}(r;\lambda)+\eta] \mathbf{v}
=: \mathcal{A}^\eta(r;\lambda) \mathbf{v}.
\end{equation}
We see that the asymptotic far-field system belonging to (\ref{eta:ff}) is given by (\ref{eta:wt}). In the core region, we use again the logarithmic radial time $s=\log r$ and see that (\ref{eta:nowt}) becomes
\begin{align*}
u_s = & v + \eta\rme^s u \\ \nonumber
v_s = & -\partial_{\psi\psi} u + \eta\rme^s v - \rme^{2s} D^{-1}
[\omega_*\partial_{\psi} u + f^\prime(u_*(\rme^s,\psi)) u - \lambda u]
\end{align*}
which we write, using the notation from (\ref{e:sdabs0}), as
\begin{equation}\label{eta:core}
\mathbf{v}_s = [\mathcal{A}_\mathrm{core}(s;\lambda) + \eta\rme^s] \mathbf{v}
=: \mathcal{A}_\mathrm{core}^\eta(s;\lambda) \mathbf{v}.
\end{equation}
We observe that the formal limiting problem of (\ref{eta:core}) for $s\to-\infty$ is given by the same system (\ref{e:lspdyn0inf}) as for the case $\eta=0$. We can now analyze the systems (\ref{eta:wt}), (\ref{eta:ff}), and (\ref{eta:core}) in exactly the same way as the systems (\ref{eta:nwt}),  (\ref{eta:nff}), and (\ref{e:sdabs0}). In particular, the existence results for exponential dichotomies that we established in \S\ref{s:wt} and \S\ref{s:edc}-\ref{s:eda} hold also for the new systems introduced here.

\subsection{Exponential trichotomies}\label{s:edcenter}

In this section, we discuss the spatial-dynamics formulation of the
linearization $\mathcal{L}_*-\lambda$ at a spiral wave when
$\lambda$ is in the Floquet spectrum of the asymptotic wave trains. We
will show that the equation
\begin{equation}\label{e:tdA}
\mathbf{u}_r = \mathcal{A}_\mathrm{arch}(r;\lambda) \mathbf{u} + 
\left(\begin{array}{cc} 0 & 0 \\
D^{-1} [ f^\prime(u_*(r,\cdot-k_*r-\theta(r))) - f^\prime(u_\infty) ] & 0
\end{array}\right) \mathbf{u}
=: [\mathcal{A}_\infty(\lambda) + \mathcal{C}(r)] \mathbf{u}
\end{equation}
posed on $X_r$ with
\[
\mathcal{A}_\infty(\lambda) = \left(\begin{array}{cc}
-k_*\partial_\vartheta & 1 \\
- D^{-1}[\omega_*\partial_\vartheta + f^\prime(u_\infty) - \lambda] &
-k_*\partial_\vartheta \end{array}\right), \qquad
\mathcal{C}(r) = - \left(\begin{array}{cc}
\theta^\prime(r)\partial_\vartheta & 0 \\
\frac{1}{r^2}\partial_{\vartheta\vartheta} &
\frac{1}{r}+\theta^\prime(r)\partial_\vartheta \end{array}\right).
\]
has a decomposition into exponentially decaying and growing directions
plus an additional center direction caused by the Floquet spectrum.
Throughout this section, we will make extensive use of the relation
between the operators $\hat{\mathcal{L}}_\mathrm{\rev{co}}(\nu)-\lambda_\mathrm{\rev{co}}$ and
$\mathcal{A}_\infty(\lambda)$ that we discussed in \S\ref{s:AL}.

We assume that $\lambda$ is a simple element of the Floquet
spectrum of the wave trains that has non-zero group velocity
$c_\mathrm{g,l}$. Thus, there is a simple, unique Floquet exponent
$\nu\in\rmi\R$ with $\lambda=\lambda_\mathrm{st}(\nu)$ and the associated
spectral projection of $\mathcal{A}_\infty(\lambda)$ is given by
\begin{equation}\label{e:centerproj}
P^\mathrm{c}_\infty(\lambda) =
\frac{1}{\langle\mathbf{u}^\mathrm{c}_\mathrm{ad},\mathbf{u}^\mathrm{c}\rangle}
\langle\mathbf{u}^\mathrm{c}_\mathrm{ad},\cdot\rangle \mathbf{u}^\mathrm{c},
\qquad
\mathbf{u}^\mathrm{c} = { u \choose (k_*\partial_\vartheta+\nu) u },
\qquad
\mathbf{u}^\mathrm{c}_\mathrm{ad} =
{ D(-k_*\partial_\vartheta+\nu)u_\mathrm{ad} \choose D u_\mathrm{ad} },
\end{equation}
where $[\hat{\mathcal{L}}_\mathrm{\rev{co}}(\nu)-\lambda_\mathrm{\rev{co}}(\nu)]u=0$, with
$\lambda_\mathrm{\rev{co}}(\nu)=\lambda_\mathrm{st}(\nu)+\omega_*\nu/k_*$, and
$[\hat{\mathcal{L}}_\mathrm{\rev{co}}^\mathrm{ad}(\nu)-\lambda_\mathrm{\rev{co}}(\nu)]u_\mathrm{ad}=0$.
We also define $P^\mathrm{h}_\infty(\lambda)=\id-P^\mathrm{c}_\infty(\lambda)$. Note that the linear group velocity $c_\mathrm{g,l}$ of
$\lambda_\mathrm{st}(\nu)$ is given by
\begin{equation}\label{e:cgl}
c_\mathrm{g,l} =
- \frac{2\langle u_\mathrm{ad},Dv\rangle}{\langle u_\mathrm{ad},u\rangle},
\end{equation}
as can be seen by differentiating (\ref{e:cgcomp}) with respect to
$\nu$ and taking the $L^2$-scalar product with $u_\mathrm{ad}$, and the projection $P^\mathrm{c}_\infty$ is therefore well defined since we assumed that $c_\mathrm{g,l}\neq0$.

Proceeding as in \S\ref{s:edff}, we see that the limiting equation
\begin{equation}\label{e:tdinf}
\mathbf{u}_r = \mathcal{A}_\infty(\lambda) \mathbf{u}
\end{equation}
has an exponential trichotomy on $X_r$ with three complementary projections $P^\mathrm{s}_\infty(\lambda)$, $P^\mathrm{u}_\infty(\lambda)$, and $P^\mathrm{c}_\infty(\lambda)$ that project onto the stable, the unstable and the center part of the spectrum of $\mathcal{A}_\infty(\lambda)$. The center subspace is one-dimensional and can be characterized as the space of initial conditions whose solutions grow at most exponentially in forward and backward ``time'' $r$ with an exponential rate $\eta>0$ which we can choose to be smaller than the decay rates in the stable and unstable subspaces. Solving (\ref{e:tdA}) using exponential weights and invoking Proposition~\ref{p:edff}, we see that (\ref{e:tdA}) also has an exponential trichotomy with projections that converge to the projections of (\ref{e:tdinf}) as $r\to\infty$.

Next, we observe that the operators
$\mathcal{C}(r)P^\mathrm{c}_\infty(\lambda)$ and
$P^\mathrm{c}_\infty(\lambda)\mathcal{C}(r)$ are bounded with norm
$\rmO(r^{-1})$. This is clear for the first operator, while the second
operator can be expressed as
\[
P^\mathrm{c}_\infty(\lambda) \mathcal{C}(r) \mathbf{w} =
\frac{1}{\langle\mathbf{u}^\mathrm{c}_\mathrm{ad},\mathbf{u}^\mathrm{c}\rangle}
\langle\mathcal{C}(r)\mathbf{u}^\mathrm{c}_\mathrm{ad},\mathbf{w}\rangle
\mathbf{u}^\mathrm{c}
\]
whenever $\mathbf{w}$ is in $X^1$, which proves the claim. Hence, the
equation
\begin{equation}\label{e:tddec}
\mathbf{u}_r = [\mathcal{A}_\infty(\lambda)
+ P^\mathrm{c}_\infty(\lambda)\mathcal{C}(r)P^\mathrm{c}_\infty(\lambda)
+ P^\mathrm{h}_\infty(\lambda)\mathcal{C}(r)P^\mathrm{h}_\infty(\lambda)] \mathbf{u}
\end{equation}
is a small perturbation of (\ref{e:tdA}), and
Proposition~\ref{p:edrobust} shows that (\ref{e:tddec}) has an
exponential trichotomy also with projections close to those of
(\ref{e:tdinf}). Equation (\ref{e:tddec}) can be written as
\begin{align*}
\mathbf{v}^\mathrm{h}_r = &  [\mathcal{A}_\infty(\lambda)
+ P^\mathrm{h}_\infty(\lambda)\mathcal{C}(r)] \mathbf{v}^\mathrm{h} \\
\mathbf{v}^\mathrm{c}_r = & 
[\nu + P^\mathrm{c}_\infty(\lambda)\mathcal{C}(r)] \mathbf{v}^\mathrm{c},
\end{align*}
where $\mathbf{v}^\mathrm{h}\in\Rg(P^\mathrm{h}_\infty(\lambda))$ and
$\mathbf{v}^\mathrm{c}\in\Rg(P^\mathrm{c}_\infty(\lambda))$. Recall that
$\nu\in\rmi\R$ is the Floquet exponent associated with $\lambda$. Since
this system is decoupled and because the second equation clearly
corresponds to the center direction, we can conclude that the first
equation has an exponential dichotomy that accounts for the remaining
strongly stable and unstable directions. Thus, we have shown the
following result.

\begin{Lemma}\label{r:et}
The equation
\[
\mathbf{w}_r =
[\mathcal{A}_\infty(\lambda) + P^\mathrm{h}_\infty(\lambda)\mathcal{C}(r)]\mathbf{w},
\qquad
\mathbf{w}\in\Rg(P^\mathrm{h}_\infty(\lambda))
\]
has an exponential dichotomy on $X_r$ on the interval $[R_*,\infty)$.
\end{Lemma}

For later reference, we remark that, when setting $\lambda=\nu=0$ in
(\ref{e:centerproj}), we have
\[
\mathbf{u}^\mathrm{c} = \mathbf{u}_\infty^\prime =
{ \partial_\vartheta u_\infty \choose
k_*\partial_{\vartheta\vartheta} u_\infty },
\qquad
\mathbf{u}^\mathrm{c}_\mathrm{ad} =
{ -k_* D\partial_\vartheta u_\mathrm{ad} \choose D u_\mathrm{ad} },
\]
where $\mathcal{L}_\mathrm{co}\partial_\vartheta u_\infty=0$ and
$\mathcal{L}^\mathrm{ad}_\mathrm{co}u_\mathrm{ad}=0$.

%%%%%%%%%%%%%%%%%%%%%%%%%%%%%%%%%%%%%%%%%%%%%%%%%%%%%%%%%%%%%%%%%%%%%%%%%

\section{Fredholm properties}\label{s:fred}

In this section, we prove Theorem~\ref{t:fm} and Proposition~\ref{p:extpt}, \rev{which characterize the Fredholm boundaries and the regions of constant Fredholm index for the linearization at a planar spiral wave. In \S\ref{s:edcomp}, we relate the relative Morse indices of the asymptotic wave trains to the Fredholm indices of maps that involve the exponential-dichotomy projections in the core and far field. In \S\ref{s:edinv}, we show that the existence of exponential dichotomies for the spatial eigenvalue problem implies Fredholm properties of the linearization at a spiral wave, which will complete the proof of Theorem~\ref{t:fm}. We then use these results in \S\ref{ss:extpt} to prove Proposition~\ref{p:extpt}.}

\subsection{Fredholm and Morse indices revisited}\label{s:edcomp}

In preparation of the proof of Theorem~\ref{t:fm}, we relate the relative Morse index of the asymptotic wave train, \rev{which we defined in \S\ref{s:wts.morse}}, to Fredholm properties of the exponential dichotomies of the spiral wave. \rev{Assume that $\lambda$ is not in the Floquet spectrum of the asymptotic wave trains. In Proposition~\ref{p:mor}, we showed that the relative Morse index $i_\mathrm{M}(\lambda)$ is equal to the Fredholm indices of the Fredholm operators}
\[
P^\mathrm{u}_\mathrm{wt}(\lambda_\mathrm{inv}): \quad
\Rg(P^\mathrm{u}_\mathrm{wt}(\lambda)) \longrightarrow
\Rg(P^\mathrm{u}_\mathrm{wt}(\lambda_\mathrm{inv}))
\]
and
\[
\iota: \quad
\Rg(P^\mathrm{u}_\mathrm{wt}(\lambda)) \times
\Rg(P^\mathrm{s}_\mathrm{wt}(\lambda_\mathrm{inv})) \longrightarrow X, \quad
(\mathbf{u}^\mathrm{u},\mathbf{u}^\mathrm{s}) \longmapsto \mathbf{u}^\mathrm{u}+\mathbf{u}^\mathrm{s},
\]
where $P^\mathrm{u}_\mathrm{wt}(\lambda)$ is the projection of the exponential dichotomy associated with the wave trains defined in Proposition~\ref{p:mor}. \rev{In Propositions~\ref{p:dc} and~\ref{p:edff}, we established the existence of exponential dichotomies $\widehat{P}_-^\mathrm{s/u}(r;\lambda)$ and $P_+^\mathrm{s/u}(r;\lambda)$ on $X_r$ in the core and the far field, defined for $r\leq R$ and $r\geq R$, respectively, of the linearization at the spiral wave, where we used the notation introduced in \S\ref{s:ced}.}

\begin{Proposition}\label{p:morse}
If $\lambda$ is not in the Floquet spectrum of the asymptotic wave
train, then the maps
\begin{align*}
\iota_\mathrm{spiral}(\lambda): &\quad
\Rg(P^\mathrm{s}_+(R;\lambda)) \times
\Rg(\widehat{P}^\mathrm{u}_-(R;\lambda)) \longrightarrow X_R, \quad
(\mathbf{u}^\mathrm{s}_+,\mathbf{u}^\mathrm{u}_-) \longmapsto \mathbf{u}^\mathrm{s}_++\mathbf{u}^\mathrm{u}_-\\
\tilde{\iota}_\mathrm{spiral}(\lambda): &\quad
\Rg(P^\mathrm{u}_+(R;\lambda)) \times
\Rg(\widehat{P}^\mathrm{s}_-(R;\lambda)) \longrightarrow X_R, \quad
\rev{(\mathbf{u}^\mathrm{u}_+,\mathbf{u}^\mathrm{s}_-)} \longmapsto \mathbf{u}^\mathrm{u}_++\mathbf{u}^\mathrm{s}_-
\end{align*}
are Fredholm, and their indices are given by 
\[
\ind(\iota_\mathrm{spiral}(\lambda)) = - \ind(\tilde{\iota}_\mathrm{spiral}(\lambda)) = -i_\mathrm{M}(\lambda).
\]
\end{Proposition}

\begin{Proof}
\rev{We will use the following argument repeatedly for} different pairs of projections. Suppose that $P_1$ and $Q$ are projections in $\mathrm{L}(X)$ with the associated map $\iota_1$ given by
\[
\iota_1: \quad
\Rg(P_1)\times\Rg(Q) \longrightarrow X, \quad
(\mathbf{u}_1,\mathbf{u}_2) \longmapsto \mathbf{u}_1+\mathbf{u}_2.
\]
Let $P_2$ be another projection in $\mathrm{L}(X)$ together with the map $\iota_2$
\[
\iota_2: \quad
\Rg(P_2)\times\Rg(Q) \longrightarrow X, \quad
(\mathbf{u}_1,\mathbf{u}_2) \longmapsto \mathbf{u}_1+\mathbf{u}_2
\]
and note that
\begin{eqnarray*}
\iota_1(\mathbf{u}_1,\mathbf{u}_2) = P_1\mathbf{u}_1+\mathbf{u}_2 & = &
P_2\mathbf{u}_1+\mathbf{u}_2 + (P_1-P_2)\mathbf{u}_1
= P_2|_{\Rg(P_1)}\mathbf{u}_1+\mathbf{u}_2 + (P_1-P_2)\mathbf{u}_1 \\ & = &
\iota_2 \circ \left(P_2|_{\Rg(P_1)}\times\id\right)(\mathbf{u}_1,\mathbf{u}_2) + (P_1-P_2)\mathbf{u}_1.
\end{eqnarray*}
We can then conclude that if $\iota_1$ is Fredholm with index $i_1$, the map $P_2:\Rg(P_1)\to\Rg(P_2)$ is Fredholm with index $i_{12}$, and the difference $P_1-P_2:\Rg(P_1)\to X$ is compact (or so small that the map $(\mathbf{u}_1,\mathbf{u}_2) \mapsto\iota_1(\mathbf{u}_1,\mathbf{u}_2) +(P_2-P_1)\mathbf{u}_1$ is still Fredholm with index $i_1$), then $\iota_2$ is Fredholm with index $i_2=i_1-i_{12}$.

We first focus on Fredholm properties of $\tilde{\iota}_\mathrm{spiral}$. We fix $\lambda$ and omit the dependence of the projections on $\lambda$. To relate the Morse index $i_\mathrm{M}$ and the Fredholm index of the pair $(P^\mathrm{u}_+(R),\widehat{P}^\mathrm{s}_-(R))$, we construct a sequence of pairs of projections and account for the changes of the Fredholm index when switching from one pair to the next. To make the notation less awkward, we give only the projections instead of the associated maps $\iota$. Finally, to compare projections in the core and far field, we use the coordinates and isomorphisms discussed in \S\ref{s:ced} that allow us to consider the relevant projections on the common space $L^2\times L^2$ instead of on $X_R$.

We begin with the unstable projection of the reference equation
\[
D u_{rr} + \omega \partial_\psi u = \lambda_\mathrm{\rev{inv}} u
\]
in the far field and the stable projection of the reference equation
\[
u_{ss} + u_{\psi\psi} = 0
\]
in the core. Using explicit computations in angular Fourier space and pulling the ranges of the resulting projections back to $L^2\times L^2$ using the isomorphisms from Lemma~\ref{l:xy} and \S\ref{s:ced}, it is not difficult to see that the Fredholm index of the resulting pair $(\mathcal{P}^\mathrm{u}_{+,\mathrm{ref}},\widehat{\mathcal{P}}^\mathrm{s}_{-,\mathrm{ref}})$ is zero. Next, we consider the pair $(\mathcal{P}^\mathrm{u}_\infty,\widehat{\mathcal{P}}^\mathrm{s}_{-,\mathrm{ref}})$. Since the Fredholm index of
\[
\mathcal{P}^\mathrm{u}_\infty: \quad
\Rg(\mathcal{P}^\mathrm{u}_\infty) \longrightarrow
\Rg(\mathcal{P}^\mathrm{u}_{+,\mathrm{ref}})
\]
is, by definition, equal to $i_\mathrm{M}$, and the difference of $\mathcal{P}^\mathrm{u}_\infty$ and $\mathcal{P}^\mathrm{u}_{+,\mathrm{ref}}$ is compact due to Proposition~\ref{p:mor}(ii), we see that the Fredholm index of the pair $(\mathcal{P}^\mathrm{u}_\infty,\widehat{\mathcal{P}}^\mathrm{s}_{-,\mathrm{ref}})$ is equal to $i_\mathrm{M}$. Closeness of projections proved in Proposition~\ref{p:edff} shows that switching to the pair $(\mathcal{P}^\mathrm{u}_+(R),\widehat{\mathcal{P}}^\mathrm{s}_{-,\mathrm{ref}})$ does not change the Fredholm index. \rev{Our final step consists of replacing $\widehat{\mathcal{P}}^\mathrm{s}_{-,\mathrm{ref}}$ by $\widehat{P}^\mathrm{s}_-(R)$. To do so, we claim that the operators
\begin{equation}\label{e:60}
\widehat{P}^\mathrm{s}_-(r):\Rg(\widehat{\mathcal{P}}^\mathrm{s}_{-,\mathrm{ref}})\longrightarrow\Rg(\widehat{P}^\mathrm{s}_-(r))
\end{equation}
are Fredholm of index zero for all $0<r\leq R$. For $0<r\ll1$, this claim follows from the convergence of the stable core projections to the stable reference projection as $r\to0$ that we established in \S\ref{s:edc}. We can then apply Fourier projections as in \cite[\S4]{ss-fred} to show that the operator in (\ref{e:60}) is a Fredholm operator of the same index independently of $r$, which establishes the claim for all $r$, and that the difference $\widehat{P}^\mathrm{s}_-(r)-\widehat{\mathcal{P}}^\mathrm{s}_{-,\mathrm{ref}}$ is compact for all $r$. Hence, appealing to our general argument that we outlined at the beginning of the proof, we can indeed} replace the pair $(\mathcal{P}^\mathrm{u}_+(R),\widehat{\mathcal{P}}^\mathrm{s}_{-,\mathrm{ref}})$ by $(\mathcal{P}^\mathrm{u}_+(R),\widehat{P}^\mathrm{s}_-(R))$ without changing the index. This shows that the Fredholm index of
\[
\Rg(P^\mathrm{u}_+(R;\lambda)) \times
\Rg(\rev{\widehat{P}^\mathrm{s}_-(R;\lambda)}) \longrightarrow X, \quad
(\mathbf{u}^\mathrm{u},\mathbf{u}^\mathrm{s}) \longmapsto \mathbf{u}^\mathrm{u}+\mathbf{u}^\mathrm{s}
\]
is indeed equal to $i_\mathrm{M}(\lambda)$ as claimed.

The same considerations apply to the map $\iota_\mathrm{spiral}$, where the Fredholm index is now determined by the map
\[
\mathcal{P}^\mathrm{s}_\infty: \quad
\Rg(\mathcal{P}^\mathrm{s}_\infty) \longrightarrow
\Rg(\mathcal{P}^\mathrm{s}_{+,\mathrm{ref}}).
\]
Since the stable subspace for the linearization at wave trains is simply obtained by reversing spatial time, the relative Morse index changes sign and we find that the Fredholm index is $-i_\mathrm{M}$. This concludes the proof for both $\iota_\mathrm{spiral}$ and $\tilde{\iota}_\mathrm{spiral}$.
\end{Proof}

\subsection{Exponential dichotomies imply Fredholm properties}\label{s:edinv}

We prove Theorem~\ref{t:fm}. Note that it is a consequence of, for instance, \cite[Lemma~6.5]{ssw2} that $\mathcal{L}_*-\lambda$ is not Fredholm whenever $\lambda$ is in the Floquet spectrum of the asymptotic wave trains. \rev{Thus, it suffices to show that $\mathcal{L}_*-\lambda$ is Fredholm whenever $\lambda$ is not in the Floquet spectrum of the asymptotic wave trains. To prove this, we follow the \rev{strategy of the} proofs in \cite[\S5.2]{ss-fred} with modifications in the regime $s\to-\infty$.}

Since $\lambda$ is not in the Floquet spectrum of the wave trains, there is an $R\gg1$ so that \rev{the far-field} equation (\ref{e:lspdyn})
\begin{align}\label{e:fflin}
u_r = &  v \\ \nonumber
v_r = &  - \frac{1}{r} v - \frac{1}{r^2} \partial_{\psi\psi} u
- D^{-1}[\omega_*\partial_{\psi} u + f^\prime(u_*(r,\psi)) u - \lambda u]
\end{align}
has an exponential dichotomy \rev{for $r\geq R$ with projections $P^\mathrm{s/u}_+(r;\lambda)$ defined on $X_r$}. The equation in the core region (\ref{e:lspdyn0})
\begin{align}\label{e:crlin}
u_s = &  w \\ \nonumber
w_s = &  -\partial_{\psi\psi} u - \rme^{2s} D^{-1}[
\omega_*\partial_{\psi} u + f^\prime(u_*(\rme^s,\psi)) u - \lambda u]
\end{align}
with $s=\log r$ always has an exponential dichotomy and, following \S\ref{s:ced}, we denote the associated projections by $P^\mathrm{s/u}_-(s;\lambda)$ on $X$ and by $\widehat{P}^\mathrm{s/u}_-(r;\lambda)$ on $X_r$. Combining the spatial dynamics equations in the core and far field, and using the time variable $r=\rme^s$ for $r<R$, we obtain an abstract ordinary differential equation
\begin{equation}\label{e:adtime}
\mathbf{u}_r = \mathcal{A}_\lambda(r) \mathbf{u},
\end{equation}
which coincides with (\ref{e:fflin}) for $r\geq R$ and with (\ref{e:crlin}) for $s=\log r\leq\log R$. We say that (\ref{e:adtime}) has an exponential dichotomy on $\R$ if (\ref{e:fflin}) has an exponential dichotomy on $X_r$ for $r\geq R$ and if the unstable subspace $\Rg(\rev{\widehat{P}^\mathrm{u}_-(R;\lambda)})$ of the exponential dichotomy in the core region and the stable subspace $\Rg(P^\mathrm{s}_+(R;\lambda))$ in the far field span the space $X_R$ at radial time $R$ and have trivial intersection.

Assume that $u=u(r,\psi)$ belongs to the null space of $\mathcal{L}_*-\lambda$, then the function $(u,u_r)$ is a bounded solution to the spatial-dynamics formulation (\ref{e:adtime}). \rev{Using Lemmas~\ref{l:abddff} and~\ref{l:abddcore}}, it follows that $(u(R,\psi),u_r(R,\psi))$ belongs to $\Rg(\rev{\widehat{P}^\mathrm{u}_-(R;\lambda)})\cap\Rg(P^\mathrm{s}_+(R;\lambda))$. \rev{Since different bounded solutions cannot share the same initial data $(u(R,\psi),u_r(R,\psi))$ by \cite[Theorem~2.5]{mz}, Proposition~\ref{p:morse} implies that the null space of $\mathcal{L}_*-\lambda$ is finite-dimensional.} Applying the above arguments to the $L^2$-adjoint of $\mathcal{L}_*-\lambda$ shows that the codimension of the closure of the range of $\mathcal{L}_*-\lambda$ is also finite-dimensional.

\rev{To complete the proof of Theorem~\ref{t:fm}, we need to (i) show that the range of $\mathcal{L}_*-\lambda$ is closed since the operator}
\[
\mathcal{L}_* - \lambda = D \Delta + \omega_* \partial_\psi + f^\prime(u_*(r,\psi)) - \lambda
\]
\rev{is then Fredholm and (ii) verify the statement about its index. We claim that (ii) follows from (i). Indeed, if $\mathcal{L}_*-\lambda$ is Fredholm,} we replace the term $f^\prime(u_*(r,\psi))u$ in the above expression for $\mathcal{L}_*-\lambda$ with the Galerkin  approximation $Q_m(f^\prime(u_*(r,\psi))u)$ where $Q_m$ denotes the orthogonal projection onto the first $m$ vector-valued Fourier modes.  Considered as operators from their common domain into $L^2$, the resulting two operators are close for $m$ sufficiently large. In particular, the operator
\begin{equation}\label{e:gao}
u \longmapsto D \Delta u + \omega_* \partial_\psi u + Q_m \left( f^\prime(u_*(r,\psi)) u \right) - \lambda u
\end{equation}
is also Fredholm and has the same index as $\mathcal{L}_*-\lambda$. The spatial-dynamics formulation for the Galerkin-approximated operator is upper triangular, and it is not difficult to see that the relative Morse index of the wave trains and the Fredholm index of the operator (\ref{e:gao}) are as in (\ref{e:fred=morse}) since the relevant system is finite-dimensional. \rev{Thus, the statement about the Fredholm index in Theorem~\ref{t:fm} follows once we know that the range of $\mathcal{L}_*-\lambda$ is closed, and we focus now on proving closedness of the range.}

Suppose therefore that
$(\mathcal{L}_*-\lambda)u_\ell=f_\ell\in L^2(\R^2,\C^N)$ where
$f_\ell\to f$ in $L^2$ as $\ell\to\infty$. We need to prove that, for
a suitable choice of the $u_\ell$, the sequence $u_\ell$ converges to
$u$ in $L^2$ which, by closedness of $\mathcal{L}_*$, would imply that
$(\mathcal{L}_*-\lambda)u=f\in L^2(\R^2)$ so that the range is closed.
To prove convergence of the $u_\ell$, we use the spatial-dynamics
formulation (\ref{e:adtime}).

Recall the definitions (\ref{e:spdnorm}) and (\ref{e:spdnorm1}) of the
spaces $X$, $X^1$ and $X_r$, $X_r^1$, \rev{respectively}. We define the function spaces
\begin{align*}
\mathcal{X} & :=  \{ (\mathbf{u}_-,\mathbf{u}_+) \in
L^2_\mathrm{loc}((-\infty,\log R],X) \times
L^2_\mathrm{loc}([R,\infty),X_r) ;\;
\|\mathbf{u}\|_{\mathcal{X}}<\infty \}, \\
\|(\mathbf{u}_-,\mathbf{u}_+)\|^2_{\mathcal{X}} & := 
\|\rme^s\mathbf{u}_-\|^2_{L^2((-\infty,\log R],X)}
+ \|r^{1/2}\mathbf{u}_+\|^2_{L^2([R,\infty),X_r)}
\end{align*}
and
\begin{align*}
\mathcal{X}^1  := & \{ (\mathbf{u}_-,\mathbf{u}_+) \in
\left( L^2_\mathrm{loc}((-\infty,\log R],X^1) \times
L^2_\mathrm{loc}([R,\infty),X^1_r) \right) \cap \\ & 
\left( H^1_\mathrm{loc}((-\infty,\log R],X) \times
H^1_\mathrm{loc}([R,\infty),X_r) \right) ;\; 
\|\mathbf{u}\|_{\mathcal{X}^1}<\infty \mbox{ and }
\mathbf{j}(R)\mathbf{u}_-(\log R)=\mathbf{u}_+(R) \} \\
\|(\mathbf{u}_-,\mathbf{u}_+)\|^2_{\mathcal{X}^1} := & 
\|\rme^s\mathbf{u}_-\|^2_{H^1((-\infty,\log R],X)}
+ \|\rme^s\mathbf{u}_-\|^2_{L^2((-\infty,\log R],X^1)} \\  &
+ \|r^{1/2}\mathbf{u}_+\|^2_{H^1([R,\infty),X_r)}
+ \|r^{1/2}\mathbf{u}_+\|^2_{L^2([R,\infty),X_r^1)},
\end{align*}
where $\mathbf{j}(R)$ was \rev{introduced in} (\ref{e:jr}). We then define the operator $\mathcal{T}:\mathcal{X}^1\to\mathcal{X}$
by $\mathcal{T}(\mathbf{u}_-,\mathbf{u}_+)=
(\mathcal{T}_-\mathbf{u}_-,\mathcal{T}_+\mathbf{u}_+)$ where 
\begin{align*}
\mathcal{T}_- \mathbf{u}_- := &
\left(\begin{array}{c} \displaystyle
\frac{\rmd}{\rmd s} u_- - v_- \\[1.5ex] \displaystyle
\frac{\rmd}{\rmd s} v_- + \partial_{\psi\psi} u_- + \rme^{2s}
D^{-1}[\omega_*\partial_{\psi} u_- + f^\prime(u_*(\rme^s,\psi)) u_- - \lambda u_-]
\end{array}\right), \\
\mathcal{T}_+ \mathbf{u}_+ := &
\left(\begin{array}{c} \displaystyle
\frac{\rmd}{\rmd r} u_+ - v_+ \\[1.5ex] \displaystyle
\frac{\rmd}{\rmd r} v_+ + \frac{1}{r} v_+
+ \frac{1}{r^2} \partial_{\psi\psi} u_+
+ D^{-1}[\omega_*\partial_{\psi} u_+ + f^\prime(u_*(r,\psi)) u_+ - \lambda u_+]
\end{array}\right),
\end{align*}
and $\mathbf{u}_\pm=(u_\pm,v_\pm)$. It is not difficult to check that
$\mathcal{T}$ is closed when considered as an unbounded operator on
$\mathcal{X}$, \rev{since} the operator is a bounded perturbation of the
Laplacian on the half cylinder $(-\infty,\log R]\times S^1$ and of
$D\Delta+\omega_*\partial_\psi$ on $[R,+\infty)$ when rewritten as
spatial dynamical systems.

We return to the sequences $u_\ell\to u$ and $f_\ell\to f$ in
$L^2(\R^2,\C^N)$. We define
\[
\mathbf{f}_{\ell,+} = { 0 \choose f_\ell(r,\psi) }, \quad
\mathbf{f}_{\ell,-} = { 0 \choose \rme^{2s}f_\ell(\rme^s,\psi) }, \quad
\mathbf{u}_{\ell,+} = { u_{\ell}(r,\psi) \choose
	\partial_ru_\ell(r,\psi)}, \quad
\mathbf{u}_{\ell,-} = { u_\ell(\rme^s,\psi) \choose
	\partial_s[u_\ell(\rme^s,\psi)] }.
\]
We claim that $\mathbf{f}_\ell\in\mathcal{X}$ and
$\mathbf{u}_\ell\in\mathcal{X}^1$. This claim can be easily checked by
transforming the $L^2$ and the $H^2$-norm in the plane into polar
coordinates $(r,\psi)$. Indeed, by Fubini's theorem, the
$\mathcal{X}$-norm of the second component of $\mathbf{u}_-$ and
$\mathbf{u}_+$ is precisely the $L^2$-norm in the plane written in polar
coordinates. Similarly, bounds on the norms of $\partial_{rr}u$,
$r^{-1}\partial_{r\psi}u$, and $r^{-2}\partial_{\psi\psi} u$ in
$L^2(\R^2,\C^N)$ imply $L^2$-bounds on the norms of
$\rme^s\partial_{ss}u_-(s,\psi)$, $\rme^s\partial_{s\psi}u_-(s,\psi)$
and $\rme^{-s}\partial_{\psi\psi}u_-(s,\psi)$. The
$\mathcal{X}^1$-norm of the $\mathbf{u}_+$-component is equivalent to
the norm induced by the domain of $\Delta+\partial_\psi$ on
$L^2(\R^2,\C^N)$. This proves our claim that
$\mathbf{f}_\ell\in\mathcal{X}$ and $\mathbf{u}_\ell\in\mathcal{X}^1$. In
particular, we have $\mathcal{T}\mathbf{u}_\ell=\mathbf{f}_\ell$ for all
$\ell$.

A possibly different solution to
$\mathcal{T}\tilde{\mathbf{u}}_\ell=\mathbf{f}_\ell$ is given by the
variation-of-constant formula
\begin{eqnarray}\label{eq:int}
\tilde{\mathbf{u}}_{\ell,-}(s) & = & 
\Phi_-^\mathrm{u}(s;\log R)\mathbf{v}_{\ell,-}
+ \int_{\log R}^s\Phi_-^\mathrm{u}(s;\zeta)\mathbf{f}_\ell(\zeta)\,\rmd\zeta
+ \int_{-\infty}^s\Phi_-^\mathrm{s}(s;\zeta)\mathbf{f}_\ell(\zeta)\,\rmd\zeta,
\qquad s\leq\log R \nonumber \\
\tilde{\mathbf{u}}_{\ell,+}(r) & = & 
\Phi_+^\mathrm{s}(r;R)\mathbf{v}_{\ell,+}
+ \int_R^r\Phi_+^\mathrm{s}(r;\zeta)\mathbf{f}_\ell(\zeta)\,\rmd\zeta
+ \int_\infty^r\Phi_+^\mathrm{u}(r;\zeta)\mathbf{f}_\ell(\zeta)\,\rmd\zeta, 
\qquad r\geq R,
\end{eqnarray}
where the elements $\mathbf{v}_{\ell,\pm}$ are defined by
\[
\mathbf{v}_{\ell,+} = P_+^\mathrm{s}(R) \mathbf{u}_{\ell,+}(R), \qquad
\mathbf{v}_{\ell,-} = P_-^\mathrm{u}(\log R) \mathbf{u}_{\ell,-}(\log R)
\]
and the evolution operators $\Phi^\mathrm{s,u}_-$ and $\Phi^\mathrm{s,u}_+$ that appear in (\ref{eq:int}) are the exponential dichotomies in the core and far-field regions that we constructed in \S\ref{s:edc} and \S\ref{s:edff}, respectively. Using the properties of $\Phi_\pm^\mathrm{u/s}$, we see that the integrals in (\ref{eq:int}) converge absolutely since the lack of exponential decay in the dichotomies on $(-\infty,\log R]$ is compensated for by the exponential decay of the right-hand side $\mathbf{f}$.

We first prove that $\tilde{\mathbf{u}}_\ell=\mathbf{u}_\ell$ for all $\ell$. The difference $\tilde{\mathbf{u}}_{\ell,+}(r)-\mathbf{u}_{\ell,+}(r)$ is uniformly bounded for $r\geq R$ and satisfies the homogeneous equation $\mathcal{T}_+\mathbf{u}_+=0$ for $r\geq R$ with $\tilde{\mathbf{u}}_{\ell,+}(R)-\mathbf{u}_{\ell,+}(R)\in\Ns(P_+^\mathrm{s}(R))$. Lemma~\ref{l:abddff} readily implies that $\tilde{\mathbf{u}}_{\ell,+}(r)=\mathbf{u}_{\ell,+}(r)$ for $r\geq R$. The same argument combined with Lemma~\ref{l:abddcore} shows that $\tilde{\mathbf{u}}_{\ell,-}(s)=\mathbf{u}_{\ell,-}(s)$ for $s\leq\log R$. Thus, we have $\tilde{\mathbf{u}}_\ell=\mathbf{u}_\ell$ as claimed.

Since $\tilde{\mathbf{u}}_\ell=\mathbf{u}_\ell$, we know that $\mathbf{u}_\ell$ satisfies (\ref{eq:int}). Setting $s=\log R$ and $r=R$ in (\ref{eq:int}), and using that $\mathbf{j}(R)\mathbf{u}_{\ell,-}(\log R)=\mathbf{u}_{\ell,+}(R)$, we see that
\begin{equation}\label{e:61}
\mathbf{j}(R)\mathbf{v}_{\ell,-} + \mathbf{j}(R)\int_{-\infty}^{\log R}\Phi_-^\mathrm{s}(\log R;\zeta)\mathbf{f}_\ell(\zeta)\,\rmd\zeta = \mathbf{v}_{\ell,+} + \int_\infty^R\Phi_+^\mathrm{u}(R;\zeta)\mathbf{f}_\ell(\zeta)\,\rmd\zeta.
\end{equation}
Setting $\widehat{\mathbf{v}}_{\ell,-}:=-\mathbf{j}(R)\mathbf{v}_{\ell,-}\in\Rg(\widehat{P}^\mathrm{u}_-(R))$, we can write (\ref{e:61}) as
\[
\mathbf{v}_{\ell,+} + \widehat{\mathbf{v}}_{\ell,-} = 
\mathbf{j}(R)\int_{-\infty}^{\log R}\Phi_-^\mathrm{s}(\log R;\zeta)\mathbf{f}_\ell(\zeta)\,\rmd\zeta
- \int_\infty^R\Phi_+^\mathrm{u}(R;\zeta)\mathbf{f}_\ell(\zeta)\,\rmd\zeta
\]
or, equivalently, as
\begin{equation}\label{e:62}
\iota_\mathrm{spiral}(\mathbf{v}_{\ell,+},\widehat{\mathbf{v}}_{\ell,-}) =
\mathbf{j}(R)\int_{-\infty}^{\log R}\Phi_-^\mathrm{s}(\log R;\zeta)\mathbf{f}_\ell(\zeta)\,\rmd\zeta
- \int_\infty^R\Phi_+^\mathrm{u}(R;\zeta)\mathbf{f}_\ell(\zeta)\,\rmd\zeta
\end{equation}
where $(\mathbf{v}_{\ell,+},\widehat{\mathbf{v}}_{\ell,-})\in\Rg(P^\mathrm{s}_+(R))\times\Rg(\widehat{P}^\mathrm{u}_-(R))$. Since the right-hand side of (\ref{e:62}) lies in $\Rg(\iota_\mathrm{spiral})$ for all $\ell$ and converges in $X_R$ as $\ell\to\infty$, and the map $\iota_\mathrm{spiral}$ is Fredholm by Proposition~\ref{p:morse}, we conclude that the sequence $(\mathbf{v}_{\ell,+},\widehat{\mathbf{v}}_{\ell,-})$ converges in $\Rg(P^\mathrm{s}_+(R))\times\Rg(\widehat{P}^\mathrm{u}_-(R))$ upon subtracting appropriate elements in the null space of $\iota_\mathrm{spiral}$.

Hence, we have shown that the right-hand side of (\ref{eq:int}) converges for $\ell\to\infty$, and we conclude that $\mathbf{u}_\ell=\tilde{\mathbf{u}}_\ell$ converges to an element $\mathbf{u}$ in $\mathcal{X}$. Restriction to the first component $u_\pm$ of $\mathbf{u}_\pm$ shows that $u_\ell\to u$ in $L^2(\R^2,\C^N)$. Inspecting the integral equation (\ref{eq:int}) for the limit $\mathbf{u}$, we see that $u\in H^2$. This proves that the range of $\mathcal{L}_*-\lambda$ is closed and, together with the previous observations, completes the proof of Theorem~\ref{t:fm}.

\subsection{Proof of Proposition~\ref{p:extpt}}\label{ss:extpt}

To prove Proposition~\ref{p:extpt}, we note that the operator $\mathcal{L}_*-\lambda$ posed on $L^2_\eta(\R^2,\C^N)$ corresponds to the spatial dynamical systems (\ref{eta:ff}) and (\ref{eta:core}) for which we constructed exponential dichotomies in \S\ref{s:edweights}. Applying the results established in the previous sections to the weighted systems (\ref{eta:wt}), (\ref{eta:ff}), and (\ref{eta:core}) therefore completes the proof of Proposition~\ref{p:extpt}.

%%%%%%%%%%%%%%%%%%%%%%%%%%%%%%%%%%%%%%%%%%%%%%%%%%%%%%%%%%%%%%%%%%%%%%%%%

\section{Robustness and asymptotics of spiral waves}\label{s:rob}

\begin{figure}
 \centering\includegraphics[width=0.7\textwidth]{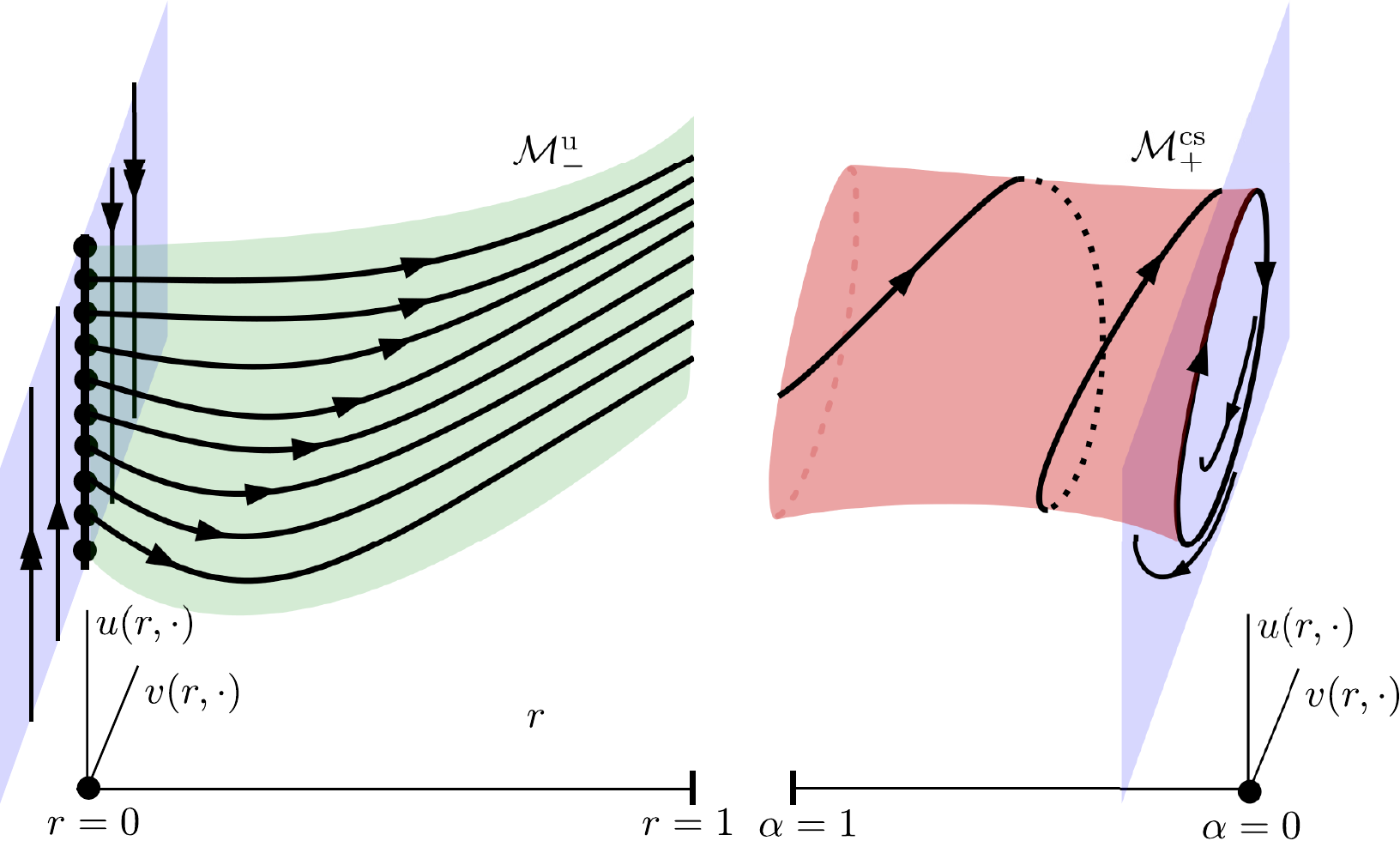}
 \caption{Illustration of the construction of robust spiral waves. In the region $0\leq r\leq 1$, we use dynamics in $\tau=\log r$ to construct the manifold $\mathcal{M}^\mathrm{u}_-$  of solutions that are bounded as $r\to 0$. in the region $1\geq \alpha=\frac{1}{r}\geq 0$, we construct the manifold  $\mathcal{M}^\mathrm{cs}_+$ of solutions that are asymptotic to the wave train solutions at $\alpha=0$. The robust intersection, as the parameter $\omega_*$ is varied, gives the spiral wave solution. The schematic is shown in $(2+1)$-dimensional phase space and should be thought as amended by infinitely many stable and unstable (and equally many) directions in the $(u,v)$-direction. 
 }\label{f:schematic_spatial_dynamics}
\end{figure}

The goal \rev{of} this section is to prove Proposition~\ref{p:exppha} and Theorem~\ref{t:rob} \rev{about robustness and far-field expansions of planar spiral waves. As in \cite{s-siam,ss-sup,ss-spst}, our strategy is to view spiral waves as heteroclinic orbits in the radial variable $r$, and we now describe this idea in detail and illustrate it further in Figure~\ref{f:schematic_spatial_dynamics}. First, we cast the steady-state equation (\ref{e:rds})
\begin{equation}\label{e:rdsn}
0 = D \left(\partial_{rr}+\frac{1}{r}\partial_r + \frac{1}{r^2}\partial_{\psi\psi}\right) u + \omega \partial_\psi u + f(u;\mu), \qquad u = u(r,\psi)\in\R^N
\end{equation}
for spiral waves as the} dynamical system
\begin{align}\label{e:spdynpa}
u_r = &  v \\ \nonumber
v_r = &  - \frac{1}{r} v - \frac{1}{r^2} \partial_{\psi\psi} u
- D^{-1}[\omega \partial_{\psi} u + f(u;\mu)]
\end{align}
in the spatial variable $r$. We consider (\ref{e:spdynpa}) in the \rev{phase space $X=H^1(S^1,\R^N)\times L^2(S^1,\R^N)$} with norms defined in (\ref{e:spdnorm}). Throughout this section, we assume that (\ref{e:rdsn}) with $\mu=0$ and $\omega=\omega_*\neq0$ admits a smooth Archimedean spiral wave $u_*(r,\psi)$ that emits a spectrally stable wave train $u_\infty$ with non-zero wavenumber $k_*\neq0$. Most of the work in this section is concerned with constructing nonlinear analogues of the stable and unstable subspaces for the linear dichotomies. These nonlinear analogues are infinite-dimensional manifolds $\mathcal{M}_-^\mathrm{u}$ and $\mathcal{M}^\mathrm{cs}_+$, which contain solutions that are bounded as $r\to0$ and asymptotic to wave trains as $r\to\infty$, respectively. The intersection of these two infinite-dimensional manifolds \rev{captures the spiral-wave solutions we are interested in as heteroclinic orbits}. A schematic of this approach is illustrated in Figure~\ref{f:schematic_spatial_dynamics}.

First, we discuss the existence of solutions near $u_*$ for $r\leq R$
and $\mu$ close to zero.

\begin{Proposition}\label{p:core}
For any fixed choice of $R$ with $0<R<\infty$, there exists a smooth
manifold $\mathcal{M}^\mathrm{u}_-(\mu,\omega)\subset X$, which
depends smoothly on $\mu$ and $\omega$ for $\mu$ close to zero and
$\omega$ close to $\omega_*$, such that
$\mathcal{M}^\mathrm{u}_-(\mu,\omega)$ consists precisely of all
boundary data of smooth solutions to (\ref{e:spwp}) in $|x|\leq R$
that are close to the original spiral wave (in particular,
$(u_*(R,\cdot),\partial_ru_*(R,\cdot))\in
\mathcal{M}^\mathrm{u}_-(0,\omega_*)$).
More precisely, there is an $\varepsilon>0$ such that any solution
$u(r,\psi)\in H^2(|x|\leq R)$ of (\ref{e:spwp}) with $u(r,\psi)$
$\varepsilon$-close to $u_*(r,\psi)$ in $H^2(|x|\leq R)$ satisfies
$(u(R,\cdot),\partial_ru(R,\cdot))\in\mathcal{M}^\mathrm{u}_-(\mu,\omega)$.
Conversely, for any
$(u(R,\cdot),\partial_ru(R,\cdot))\in\mathcal{M}^\mathrm{u}_-(\mu,\omega)$,
there actually exists a solution $u(r,\psi)$ of (\ref{e:spwp}) that is
$\varepsilon$-close to $u_*(r,\psi)$ in $H^2(|x|\leq R)$ and satisfies
the given Dirichlet and Neumann boundary values. These solutions
depend continuously on $\mu$ and the boundary values. The tangent
space of $\mathcal{M}^\mathrm{u}_-(0,\omega_*)$ in
$(u_*(R,\cdot),\partial_ru_*(R,\cdot))$ is given by
$\Rg(P_-^\mathrm{u}(R))$ defined in \eqref{e:cdich}.
\end{Proposition}

\begin{Proof}
The manifold is the union of the strong unstable fibers of the subspace \rev{$\{(u_0,0);\; u_0\in\R^N\}$} of constant functions in the center space $E_{-\infty}^\mathrm{c}$ at $s=-\infty$, where $s=\log r$ is the rescaled logarithmic radial variable introduced in \S\ref{s:edc}. Using the exponential dichotomies in the core region, we can construct this manifold by applying the uniform contraction mapping principle to the fixed-point equation
\[
\begin{pmatrix} u \\ w \end{pmatrix}(s) =
\Phi_-^\mathrm{u}(s;\log R) \begin{pmatrix} u^\mathrm{u}_0 \\ w^\mathrm{u}_0 \end{pmatrix}
+ \int_{\log R}^s \Phi_-^\mathrm{u}(s;\zeta) G(u(\zeta),\zeta,\mu)\,\rmd\zeta
+ \int_{-\infty}^s \Phi_-^\mathrm{s}(s;\zeta) G(u(\zeta),\zeta,\mu)\,\rmd\zeta, \quad s\leq\log R,
\]
where $(u^\mathrm{u}_0,w^\mathrm{u}_0)\in\Rg(P^\mathrm{u}_-(\log R))$ and
\[
G(u,s,\mu) := \rme^{2s} \begin{pmatrix} 0 \\ f(u_*(\rme^s,\psi)+u,\mu) - f(u_*(\rme^s,\psi),0) - f^\prime(u_*(\rme^s,\psi),0) u \end{pmatrix} = \rme^{2s} \rmO(|u|^2+|\mu|).
\]
Smoothness of the strong unstable fibers and smooth dependence on the asymptotic value and the parameter $\mu$ follow from the uniform contraction mapping principle. Smoothness in $\omega$ can be shown similarly upon dividing (\ref{e:rdsn}) by $\omega$ and rescaling $r$ so that (\ref{e:rdsn}) depends on $\omega$ only through the nonlinearity.
\end{Proof}

The following result deals with solutions in the far field $r\gg1$. Its proof is considerably more complicated and will occupy the remainder of this section. \rev{The main challenge is that spiral waves converge only algebraically with order $1/r$ in Archimedean coordinates as $r\to\infty$. In contrast, the scaling $s=\log r$ for the core region ensures that the coefficients of the core equation converge exponentially as $s\to-\infty$, which simplifies the analysis tremendously.}

\begin{Proposition}\label{p:stmaff}
Choose an $\varepsilon>0$, then for all $\mu$ sufficiently close to zero and any $R>0$ sufficiently large there exists a smooth manifold $\mathcal{M}^\mathrm{cs}_+(\mu,\omega)\subset X$ \rev{that depends smoothly on $(\mu,\omega)$ and that} contains precisely the boundary data of smooth and bounded solutions to (\ref{e:spwp}) in $|x|\geq R$ that are \rev{$\varepsilon$-close} to the original spiral wave. In particular, $(u_*(R,\cdot),\partial_ru_*(R,\cdot))\in\mathcal{M}^\mathrm{cs}_+(0,\omega_*)$. More precisely, we have the following.
\begin{compactitem}
\item Let $u(r,\psi)\in H^2(|x|\geq R)$ be a solution to
(\ref{e:spwp}) such that
\begin{equation}\label{e:asypha1}
|(u(r,\cdot) - u_\infty(k_*r+\theta(r)+\cdot),
\partial_r u(r,\cdot) - \partial_r u_\infty(k_*r+\theta(r)+\cdot))|_{X_r}
\to 0
\end{equation}
as $r\to\infty$ for some phase function $\theta(r)$, where $u_\infty$ denotes the $\mu$-dependent wave train with frequency $\omega$, and
\[
|(u(R,\cdot) - u_*(R,\cdot),
\partial_r u(R,\cdot) - \partial_r u_*(R,\cdot))|_{X} < \varepsilon,
\]
then $(u(R,\cdot),\partial_r u(R,\cdot))\in\mathcal{M}^\mathrm{cs}_+(\mu,\omega)$.
\item Conversely, for any $(u(R,\cdot),\partial_ru(R,\cdot))\in\mathcal{M}^\mathrm{cs}_+(\mu,\omega)$, there exist a solution $u(r,\psi)$ of (\ref{e:spwp}) in $r\geq R$ with the given Dirichlet and Neumann boundary values and a smooth phase $\theta(r)$ that satisfies $\theta^\prime(r)\to0$ as $r\to\infty$ such that $u(r,\psi)$ is asymptotic to the profile $u_\infty(kr+\theta(r)+\psi)$, where the wavenumber $k$ is determined implicitly through the nonlinear dispersion relation $\omega_*(k)=\omega$ of the wave trains, and (\ref{e:asypha1}) holds. The solution $(u,\partial_r u)(r)\in X_r$ and the phase $\theta(r)$ depend smoothly on $(\mu,\omega)$ and the boundary values in $\mathcal{M}^\mathrm{cs}_+(\mu,\omega)$.
\item The tangent space of $\mathcal{M}^\mathrm{cs}_+(0,\omega_*)$ in $(u_*(R,\cdot),\partial_r u_*(R,\cdot))$ is \rev{given by} $\Rg(P_+^\mathrm{cs}(R))$ as defined in Proposition~\ref{p:edff}.
%\item The tangent space of $\mathcal{M}^\mathrm{cs}_+(0,\omega_*)$ in $(u_*(R,\cdot),\partial_r u_*(R,\cdot))$ is $\varepsilon$-close to $\Rg(P_+^\mathrm{cs}(R))$ as defined in Proposition~\ref{p:edff} and Definition~\ref{d:dich}.
\end{compactitem}
\end{Proposition}

\begin{Proof}
Throughout the proof, we fix $\mu$, since smooth dependence of the
manifold and solutions on $\mu$ will become clear from the proof. We
have smooth dependence of these objects on $\omega$ since we may divide
(\ref{e:rdsn}) by $\omega$ and then rescale $r$ so that (\ref{e:rdsn})
depends on $\omega$ only through the smooth nonlinearity.

Recall the steady-state equation
\begin{align*}
u_r = &  v \\
v_r = &  - \frac{1}{r} v - \frac{1}{r^2} \partial_{\psi\psi} u
- D^{-1}[\omega_*\partial_{\psi} u + f(u)].
\end{align*}
Introducing the Archimedean coordinate
$\vartheta=k_*r+\theta_0\log r+\psi$ with $\theta_0$ to be determined,
we obtain the equation
\begin{align}\label{e:asyabs}
u_r = &  -\left(k_*+\frac{\theta_0}{r}\right)\partial_\vartheta u + v
\\ \nonumber
v_r = &  -\left(k_*+\frac{\theta_0}{r}\right)\partial_\vartheta v
- \frac{1}{r} v - \frac{1}{r^2} \partial_{\vartheta\vartheta} u
- D^{-1}[\omega_*\partial_\vartheta u + f(u)]
\end{align}
for which we seek solutions $\mathbf{u}(r,\vartheta)=(u,v)(r,\vartheta)$.

\paragraph{Formal expansion:}
Before we embark on a rigorous analysis of (\ref{e:asyabs}), we seek
formal solutions of the form
\[
u(r,\vartheta) = u_\infty(\vartheta) + \frac{1}{r} u_1(\vartheta)
+ \rmO\left(\frac{1}{r^2}\right).
\]
In the following formal analysis, we shall neglect all terms that are
formally of order $\rmO(r^{-2})$. Thus, from the first equation in
(\ref{e:asyabs}), we get
\[
v = k_* u_\infty^\prime
+ \frac{1}{r} (\theta_0 u_\infty^\prime + k_* u_1^\prime)
+ \rmO\left(\frac{1}{r^2}\right),
\]
where $u^\prime=u_\vartheta$. Substituting this into the second
equation in (\ref{e:asyabs}), we obtain, after some calculations, the
equation
\begin{equation}\label{e:ss}
k_*^2 D u_\infty^{\prime\prime} + \omega_* u_\infty^\prime
+ f(u_\infty) = 0
\end{equation}
at order $\rmO(1)$ and the equation
\begin{equation}\label{e:formal1}
k_*^2 D u_1^{\prime\prime} + \omega_* u_1^\prime
+ f^\prime(u_\infty) u_1 =
- k_* D (2\theta_0 u_\infty^{\prime\prime} + u_\infty^\prime)
\end{equation}
at order $\rmO(r^{-1})$. We can solve this equation for $u_1$ if and
only if the right-hand side is in the range of the operator
\rev{$\hat{\mathcal{L}}_\mathrm{co}(0)$}, the linearization of the one-dimensional
reaction-diffusion system at the wave train $u_\infty$. Thus, we
need the compatibility condition
\[
\langle u_\mathrm{ad},2\theta_0 D u_\infty^{\prime\prime}
+ D u_\infty^\prime \rangle = 0,
\]
which gives
\begin{equation}\label{e:theta1}
\theta_0 = - \frac{\langle u_\mathrm{ad},D u_\infty^\prime \rangle}%
{2\langle u_\mathrm{ad},D u_\infty^{\prime\prime} \rangle}
= \frac{k_* d_\perp}{c_\mathrm{g}},
\end{equation}
upon using (\ref{e:ndr}) and (\ref{e:dperp}). Substituting this
expression into (\ref{e:formal1}), and using (\ref{e:nunu}) and
(\ref{e:uk}), we see that
\begin{equation}\label{e:u1}
u_1 = a_1 u_\infty^\prime + u_\mathrm{h}, \qquad
u_\mathrm{h} = \theta_0 \partial_k u_\infty - \frac{k_*}{2} u_{\nu\nu},
\end{equation}
where $a_1\in\R$ is arbitrary. Before we proceed, we remark that
\begin{equation}\label{e:comp}
\langle u_\mathrm{ad},2k_*\theta_0 D u_\infty^{\prime\prime\prime}
+ k_* D u_\infty^{\prime\prime}
+ f^{\prime\prime}(u_\infty)[u_\infty^\prime,u_1]\rangle = 0,
\end{equation}
for any $a_1\in\R$, where $u_1$ is given by (\ref{e:u1}). Indeed, taking
the derivative of (\ref{e:formal1}) with respect to $\vartheta$, we
see that
\[
\rev{\mathcal{L}_\mathrm{co}} u_1^{\prime} =
- k_* D (2\theta_0 u_\infty^{\prime\prime\prime} + u_\infty^{\prime\prime})
- f^{\prime\prime}(u_\infty)[u_\infty^\prime,u_1],
\]
so that the right-hand side is in the range of the operator
\rev{$\mathcal{L}_\mathrm{co}$}, which proves (\ref{e:comp}). The same arguments show
that
\begin{equation}\label{e:fdp}
\langle u_\mathrm{ad},
f^{\prime\prime}(u_\infty)[u_\infty^\prime]^2 \rangle = 0,
\end{equation}
upon taking two derivatives of (\ref{e:ss}) with respect to
$\vartheta$.

\paragraph{Rigorous analysis:}
Now that we know the formal solution up to order $\rmO(r^{-2})$, we
begin with the rigorous analysis of (\ref{e:asyabs})
\begin{align*}
u_r = &  -\left(k_*+\frac{\theta_0}{r}\right)\partial_\vartheta u + v \\
v_r = &  -\left(k_*+\frac{\theta_0}{r}\right)\partial_\vartheta v
- \frac{1}{r} v - \frac{1}{r^2} \partial_{\vartheta\vartheta} u
- D^{-1}[\omega_*\partial_\vartheta u + f(u)].
\end{align*}
As in (\ref{e:theta1}), we set $\theta_0=k_*d_\perp/c_\mathrm{g}$, and
seek solutions $\mathbf{u}(r,\vartheta)=(u,v)(r,\vartheta)$ of
(\ref{e:asyabs}) of the form
\begin{equation}\label{e:ffans}
\mathbf{u}(r,\vartheta) =
\mathbf{u}_\infty(\vartheta) + \tilde{\mathbf{w}}(r,\vartheta),
\end{equation}
where $\mathbf{u}_\infty=(u_\infty,k_*u_\infty^\prime)$. Note that
$\mathbf{u}_\infty=(u_\infty(\vartheta),ku_\infty^\prime(\vartheta))$ is
an $r$-independent solution to the asymptotic equation which is
obtained formally by setting $r=\infty$ in (\ref{e:asyabs}). We
substitute the ansatz (\ref{e:ffans}) into (\ref{e:asyabs}) and obtain
\begin{equation}\label{e:ddec}
\tilde{\mathbf{w}}_r =
[\mathcal{A}_\infty + \mathcal{C}(r)] \tilde{\mathbf{w}}
+ \mathcal{C}(r) \mathbf{u}_\infty + \mathcal{G}(\tilde{\mathbf{w}}),
\end{equation}
where
\[
\mathcal{A}_\infty = \left(\begin{array}{cc}
-k_*\partial_\vartheta & 1 \\
- D^{-1}[\omega_*\partial_\vartheta + f^\prime(u_\infty)] &
-k_*\partial_\vartheta \end{array}\right), \qquad
\mathcal{C}(r) = - \frac{1}{r} \left(\begin{array}{cc}
\theta_0\partial_\vartheta & 0 \\
\frac{1}{r}\partial_{\vartheta\vartheta} &
1+\theta_0\partial_\vartheta \end{array}\right)
\]
and
\[
\mathcal{G}(\mathbf{w}) = \mathcal{G}(w_1,w_2) =
\left(\begin{array}{c}
0 \\ -D^{-1}[f(u_\infty+w_1) - f(u_\infty) - f^\prime(u_\infty)w_1]
\end{array}\right),
\]
so that $\mathcal{G}(\mathbf{w})=\rmO(|\mathbf{w}|^2_X)$.

Recall that, by assumption, the center eigenspace of
$\mathcal{A}_\infty$ is one-dimensional and spanned by
$\mathbf{u}_\infty^\prime$. We use the center spectral projection
$P^\mathrm{c}_\infty$ of $\mathcal{A}_\infty$, see \S\ref{s:AL}
and \S\ref{s:edcenter}, and write
\begin{equation}\label{e:ffans2}
\tilde{\mathbf{w}}(r) =
a(r) \mathbf{u}_\infty^\prime + \mathbf{w}(r) + \frac{1}{r} \mathbf{u}_1,
\end{equation}
where $a(r)\in\R$ and
\[
\mathbf{u}_1 = { u_1 \choose k_* u_1^\prime + \theta_0 u_\infty^\prime }.
\]
We require that $\mathbf{w}(r)\in\Rg(P^\mathrm{h}_\infty)$ where
$P^\mathrm{h}_\infty=\id-P^\mathrm{c}_\infty$. Substituting
this ansatz into (\ref{e:ddec}), we obtain
\[
a_r \mathbf{u}_\infty^\prime + \mathbf{w}_r =
[\mathcal{A}_\infty+\mathcal{C}(r)] (a\mathbf{u}_\infty^\prime+\mathbf{w})
+ \mathcal{C}(r) \mathbf{u}_\infty
+ \frac{1}{r} [\mathcal{A}_\infty+\mathcal{C}(r)] \mathbf{u}_1
+ \frac{1}{r^2} \mathbf{u}_1
+ \mathcal{G}(a\mathbf{u}_\infty^\prime+\mathbf{w}+\mathbf{u}_1/r).
\]
Using the definition of $\mathbf{u}_1$, we see that
\[
\mathcal{C}(r) \mathbf{u}_\infty
+ \frac{1}{r} \mathcal{A}_\infty \mathbf{u}_1 =
\frac{1}{r^2} \mathcal{C}_2 \mathbf{u}_\infty, \qquad
\mathcal{C}_2 = \left(\begin{array}{cc} 0 & 0 \\
-\partial_{\vartheta\vartheta} & 0 \end{array}\right),
\]
which gives the system
\[
a_r \mathbf{u}_\infty^\prime + \mathbf{w}_r =
[\mathcal{A}_\infty+\mathcal{C}(r)] (a\mathbf{u}_\infty^\prime+\mathbf{w})
+ \frac{1}{r^2} \mathcal{C}_2 \mathbf{u}_\infty
+ \frac{1}{r} \mathcal{C}(r) \mathbf{u}_1 + \frac{1}{r^2} \mathbf{u}_1
+ \mathcal{G}(a\mathbf{u}_\infty^\prime+\mathbf{w}+\mathbf{u}_1/r),
\]
or
\[
a_r \mathbf{u}_\infty^\prime + \mathbf{w}_r =
[\mathcal{A}_\infty+\mathcal{C}(r)] (a\mathbf{u}_\infty^\prime+\mathbf{w})
+ \mathcal{R}_1(r)
+ \mathcal{G}(a\mathbf{u}_\infty^\prime+\mathbf{w}+\mathbf{u}_1/r),
\]
upon setting
\begin{equation}\label{e:r1}
\mathcal{R}_1(r) =
\frac{1}{r^2} \mathcal{C}_2 \mathbf{u}_\infty
+ \frac{1}{r} \mathcal{C}(r) \mathbf{u}_1 + \frac{1}{r^2} \mathbf{u}_1
= \rmO\left(\frac{1}{r^2}\right).
\end{equation}
Next, we project onto the center and the hyperbolic part using the
spectral projections $P_\infty^\mathrm{c}$ and $P_\infty^\mathrm{h}$
and obtain
\begin{align*}
a_r \mathbf{u}_\infty^\prime = &  P^\mathrm{c}_\infty\left[
\mathcal{C}(r) a \mathbf{u}_\infty^\prime + \mathcal{C}(r)\mathbf{w}
+ \mathcal{R}_1(r)
+ \mathcal{G}(a\mathbf{u}_\infty^\prime+\mathbf{w}+\mathbf{u}_1/r)
\right] \\
\mathbf{w}_r = & 
[\mathcal{A}_\infty+P^\mathrm{h}_\infty\mathcal{C}(r)]\mathbf{w}
+ P^\mathrm{h}_\infty\left[ \mathcal{C}(r) a \mathbf{u}_\infty^\prime
+ \mathcal{R}_1(r)
+ \mathcal{G}(a\mathbf{u}_\infty^\prime+\mathbf{w}+\mathbf{u}_1/r)
\right].
\end{align*}
We rewrite the equation for $a$ using the explicit form of the
projection $P_\infty^\mathrm{c}$ from \S\ref{s:edcenter} which
gives
\begin{align*}
a_r = &  \left\langle \mathbf{u}_\mathrm{ad},
\mathcal{C}(r) a \mathbf{u}_\infty^\prime
+ \mathcal{C}(r)\mathbf{w} + \mathcal{R}_1(r)
+ \mathcal{G}(a\mathbf{u}_\infty^\prime+\mathbf{w}+\mathbf{u}_1/r)
\right\rangle \\ = & 
a\langle\mathbf{u}_\mathrm{ad},\mathcal{C}(r)\mathbf{u}_\infty^\prime\rangle
+ \langle\mathcal{C}(r)^*\mathbf{u}_\mathrm{ad},\mathbf{w}\rangle
+ \langle\mathbf{u}_\mathrm{ad},\mathcal{R}_1(r)\rangle
+ \langle u_\mathrm{ad},
D \mathcal{G}_2(a\mathbf{u}_\infty^\prime+\mathbf{w}+\mathbf{u}_1/r)\rangle.
\end{align*}
We can write the second component of the nonlinearity as
\[
\mathcal{G}_2(a\mathbf{u}_\infty^\prime+\mathbf{w}+\mathbf{u}_1/r) =
- D^{-1} \left[
\frac{1}{2} f^{\prime\prime}(u_\infty)[au_\infty^\prime+u_1/r]^2
- g_1(a,w_1,r) w_1 - g_2(a,r) \right],
\]
where
\begin{equation}\label{e:g12}
g_1(a,w_1,r) = \rmO\left(|a|+|w_1|+1/r\right), \qquad
g_2(a,r) = \rmO\left( (|a|+1/r)^3 \right).
\end{equation}
Hence, we find
\begin{align*}
a_r = & 
a\langle\mathbf{u}_\mathrm{ad},\mathcal{C}(r)\mathbf{u}_\infty^\prime\rangle
+ \langle\mathcal{C}(r)^*\mathbf{u}_\mathrm{ad},\mathbf{w}\rangle
+ \langle\mathbf{u}_\mathrm{ad},\mathcal{R}_1(r)\rangle \\  &
- \frac{1}{2} \langle u_\mathrm{ad},
f^{\prime\prime}(u_\infty)[au_\infty^\prime+u_1/r]^2\rangle
+ \langle u_\mathrm{ad},g_1(a,w_1,r)w_1+g_2(a,r) \rangle \rev{.}
\end{align*}
Using the definition
\begin{equation}\label{e:r2}
\mathcal{R}_2(r) =
\langle\mathbf{u}_\mathrm{ad},\mathcal{R}_1(r)\rangle
- \frac{1}{2r^2} \langle u_\mathrm{ad},
f^{\prime\prime}(u_\infty)[u_1]^2\rangle
= \rmO\left(\frac{1}{r^2}\right)
\end{equation}
and exploiting (\ref{e:fdp}), we find
\begin{align*}
a_r = & 
a \langle\mathbf{u}_\mathrm{ad},\mathcal{C}(r)\mathbf{u}_\infty^\prime\rangle
+ \frac{a}{r} \langle u_\mathrm{ad},
f^{\prime\prime}(u_\infty)[u_\infty^\prime,u_1]\rangle \\ & 
+ \langle\mathcal{C}(r)^*\mathbf{u}_\mathrm{ad},\mathbf{w}\rangle
+ \mathcal{R}_2(r)
+ \langle u_\mathrm{ad},g_1(a,w_1,r)w_1+g_2(a,r) \rangle.
\end{align*}
Writing out the first scalar product, and using the identity
(\ref{e:comp}), we see that the $a/r$ terms actually vanish, so that
we obtain the final equation
\begin{align}
a_r = & 
-\frac{a}{r^2} \langle u_\mathrm{ad},u_\infty^{\prime\prime}\rangle
+ \langle\mathcal{C}(r)^*\mathbf{u}_\mathrm{ad},\mathbf{w}\rangle
+ \mathcal{R}_2(r)
+ \langle u_\mathrm{ad},g_1(a,w_1,r)w_1+g_2(a,r) \rangle
\label{e:aweqn1} \\ \label{e:aweqn2}
\mathbf{w}_r = & 
[\mathcal{A}_\infty+P^\mathrm{h}_\infty\mathcal{C}(r)]\mathbf{w}
+ P^\mathrm{h}_\infty\left[ \mathcal{C}(r) a \mathbf{u}_\infty^\prime
+ \mathcal{R}_1(r)
+ \mathcal{G}(a\mathbf{u}_\infty^\prime+\mathbf{w}+\mathbf{u}_1/r)
\right].
\end{align}
While we could proceed from here on and solve (\ref{e:aweqn1})-(\ref{e:aweqn2}) directly using Banach's fixed-point theorem applied to a corresponding integral equation, we will first simplify the equation further using normal-form transformations as this will help us obtain the higher-order expansions stated in (\ref{e:expand}). To do so, we note that the definitions of the remainders $\mathcal{R}_\ell$ and the nonlinearities $g_\ell$ and $\mathcal{G}$ imply that all terms appearing in (\ref{e:aweqn1})-(\ref{e:aweqn2}) admit a formal expansion in terms of $(1/r,a,\mathbf{w})$. Possibly after modifying the remainder terms $\mathcal{R}_j$, we can also assume that the nonlinearities $g_2$ and $\mathcal{G}$ vanish at $(a,\mathbf{w})=0$. We claim that we can perform a sequence of subsequent transformations
\[
a\mapsto a+\rmO(r^{-j+1}),\qquad \mathbf{w}\mapsto \mathbf{w}+\rmO(r^{-j})
\]
for $j=2,\ldots,K+1$ so that the system (\ref{e:aweqn1})-(\ref{e:aweqn2}) is transformed into a system of the same form, but with remainders $\mathcal{R}_\ell=\rmO(r^{-(K+2)})$. To see this, we proceed inductively and assume  $\mathcal{R}_\ell=\mathcal{R}_{\ell,j}r^{-j}+\rmO(r^{-(j+1)})$. First, the substitution
\begin{equation}\label{e:subsn1}
a_\mathrm{new}=a+\mathcal{R}_{2,j}\frac{r^{-j+1}}{-j+1}
\end{equation}
preserves the general form of the equation for $a_r$ but eliminates terms of order $r^{-j}$ in the inhomogeneous terms $\mathcal{R}_\ell$. In fact, in the first equation only the terms $a_r=\mathcal{R}_2(r)$ yield terms of order $r^{-j}$ after the substitution (\ref{e:subsn1}), so that the choice (\ref{e:subsn1}) for $a_\mathrm{new}$ cancels those terms.
We now substitute this new variable $a_\mathrm{new}$ into the equation for $\mathbf{w}_r$ and collect inhomogeneous terms (terms that vanish for $a=0, \mathbf{w}=0$) in the new remainder $\tilde{\mathcal{R}}_1$. Note that the terms $\mathcal{C}(r)a \mathbf{u}_\infty^\prime$  and $\mathcal{G}(a \mathbf{u}_\infty^\prime\mathbf{u}_1/r)$ contribute 
a new term at order $r^{-j}$, but there are no inhomogeneous terms of lower order. We next remove the term $\tilde{\mathcal{R}}_{1,j}r^{-j}$ using the substitution
\begin{equation}\label{e:subsn2}
\mathbf{w}_\mathrm{new}=\mathbf{w}+\mathcal{A}_\infty^{-1}P_\infty^\mathrm{h}\tilde{\mathcal{R}}_{1,j}r^{-j}. 
\end{equation}
In the equation for $\mathbf{w}_r$, only $\mathcal{A}_\infty\mathbf{w}+P_\infty^\mathrm{h}\mathcal{R}_1(r)$  yield terms of order $r^{-j}$ after the substitution (\ref{e:subsn2}), and those contributions cancel due to the choice of transformation in (\ref{e:subsn2}). Also note that the coefficients $\mathcal{R}_{\ell,j}$ are smooth, a property that is preserved under the transformation (\ref{e:subsn1}-\ref{e:subsn2}). Repeating this change of coordinates shows that  we can assume from now on that $\mathcal{R}_j=\rmO(r^{-(K+2)})$ in (\ref{e:aweqn1})-(\ref{e:aweqn2}) and that nonlinear terms vanish for $(a,\mathbf{w})=0$. 

Our next goal is to derive an integral equation and solve the system with a fixed point argument in appropriate function spaces. We therefore note that Lemma~\ref{r:et} implies
that the principal part
\begin{equation}\label{e:hyppart}
\mathbf{w}_r =
[\mathcal{A}_\infty+P^\mathrm{h}_\infty\mathcal{C}(r)]\mathbf{w}, \qquad
\mathbf{w}\in\Rg(P^\mathrm{h}_\infty)
\end{equation}
of (\ref{e:aweqn2}) has an exponential dichotomy on $X_r$ which we
denote by $\Phi^\mathrm{ss}(r,s)$ and $\Phi^\mathrm{uu}(r,s)$. The
desired integral equation is given by
\begin{align}
a(r) = &  \int_{-\infty}^r \left[
-\frac{a(s)}{s^2} \langle u_\mathrm{ad},u_\infty^{\prime\prime}\rangle
+ \langle\mathcal{C}(s)^*\mathbf{u}_\mathrm{ad},\mathbf{w}(s)\rangle
+ \mathcal{R}_2(s) \right.
\nonumber \\ \label{e:awinteqn}  & \qquad\quad \left.
+ \langle u_\mathrm{ad},g_1(a(s),w_1(s),s)w_1(s)+g_2(a(s),s)\rangle
\right] \rmd s \\ \nonumber
\mathbf{w}(r) = &  \Phi^\mathrm{ss}(r,R) \mathbf{w}^\mathrm{ss}_0
+ \int_R^r \Phi^\mathrm{ss}(r,s) \mathbf{f}(a(s),\mathbf{w}(s),s)\,\rmd s
+ \int_\infty^r \Phi^\mathrm{uu}(r,s) \mathbf{f}(a(s),\mathbf{w}(s),s)\,\rmd s,
\end{align}
where we take $r\in[R,\infty)$, where $\mathbf{w}^\mathrm{ss}_0$ lies in
the stable subspace $\Rg(P^\mathrm{ss}(R))$, and where
\[
\mathbf{f}(a,\mathbf{w},r) = P^\mathrm{h}_\infty\left[
\mathcal{C}(r) a \mathbf{u}_\infty^\prime + \mathcal{R}_1(r)
+ \mathcal{G}(a\mathbf{u}_\infty^\prime+\mathbf{w}+\mathbf{u}_1/r) \right].
\]
We regard (\ref{e:awinteqn}) as a fixed-point equation
\begin{equation} \label{e:fpe}
{ a \choose \mathbf{w} } =
{ \mathcal{F}_1(a,\mathbf{w}) \choose \mathcal{F}_2(a,\mathbf{w}) }, \qquad
(a,\mathbf{w}) \in \mathcal{X}_\varepsilon,
\end{equation}
with parameter $\mathbf{w}^\mathrm{ss}_0$ on the space
\[
\mathcal{X}_\varepsilon := \left\{
(a,\mathbf{w}) \in C^0([R,\infty),\R\times X) ;\;
\|a\| := \sup_{r\geq R} r^{K+\varepsilon}|a(r)| < \infty ,\;
\|\mathbf{w}\| := \sup_{r\geq R} r^{K+1+\varepsilon}
|\mathbf{w}(r)|_{X_r} < \infty
\right\}
\]
equipped with the norm $\|a\|+\|\mathbf{w}\|$, where $\varepsilon\in(0,1)$.
Using the norm on $\mathcal{X}_\varepsilon$ as well as the estimates
(\ref{e:r1}), (\ref{e:g12}) and (\ref{e:r2}), it is not difficult to
check that there is constant $C>0$ such that
\begin{align*}
\|\mathcal{F}_1(a,\mathbf{w})\| & \leq 
C \left[ 1 + \frac{1}{R^\varepsilon}(\|a\|+\|\mathbf{w}\|) \right] \\
\|\mathcal{F}_2(a,\mathbf{w})\| & \leq  K |w^\mathrm{ss}_0|
+ \frac{C}{R} \left[ 1 + \|a\|+\|\mathbf{w}\| \right] \\
\|\mathrm{D}_{(a,\mathbf{w})}\mathcal{F}_1(a,\mathbf{w})\| & \leq 
\frac{C}{R^\varepsilon},
\end{align*}
where $K$ denotes the constant of the exponential dichotomy of
(\ref{e:hyppart}). Indeed, the exponential decay estimates for the
evolution operators $\Phi^{\mathrm{ss}}$ and $\Phi^{\mathrm{uu}}$
imply that the integral operators appearing in $\mathcal{F}_2$
reproduce algebraic weights. Also, due to the embedding
$H^{\frac12}\hookrightarrow L^p$ for any $p<\infty$, the nonlinearities
$g_1$ and $g_2$ define smooth maps from $\mathcal{X}_\varepsilon$ into
itself provided $f$ satisfies certain polynomial growth conditions
which hold after the standard cut-off close to $\mathbf{w}=0$.
Alternatively, we may invoke Remark~\ref{r:halpha} and consider the
equation on a space $H^{\alpha+1/2}\times H^\alpha$ for some
$\alpha>0$ so that $u\in C^0$.

The estimates for $\mathcal{F}$ show that the right-hand side of
(\ref{e:fpe}) is a uniform contraction, which maps the closed subset
\[
\mathcal{Z} = \left\{ (a,\mathbf{w}) \in \mathcal{X}_\varepsilon ;\;
\|a\| < 2C ,\; \|\mathbf{w}\| < \delta \right\}
\]
into itself for any $\mathbf{w}^\mathrm{ss}_0$ with
$|\mathbf{w}^\mathrm{ss}_0|_{X_R}<\delta/2$ and for any $R$ larger than
some $R_*\gg1$ and any $\delta>0$ sufficiently small. Therefore, for
any such $w^\mathrm{ss}_0$, there exists a unique fixed point
$(a,\mathbf{w})$ of (\ref{e:awinteqn}) in $\mathcal{Z}$ that depends smoothly on
$\mathbf{w}^\mathrm{ss}_0$. Exploiting the norm in
$\mathcal{X}_\varepsilon$, we see that $a(r)$ decays with rate
$1/r^{K+\varepsilon}$, while $\mathbf{w}(r)$ decays like $1/r^{K+1+\varepsilon}$ as
$r\to\infty$. 

The family of traces $\mathbf{w}(R)$, considered as a function of
$\mathbf{w}^\mathrm{ss}_0$, describes a graph over
$\Rg(P^\mathrm{ss}(R))$. To describe the manifold
$\mathcal{M}^\mathrm{cs}_+$ as a graph over $\Rg(P^\mathrm{cs}(R))$,
we replace the term $\mathbf{u}_\infty(\vartheta)$ in our ansatz
(\ref{e:ffans}) by $\mathbf{u}_\infty(\vartheta+a_\infty)$ and treat
the asymptotic phase $a_\infty\in\R$ as a parameter (in addition to
$\mathbf{w}^\mathrm{ss}_0$). The right-hand side of the fixed-point
equation (\ref{e:awinteqn}) is then a contraction uniformly in
$(a_\infty,\mathbf{w}^\mathrm{ss}_0)$, and the resulting fixed points
depend smoothly on $(a_\infty,\mathbf{w}^\mathrm{ss}_0)$. This
eventually proves the existence and characterization of the manifold
$\mathcal{M}^\mathrm{cs}_+$ as a graph over $\Rg(P^\mathrm{cs}(R))$ as
desired. As mentioned at the beginning of the proof, smooth dependence
on the external parameter $\mu$ and on the frequency $\omega$ follows
in the same fashion.
\end{Proof}

\begin{Proof}[~of Proposition~\ref{p:exppha}.]
During the proof of the preceding Proposition~\ref{p:stmaff}, we
actually derived the expansion for the solution $u_*$. Reverting the normal form transformations, we find an expansion for $\mathbf{w}$ and $a$ up to any finite order and can then use (\ref{e:ffans2}) and (\ref{e:ffans}) to derive an expansion for $\mathbf{u}$. Next, interpreting $a(r)$ as a phase correction, $\mathbf{u}_\infty+a(r)\mathbf{u}_\infty'=\mathbf{u}_\infty(\cdot+a(r))+\rmO(a(r)^2)$, we find an expansion for $u_*$ as in (\ref{e:expand}). Finally, from the proof of Proposition~\ref{p:stmaff}, we also find the leading-order expansion
\begin{align*}
u_*(r,\psi) = &  u_\infty(k_*r+\theta_*(r)+\psi)
+ \frac{1}{r} u_1(k_*r+\theta_*(r)+\psi)
+ \rmO\left(\frac{1}{r^{2}}\right) \\
\theta_*(r) = &  \frac{k_*d_\perp}{c_\mathrm{g}} \log r
+ \rmO\left(\frac{1}{r}\right) \\
u_1(\vartheta) = &  k_* \left( \frac{d_\perp}{c_\mathrm{g}}
\partial_k u_\infty - \frac{1}{2} u_{\nu_\perp\nu_\perp} \right)
\end{align*}
as claimed in Proposition~\ref{p:exppha}.
\end{Proof}

\begin{Proof}[~of Theorem~\ref{t:rob}.]
We assumed that the spiral wave $u_*$ is transverse (see Definition~\ref{d:transverse}) and therefore know that the generalized kernel of the linearization $\mathcal{L}_*$ about $u_*$ posed on the exponentially weighted space $L^2_\eta$ with $\eta\in J_0(0)=(-\Re\nu_0(0),0)\in\R^-$ is one-dimensional and spanned by $\partial_\psi u_*(r,\varphi)$. We conclude that the tangent spaces to $\mathcal{M}^\mathrm{u}_-(0)$ and $\mathcal{M}^\mathrm{cs}_+(0)$ intersect in a one-dimensional subspace spanned by $\partial_\psi(u_*(R,\cdot),\partial_r u_*(R,\cdot))$. Using the results proved in \S\ref{s:edweights}, we know that the complement of the sum of the tangent spaces is also one-dimensional and spanned by $\mathbf{u}^\perp\in X$, say. To prove persistence of the intersection, we have to compute the derivative of the manifolds $\mathcal{M}^\mathrm{cs,u}_\pm(0)$ with respect to $\omega$ and show that the projection onto $\mathbf{u}^\perp$ of the difference does not vanish. Indeed, this would prove that the linearized equation is onto if we include the parameter $\omega$ as an independent variable. We argue by contradiction and assume that this difference is contained in the sum of the tangent spaces. Using the adjoint evolution operators, which exist due to the results in \S\ref{s:eda}, we see that the function $(0,D^{-1}\partial_\psi u_*(r,\psi))$, the derivative of the difference of the two invariant manifolds with respect to $\omega$, is contained in the range of the operator $\mathcal{T}$ from \S\ref{s:edinv}. Using regularity properties of solutions to $\mathcal{T}\mathbf{u}=\mathbf{f}$ for smooth right-hand sides $\mathbf{f}$, we see that the first component $u$ of $\mathbf{u}$ is a classical solution to $\mathcal{L}_*u=\partial_\psi u_*$. This contradicts the assumption that the generalized kernel of $\mathcal{L}_*$ considered in $L^2_\eta$ with $\eta\in J_0(0)$ has dimension one.
\end{Proof}
    
\begin{Remark}\label{r:spiralsinks}
In the proof of Theorem~\ref{t:rob}, we have seen that spirals that
emit spectrally stable wave trains actually select the frequency of
rotation (and a wavenumber via the inverse nonlinear dispersion
relation). If we had assumed that the group velocity of the spectrally
stable wave trains in the far field is negative, we would have found
spiral waves for an open interval of frequencies
which are selected by the wavenumber of the wave trains that transport
towards the core. These spiral sinks have been found in the
complex Ginzburg--Landau equation \cite{gre}.
\end{Remark}

%%%%%%%%%%%%%%%%%%%%%%%%%%%%%%%%%%%%%%%%%%%%%%%%%%%%%%%%%%%%%%%%%%%%%%%%%

\section{Shape of eigenfunctions, and transverse instabilities}\label{s:ps}

\rev{In \S\ref{s81}, we investigate the spatial shape of eigenfunctions $u$, which satisfy $\mathcal{L}_* u=\lambda u$, and prove the far-field expansions of their profiles in terms of spatial eigenvalues that we formulated in \S\ref{s:mr.exp}.  We focus on the proof of Proposition~\ref{p:asyeig} and note that Proposition~\ref{p:evalgrowth} is an immediate consequence of the spatial-dynamics formulation of the eigenvalue problem and the existence of exponential dichotomies in the far field that we introduced in \S\ref{s:ed}. In \S\ref{s82}, we prove Lemma~\ref{l:spiralperp}, which states that transverse instabilities of the asymptotic wave train prevent the spectral mapping theorem from holding.}

\subsection{Proof of Proposition~\ref{p:asyeig}}\label{s81}

We expand the eigenvalue problem
\begin{align}
u_r = &  - (k_*+\theta_*^\prime(r)) \partial_\vartheta u + v
\nonumber \\ \label{e:evpcon}
v_r = &  - (k_*+\theta_*^\prime(r)) \partial_\vartheta v
- \frac{1}{r} v - \frac{1}{r^2}\partial_{\vartheta\vartheta} u
\\ \nonumber & 
- D^{-1} \left[\omega_*\partial_\vartheta u
+ f^\prime(u_\infty(\vartheta)) u - \lambda u
+ \frac{1}{r} f^{\prime\prime}(u_\infty(\vartheta))[u_1(\vartheta),u]
+ \rmO(r^{-2}) u \right],
\end{align}
at $r=\infty$, where $\theta_*(r)$ is the asymptotic phase of the
spiral wave relative to the emitted wave trains and $u_1$ is the
first-order correction of the profile of the spiral wave; see
Proposition~\ref{p:exppha}. We rewrite this equation as an abstract
equation
\begin{equation}\label{e:evpabs}
\mathbf{u}_r = [\mathcal{A}_\infty + \mathcal{C}(r)] \mathbf{u},
\end{equation}
where
\begin{align}\label{e:acdef}
\mathcal{A}_\infty = &  \left(\begin{array}{cc}
-k_*\partial_\vartheta & 1 \\
- D^{-1}[\omega_*\partial_\vartheta + f^\prime(u_\infty) - \lambda] &
-k_*\partial_\vartheta \end{array}\right), \\ \nonumber
\mathcal{C}(r) = &  - \left(\begin{array}{cc}
\theta_*^\prime(r)\partial_\vartheta & 0 \\
\frac{1}{r^2}\partial_{\vartheta\vartheta} &
\frac{1}{r}+\theta_*^\prime(r)\partial_\vartheta \end{array}\right)
- \frac{1}{r} \left(\begin{array}{cc}
0 & 0 \\ D^{-1} f^{\prime\prime}(u_\infty)[u_1,\cdot] + \rmO(1/r)
& 0 \end{array}\right).
\end{align}
Denote by $P^\mathrm{c}_\infty$ and $P^\mathrm{h}_\infty$ the
complementary spectral projections onto the center and the hyperbolic
subspace, respectively, of the asymptotic equation
\[
\mathbf{u}_r = \mathcal{A}_\infty \mathbf{u},
\]
which represents the eigenvalue problem of the wave trains. Let
$\nu\in\rmi\R$ be the, by assumption unique, Floquet exponent with
$\lambda=\lambda_\mathrm{st}(\nu)$. By hypothesis, $\Rg(P^\mathrm{c}_\infty)$ is
one-dimensional. The results in \S\ref{s:edcenter} show that it
is spanned by
\[
\mathbf{u}^\mathrm{c} = { u \choose (k_*\partial_\vartheta+\nu) u } ,
\]
where $[\hat{\mathcal{L}}_\mathrm{\rev{co}}(\nu)-\lambda_\mathrm{\rev{co}}(\nu)]u=0$ and
$\lambda_\mathrm{\rev{co}}(\nu)=\lambda_\mathrm{st}(\nu)+\omega_*\nu/k_*$.
The results in \S\ref{s:edcenter} also show that the projection
$P^\mathrm{c}_\infty$ is given by
\[
P^\mathrm{c}_\infty =
\frac{1}{\langle\rev{\mathbf{u}_\mathrm{ad}},
\mathbf{u}^\mathrm{c}\rangle}
\langle\rev{\mathbf{u}_\mathrm{ad}},\cdot\rangle
\mathbf{u}^\mathrm{c}, \qquad
\rev{\mathbf{u}_\mathrm{ad}} =
{ D(-k_*\partial_\vartheta+\nu)u_\mathrm{ad} \choose D u_\mathrm{ad} },
\]
where
$[\hat{\mathcal{L}}_\mathrm{\rev{co}}^\mathrm{ad}(\nu)-\lambda_\mathrm{\rev{co}}(\nu)]u_\mathrm{ad}=0$.
Thus, given the eigenfunction $\mathbf{u}$ of the spiral wave, we write
\[
\mathbf{u} = P^\mathrm{c}_\infty \mathbf{u} + P^\mathrm{h}_\infty \mathbf{u}
=: a(r) \mathbf{u}^\mathrm{c} + \mathbf{w}
\]
so that (\ref{e:evpabs}) becomes
\begin{align}
a_r \mathbf{u}^\mathrm{c} = & 
a [\nu + P^\mathrm{c}_\infty \mathcal{C}(r)] \mathbf{u}^\mathrm{c}
+ P^\mathrm{c}_\infty \mathcal{C}(r) \mathbf{w}
\label{e:ffefa} \\ \label{e:ffefw}
\mathbf{w}_r = & 
[\mathcal{A}_\infty + P^\mathrm{h}_\infty \mathcal{C}(r)] \mathbf{w}
+ a P^\mathrm{h}_\infty \mathcal{C}(r) \mathbf{u}^\mathrm{c}.
\end{align}
Since the second summand in the definition (\ref{e:acdef}) of
$\mathcal{C}(r)$ converges to zero in norm as $r\to\infty$, we can
apply Proposition~\ref{p:edrobust} and Lemma~\ref{r:et} which show
that the equation
\[
\mathbf{w}_r =
[\mathcal{A}_\infty + P^\mathrm{h}_\infty \mathcal{C}(r)] \mathbf{w}
\]
has an exponential dichotomy on $[R,\infty)$ for $R\gg1$ sufficiently
large. Thus, (\ref{e:ffefw}) is equivalent to
\begin{equation}\label{e:ffefwint}
\mathbf{w}(r) = \Phi^\mathrm{ss}(r,R) \mathbf{w}^\mathrm{ss}_0
+ \int_R^r a(s) \Phi^\mathrm{ss}(r,s)
P^\mathrm{h}_\infty \mathcal{C}(s) \mathbf{u}^\mathrm{c}\,\rmd s
+ \int_\infty^r a(s) \Phi^\mathrm{uu}(r,s)
P^\mathrm{h}_\infty \mathcal{C}(s) \mathbf{u}^\mathrm{c}\,\rmd s.
\end{equation}
To analyze the equation (\ref{e:ffefa}) for $a(r)$, we first neglect
the coupling term involving $\mathbf{w}$. The remaining equation is
\[
a_r = \left[ \nu +
\frac{\langle\mathbf{u}_\mathrm{ad},
\mathcal{C}(r)\mathbf{u}^\mathrm{c}\rangle}%
{\langle\mathbf{u}_\mathrm{ad},\mathbf{u}^\mathrm{c}\rangle} \right] a.
\]
This expression can be evaluated as in the proof of
Proposition~\ref{p:stmaff}, and we obtain
\begin{equation}\label{e:aflow}
a_r =
\left[ \nu - \frac{1}{r}\; \frac{\langle u_\mathrm{ad},
[2\theta_0\partial_\vartheta+1] D v
+ f^{\prime\prime}(u_\infty)[u_1,u]\rangle}{2\langle u_\mathrm{ad},Dv\rangle}
+ \rmO\left(\frac{1}{r^2}\right) \right] a,
\end{equation}
where $\theta_0$ is given by $\theta_0=k_*d_\perp/c_\mathrm{g}$, see
(\ref{e:theta1}). Recall from (\ref{e:cgl}) that the linear group
velocity of $\lambda_\mathrm{st}(\nu)$ is given by
\[
c_\mathrm{g,l} =
-\frac{2\langle u_\mathrm{ad},Dv\rangle}{\langle u_\mathrm{ad},u\rangle}.
\]
In particular, the denominator in (\ref{e:aflow}) is non-zero.
Integrating (\ref{e:aflow}) gives
\begin{equation}\label{e:ffa}
a(r) = a_0 r^\alpha \rme^{\nu r}
\left[1+\rmO\left(\frac{1}{r}\right)\right],
\end{equation}
for some $a_0\in\C$ where
\[
\alpha = \frac{\langle u_\mathrm{ad},
[(2k_*d_\perp/c_\mathrm{g})\partial_\vartheta+1] D v
+ f^{\prime\prime}(u_\infty)[u_1,u]\rangle}%
{c_\mathrm{g,l}\,\langle u_\mathrm{ad},u\rangle}.
\]
We claim that $a_0\neq0$. Indeed, if $a_0$ were zero, the expansion of
the eigenfunction $u(r,\psi)$ of the spiral wave would only involve
the solution $\mathbf{w}$ of the hyperbolic part. As a consequence, due
to (\ref{e:ffefwint}), $u(r,\psi)$ would decay exponentially as
$r\to\infty$ so that the null space of $\mathcal{L}_*-\lambda$ would
be non-trivial in $L^2_\eta$ for any sufficiently small $\eta>0$. This
contradicts our hypotheses. Thus, $a_0\neq0$.

Using (\ref{e:ffa}), we can therefore also integrate equation
(\ref{e:ffefa}). Putting the resulting equation and (\ref{e:ffefwint})
together, we see that (\ref{e:ffefa})-(\ref{e:ffefw}) is equivalent to
\begin{align}\label{e:efinteqn}
a(r) = &  a_0 r^\alpha \rme^{\nu r}
+ \int_\infty^r \frac{r^\alpha}{s^\alpha} \rme^{\nu(r-s)}
\left[ \rmO\left(\frac{1}{s^2}\right) a(s)
+ \frac{\langle\mathcal{C}(s)^*\mathbf{u}_\mathrm{ad},\mathbf{w}(s)\rangle}%
{\langle\mathbf{u}_\mathrm{ad},\mathbf{u}^\mathrm{c}\rangle} \right] \,\rmd s
\\ \nonumber
\mathbf{w}(r) = &  \Phi^\mathrm{ss}(r,R) \mathbf{w}^\mathrm{ss}_0
+ \int_R^r a(s) \Phi^\mathrm{ss}(r,s)
P^\mathrm{h}_\infty \mathcal{C}(s) \mathbf{u}^\mathrm{c}\,\rmd s
+ \int_\infty^r a(s) \Phi^\mathrm{uu}(r,s)
P^\mathrm{h}_\infty \mathcal{C}(s) \mathbf{u}^\mathrm{c}\,\rmd s,
\end{align}
where the $\rmO(1/r^2)$ term in the equation for $a$ coincides with
the corresponding term in (\ref{e:aflow}). Note that
\begin{equation}\label{e:efffpest}
\left|\frac{\langle\mathcal{C}(r)^*\mathbf{u}_\mathrm{ad},\mathbf{w}\rangle}%
{\langle\mathbf{u}_\mathrm{ad},\mathbf{u}^\mathrm{c}\rangle}\right|
\leq C \frac{|\mathbf{w}|_X}{r}, \qquad
\left|P^\mathrm{h}_\infty\mathcal{C}(r)\mathbf{u}^\mathrm{c}\right|_X
\leq \frac{C}{r}
\end{equation}
for some constant $C>0$ that is independent of $r$.

We regard (\ref{e:efinteqn}) as the fixed-point equation
\[
{ a \choose \mathbf{w} } =
{ \mathcal{F}_1(a,\mathbf{w}) \choose \mathcal{F}_2(a,\mathbf{w}) }, \qquad
(a,\mathbf{w}) \in \mathcal{X}_\varepsilon
\]
on the space
\[
\mathcal{X}_\varepsilon = \left\{
(a,\mathbf{w}) \in C^0([R,\infty),\R\times X) ;\;
\|a\| := \sup_{r\geq R} r^{-\alpha} |a(r)| < \infty ,\;
\|\mathbf{w}\| := \sup_{r\geq R} r^{1-\alpha-\varepsilon}
|\mathbf{w}(r)|_{X_r} < \infty \right\},
\]
equipped with the norm $\|a\|+\|\mathbf{w}\|$, where $\varepsilon\in(0,1)$
is fixed. Using that
\[
\int_1^r \left(\frac{r}{s}\right)^\alpha \rme^{-(r-s)} \,\rmd s \to 1
\quad\mbox{as}\quad r\to\infty
\]
for any $\alpha\in\R$ (which follows from the fact that the integral
on the left-hand side satisfies the differential equation
$b_r=1+(\alpha/r-1)b$), and exploiting the estimate
(\ref{e:efffpest}), it is not difficult to check that there is a
constant $C>0$ such that
\begin{align*}
\|\mathcal{F}_1(a,\mathbf{w})\| & \leq 
|a_0| + C\left[\frac{\|a\|}{R}+\frac{\|w\|}{R^{1-\varepsilon}}\right] \\
\|\mathcal{F}_2(a,\mathbf{w})\| & \leq 
C\left[|\mathbf{w}^\mathrm{ss}_0|_{X_R}+\frac{\|a\|}{R^\varepsilon}\right].
\end{align*}
Thus, there exists a unique solution of (\ref{e:efffpest}) provided we
choose $R\gg1$ sufficiently large. This solution is given by
\[
a(r) = a_0 r^\alpha \rme^{\nu r}
\left[1+\rmO\left(\frac{1}{r^{1-\varepsilon}}\right)\right], \qquad
\mathbf{w}(r) = \rmO\left(\frac{1}{r^{1+\alpha}}\right).
\]
Now that we have obtained a solution in the weighted space, we can
substitute $\mathbf{w}(r)$ back into the integral equation for $a$, and
we see that, in fact,
\[
a(r) = a_0 r^\alpha \rme^{\nu r}
\left[1+\rmO\left(\frac{1}{r}\right)\right].
\]
This completes the proof of Proposition~\ref{p:asyeig}.

\subsection{Proof of Lemma~\ref{l:spiralperp}}\label{s82}

Decay estimates for strongly continuous semigroups are tied to uniform estimates for the resolvent of their generator on vertical lines. It follows from \rev{\cite[Theorem~II.1.10(iii)]{engelnagel}} that it suffices to find $\lambda_n\in\C$ with $\Re\lambda_n=\Re\lambda_*$ and nonzero elements $u_n\in L^2(\R^2,\C^N)$ such that
\[
\frac{1}{|u_n|_{L^2}} |(\mathcal{L}_*-\lambda_n) u_n|_{L^2} \longrightarrow 0 \mbox{ as } n\longrightarrow\infty
\]
to prove the statement of the lemma. We first choose a smooth cutoff function $\chi(x)$ such that
\[
\chi(x) \left\{ \begin{array}{lclcl}
& = &   1 && |x|\leq1\\
& \in & [0,1] && 1\leq |x|\leq2\\
& = &   0 && |x|\geq2.\\
\end{array}\right.
\]
Next, we let
\[
u_n(r,\psi) := \rme^{\rmi n^2\psi} v_\infty(kr+\theta_*(r)+\psi) \chi\left(\frac{\gamma r-n^2}{n}\right), \qquad
\lambda_n:=\lambda_*+\rmi\omega_* n^2.
\]
Note that $\Re\lambda_n=\Re\lambda_*$ for all $n$ as required and
\[
\frac{n^2}{r} = \gamma + \rmO(1/n) \mbox{ whenever } \chi\left(\frac{\gamma r-n^2}{n}\right)\neq0
\]
uniformly in $r\geq0$. Furthermore, the support of $u_n$ lies in an annulus of diameter $2n/\gamma$ centered at $r=n^2/\gamma$. It follows that there is a constant $C_0>0$ so that
\[
|u_n|_{L^2(\R^2,\C^N)} \geq C_0 n^{\frac32} |v_\infty|_{L^2(S^1,\C^N)}
\]
for all $n\gg1$. Next, using (\ref{e:expand}), we find
\begin{eqnarray*}
(\mathcal{L}_*-\lambda_n) u_n & = &
\left[ D\left(\partial_{rr} + \frac{1}{r} \partial_r + \frac{1}{r^2} \partial_{\psi\psi}\right) + \omega_* \partial_\psi + f^\prime(u_*(r,\psi)) - \lambda_n \right] u_n(r,\psi) \\ & = &
\rme^{\rmi n^2\psi} \left[ D\left( k^2 v_\infty^{\prime\prime} \chi + \rmO(1/n) - \frac{n^4}{r^2} v_\infty \chi \right) + \rmi \omega_* n^2 v_\infty \chi + \omega_* v_\infty^\prime \chi + \rmO(1/n) \right. \\ && \left.
+ f^\prime(u_\infty(kr+\theta_*(r)+\psi)+\rmO(1/n^2)) v_\infty \chi - (\lambda_* +\rmi\omega_* n^2) v_\infty \chi \right] \\ & = &
\rme^{\rmi n^2\psi} \left[ \chi \left( D(k^2 v_\infty^{\prime\prime} - \gamma^2 v_\infty) + \omega_* v_\infty^\prime + f^\prime(u_\infty(kr+\theta_*(r)+\psi)) v_\infty - \lambda_* v_\infty \right) + \rmO(1/n) \right] \\ & = &
\rmO(1/n),
\end{eqnarray*}
where we used (\ref{e:tr}) to obtain the last identity. Since $(\mathcal{L}_*-\lambda_n) u_n$ has the same support as $u_n$, we see that there is a constant $C_1>0$ so that
\[
|(\mathcal{L}_*-\lambda_n) u_n|_{L^2(\R^2,\C^N)} \leq C_1 n^{\frac12} |v_\infty|_{L^2(S^1,\C^N)},
\]
for all $n\gg1$. We conclude that
\[
\frac{1}{|u_n|_{L^2}} |(\mathcal{L}_*-\lambda_n) u_n|_{L^2} \leq \frac{C_1}{C_0 n},
\]
which completes the proof of Lemma~\ref{l:spiralperp}.

%%%%%%%%%%%%%%%%%%%%%%%%%%%%%%%%%%%%%%%%%%%%%%%%%%%%%%%%%%%%%%%%%%%%%%%%%

\section{Spiral waves on large finite disks}\label{s:trunc}

In this section, we prove Theorem~\ref{t:trunc}, \rev{which states that planar spiral waves persist under domain truncation to large bounded disks provided that the boundary conditions can be accommodated via boundary sinks.}
First, we prepare the
actual proof by discussing in \S\ref{s:truncbs} the
boundary \rev{sinks} whose existence we assumed in
Theorem~\ref{t:trunc}. In \S\ref{s:truncc}, we construct
solutions to the spatial-dynamics formulation
\begin{align*}
u_r = &  -[k+\theta^\prime(r)]\partial_\vartheta u + v \\
v_r = &  -[k+\theta^\prime(r)]\partial_\vartheta v
- \frac{1}{r} v - \frac{1}{r^2} \partial_{\vartheta\vartheta} u
- D^{-1}[\omega\partial_\vartheta u + f(u)]
\end{align*}
separately in the core region, the far-field region and the
boundary-layer region. These solutions are then matched in
\S\ref{s:truncm} in the transitions zones between core, far
field, and the boundary \rev{sink}.

\subsection{Boundary sinks}\label{s:truncbs}

Recall that we assumed that there is a solution
\rev{$u(x,t)=u_\mathrm{bs}(x,\omega_*t)$} of the reaction-diffusion
equation
\begin{align}\label{e:1drds}
u_t = D u_{xx} + f(u), & \qquad  x\in(-\infty,0), \\ \nonumber
(u,u_x)(0,t) \in E^\mathrm{bc}_0, &  \qquad t>0,
\end{align}
where $u_\mathrm{bs}(x,\tau)$ is $2\pi$-periodic in $\tau$ with
\[
|u_\mathrm{bs}(x,\cdot) - u_\infty(k_*x-\cdot)|_{H^1(S^1)}\to0
\quad\mbox{ as }\quad x\to-\infty
\]
and $E^\mathrm{bc}_0\subset\R^{2N}$ is an $N$-dimensional subspace.
We denote by $\Phi_t(u_0)$ the semiflow associated with
(\ref{e:1drds}) on $H^1(\R^-,\R^N)$. Since we assumed that the
asymptotic wave trains are spectrally stable and have positive group
velocity, we know that the Fredholm index of the linearization
$\Psi_\mathrm{bs}=\mathrm{D}\Phi_{2\pi/\omega_*}(u_\mathrm{bs}(\cdot,0))$ is
$+1$ in the region to the left of the Floquet spectrum at
$\lambda=0$. We also assumed that \rev{the linearization of (\ref{e:1drds}) about the boundary sink $u_\mathrm{bs}(x,\omega_*t)$ does not have a solution that decays to zero exponentially as $x\to-\infty$.}

Next, we interpret these hypotheses in terms of the spatial-dynamics
formulation
\begin{align}\label{e:ty}
u_x = &  v \\ \nonumber
v_x = &  - D^{-1}[-\omega\partial_\tau u + f(u)],
\end{align}
where $\mathbf{u}(x)=(u,v)(x)\in Y=H^{\frac12}(S^1)\times L^2(S^1)$ for
all $x\in(-\infty,0)$ with the boundary condition
\begin{equation}\label{e:tybc}
\mathbf{u}(0) \in E^\mathrm{bc}_Y :=
\{ (u,v)\in Y;\; (u(\tau),v(\tau))\in E^\mathrm{bc}_0 \;\forall\tau \}.
\end{equation}
Note that
$\mathbf{u}_\mathrm{bs}:=(u_\mathrm{bs},\partial_x u_\mathrm{bs})$ is
a solution to (\ref{e:ty})-(\ref{e:tybc}) for $\omega=\omega_*$.
Furthermore, the assumptions on \rev{spectrum and group velocity of the asymptotic wave train} imply that the
linearization
\begin{align*}
u_x = &  v \\
v_x = & 
- D^{-1}[-\omega_*\partial_\tau u + f^\prime(u_\mathrm{bs})u]
\end{align*}
of (\ref{e:ty}) at $\mathbf{u}_\mathrm{bs}$ has an exponential
dichotomy on $Y$ with strong unstable projections $P^\mathrm{uu}(x)$ and
center-unstable $P^\mathrm{cu}(x)$ both defined for $x\leq0$.
In addition, \rev{the assumption that the linearization of (\ref{e:1drds}) about $u_\mathrm{bs}(x,\omega_*t)$ does not have an exponentially decaying solution implies the transversality conditions}
\begin{equation}\label{e:tpf}
\Rg(P^\mathrm{cu}(0)) \pitchfork E^\mathrm{bc}_Y =
\{\partial_\tau\mathbf{u}_\mathrm{bs}\}, \qquad
\Rg(P^\mathrm{uu}(0)) \pitchfork E^\mathrm{bc}_Y = \{0\}.
\end{equation}
\rev{We show in the next lemma that these transversality properties imply that the boundary sink, whose existence we assumed only for the fixed temporal frequency $\omega_*$, is robust so that it persists when we change $\omega$ from $\omega_*$ to nearby values.}

\begin{Lemma}\label{l:tr}
Up to the time-shift symmetry, equations (\ref{e:ty})-(\ref{e:tybc})
have a locally unique solution $\mathbf{u}_\mathrm{bs}(x;\omega)$ for each
$\omega$ close to $\omega_*$. Furthermore, there is a $\kappa>0$ and a
constant $C$ such that
\[
|\mathbf{u}_\mathrm{bs}(x,\cdot;\omega)
-\mathbf{u}_\infty(kx-\cdot;\omega)|_{Y}
\leq C \rme^{-\kappa|x|}, \qquad x\leq0,
\]
the solutions $\mathbf{u}_\mathrm{bs}(x;\omega)$ depend smoothly on                                                                                                                                                                                                                                                                 
$\omega$, and the linearization of (\ref{e:ty}) at each
$\mathbf{u}_\mathrm{bs}(x;\omega)$ has an exponential dichotomy on $Y$ that
satisfies (\ref{e:tpf}).
\end{Lemma}

\begin{Proof}
The proof involves two steps. First, we compute the strong unstable
fibers of the asymptotic wave trains
$\mathbf{u}_\infty(kx-\omega(k)t;k)$ (note that we can switch forth
and back between parametrizing solutions via $k$ or $\omega$ since the
group velocity $\omega^\prime(k_*)$ is not zero). In the second step, we
match the strong unstable fibers and the boundary condition using the
boundary-sink solution $\mathbf{u}_\mathrm{bs}(x,\tau)$.
\rev{The existence of exponential dichotomies of the linearization about each of the boundary sinks and the transversality property (\ref{e:tpf}) follow from the robustness theorem for exponential dichotomies.}

After rescaling $x\to\sqrt{\omega}x$,  the original reaction-diffusion
equation (\ref{e:ty}) reads
\begin{align}\label{e:tuu}
u_x = &  v \\ \nonumber
v_x = &  - D^{-1}\left[-\partial_\tau u + \frac{1}{\omega} f(u)\right].
\end{align}
We write
$\mathbf{u}(x,\tau)=\mathbf{u}_\infty(kx-\tau;\omega)+\mathbf{w}(x,\tau)$,
then $\mathbf{w}(x)$ converges to zero as $x\to-\infty$ if and only
if it satisfies the integral equation
\begin{equation}\label{e:wuu}
\mathbf{w}(x) = \Phi^\mathrm{uu}(x,-R)\mathbf{w}^\mathrm{uu}_0
+ \int_{-R}^x \Phi^\mathrm{uu}(x,\xi)\mathcal{G}(\xi,\mathbf{w}(\xi))\,\rmd \xi
+ \int_{-\infty}^x \Phi^\mathrm{cs}(x,\xi)\mathcal{G}(\xi,\mathbf{w}(\xi))\,\rmd \xi
\end{equation}
on $Y$ for $x\in(-\infty,-R]$, where
\[
\mathcal{G}(x,\mathbf{w}) = \frac{1}{\omega}
{ 0 \choose -D^{-1}[f(u_\infty(x;\omega)+w_1)-f(u_\infty(x;\omega))
-f^\prime(u_\infty(x;\omega))w_1] }
\]
and $\Phi^\mathrm{uu}$ and $\Phi^\mathrm{cs}$ denote the
$\omega$-dependent exponential dichotomies of the linearization of
(\ref{e:tuu}) at the wave trains $\mathbf{u}_\infty(\cdot;\omega)$.
We denote by $\kappa$ the exponential rate associated with this
dichotomy. We can solve (\ref{e:wuu}) on an appropriate function space
with exponential weight $\rme^{\kappa|x|}$, using a contraction mapping
theorem and choosing $R\gg 1$ sufficiently large. Since the exponential dichotomies \cite{pss} as well as the
other terms in (\ref{e:wuu}) depend smoothly on $\omega$, so do the
strong unstable fibers which are the fixed points of (\ref{e:wuu}).

The second step is carried out analogously by writing down an integral
equation on $[-R,0]$ and using the exponential dichotomies of the
linearization of (\ref{e:tuu}) at the sink
$\mathbf{u}_\mathrm{bs}$. The fact that the subspace $E^\mathrm{bc}_Y$
and the range of the center-unstable projection of the sink intersect
transversely by (\ref{e:tpf}) allows us to then solve the resulting
integral equation for all $\omega$ close to $\omega_*$. We omit the
details.
\end{Proof} 

% boundary conditions on Y and on X?
% what does ``pointwise'' mean?
% assume that E^bc_1 is closed in Y and in X?

Next, we transform the above solutions and statements into Archimedean
coordinates. Thus, we define a new function
$\hat{\mathbf{u}}_\mathrm{bs}(x,\vartheta)$ by
\[
\hat{\mathbf{u}}_\mathrm{bs}(x,\vartheta) :=
\mathbf{u}_\mathrm{bs}(x,k_*x-\vartheta), \qquad\mbox{i.e.}\qquad
\mathbf{u}_\mathrm{bs}(x,\tau) = 
\hat{\mathbf{u}}_\mathrm{bs}(x,k_*x-\tau),
\]
so that
\begin{equation}\label{e:tblcc}
|\hat{\mathbf{u}}_\mathrm{bs}(x,\vartheta;\omega)
-\mathbf{u}_\infty(\vartheta;\omega)| \leq C \rme^{-\kappa|x|}
\end{equation}
as $x\to-\infty$. The boundary conditions remain unchanged since $\mathbf{u}(0)\in E^\mathrm{bc}_Y$ means that $\mathbf{u}(0,\tau)\in E^\mathrm{bc}_0$ pointwise in $\tau$. Hence, dropping the \rev{hats} and using $\rho$ instead of $x$, we see that
$\mathbf{u}_\mathrm{bs}(\rho;\omega)$ satisfies the system
\begin{align}\label{e:trho}
\partial_\rho u = &  -k_*\partial_\vartheta u + v \\ \nonumber
\partial_\rho v = &  -k_*\partial_\vartheta v
- D^{-1}[\omega\partial_\vartheta u + f(u)]
\end{align}
for $\omega$ close to $\omega_*$, where $\rho\in(-\infty,0)$ and
$\mathbf{u}(\rho)=(u,v)(\rho)\in Y$ satisfies
$\mathbf{u}(0)\in E^\mathrm{bc}_Y$. Since the only difference between
(\ref{e:ty}) and (\ref{e:trho}) is the appearance in (\ref{e:trho}) of
the generator of the shift in $\vartheta$, we still have
transversality of $E^\mathrm{bc}_Y$ and the range
$\Rg(P^\mathrm{uu}_\mathrm{bs}(0))$ of the exponential dichotomy of
the linearization of (\ref{e:trho}) at $\mathbf{u}_\mathrm{bs}$
(this can be proved as in \S\ref{s:equiv} using the detour via
the corresponding operators $\mathcal{T}$).

\subsection{Construction of core, far-field and boundary-layer solutions}
\label{s:truncc}

We can now address the existence of solutions to the equation
\begin{align}\label{e:tsp}
u_r = &  -[k+\theta^\prime(r)]\partial_\vartheta u + v
\\ \nonumber
v_r = &  -[k+\theta^\prime(r)]\partial_\vartheta v
- \frac{1}{r} v - \frac{1}{r^2} \partial_{\vartheta\vartheta} u
- D^{-1}[\omega\partial_\vartheta u + f(u)],
\end{align}
where $\mathbf{u}(r)=(u,v)(r)\in X_r$ for $r\in(0,R)$ with the
boundary condition
\begin{equation}\label{e:tbc}
\mathbf{u}(R) = (u,v)(R) \in E^\mathrm{bc}.
\end{equation}
Solutions to (\ref{e:tsp})-(\ref{e:tbc}) correspond to
rigidly-rotating spiral waves on the disk of radius $R$.

\begin{figure}
\centering\includegraphics[width=.7\textwidth]{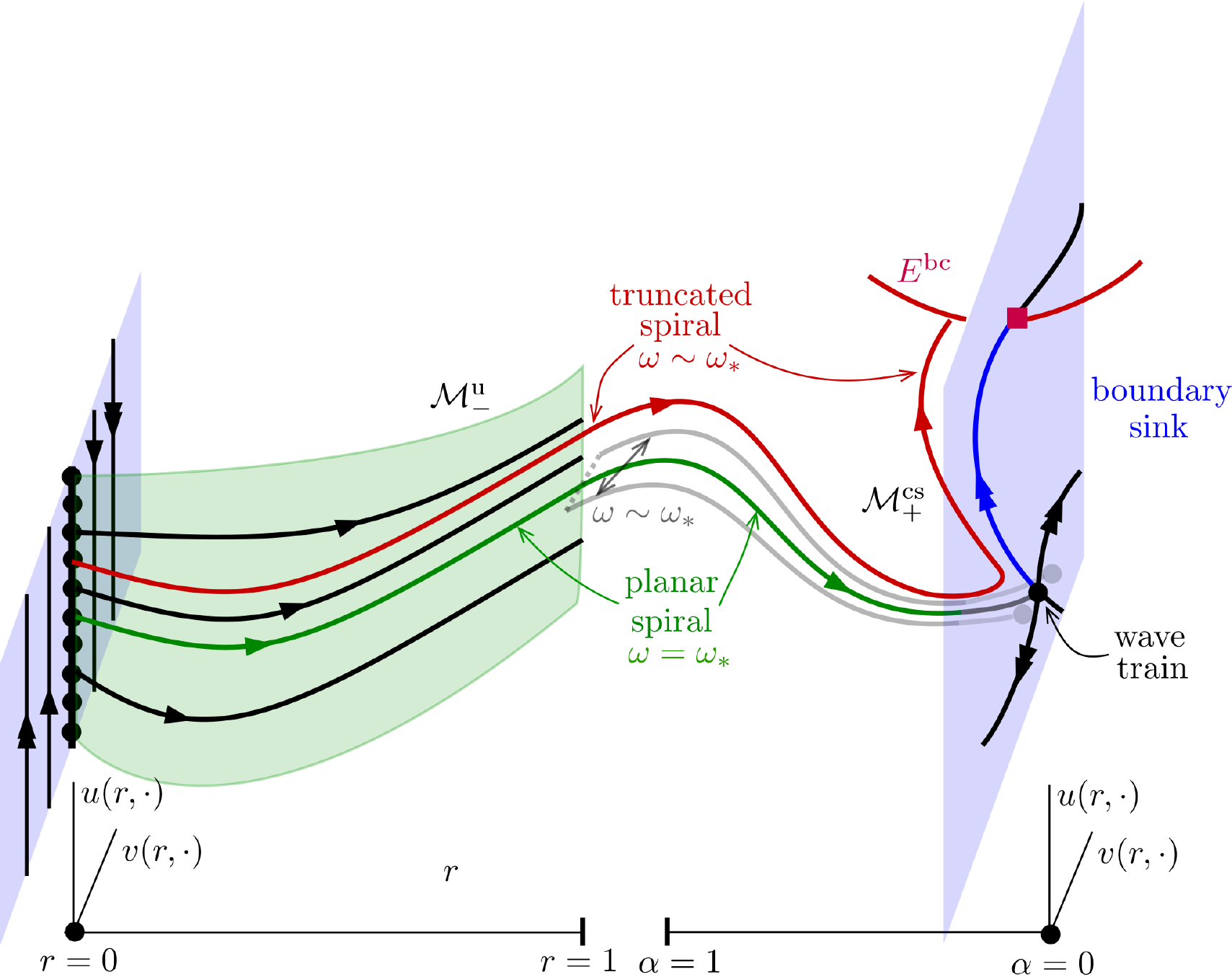}
\caption{Gluing ingredients for the construction of spiral waves in finite-size disks. Planar spiral (green) consisting of core region and far-field region, constructed as a transverse heteroclinic as $\omega$ varies near $\omega_*$ to the wave train solution at $r=\infty$, $\alpha=0$. In $\alpha=0$, the boundary sink is contained in the unstable manifold of wave trains as a transverse intersection of the unstable manifold of wave trains (2-dimensional in picture) with the boundary condition $E^\mathrm{bc}$ (0-dimensional in picture). A truncated spiral follows the green and blue curve to intersect the extension of $E^\mathrm{bc}$ outside of $\alpha=0$ at $\alpha=\frac{1}{R}>0$.}\label{f:spiral_bdy_sink}
\end{figure}

To show the existence of these spirals, we will, for some fixed
$R_*\gg1$, construct solutions to (\ref{e:tsp})-(\ref{e:tbc}) in the
core region $(0,R_*)$, the far field $(R_*,R-\kappa^{-1}\log R)$, and
the boundary layer $(R-\kappa^{-1}\log R,R)$ for $R\gg1$. These
solutions are then matched at $r=R_*$ and $r=R-\kappa^{-1}\log R$. The
constant $R_*$ will be chosen as in Proposition~\ref{p:stmaff}, while
$\kappa>0$ is as in Lemma~\ref{l:tr}.
A sketch of the construction in phase space is shown in Figure~\ref{f:spiral_bdy_sink}.

\paragraph{Core region.}

We begin by discussing (\ref{e:tsp}) in the core region $(0,R_*)$. Proposition~\ref{p:core} shows that the relevant solutions to (\ref{e:tsp}) on $(0,R_*)$ are those that have initial data in the manifold $\mathcal{M}^\mathrm{u}_-(\omega)$. \rev{Since we can write $\mathcal{M}^\mathrm{u}_-(\omega)$ near $u_*(R_*)$ as a graph of a map from $\Rg(\widehat{P}^\mathrm{u}_-(R_*))$ into $\Rg(\widehat{P}^\mathrm{s}_-(R_*))$, we can parametrize the elements of $\mathcal{M}^\mathrm{u}_-(\omega)$ as functions of $\mathbf{w}_\mathrm{core}^\mathrm{u}\in\Rg(\widehat{P}^\mathrm{u}_-(R_*))$ via}
\begin{equation}\label{e:tcorepar}
\mathbf{u}_\mathrm{core}(\rev{R}_*;\omega,\mathbf{w}_\mathrm{core}^\mathrm{u})
\in\mathcal{M}^\mathrm{u}_-(\omega), \qquad
\mathbf{u}_\mathrm{core}(R_*;\omega,\mathbf{w}_\mathrm{core}^\mathrm{u})
- \mathbf{u}_*(R_*) - \mathbf{w}_\mathrm{core}^\mathrm{u}
\in\Rg(\widehat{P}^\mathrm{s}_-(R_*)).
\end{equation}

\paragraph{Far-field region.}

Next, we consider solutions to (\ref{e:tsp}) in the far field for
$r\in(R_*,R-\kappa^{-1}\log R)$. Proposition~\ref{p:stmaff} shows that
the center-stable manifold $\mathcal{M}^\mathrm{cs}_+(\omega)$ of the
asymptotic wave trains is smooth in $\omega$ and can be parametrized
as \rev{the graph of a map from $\Rg(P^\mathrm{cs}_+(R_*))$ into $\Rg(P^\mathrm{uu}_+(R_*))$. We can then use a rotation of the planar spiral wave $u_*(R_*,\cdot)$ by an angle $\alpha$ to parametrize the center direction in $\Rg(P^\mathrm{cs}_+(R_*))$ and use vectors $\mathbf{w}_\mathrm{ff}^\mathrm{ss}\in\Rg(P^\mathrm{ss}_+(R_*))$ close to zero to parametrize the remaining strong stable directions.}
Thus, for $R_*\leq r\leq R-\kappa^{-1}\log R$, we \rev{can} write
\begin{equation}\label{e:tffans}
\mathbf{u}(r) =
\mathbf{u}_\mathrm{ff}(r;\omega,\alpha,\mathbf{w}_\mathrm{ff}^\mathrm{ss})
+ \mathbf{w}_\mathrm{ff}(r),
\end{equation}
where
$\mathbf{u}_\mathrm{ff}(r;\omega,\alpha,\mathbf{w}_\mathrm{ff}^\mathrm{ss})$
denotes the solution of (\ref{e:tsp}) on $[R_*,\infty)$ with
\begin{equation}\label{e:tffpar}
\mathbf{u}_\mathrm{ff}(R_*;\omega,\alpha,\mathbf{w}_\mathrm{ff}^\mathrm{ss})
\in\mathcal{M}^\mathrm{cs}_+(\omega), \qquad
\mathbf{u}_\mathrm{ff}(R_*;\omega,\alpha,\mathbf{w}_\mathrm{ff}^\mathrm{ss})
- \mathbf{u}_*(R_*,\cdot+\alpha) - \mathbf{w}_\mathrm{ff}^\mathrm{ss}
\in\Rg(P^\mathrm{uu}_+(R_*)),
\end{equation}
\rev{for $\mathbf{w}_\mathrm{ff}^\mathrm{ss}\in\Rg(P^\mathrm{ss}_+(R_*))$}. We
see that $\mathbf{u}(r)$ is a solution to (\ref{e:tsp}) if and only
if $\mathbf{w}_\mathrm{ff}(r)$ satisfies
\begin{equation}\label{e:tff}
\mathbf{w}_r = [\mathcal{A}(r) + \mathcal{C}(r)] \mathbf{w}
+ \mathcal{G}(r,\mathbf{w}),
\end{equation}
with
\begin{align}
\mathcal{A}(r) = & 
\left(\begin{array}{cc} -k\partial_\vartheta & 1 \\
- D^{-1}[\omega\partial_\vartheta + f^\prime(u_\mathrm{ff}(r))] &
-k\partial_\vartheta \end{array}\right)
\nonumber \\ \label{e:tffdef}
\mathcal{C}(r) = & 
- \left(\begin{array}{cc} \theta^\prime(r)\partial_\vartheta & 0 \\
\frac{1}{r^2}\partial_{\vartheta\vartheta} &
\frac{1}{r}+\theta^\prime(r)\partial_\vartheta \end{array}\right)
\\ \nonumber
\mathcal{G}(r,\mathbf{w}) = & 
{ 0 \choose -D^{-1}[f(u_\mathrm{ff}(r)+w_1) - f(u_\mathrm{ff}(r))
- f^\prime(u_\mathrm{ff}(r))w_1] } \;=\; \rmO(|\mathbf{w}|_{X_r}^2),
\end{align}
where
$\mathbf{u}_\mathrm{ff}(r)=
\mathbf{u}_\mathrm{ff}(r;\omega,\alpha,\mathbf{w}_\mathrm{ff}^\mathrm{ss})$
is the solution discussed in (\ref{e:tffpar}). Note that
$\mathbf{w}(r)=0$ is always a solution of (\ref{e:tffdef}) since it is
the deviation from the actual solution $\mathbf{u}_\mathrm{ff}(r)$. On
account of Propositions~\ref{p:edff} and~\ref{p:edrobust}, the linear equation
\[
\mathbf{w}_r = [\mathcal{A}(r) + \mathcal{C}(r)] \mathbf{w},
\]
has an exponential dichotomy on $[R_*,\infty)$ for some sufficiently
large $R_*\gg1$ uniformly in $(\omega,\alpha,\mathbf{w}_\mathrm{ff}^\mathrm{ss})$ close to zero (note that these dichotomies depend on
$(\omega,\alpha,\mathbf{w}_\mathrm{ff}^\mathrm{ss})$, but we will
suppress this dependence in our notation). We have the estimates
\begin{align}\label{e:tffed}
\|\Phi^\mathrm{cs}_\mathrm{ff}(r,s)\|_{\mathrm{L}(X_r,X_s)} & \leq  C \rme^{\delta|r-s|},\ \ 
 \qquad  r\geq s\geq R_* ,\\ \nonumber
\|\Phi^\mathrm{uu}_\mathrm{ff}(r,s)\|_{\mathrm{L}(X_r,X_s)} & \leq  C \rme^{-\kappa|r-s|},\qquad 
 s\geq r\geq R_*
\end{align}
for some $\kappa>0$ and some arbitrarily small $\delta>0$. Using the
dichotomies, we can convert (\ref{e:tff}) into the integral equation
\[
\mathbf{w}(r) =
\Phi^\mathrm{uu}_\mathrm{ff}(r,R-\kappa^{-1}\log R)\mathbf{w}_\mathbf{ff}^\mathrm{uu}
+ \int_{R_*}^r \Phi^\mathrm{cs}_\mathrm{ff}(r,s)\mathcal{G}(s,\mathbf{w}(s))\,\rmd s
+ \int_{R-\kappa^{-1}\log R}^r \Phi^\mathrm{uu}_\mathrm{ff}(r,s)
\mathcal{G}(s,\mathbf{w}(s))\,\rmd s,
\]
where $r\in[R_*,R-\kappa^{-1}\log R]$. Using the estimates
(\ref{e:tffed}) for the exponential dichotomies and the fact that the
nonlinearity $\mathcal{G}$ defined in (\ref{e:tffdef}) vanishes with order
$\rmO(|\mathbf{w}|^2)$, it is not hard to see that the integral equation can be
solved using a contraction mapping theorem in the function space
\[
\mathcal{X}_\mathrm{ff} =
\left\{ \mathbf{w}\in C^0([R_*,R-\kappa^{-1}\log R],X);\;
\|\mathbf{w}\| = \sup_{r\in[R_*,R-\kappa^{-1}\log R]}
R^{-1}\rme^{\kappa(R-r)} |\mathbf{w}(r)|_{X_r} < \infty \right\}
\]
uniformly in $R$ and for any $\mathbf{w}_\mathbf{ff}^\mathrm{uu}$
sufficiently close to zero. The resulting solutions are smooth in
$(\omega,\alpha,\mathbf{w}_\mathbf{ff}^\mathrm{ss},\mathbf{w}_\mathbf{ff}^\mathrm{uu})$
near $(\omega_*,0,0,0)$, and they satisfy
\begin{equation}\label{est:ff}
\|\mathbf{w}\|=\rmO(|\mathbf{w}_\mathbf{ff}^\mathrm{uu}|).
\end{equation}
For later reference, we evaluate these solutions at the end points $r=R_*$ and
at $r=R-\kappa^{-1}\log R$ of their interval of existence:
\begin{align}
\mathbf{w}_\mathrm{ff}(R_*) = & 
\Phi^\mathrm{uu}_\mathrm{ff}(R_*,R-\kappa^{-1}\log R)\mathbf{w}_\mathbf{ff}^\mathrm{uu}
+ \int_{R-\kappa^{-1}\log R}^{R_*} \Phi^\mathrm{uu}_\mathrm{ff}(R_*,r)
\mathcal{G}(r,\mathbf{w}(r))\,\rmd r
\nonumber \\ \label{e:tffleft} = & 
\rmO\left(R\rme^{-\kappa(R-R_*)}
|\mathbf{w}_\mathbf{ff}^\mathrm{uu}|\right), \\
\mathbf{w}_\mathrm{ff}(R-\kappa^{-1}\log R) = & 
P_\mathrm{ff}^\mathrm{uu}(R-\kappa^{-1}\log R)
\mathbf{w}_\mathbf{ff}^\mathrm{uu}
+ \int_{R_*}^{R-\kappa^{-1}\log R} \Phi^\mathrm{cs}_\mathrm{ff}(R-\kappa^{-1}\log R,r)
\mathcal{G}(r,\mathbf{w}(r))\,\rmd r
\nonumber \\ \label{e:tffright} = & 
P_\mathrm{ff}^\mathrm{uu}(R-\kappa^{-1}\log R)
\mathbf{w}_\mathbf{ff}^\mathrm{uu}
+ \rmO\left(|\mathbf{w}_\mathbf{ff}^\mathrm{uu}|^2\right).
\end{align}

\paragraph{Boundary-layer region.}

In the last step, we consider solutions to (\ref{e:tsp})-(\ref{e:tbc})
in the boundary-layer region where $r\in[R-\kappa^{-1}\log R,R]$. It
is convenient to use the independent variable $\rho=r-R$ instead of
$r=\rho+R$. In this variable, (\ref{e:tsp})-(\ref{e:tbc}) become
\begin{align}\label{e:tbleqn}
u_\rho = &  -[k+\theta^\prime(\rho+R)]\partial_\vartheta u + v
\\ \nonumber
v_\rho = &  -[k+\theta^\prime(\rho+R)]\partial_\vartheta v
- \frac{v}{\rho+R} - \frac{\partial_{\vartheta\vartheta} u}{(\rho+R)^2} 
- D^{-1}[\omega\partial_\vartheta u + f(u)],
\end{align}
where $\rho\in(-\kappa^{-1}\log R,0)$, and 
\begin{equation}\label{e:tblbc}
(u,v)(0) \in E^\mathrm{bc}.
\end{equation}
We seek solutions to (\ref{e:tbleqn})--(\ref{e:tblbc}) using the
ansatz
\begin{equation}\label{e:tblans}
\mathbf{u}(\rho) =
\mathbf{u}_\mathrm{bs}(\rho;\omega) + \mathbf{w}_\mathrm{bs}(\rho),
\end{equation}
where $\mathbf{u}_\mathrm{bs}(\rho;\omega)$ is the boundary-layer
solution of (\ref{e:trho}). In particular, we see that
$\mathbf{u}_\mathrm{bs}(\rho;\omega)$ satisfies (\ref{e:tblbc}) and
also (\ref{e:tbleqn}) if we formally set $R=\infty$. Substituting the
ansatz (\ref{e:tblans}) into (\ref{e:tsp}), we obtain
\begin{equation}\label{e:tbs}
\mathbf{w}_\rho = [\mathcal{A}_\mathrm{bs}(\rho) + \mathcal{C}(\rho)] \mathbf{w}
+ \mathcal{C}(\rho) \mathbf{u}_\mathrm{bs}(\rho)
+ \mathcal{G}(\rho,\mathbf{w})
\end{equation}
with $\rho\in(-\kappa^{-1}\log R,0)$, where
\begin{align}
\mathcal{A}_\mathrm{bs}(\rho) = & 
\left(\begin{array}{cc} -k\partial_\vartheta & 1 \\
- D^{-1}[\omega\partial_\vartheta + f^\prime(u_\mathrm{bs}(\rho))] &
-k\partial_\vartheta \end{array}\right),
\nonumber \\ \label{e:tbldef}
\mathcal{C}(\rho) = & 
- \left(\begin{array}{cc} \theta^\prime(\rho+R)\partial_\vartheta & 0 \\
\frac{1}{(\rho+R)^2}\partial_{\vartheta\vartheta} &
\frac{1}{\rho+R}+\theta^\prime(\rho+R)\partial_\vartheta \end{array}\right),
\\ \nonumber
\mathcal{G}(\rho,\mathbf{w}) = & 
{ 0 \choose -D^{-1}[f(u_\mathrm{bs}(\rho)+w_1)
- f(u_\mathrm{bs}(\rho)) - f^\prime(u_\mathrm{bs}(\rho))w_1] }
\;=\; \rmO(|\mathbf{w}|^2),
\end{align}
and
$\mathbf{u}_\mathrm{bs}(\rho)=\mathbf{u}_\mathrm{bs}(\rho;\omega)$.
Our strategy for solving (\ref{e:tbs}) is the same as before: we show
that the linear equation
\begin{equation}\label{e:tbll}
\mathbf{w}_\rho = [\mathcal{A}_\mathrm{bs}(\rho) + \mathcal{C}(\rho)] \mathbf{w}
\end{equation}
has an exponential dichotomy on $X_{\rho+R}$, uniformly in $\omega$ and $R$, then use
the dichotomy to convert (\ref{e:tbs}) to an integral equation, and
lastly solve the integral equation in a function space with
appropriate weights. The existence of exponential dichotomies for
(\ref{e:tbll}) on $[-\kappa^{-1}\log R,0]$ that satisfy the estimates
\begin{align}
\|\Phi^\mathrm{cs}_\mathrm{bs}(\rho,\sigma)\| & \leq  C \rme^{\delta|\rho-\sigma|}, \ \qquad 
 -\kappa^{-1}\log R\leq\sigma\leq\rho\leq0, \nonumber
\\ \label{e:tbled}
\|\Phi^\mathrm{ss}_\mathrm{bs}(\rho,\sigma)\| & \leq  C \rme^{-\kappa|\rho-\sigma|},\qquad 
 -\kappa^{-1}\log R\leq\sigma\leq\rho\leq0,
\\ \nonumber
\|\Phi^\mathrm{uu}_\mathrm{bs}(\rho,\sigma)\| & \leq  C \rme^{-\kappa|\rho-\sigma|},\qquad 
-\kappa^{-1}\log R\leq\rho\leq\sigma\leq0,
\end{align}
for some $\kappa>0$ and some arbitrarily small $\delta>0$ can be established following the arguments in the proof of Proposition~\ref{p:edff} upon freezing the $\rho$-dependent coefficients that appear in the definition (\ref{e:tbldef}) of $\mathcal{A}_\mathrm{bs}(\rho)$ and $\mathcal{C}(\rho)$ at their value at $\rho=-\kappa^{-1}\log R$ for $\rho\leq-\kappa^{-1}\log R$. Equation (\ref{e:tbs}) is then equivalent to
\begin{align}\label{e:tblint}
\mathbf{w}(\rho) = & 
\Phi^\mathrm{ss}_\mathrm{bs}(\rho,-\kappa^{-1}\log R)\mathbf{w}_\mathbf{bs}^\mathrm{ss}
+ \int_{-\kappa^{-1}\log R}^\rho \Phi^\mathrm{cs}_\mathrm{bs}(\rho,\sigma)
\left[\mathcal{C}(\sigma)\mathbf{u}_\mathrm{bs}(\sigma)
+\mathcal{G}(\sigma,\mathbf{w}(\sigma))\right]\,\rmd\sigma
\\ & \nonumber
+ \Phi^\mathrm{uu}_\mathrm{bs}(\rho,0)\mathbf{w}_\mathbf{bs}^\mathrm{uu}
+ \int_0^\rho \Phi^\mathrm{uu}_\mathrm{bs}(\rho,\sigma)
\left[\mathcal{C}(\sigma)\mathbf{u}_\mathrm{bs}(\sigma)
+\mathcal{G}(\sigma,\mathbf{w}(\sigma))\right]\,\rmd\sigma,
\end{align}
where $\rho\in[-\kappa^{-1}\log R,0]$. If we consider (\ref{e:tblint})
as a fixed-point equation on the space
\[
\mathcal{X}_\mathrm{bs} =
\left\{ \mathbf{w}\in C^0([-\kappa^{-1}\log R,0],X);\;
\|\mathbf{w}\| = \sup_{\rho\in[-\kappa^{-1}\log R,0]}
|\mathbf{w}(\rho)|_{X_{\rho+R}} < \infty \right\}
\]
and fix $\varepsilon$ so that $1>\varepsilon>\delta>0$, then there is a
constant $K>0$ such that (\ref{e:tblint}) has a unique solution for
every $R$ sufficiently large in the ball
$\|\mathbf{w}\|\leq KR^{-1+\varepsilon}$ for every $\omega$ close to
$\omega_*$ and every $\mathbf{w}_\mathbf{bs}^\mathrm{ss}$ and
$\mathbf{w}_\mathbf{bs}^\mathrm{uu}$ of norm less than
$R^{-1+\varepsilon}$. This claim follows upon exploiting the estimates
(\ref{e:tbled}) for the exponential dichotomies and the fact that the
nonlinearity $\mathcal{G}$ (\ref{e:tbldef}) vanishes with order
$\rmO(|\mathbf{w}|^2)$. Furthermore, the resulting solutions are
smooth in
$(\omega,\mathbf{w}_\mathbf{bs}^\mathrm{ss},
\mathbf{w}_\mathbf{bs}^\mathrm{uu})$,
and they satisfy
\begin{equation}\label{est:west}
\|\mathbf{w}\| = \rmO\left(|\mathbf{w}_\mathbf{bs}^\mathrm{ss}|
+|\mathbf{w}_\mathbf{bs}^\mathrm{uu}|+R^{-1+\delta}\right).
\end{equation}
Evaluating the solution at $\rho=-\kappa^{-1}\log R$ and at $\rho=0$
gives
\begin{align}
\mathbf{w}_\mathrm{bs}(-\kappa^{-1}\log R) = & 
P_\mathbf{bs}^\mathrm{ss}(-\kappa^{-1}\log R)
\mathbf{w}_\mathbf{bs}^\mathrm{ss}
+ \Phi^\mathrm{uu}_\mathrm{bs}(-\kappa^{-1}\log R,0)\mathbf{w}_\mathbf{bs}^\mathrm{uu}
\label{e:tblleft} \\ & \nonumber
+ \int_0^{-\kappa^{-1}\log R} \Phi^\mathrm{uu}_\mathrm{bs}(-\kappa^{-1}\log R,\rho)
\left[\mathcal{C}(\rho)\mathbf{u}_\mathrm{bs}(\rho)
+\mathcal{G}(\rho,\mathbf{w}(\rho))\right]\,\rmd\rho
\\ = &  \nonumber
P_\mathbf{bs}^\mathrm{ss}(-\kappa^{-1}\log R)
\mathbf{w}_\mathbf{bs}^\mathrm{ss}
+ \rmO\left( \frac{1}{R^{1-\delta}}
+\frac{|\mathbf{w}_\mathbf{bs}^\mathrm{uu}|}{R}
+|\log R|\left[ |\mathbf{w}_\mathbf{bs}^\mathrm{ss}|^2
+|\mathbf{w}_\mathbf{bs}^\mathrm{uu}|^2\right] \right),
\\
\mathbf{w}_\mathrm{bs}(0) = & 
P_\mathbf{bs}^\mathrm{uu}(0)\mathbf{w}_\mathbf{bs}^\mathrm{uu}
+ \Phi^\mathrm{ss}_\mathrm{bs}(0,-\kappa^{-1}\log R)\mathbf{w}_\mathbf{bs}^\mathrm{ss}
\label{e:tblright} \\ & \nonumber
+ \int_{-\kappa^{-1}\log R}^0 \Phi^\mathrm{cs}_\mathrm{bs}(0,\rho)
\left[\mathcal{C}(\rho)\mathbf{u}_\mathrm{bs}(\rho)
+\mathcal{G}(\rho,\mathbf{w}(\rho))\right]\,\rmd\rho
\\ = &  \nonumber
P_\mathbf{bs}^\mathrm{uu}(0)\mathbf{w}_\mathbf{bs}^\mathrm{uu}
+ \rmO\left( \frac{1}{R^{1-\delta}}
+\frac{|\mathbf{w}_\mathbf{bs}^\mathrm{ss}|}{R}
+|\log R|\left[ |\mathbf{w}_\mathbf{bs}^\mathrm{ss}|^2
+|\mathbf{w}_\mathbf{bs}^\mathrm{uu}|^2\right] \right).
\end{align}
This completes the construction of the solutions in the core, the
far-field, and the boundary-layer region.

\subsection{Matching of core, far-field, and boundary-layer solutions}
\label{s:truncm}

It remains to match the solutions obtained in the last section at
$r=R_*$, $r=R-\kappa^{-1}\log R$, and $r=R$. Using the coordinate transformations established in \S\ref{s:ced}, we push the matching conditions to the space $L^2\times L^2$ and solve them in this space as this allows us to compare the various projections with the $r$-independent projections of the asymptotic wave train. Note that the estimates we established above remain valid when we transform the equations to $L^2\times L^2$.

\paragraph{Matching far-field and boundary-layer solutions.}
First, we match the far-field solution at $r=R-\kappa^{-1}\log R$ and the
boundary-layer solutions at $\rho=-\kappa^{-1}\log R$. From
(\ref{e:tffans}) and (\ref{e:tblans}) we obtain the equation
\[
\mathbf{u}_\mathrm{ff}(R-\kappa^{-1}\log R;\omega,\alpha,
\mathbf{w}_\mathrm{ff}^\mathrm{ss})
+ \mathbf{w}_\mathrm{ff}(R-\kappa^{-1}\log R)
= \mathbf{u}_\mathrm{bs}(-\kappa^{-1}\log R;\omega)
+ \mathbf{w}_\mathrm{bs}(-\kappa^{-1}\log R).
\]
Note that Proposition~\ref{p:stmaff} implies that
\[
\mathbf{u}_\mathrm{ff}(R-\kappa^{-1}\log R;\omega,\alpha,
\mathbf{w}_\mathrm{ff}^\mathrm{ss})
= \mathbf{u}_\infty(\cdot;\omega)
+ \mathbf{u}_\infty^\prime(\cdot;\omega) \alpha
+ \rmO\left(\alpha^2+\frac{1}{R}\right),
\]
while (\ref{e:tblcc}) shows that
\[
\mathbf{u}_\mathrm{bs}(-\kappa^{-1}\log R;\omega)
= \mathbf{u}_\infty(\cdot;\omega)
+ \rmO\left(\frac{1}{R}\right).
\]
Using these facts together with (\ref{e:tffright}) and
(\ref{e:tblleft}), we arrive at the equation
\begin{align}\label{e:tmffbs}
 \MoveEqLeft[8] \mathbf{u}_\infty^\prime(\cdot;\omega) \alpha
+ \mathcal{P}_\mathrm{ff}^\mathrm{uu}(R-\kappa^{-1}\log R)
\mathbf{w}_\mathbf{ff}^\mathrm{uu}
- \mathcal{P}_\mathbf{bs}^\mathrm{ss}(-\kappa^{-1}\log R)
\mathbf{w}_\mathbf{bs}^\mathrm{ss},   \nonumber\\ 
&=  
\rmO\left( \frac{1}{R^{1-\delta}} + \alpha^2
+ |\mathbf{w}_\mathbf{ff}^\mathrm{uu}|^2
+ \frac{|\mathbf{w}_\mathbf{bs}^\mathrm{uu}|}{R}
+ |\log R|\left[ |\mathbf{w}_\mathbf{bs}^\mathrm{ss}|^2
+ |\mathbf{w}_\mathbf{bs}^\mathrm{uu}|^2\right] \right)
\end{align}
where
\[
\mathbf{w}_\mathbf{ff}^\mathrm{uu}
\in\Rg(\mathcal{P}^\mathrm{uu}_\infty(\omega_*)), \qquad
\mathbf{w}_\mathbf{bs}^\mathrm{ss}
\in\Rg(\mathcal{P}^\mathrm{ss}_\infty(\omega_*)),
\]
with $|\mathbf{w}_\mathbf{bs}^\mathrm{ss}|\leq R^{-1+\varepsilon}$. We will solve (\ref{e:tmffbs}) below.

\paragraph{Matching boundary conditions.}
Next, we need to satisfy the boundary conditions at $r=R$ or,
alternatively, at $\rho=0$. Using the ansatz (\ref{e:tblans}) and
exploiting the fact that
$\mathbf{u}_\mathrm{bs}(0;\omega)\in E^\mathrm{bc}$ satisfies the
boundary condition, it remains to solve
$\mathbf{w}_\mathrm{bs}(0)\in E^\mathrm{bc}$. It is convenient to
introduce the projection $\mathcal{P}^\mathrm{bc}$ defined via
\[
\Rg(\mathcal{P}^\mathrm{bc}) = \Rg(\mathcal{P}_\mathbf{bs}^\mathrm{uu}(0;\omega_*)),
\qquad
\Ns(\mathcal{P}^\mathrm{bc}) = E^\mathrm{bc}.
\]
Note that $\mathcal{P}^\mathrm{bc}$ is well defined on account of the results in
\S\ref{s:truncbs} and bounded uniformly in $R$. Using this
projection, we see that $\mathbf{w}_\mathrm{bs}(0)\in E^\mathrm{bc}$
if and only if $\mathcal{P}^\mathrm{bc}\mathbf{w}_\mathrm{bs}(0)=0$ which
becomes
\begin{equation}\label{e:tmblbc}
\mathcal{P}^\mathrm{bc} \left[
\mathcal{P}_\mathbf{bs}^\mathrm{uu}(0)\mathbf{w}_\mathbf{bs}^\mathrm{uu}
+ \rmO\left( \frac{1}{R^{1-\delta}}
+ \frac{|\mathbf{w}_\mathbf{bs}^\mathrm{ss}|}{R}
+ |\log R|\left[ |\mathbf{w}_\mathbf{bs}^\mathrm{ss}|^2
+ |\mathbf{w}_\mathbf{bs}^\mathrm{uu}|^2\right] \right) \right] = 0
\end{equation}
when we use (\ref{e:tblright}). Here, we can choose
$\mathbf{w}_\mathbf{bs}^\mathrm{uu}$ subject to
\[
\mathbf{w}_\mathbf{bs}^\mathrm{uu}
\in\Rg(\mathcal{P}_\mathbf{bs}^\mathrm{uu}(0;\omega_*)), \qquad
|\mathbf{w}_\mathbf{bs}^\mathrm{uu}|\leq R^{-1+\varepsilon}.
\]

\paragraph{Solving (\ref{e:tmffbs}) and (\ref{e:tmblbc}).}
Collecting the equations (\ref{e:tmffbs}) and (\ref{e:tmblbc}) we
derived so far, we obtain the system
\begin{align}
\begin{split}
0=&\mathbf{u}_\infty^\prime(\cdot;\omega) \alpha
+ \mathcal{P}_\mathrm{ff}^\mathrm{uu}(R-\kappa^{-1}\log R)
\mathbf{w}_\mathbf{ff}^\mathrm{uu}
- \mathcal{P}_\mathbf{bs}^\mathrm{ss}(-\kappa^{-1}\log R)
\mathbf{w}_\mathbf{bs}^\mathrm{ss} \\ 
&+ \rmO\left( \frac{1}{R^{1-\delta}} + \alpha^2
+ |\mathbf{w}_\mathbf{ff}^\mathrm{uu}|^2
+ \frac{|\mathbf{w}_\mathbf{bs}^\mathrm{uu}|}{R}
+ |\log R|\left[ |\mathbf{w}_\mathbf{bs}^\mathrm{ss}|^2
+ |\mathbf{w}_\mathbf{bs}^\mathrm{uu}|^2\right] \right) \\
0=&\mathcal{P}^\mathrm{bc} \left[
\mathcal{P}_\mathbf{bs}^\mathrm{uu}(0)\mathbf{w}_\mathbf{bs}^\mathrm{uu}
+ \rmO\left( \frac{1}{R^{1-\delta}}
+ \frac{|\mathbf{w}_\mathbf{bs}^\mathrm{ss}|}{R}
+ |\log R|\left[ |\mathbf{w}_\mathbf{bs}^\mathrm{ss}|^2
+ |\mathbf{w}_\mathbf{bs}^\mathrm{uu}|^2\right] \right) \right] ,
\end{split}\label{e:tm1}
\end{align}
where
\[
\mathbf{w}_\mathbf{ff}^\mathrm{uu}
\in\Rg(\mathcal{P}^\mathrm{uu}_\infty(\omega_*)), \qquad
\mathbf{w}_\mathbf{bs}^\mathrm{ss}
\in\Rg(\mathcal{P}^\mathrm{uu}_\infty(\omega_*)), \qquad
\mathbf{w}_\mathbf{bs}^\mathrm{uu}
\in\Rg(\mathcal{P}_\mathbf{bs}^\mathrm{uu}(0;\omega_*)),
\]
and $\alpha$ need to be close to zero with $|\mathbf{w}_\mathbf{bs}^\mathrm{ss}|\leq R^{-1+\varepsilon}$ and $|\mathbf{w}_\mathbf{bs}^\mathrm{uu}|\leq R^{-1+\varepsilon}$.
The remainder terms, and their derivatives with respect to
$(\mathbf{w}_\mathbf{ff}^\mathrm{uu},\mathbf{w}_\mathbf{ff}^\mathrm{ss},
\mathbf{w}_\mathbf{bs}^\mathrm{uu},\mathbf{w}_\mathbf{bs}^\mathrm{ss})$
and $\omega$, are uniform in $R$, $\omega$ and
$\mathbf{w}_\mathbf{ff}^\mathrm{ss}$. Note that the projections
$\mathcal{P}_\mathbf{bs}^\mathrm{uu}(0)$,
$\mathcal{P}_\mathrm{ff}^\mathrm{uu}(R-\kappa^{-1}\log R)$ and
$\mathcal{P}_\mathbf{bs}^\mathrm{ss}(-\kappa^{-1}\log R)$ are smooth in $\omega$
and that $\mathcal{P}_\mathrm{ff}^\mathrm{uu}(r)$ and
$\mathcal{P}_\mathbf{bs}^\mathrm{uu}(-\rho)$ are close to
$\mathcal{P}^\mathrm{uu}_\infty(\omega_*)$ for $r$ and $\rho$ sufficiently
large and $\omega$ close to $\omega_*$. We also recall that
$E^\mathrm{bc}\oplus\Rg(\mathcal{P}_\mathbf{bs}^\mathrm{uu}(0;\omega_*))=X$.
Using that $1>\varepsilon>\delta>0$, it is then not difficult to solve
(\ref{e:tm1}) for
$(\mathbf{w}_\mathbf{ff}^\mathrm{uu},\mathbf{w}_\mathbf{bs}^\mathrm{uu},
\mathbf{w}_\mathbf{bs}^\mathrm{ss},\alpha)$ as a smooth function of
$\omega$ and $\mathbf{w}_\mathbf{ff}^\mathrm{ss}$. Furthermore, for
some constant $C>0$, we have
\begin{equation}\label{e:tmffblest}
|\alpha|
+ |\mathbf{w}_\mathbf{ff}^\mathrm{uu}|
+ |\mathbf{w}_\mathbf{bs}^\mathrm{ss}|
+ |\mathbf{w}_\mathbf{bs}^\mathrm{uu}|\leq \frac{C}{R^{1-\delta}}
\end{equation}
uniformly in $\omega$ and $\mathbf{w}_\mathbf{ff}^\mathrm{ss}$. The
above estimate is also true for derivatives with respect to $\omega$
and $\mathbf{w}_\mathbf{ff}^\mathrm{ss}$.

\paragraph{Matching core and far-field solutions.}

Lastly, using (\ref{e:tffans}), we see that the matching condition of
the core and the far-field solution at $r=R_*$ is given by
\[
\mathbf{u}_\mathrm{core}(R_*;\omega,\mathbf{w}_\mathrm{core}^\mathrm{u})
= \mathbf{u}_\mathrm{ff}(R_*;\omega,\alpha(\omega,\mathbf{w}_\mathbf{ff}^\mathrm{ss}),\mathbf{w}_\mathrm{ff}^\mathrm{ss})
+ \mathbf{w}_\mathrm{ff}(R_*),
\]
where $\alpha(\omega,\mathbf{w}_\mathbf{ff}^\mathrm{ss})$ denotes the function we obtained in the previous step. Using (\ref{e:tffleft}) and (\ref{e:tmffblest}), this equation becomes
\begin{equation}\label{e:tmcrff}
\mathbf{u}_\mathrm{core}(R_*;\omega,\mathbf{w}_\mathrm{core}^\mathrm{u})
= \mathbf{u}_\mathrm{ff}(R_*;\omega,\alpha(\omega,\mathbf{w}_\mathbf{ff}^\mathrm{ss}),\mathbf{w}_\mathrm{ff}^\mathrm{ss})
+ \rmO\left(R^{2-\delta}\rme^{-\kappa R}\right),
\end{equation}
where we can choose $\mathbf{w}_\mathrm{core}^\mathrm{u}\in\Rg(\mathcal{P}^\mathrm{u}_-(R_*))$ and $\mathbf{w}_\mathrm{ff}^\mathrm{ss}\in\Rg(\mathcal{P}^\mathrm{ss}_+(R_*))$ near zero as we wish. The arguments given in
\cite[\S4.2]{s-conv} for matching homoclinic orbits in finite
dimensions also apply to (\ref{e:tmcrff}), and we obtain that
(\ref{e:tmcrff}) has a unique solution $(\omega,\mathbf{w}_\mathrm{core}^\mathrm{u},\mathbf{w}_\mathrm{ff}^\mathrm{ss})$
for each $R\gg1$. Furthermore, \cite[Lemma~4.2]{s-conv} shows that
there is a constant $C>0$ such that
\begin{equation}\label{e:tmom}
|\omega-\omega_*|
+ |\mathbf{w}_\mathrm{core}^\mathrm{uu}|
+ |\mathbf{w}_\mathrm{ff}^\mathrm{ss}|
\leq C R^{2-\delta} \rme^{-\kappa R}.
\end{equation}
This completes the proof of the existence part of Theorem~\ref{t:trunc}. The estimates (\ref{est:trunc}) follow from the representations (\ref{e:tcorepar}), (\ref{e:tffans}), and (\ref{e:tblans}) together with the estimates (\ref{est:ff}), (\ref{est:west}),  (\ref{e:tmffblest}), and (\ref{e:tmom}).

%%%%%%%%%%%%%%%%%%%%%%%%%%%%%%%%%%%%%%%%%%%%%%%%%%%%%%%%%%%%%%%%%%%%%%%%%

\section{Spectra of spiral waves restricted to large finite disks}\label{s:abs}

In this section, we prove Theorem~\ref{t:tl}, \rev{which characterizes the spectrum of the linearization at a spiral waves restricted to a large bounded disk}. We define
\[
\mathcal{L}_{R} u = D\Delta u +\omega_* \partial_\psi u + f'(u_*(r,\psi))u
\]
on the disk $0\leq r\leq R$ with boundary conditions $au+bu_r=0$ at the boundary $r=R$, where $a,b$ are fixed constants. We denote by $\Sigma_R$ the spectrum of the operator $\mathcal{L}_R$ posed on $L^2$. Our goal is to characterize the limit of $\Sigma_R$ as $R\to\infty$. It is convenient to define the set
\[
\Sigma_\mathrm{acc} := \{\lambda\in\C\,|\, \exists R_k\to\infty, \lambda_k\in\Sigma_{R_k} \mbox{ so that } \lambda_k\to\lambda \mbox{ as } k\to\infty\}
\]
of accumulation points of $\Sigma_R$ as $R\to\infty$. We claim that $\Sigma_\mathrm{acc}$ is equal to the \emph{limiting spectral set} $\Sigma_\mathrm{st}:=\Sigma_\mathrm{abs}\cup\Sigma_\mathrm{ext}\cup\Sigma_\mathrm{bdy}$. To prove that $\Sigma_\mathrm{acc}=\Sigma_\mathrm{st}$, it suffices to show the following inclusions:
\[
\text{(i)}\ \Sigma_\mathrm{acc}\subseteq\Sigma_\mathrm{st}, \qquad
\text{(ii)}\ \Sigma_\mathrm{bdy}\subseteq\Sigma_\mathrm{acc}, \qquad
\text{(iii)}\ \Sigma_\mathrm{ext}\subseteq\Sigma_\mathrm{acc}, \qquad
\text{(iv)}\ \Sigma_\mathrm{abs}\subseteq\Sigma_\mathrm{acc}.
\]
We will establish (i) in \S\ref{s:10.1}, (ii) and (iii) in \S\ref{s:10.2}, and (iv) in \S\ref{s:10.4}. From the proofs, it will be clear that the multiplicities are preserved in the limits (ii) and (iii), while multiplicities tend to infinity in (iv).

\subsection{Excluding eigenvalues outside of the limiting spectral set}\label{s:10.1}

We prove that $\Sigma_\mathrm{acc}\subseteq\Sigma_\mathrm{st}$, thus excluding the case that eigenvalues of $\mathcal{L}_R$ can accumulate in the complement of the limiting spectral set $\Sigma_\mathrm{st}$.

\begin{Lemma}[Continuity of the resolvent under restriction]\label{l:res}
Suppose that $\lambda_*\notin\Sigma_\mathrm{st}$. Then there exist $\delta>0$ and $\bar{R}>0$ such that $B_\delta(\lambda_*)$ belongs to the resolvent set of $\mathcal{L}_R$ for all $R>\bar{R}$. Moreover, $\bar{R}(\lambda_*)$ can be chosen uniformly in compact subsets of the complement of $\Sigma_\mathrm{st}$. In particular, we conclude that $\lambda_*\notin\Sigma_\mathrm{acc}$.
\end{Lemma}

\begin{Proof}
Since $\lambda_*\notin\Sigma_\mathrm{abs}\cup\Sigma_\mathrm{ext}$, the spatial dynamical system belonging to the system $\mathcal{L}u=\lambda_* u$ associated with the linearization of the planar spiral wave admits an exponential dichotomy in appropriate weighted spaces for $r\geq0$; see \S\ref{s:edweights}. Since $\lambda_*\notin\Sigma_\mathrm{bdy}$, there exists an $\bar{R}$ so that the space $E^\mathrm{bc}$ of boundary condition in the spatial dynamics formulation is transverse to the unstable subspace of the exponential dichotomy at $r$ for each $r\geq\bar{R}$. We can therefore find exponential dichotomies that satisfy $E^\mathrm{s}(R)=E^\mathrm{bc}$ that are uniform in $R$, which proves the absence of point spectrum of $\mathcal{L}_R$ at $\lambda_*$. These arguments can be extended, uniformly in $R$, to all $\lambda$ near $\lambda_*$ by continuity of the dichotomies in $\lambda$.
\end{Proof}

\subsection{Convergence of eigenvalues to the boundary and the extended point spectrum}\label{s:10.2}

We first consider eigenvalues created by boundary conditions and prove that $\Sigma_\mathrm{bdy}\subseteq\Sigma_\mathrm{acc}$.

\begin{Lemma}[Eigenvalues induced by boundary conditions]\label{l:bspec}
Suppose that $\lambda_*\in\Sigma_\mathrm{bdy}\setminus(\Sigma_\mathrm{abs}\cup\Sigma_\mathrm{ext})$ belongs to the boundary spectrum but not to the absolute or the extended point spectrum of the spiral wave. Let $m$ be the algebraic multiplicity of $\lambda_*$ as an element of $\Sigma_\mathrm{bdy}$. Then there is a $\delta>0$ such that for all $R$ sufficiently large the truncated linearization $\mathcal{L}_R$ has precisely $m$ eigenvalues, counted with algebraic multiplicity, in $B_\delta(\lambda_*)$. Moreover, there exists a constant $C>0$ such that $|\lambda-\lambda_*|\leq CR^{-\frac{1}{3m}}$ for any eigenvalue $\lambda$ in this $\delta$-neighborhood of $\lambda_*$.
\end{Lemma}

\begin{Proof}
As in the preceding lemma, since $\lambda_*\notin\Sigma_\mathrm{abs}\cup\Sigma_\mathrm{ext}$, the spatial dynamical system belonging to the system $\mathcal{L}u=\lambda u$ associated with the linearization of the planar spiral wave admits an exponential dichotomy in appropriate weighted spaces for $r\geq0$ for each $\lambda$ near $\lambda_*$. In addition, we proved in Proposition~\ref{p:edff} that the unstable subspace $E^\mathrm{u}(R)$ of the dichotomy is $\rmO(R^{-1/3})$-close to the unstable subspace of the linearization at the asymptotic wave trains. We find eigenvalues by looking for nontrivial intersections of the unstable subspace $E^\mathrm{u}(R)$ and the boundary subspace $E^\mathrm{bc}$, which yields a linear equation with parameter $\lambda$ that is $\rmO(R^{-1/3})$-close to the equation for elements of the boundary spectrum. Using Lyapunov--Schmidt reduction, we obtain a characteristic equation for eigenvalues that is analytic in $\lambda$ of the form $\lambda^m=\rmO(R^{-1/3})$, which then gives roots as desired. Though we do not provide details, it can be shown that the multiplicity of each root obtained in this fashion is equal to the algebraic multiplicity of the corresponding eigenvalue of $\mathcal{L}_R$. 
\end{Proof}

Next, we show that elements in the extended point spectrum $\Sigma_\mathrm{ext}$ lie in $\Sigma_\mathrm{acc}$.

\begin{Lemma}[Eigenvalues induced by extended point spectrum]\label{l:extspec}
Assume that $\lambda_*\in\Sigma_\mathrm{ext}\setminus(\Sigma_\mathrm{abs}\cup\Sigma_\mathrm{bdy})$ belongs to the extended point spectrum but not to the absolute or the boundary spectrum of the spiral wave. Let $m$ be the algebraic multiplicity of $\lambda_*$; then there is a $\delta>0$ such that for all $R$ sufficiently large the truncated linearization $\mathcal{L}_R$ has precisely $m$ eigenvalues, counted with algebraic multiplicity, in $B_\delta(\lambda_*)$. Moreover, there are constants $C,\eta >0$ such $|\lambda-\lambda_*|\leq C\rme^{-\eta R}$ for any eigenvalue $\lambda$ in this $\delta$-neighborhood of $\lambda_*$. 
\end{Lemma}

\begin{Proof}
Since $\lambda_*\notin\Sigma_\mathrm{abs}$, the spatial dynamical system belonging to the system $\mathcal{L}u=\lambda u$ associated with the linearization of the planar spiral wave admits exponential dichotomies in appropriate fixed weighted spaces separately for $0\leq r\leq R_*$ and for $R_*\leq r$ for each $\lambda$ near $\lambda_*$. We restrict the dichotomy for $R_*\leq r$ to the interval $R_*\leq r\leq R$ and, using the assumption that $\lambda_*\notin\Sigma_\mathrm{bdy}$, modify the dichotomy by selecting the new stable subspace $\tilde{E}^\mathrm{s}_+(R)$ to satisfy $\tilde{E}^\mathrm{s}_+(R)=E^\mathrm{bc}$ for each $\lambda$ near $\lambda_*$. The resulting dichotomy has rates and constants that are independent of $R$ and $\lambda$ near $\lambda_*$. Furthermore, the new stable eigenspace $\tilde{E}^\mathrm{s}_+(R_*)$  at $r=R_*$ is exponentially close in $R$ to $E^\mathrm{s}_+(R_*)$. A number $\lambda$ near $\lambda_*$ is an eigenvalue of $\mathcal{L}_R$ if and only if $\tilde{E}^\mathrm{s}_+(R_*)$ and ${E}^\mathrm{u}_-(R_*)$ have a nontrivial intersection. Using Lyapunov--Schmidt reduction, this condition gives a reduced equation that is exponentially close in $R$ to the equation for eigenvalues in the extended point spectrum of the planar spiral wave. As a consequence, the eigenvalue $\lambda_*$ of multiplicity $m$ in the extended point spectrum creates precisely $m$ eigenvalues, counted with multiplicity, of $\mathcal{L}_R$ near $\lambda_*$, and the latter converge to $\lambda_*$ exponentially in $R$ as $R\to\infty$.
\end{Proof}

\subsection{Accumulation near the absolute spectrum}\label{s:10.4}

It remains to prove that $\Sigma_\mathrm{abs}\subseteq\Sigma_\mathrm{acc}$, which turns out to be more subtle than the previous cases.

Take an element $\lambda_*\in\Sigma_\mathrm{abs}$ so that $\Re\nu_{-1}(\lambda_*)=\Re\nu_0(\lambda_*)$. After choosing a new variable $\tilde{\lambda}$ via $\lambda=\lambda_*+\tilde{\lambda}$ and using an exponential weight $\Re\nu_0(\lambda_*)$, we can assume that $\lambda_*=0$ with $\nu_{-1}(0)=-\rmi/2=-\nu_0(0)$. Note that simplicity of the absolute spectrum implies that $\Re\nu_j<0$ for $j<-1$, $\Re\nu_j>0$ for $j>0$, and $\frac{\rmd\nu_{-1}}{\rmd\lambda}\neq\frac{\rmd\nu_0}{\rmd\lambda}$ at $\lambda=\lambda_*=0$. \rev{Finally, using that $\frac{\rmd\nu_{-1}}{\rmd\lambda}(0)\neq\frac{\rmd\nu_0}{\rmd\lambda}(0)$, we can make an invertible analytic change of coordinates of the $\lambda$ variable so that the absolute spectrum near $\lambda=\lambda_*=0$ is given by $\Re\lambda=0$.}

First, we describe the asymptotics of solutions in the two-dimensional center subspace associated with the spatial eigenvalues $\pm\rmi/2$. We write the spatial-dynamics formulation (\ref{e:evpcon}) associated with the eigenvalue problem $\mathcal{L}_Ru=\lambda u$ as
\begin{equation}\label{e:1041}
\mathbf{w}_r = \mathcal{A}_\lambda(r) \mathbf{w}.
\end{equation}
The associated asymptotic system, obtained by formally setting $r=\infty$ in (\ref{e:evpcon}) or (\ref{e:1041}), admits a trichotomy belonging to the splitting of spatial eigenvalues $\nu$ into center, stable, and unstable sets with associated eigenspaces $E^\mathrm{c,s,u}_\infty$. The next lemma shows that similar trichotomies exist for (\ref{e:1041}).

\begin{Proposition}\label{p:ecred}
For each $\lambda$ close to zero, the linearized equation (\ref{e:1041}) has an exponential trichotomy for $r\geq R_*$ with subspaces $E^\mathrm{c,s,u}(r)$. The center subspace $E^\mathrm{c}(r)$ has dimension two and is a graph over $E^\mathrm{c}_\infty$ that can be chosen to be of class $C^N$ in $\frac{1}{r}$ for each fixed $N<\infty$. The dynamics in $E^\mathrm{c}(r)$ projected onto $E^\mathrm{c}_\infty$ are given by the linear equation 
\begin{equation}\label{e:cred}
\mathbf{w}_\mathrm{c}' = A_\mathrm{c}(r;\lambda)\mathbf{w}_\mathrm{c}, \qquad
\mathbf{w}_\mathrm{c}\in E^\mathrm{c}_\infty
\end{equation}
and, in suitable coordinates in $E^\mathrm{c}_\infty$, 
\[
A_\mathrm{c}(r) = A_\mathrm{c}^\infty + \sum_{j=1}^N A_\mathrm{c}^j r^{-j} + \rmO(r^{-(K+1)}), \qquad
A_\mathrm{c}^\infty =
\begin{pmatrix} \nu_{-1}(\lambda) & 0 \\ 0 & \nu_0(\lambda) \end{pmatrix}.
\]
Furthermore, the subspace $E^\mathrm{c}(r)$, and the reduced equation \eqref{e:cred} are analytic in $\lambda$.
\end{Proposition}

\begin{Proof}
The proof is similar to the proof of Proposition~\ref{p:asyeig}. In the far field, for $r\geq R_*$, we can use the trichotomy for the asymptotic equation to decompose 
\begin{align}
\mathbf{w}_\mathrm{h}' = & A_\mathrm{h}(r) \mathbf{w}_\mathrm{h} + A_\mathrm{hc}(r) \mathbf{w}_\mathrm{c}
\notag \\ \label{e:hcoor}
\mathbf{w}_\mathrm{c}' = & A_\mathrm{c}(r) \mathbf{w}_\mathrm{c} + A_\mathrm{ch}(r) \mathbf{w}_\mathrm{h},
\end{align}
where we wrote $\mathbf{w}_\mathrm{h}=(\mathbf{w}_\mathrm{s},\mathbf{w}_\mathrm{u})$, and where $A_\mathrm{hc}(r),A_\mathrm{ch}(r)=\rmO(\frac{1}{r})$. Proceeding as in Proposition~\ref{p:asyeig}, we can compute the expansion 
\[
\mathbf{w}_\mathrm{h} = B(r) \mathbf{w}_\mathrm{c} + \tilde{\mathbf{w}}_\mathrm{h}, \qquad
B(r) = \sum_{j=1}^{N} B_jr^{-j}
\]
such that \eqref{e:hcoor}, written in the new variables $(\tilde{\mathbf{w}}_\mathrm{h},\mathbf{w}_\mathrm{c})$, becomes
\begin{align}
\tilde{\mathbf{w}}_\mathrm{h}' = & A_\mathrm{h}(r)\tilde{\mathbf{w}}_\mathrm{h}+\tilde{A}_\mathrm{hc}(r)\mathbf{w}_\mathrm{c}
\notag \\ \label{e:hcoor2}
\mathbf{w}_\mathrm{c}' = & \tilde{A}_\mathrm{c}(r)\mathbf{w}_\mathrm{c}+\tilde{A}_\mathrm{ch}(r)\tilde{\mathbf{w}}_\mathrm{h},
\end{align}
where $\tilde{A}_\mathrm{hc}(r)=\rmO(r^{-(K+1)})$, and where $\tilde{A}_\mathrm{c}(r)$ has an expansion in $r^{-1}$ up to order $N$. We will now argue that \eqref{e:hcoor2} has an exponential trichotomy. Artificially setting $\tilde{A}_\mathrm{hc}\equiv0$, we can construct an exponential trichotomy with $E^\mathrm{c}=\{\tilde{\mathbf{w}}_\mathrm{h}=0\}$. Quantitative robustness of exponential trichotomies then guarantees that the equation with $\tilde{A}_\mathrm{hc}(r)=\rmO(r^{-(K+1)})$ also admits an exponential trichotomy with subspace $E^\mathrm{c}(r)=\rmO(r^{-(K+1)})$. Diagonalizing the linear part within this subspace implies the claim of the proposition.
\end{Proof}

The key step in our analysis of the far-field asymptotics of (\ref{e:cred}) is a change of coordinates that allows us to diagonalize the system in $E^\mathrm{c}$ uniformly in $r$ and $\lambda$. This result is presented in the following proposition. 

\begin{Proposition}\label{p:absspecdiag}
For each fixed natural number $M$, there exists a linear, $(r,\lambda)$-dependent change of coordinates in $E^\mathrm{c}_\infty$ that is $C^{M+1}$ in $\lambda$ and has an expansion up to order $M$ in $r^{-1}$ such that the reduced equation \eqref{e:cred} is of the form
\[
\mathbf{w}_\mathrm{c}'=A_\mathrm{c}(r,\lambda)\mathbf{w}_\mathrm{c}, \quad
A_\mathrm{c}(\lambda,r)=
\begin{pmatrix}
\nu^1(\lambda,r)&0\\0&\nu^2(\lambda,r)
\end{pmatrix},\quad
\nu^j(\lambda,r)=\nu^j_\infty(\lambda)+\sum_{\ell=1}^{M}\nu^j_\ell(\lambda)r^{-\ell} + \rmO(r^{-(M+1)}),
\]
where $\nu^j_\infty(\lambda)$ are the eigenvalues $\nu_{-1}(\lambda)$ and $\nu_0(\lambda)$ from Proposition~\ref{p:ecred}.
\end{Proposition}

\begin{Proof}
To prove the statement, we need to continue the asymptotic eigenvectors that belong to the asymptotic matrix at $r=\infty$ on the center space associated with the eigenvalues $\nu^1$ and $\nu^2$ to finite values of $r$. The proof is divided into several steps.

\paragraph{Step 1.}
We perform a sequence of near-identity transformation of the form $\id+r^{-j}B_j(\lambda)$ so that $A_\mathrm{c}(r,\lambda)$ is diagonal up to terms of order $r^{-(K+1)}$. This can be readily accomplished since diagonalizing at order $r^{-j}$ introduces error terms only of order $r^{-(j+1)}$.

\paragraph{Step 2.}
We introduce projective coordinates $z=w_\mathrm{c}^1/w_\mathrm{c}^2$, thus reducing the linear two-dimensional equation to an equation on the complex Grassmannian $\bar{\C}\sim S^2$. Setting $\alpha:=r^{-1}$, we arrive at the equation
\[
z'=(\nu^1(\lambda,r)-\nu^2(\lambda,r))z+\rmO(\alpha^{M+1})+z^2\rmO(\alpha^{M+1}), \quad
\nu^1(\lambda,r)-\nu^2(\lambda,r)=\rmi + (\nu^1_\lambda(0)-\nu^2_\lambda(0))\lambda +\rmO(\lambda^2) + \rmO(\alpha).
\]
\rev{Since the analytic coordinate transformation for $\lambda$ we discussed at the beginning of this section ensures that the absolute spectrum through $\lambda_*=0$ is given by the line $\Re\lambda=0$}, the equation for $z$ becomes
\begin{align*}
z'&=\rmi z + \lambda z + \sum_{j=1}^{M}a_j \alpha^j z+ \rmO(\alpha^{M+1})+z^2\rmO(\alpha^{M+1}) \notag \\ 
\alpha'&=-\alpha^2 \\ \notag
\lambda'&=0.
\end{align*}
\rev{Note that} we included the equations for $\alpha$ and $\lambda$ for later use.

\paragraph{Step 3.}
Next, using the corotating frame $z=\rme^{\rmi(1+\Im\lambda)r}\tilde{z}$ and setting $\lambda_\mathrm{r}:=\Re\lambda$, we find, after dropping the tildes, that $z$ satisfies the system
\begin{align}
z' = &  \lambda_\mathrm{r} z + \sum_{j=1}^{M}a_j \alpha^j z+ \rmO(\alpha^{M+1})+ z\rmO(\alpha^{M+1})+z^2\rmO(\alpha^{M+1}) \notag\\ \label{e:redproj2}
\alpha' = &  -\alpha^2 \\ \notag
\lambda' = &  0,
\end{align}
where the remainder terms now depend on $\rme^{\rmi/\alpha}$, hence inducing negative powers of $\alpha$ when we differentiate these terms in $\alpha$. Choosing $M$ sufficiently large, we conclude that the resulting system can still be differentiated arbitrarily often in $\alpha$ up to $\alpha=0$.

\paragraph{Step 4.}
The dynamics of \eqref{e:redproj2} is degenerate near the origin as the leading-order terms in the vector field are quadratic in $(z,\alpha,\lambda)$.
In order to understand the dynamics of this system, we desingularize using geometric desingularization techniques. We describe the dynamics in the neighborhood of the origin using polar coordinates, which leads to a dynamical system on $\R^+\times S^4$, where $z$ is treated as real two-dimensional variable and the parameter $\Im\lambda$ is hidden in the higher-order terms. As we will see later, it suffices to describe the dynamics near the equator $\{z=0\}$, which can be described by the following two directional coordinate charts.

\emph{1-chart:} We define the coordinates $(z_1,\lambda_1)$ via
\[
z=\alpha z_1,\quad
\lambda_\mathrm{r}=\alpha \lambda_1
\] 
and note that these parametrize our system near the $\alpha$-pole of the sphere. After \rev{introducing the independent variable $s$ with $\alpha\frac{\rmd}{\rmd r}=\frac{\rmd}{\rmd s}$ and then using again $'$ to denote $\frac{\rmd}{\rmd s}$}, equation (\ref{e:redproj2}) becomes 
\begin{eqnarray}
z_1' & = & \lambda_1 z_1 + \rev{(1+a_1)} z_1 +  \sum_{j=2}^{M}a_j \alpha^{j-1} z_1 +  \rmO(\alpha^{M+1}) + z_1\rmO(\alpha^{M}) + z_1^2 \rmO(\alpha^{M-1}) \nonumber \\ \label{e:1chart}
\rev{\alpha}' & = & -\rev{\alpha} \\ \nonumber
\lambda_1' & = &\lambda_1.
\end{eqnarray}

\emph{$2^+$-chart ($\lambda_\mathrm{r}>0$):}
To characterize the dynamics near the $z$-pole for $\lambda_\mathrm{r}>0$, we introduce the coordinates
\[
z=\lambda_\mathrm{r} z_2,\quad
\alpha=\lambda_\mathrm{r}\alpha_2.
\]
After \rev{introducing $s$ with $\lambda_\mathrm{r}\frac{\rmd}{\rmd r}=\frac{\rmd}{\rmd s}$ and then denoting $\frac{\rmd}{\rmd s}$ again by $'$}, equation (\ref{e:redproj2}) becomes
\begin{align*}
z_2'=& z_2 + a_1\alpha_2z_2 + \lambda_\mathrm{r} \sum_{j=2}^{M}a_j \alpha_2^j\lambda_\mathrm{r}^{j-2} z_2 + \rmO(\lambda_\mathrm{r}^{M+1}\alpha_2^M) \\
\alpha_2'=&-\alpha_2^2\\
\lambda_\mathrm{r}'=&0.
\end{align*}

\emph{$2^-$-chart (for $\lambda_\mathrm{r}<0$):}
Similarly, we use the variables
\[
z=-\lambda_\mathrm{r} z_2,\quad
\alpha=-\lambda_\mathrm{r}\alpha_2
\]
to parametrize the region near the $z$-pole $\lambda_\mathrm{r}<0$. After \rev{introducing $s$ with $-\lambda_\mathrm{r}\frac{\rmd}{\rmd r}=\frac{\rmd}{\rmd s}$ and then denoting $\frac{\rmd}{\rmd s}$ again by $'$}, equation (\ref{e:redproj2}) becomes
\begin{align*}
z_2'=&-z_2+a_1\alpha_2z_2 + \lambda_\mathrm{r} \sum_{j=2}^{M}a_j \alpha_2^j\lambda_\mathrm{r}^{j-2} z_2 +  \rmO(\lambda_\mathrm{r}^{M+1}\alpha_2^M) \\
\alpha_2'=&-\alpha_2^2 \\
\lambda_\mathrm{r}'=&0.
\end{align*}

% \emph{3-chart:} \[
% \alpha=z\alpha_3,\quad \lambda=z\lambda_3,
% \]
% gives after using the Euler multiplier $z\ '=\dot{\ }$, 
% \begin{align*}
%     z'=&(\lambda_3+a_1\alpha_3)z+\rmO(z^2)\\
%     \alpha_3'=&-  \alpha_3^2(a_1+1) -\lambda_3\alpha_3 + \rmO(z)\\
%     \lambda_3'=&-\lambda_3^2+a_1\alpha_3\lambda_3+\rmO(z)..
% \end{align*}

We glue these coordinate charts together near $\{\alpha_2=1\}$, which corresponds to $\{\lambda_1=\pm1\}$ for the $2^\pm$-chart, respectively. Using this information and the definitions of the charts, we see that these coordinates are related via
\[
z_1 = z_2, \qquad
\alpha=|\lambda_\mathrm{r}|.
\]
In the next step, we analyze the dynamics in the 1- and 2-chart: we refer to Figures~\ref{f:blowupeq} and~\ref{f:blowup} for illustrations of the overall dynamics.

\begin{figure}
\centering
\includegraphics[width=\textwidth]{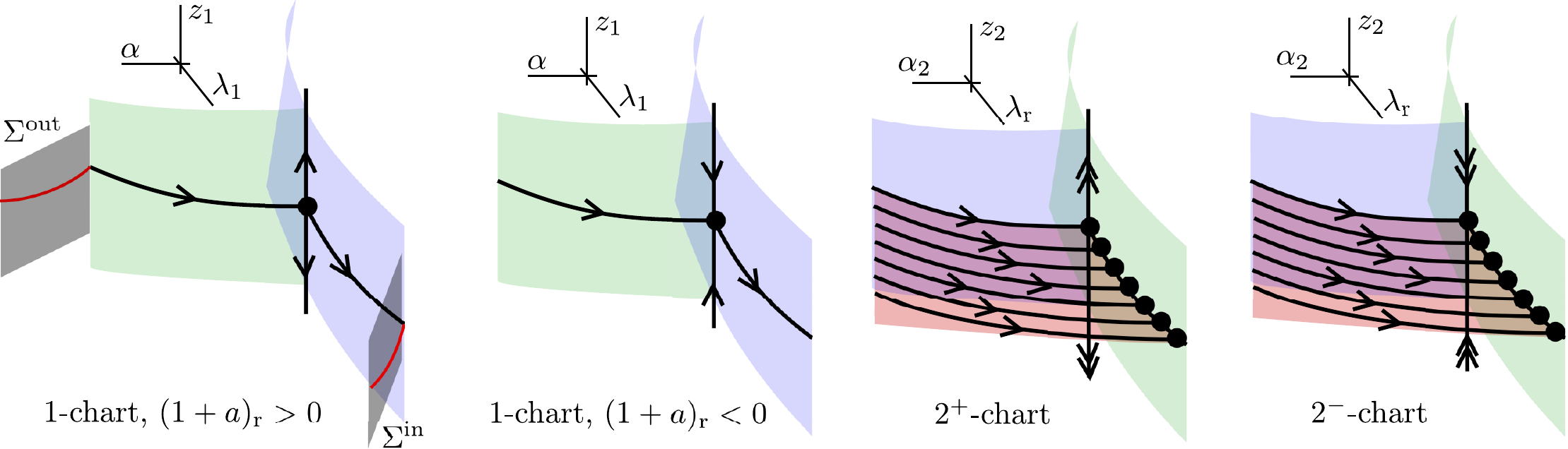}
\caption{Left two panels: We illustrate the dynamics near the origin in the 1-chart, with the sphere shaded in blue, depending on the sign of $(1+\Re a_1)$. Also included in the left panel is a schematics of the Shilnikov sections used to track the stable manifold past the singular equilibrium as described in Steps~6 and~7. Right two panels: Shown is the dynamics near the origin in the 2-charts, including the center manifold and the lines of equilibria.}\label{f:blowupeq}
\end{figure}

\paragraph{Step 5.} 
In the 1-chart, the set $\{\alpha=0\}$ corresponds to the origin in the original equation. This set is invariant and carries non-trivial dynamics due to the singular rescaling of time by $\alpha$. In this chart, the origin is an equilibrium that is stable in the direction of $\alpha$, perpendicular to the sphere, with eigenvalue $-1$. The eigenvalues within the sphere are \rev{$+1$} and $(1+a_1)$, \rev{which} depends on the leading-order correction in the expansion. The $z$-equator $\{z_1=0\}$ \rev{inside the sphere} is invariant, and solutions in this set converge to the origin in backwards time in the $1$-chart \rev{due to the eigenvalue $+1$}; see Figure~\ref{f:blowupeq}.

In \rev{each} 2-chart, there is a line of equilibria $\alpha_2=z_2=0$ that correspond to the asymptotic eigenspaces parametrized by $\lambda_\mathrm{r}$. The linearization at the equilibrium $\alpha_2=z_2=|\lambda_\mathrm{r}|=0$ inside the singular sphere has an eigenvalue zero associated with this line of equilibria, and an additional eigenvalue zero associated with the dynamics in the $\alpha_2$-direction. In addition, there is an eigenvalue $\pm1$ associated with the $z_2$-direction in the $2^\pm$-chart, respectively. The $\alpha_2$-axis is invariant with solutions converging to $\alpha_2=0$ in forward time. \rev{Recall that $\lambda_\mathrm{r}$ is real and that $\Im\lambda$ appears only as a hidden parameter in the higher-order terms of the equation for $z_2$. We will omit the hidden variable $\Im\lambda$ from our dimension counts below.}

As illustrated in Figure~\ref{f:blowup}, \rev{two heteroclinic orbits connect} the origin in the 1-chart and the origins in the $2^\pm$-charts along the $z$-equator $z_1=0$ and $z_2=0$, respectively. \rev{Eigenfunctions in the $2^\pm$-chart} correspond to trajectories that converge to the manifold of equilibria $\alpha_2=z_2=0$ as \rev{time} goes to infinity. \rev{Our goal is to construct a smooth curve in the section $\Sigma_\mathrm{out}=\{\alpha=\delta\}$ in the 1-chart that is parametrized by $\lambda_\mathrm{r}$ near zero so that the corresponding solutions converge to the manifold of equilibria at time goes to infinity in the $2^\pm$-chart depending on the sign of $\lambda_\mathrm{r}$. We accomplish this by tracking solutions in the center-stable manifolds of these equilibria back towards $\Sigma_\mathrm{out}$ and establishing smoothness in $\lambda_\mathrm{r}$ in this section. In the $2^-$-chart, all solutions converge to the manifold of equilibria as time goes to infinity. We therefore focus first on the $2^+$-chart. In this chart, we are interested in the two-dimensional center} manifold of these equilibria. The tangent space to this manifold at $\lambda_\mathrm{r}=0$ is simply the two-dimensional center eigenspace $z_2=0$. Inside this two-dimensional center manifold, the dynamics is given by $\alpha_2'=-\alpha_2^2$ and $\lambda_\mathrm{r}'=0$, and trajectories indeed correspond to the stable manifolds of the asymptotic equilibria. We are interested in tracking this manifold backward in time to and past the origin in the 1-chart to a finite value of $\alpha$. Note that the Taylor jet of this center manifold is of arbitrarily high order $\rmO(|\alpha_2\lambda_\mathrm{r}|^M)$ due to the fact that we diagonalized eigenspaces up to order $M$.

\begin{figure}
\centering
\includegraphics[width=0.8\textwidth]{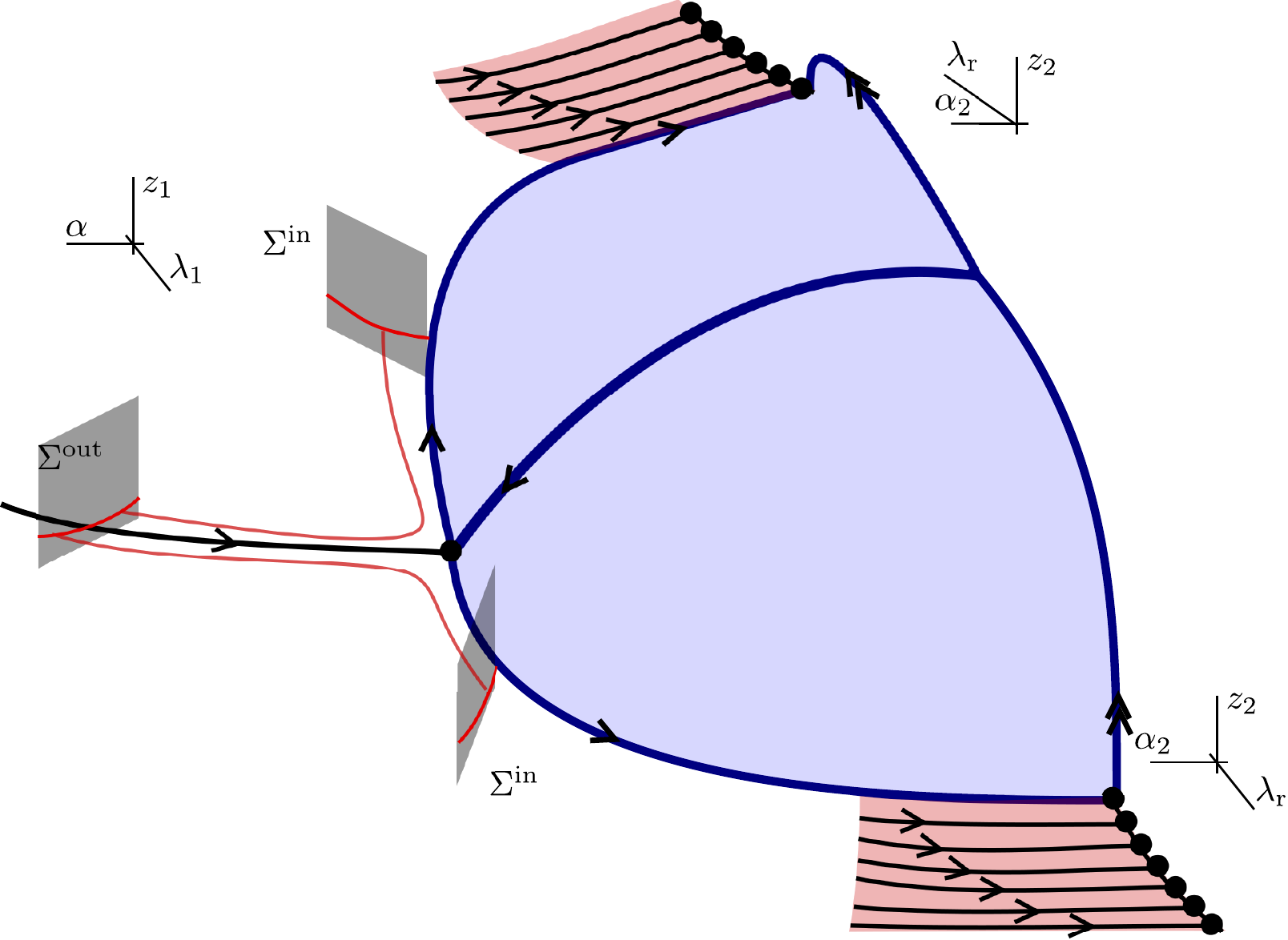}
\caption{Shown is the singular sphere with the 1- and 2-charts, singular equilibria, connecting orbits, and Shilnikov passage constructions indicated.}\label{f:blowup}
\end{figure}

\paragraph{Step 6.} 
In order to track the center manifold near the set $\{z_2=0\}$ past the 1-chart origin, we need to analyze the passage map near \rev{the origin in the 1-chart}. This is a somewhat standard Shilnikov-type problem, and we outline the results of this analysis here. The center manifold that we track is given as a graph $z_1=h(\rev{\alpha})$, with $h=\rmO(\rev{\alpha}^M)$, in a cross section $\Sigma^\mathrm{in}=\{\lambda_1=\delta\}$ of the flow. We wish to transport this manifold backwards past the equilibrium to a section $\Sigma^\mathrm{out}=\{\rev{\alpha}=\delta\}$. We add superscripts to the variables in the cross section and, due to the fact that $\lambda$ is constant and does not evolve, find that
\[
\lambda_1^\mathrm{in}\alpha^\mathrm{in} = \lambda_1^\mathrm{out}\alpha^\mathrm{out} \quad\text{ hence }\quad \lambda_1^\mathrm{out} = \alpha^\mathrm{in}.
\]
We determine $z_1^\mathrm{out}$ from the equation for $z_1'$ by integrating for a time $T$ with $\rme^{-T}=\alpha^\mathrm{in}/\delta$. Using only the linearization for illustration, we expect to arrive at
\[
z_1^\mathrm{out} = \left(\frac{\alpha^\mathrm{in}}{\delta}\right)^{a+1} \rmO\left(|\alpha^\mathrm{in}|^{M}\right) = \rmO\left(|\lambda_1^\mathrm{out}|^{M+a+1}\right).
\]
In the next step, we show how this linear calculation can be turned into a nonlinear estimate.

\paragraph{Step 7.} 
\rev{We observe that we can write the first equation in (\ref{e:1chart}) in the 1-chart as
\[
z_1^\prime = (1+a(\alpha,\lambda_1)) z_1 + \alpha^{M-1} R(z_1,\alpha,\lambda_1),
\]
where the functions $a(\alpha,\lambda_1)$ and $R(z_1,\alpha,\lambda_1)$ are smooth. We artificially augment the system (\ref{e:1chart}) by introducing the additional variable $\bar{\alpha}:=\alpha^{M-1}$ to arrive at 
\begin{align*}
z_1' = & (1+a(\alpha,\lambda_1)) z_1  +  \bar{\alpha} R(z_1,\alpha,\lambda_1) \\
{\alpha}' = & -{\alpha} \\
\lambda_1' = & \lambda_1 \\
\bar{\alpha}' = & -(M-1)\bar{\alpha}. 
\end{align*}
For $M$ large enough, the dynamics in $\bar{\alpha}$ is strongly contracting, and the system is therefore  fibered in a neighborhood of the origin over the invariant subspace $\bar{\alpha}=0$. A smooth coordinate transformation 
\[
 (\tilde z_1,\tilde \alpha,\tilde \lambda_1)=(z,\alpha,\lambda_1)+\bar{\alpha}\Psi(z_1,\alpha,\lambda_1;\bar{\alpha})
\]
will straighten out these fibers such that, after this transformation, the dynamics in $(\tilde{z}_1,\tilde{\alpha},\tilde{\lambda}_1)$ are independent of $\bar{\alpha}$, hence given through the equation at $\bar{\alpha}=0$,
\begin{align*}
\tilde{z}_1' = & (1+a(\tilde{\alpha},\tilde{\lambda}_1)) \tilde{z}_1 \\
\tilde{\alpha}' = & -\tilde{\alpha} \\
\tilde{\lambda}_1' = & \tilde{\lambda}_1.
\end{align*}
We can now solve explicitly for $\tilde{z}^\mathrm{out}$ and recover the desired estimates $\tilde{z}_1^\mathrm{out}=\rmO(|\tilde{\lambda}_1^\mathrm{out}|^{\tilde{M}})$, where $\tilde{M}$ can be chosen large provided $M$ is large, with equivalent estimates for the derivatives due to the linearity in $\tilde{z}_1$. }

\rev{In the original coordinates, $z_1^\mathrm{out}$ is therefore a smooth function of $\lambda_1^\mathrm{out}$ up to $\lambda_1^\mathrm{out}=0$, which completes the analysis of the $2^+$-chart. Next, we extend this function smoothly into the region $\lambda_1^\mathrm{out}<0$. Reversing the analysis presented above in time, we can track this manifold through the 1-chart to the $2^-$-chart, where all trajectories limit on the family of stable equilibria.}

\paragraph{Step 8.}
The preceding analysis provides $z_*(\lambda,\alpha)$ for each $\alpha$ with $0\leq\alpha\leq\alpha_*=\delta$ and shows that this expression is smooth in $\lambda$ \rev{near $\lambda=0$}. The change of coordinates $z\mapsto z-z_*(\lambda,\alpha)=:\tilde{z}$ gives, upon omitting the tilde, the new equation 
\[
z' = (\lambda+a(\lambda,\alpha)) z + \rmO(\alpha^{M})z^2,
\]
which, in particular, leaves $z=0$ invariant. Inverting $\xi:=1/z$, we find 
\[
\xi' = -(\lambda+a(\lambda,\alpha)) \xi + \rmO(\alpha^{M}).
\]
We can now repeat the construction outlined above for this new equation, which results in a function $\xi_*(\lambda,\alpha)$ with analogous properties to those of $z_*$. Subtracting this expression, we obtain an equation for $\xi$, or $z$, that is linear. Linear equations on the Grassmannian correspond to linear systems whose matrix is diagonal as claimed. This completes the proof of the proposition.
\end{Proof}

In the coordinates provided by Proposition~\ref{p:absspecdiag}, we can solve the equation inside the center eigenspace explicitly by separation of variables. In particular, there exist unique solutions that converge with asymptotics associated with the eigenvalues $\nu_{-1,0}(\lambda)$, respectively. Since the ordering of these two eigenvalues by real part is exchanged upon crossing the absolute spectrum, solutions with asymptotics given by $\nu_0(\lambda)$ give rise to eigenfunctions to one side of the absolute spectrum, while solutions with asymptotics associated with $\nu_{-1}(\lambda)$ give rise to eigenfunctions on the opposite side. With this background, we can now formulate the concept of a resonance in the absolute spectrum (see Definition~\ref{d:absgen2}) more precisely.

Recall that the system (\ref{e:evpcon}) associated with the eigenvalue problem $\mathcal{L}_Ru=\lambda u$ is given by
\begin{equation}\label{e:1043}
\mathbf{w}_r = \mathcal{A}_\lambda(r) \mathbf{w}.
\end{equation}
This equation admits exponential dichotomies on $0\leq r\leq R_*$, and the corresponding center-unstable subspace $E^\mathrm{cu}_-(R_*)$ contains all initial conditions whose associated solutions correspond to bounded solutions of $\mathcal{L}_Ru=\lambda u$ on $r\leq R_*$.

\begin{Definition}[Resonances in the absolute spectrum]\label{d:resonance}
Fix $\lambda_*$ in the simple part of the absolute spectrum.
\begin{compactitem}
\item[(i)] \emph{Resonances caused by the spiral wave:}
\rev{We say that there is a resonance in the absolute spectrum at $\lambda_*$ caused by the spiral wave if there exists a solution $\mathbf{w}(r)$ of (\ref{e:1043}) defined for $r\geq R_*$ with $\mathbf{w}(R_*)\in E^\mathrm{cu}_-(R_*)\cap E^\mathrm{cs}(R_*)$ so that, in the coordinates of Proposition~\ref{p:absspecdiag}, at least one of its two center components in $E^\mathrm{c}(r)$ vanishes for one, and hence all, values of $r$.}
\item[(ii)] \emph{Resonances caused by the boundary conditions:}
We say that there is a resonance in the absolute spectrum at $\lambda_*$ caused by the boundary conditions if $E^\mathrm{bc}\cap (E^\mathrm{c}_\infty\oplus E^\mathrm{u}_\infty)$ contains a vector whose component in $E^\mathrm{c}_\infty$ lies in $\R(1,0)^T$ or $\R(0,1)^T$.
\end{compactitem}
%\begin{compactitem}
%\item[(i)] \emph{Resonances caused by the spiral wave:}
%We say that a solution $\mathbf{w}(r)$ of (\ref{e:1043}) defined for $r\geq R_*$ with $\mathbf{w}(R_*)\in E^\mathrm{cu}_-(R_*)\cap E^\mathrm{cs}(R_*)$ at $r=R_*$ has asymptotics $\nu_0(\lambda_*)$ if its component in $E^\mathrm{c}(r)$, in the coordinates of Proposition~\ref{p:absspecdiag}, lies in $\R(0,1)^T$; we say that it has asymptotics $\nu_{-1}(\lambda_*)$ if the component in $E^\mathrm{c}(r)$ lies in $\R(1,0)^T$. We then say that there is a resonance in the absolute spectrum at $\lambda_*$ caused by the spiral wave if there exists a solution $\mathbf{w}(r)$ to (\ref{e:1043}) with $\mathbf{w}(R_*)\in E^\mathrm{cu}_-(R_*)\cap E^\mathrm{cs}(R_*)$ and its component in $E^\mathrm{c}(r)$ has asymptotics $\nu_0(\lambda_*)$ or $\nu_{-1}(\lambda_*)$ as $r\to\infty$. Note that we do allow the component in $E^\mathrm{c}$ to vanish.
%\item[(ii)] \emph{Resonances caused by the boundary conditions:}
%We say that there is a resonance in the absolute spectrum at $\lambda_*$ caused by the boundary conditions if $E^\mathrm{bc}\cap (E^\mathrm{c}_\infty\oplus E^\mathrm{u}_\infty)$ contains a vector whose component in $E^\mathrm{c}_\infty$ lies in $\R(1,0)^T$ or $\R(0,1)^T$.
%\end{compactitem}
\end{Definition}

\rev{We will exclude points in the absolute spectrum that exhibit a resonance as these points are more challenging to handle. In the context of travelling waves, the extension of the Evans function into the absolute spectrum vanishes at these points, and it is then possible that discrete eigenvalues or resonance poles emerge at these points upon adding small bounded perturbation to the underlying operator. This phenomenon was first observed in \cite{Simon} in the context of multi-dimensional Schr\"odinger operators and later found for one-dimensional travelling waves and degenerate shock waves in \cite{GZGap, KSGap, HowardNoGap}.}

\begin{Proposition}[Accumulation of eigenvalues near the absolute spectrum]\label{p:acc}
Suppose that $\lambda_*$ is a point in the simple absolute spectrum with no resonances caused by the planar spiral wave or the boundary conditions. For each fixed number $m$ there are constants $C,R_0$ such that the linearization at the spiral wave with Robin boundary conditions imposed at $|x|=R$, $R\geq R_0$,  possesses $m$ distinct eigenvalues $\lambda_j$ (with $1\leq j\leq m$) with $|\lambda_j-\lambda_*|\leq C/R$.
\end{Proposition}

\begin{Proof}
Since we assume that $\Sigma_\mathrm{abs}$ is simple at $\lambda_*$, the configuration of Morse indices of the spatial dynamical system (\ref{e:1043}) at $\lambda_*$ implies that $E^\mathrm{cu}_-(R_*)\cap E^\mathrm{cs}_+(R_*)$ has dimension at least one. We claim that the assumption that there are no resonances caused by the spiral wave implies that
\begin{equation}\label{e:106a}
\dim(E^\mathrm{cu}_-(R_*)\cap E^\mathrm{cs}_+(R_*)) = 1, \qquad
E^\mathrm{cu}_-(R_*)\cap E^\mathrm{s}_+(R_*) = \{0\}.
\end{equation}
Indeed, the second identity follows immediately as any nontrivial element in the intersection would yield a solution that decays exponentially as $r\to\infty$, thus causing a resonance as the dichotomy projection on the center space vanishes identically. The first identity follows similarly upon observing that if there are two linearly independent solutions with initial conditions in the intersection, then these solutions either span $E^\mathrm{c}_+(r)$, leading to a resonance, or one of them lies in $E^\mathrm{s}_+(r)$, yielding a contradiction to the second identity that we already proved.

Since $E^\mathrm{c}_+(R_*)$ is two-dimensional, it follows \rev{from (\ref{e:106a})} that there is a one-dimensional subspace $\rev{\tilde{V}}\subset E^\mathrm{c}_+(R_*)$ so that
\[
(E^\mathrm{cu}_-(R_*)\cap E^\mathrm{cs}_+(R_*)) \oplus E^\mathrm{s}_+(R_*) \oplus \tilde{V} = E^\mathrm{cs}_+(R_*).
\]
We now define
\[
\tilde{E}^\mathrm{cu}_+(R_*) := E^\mathrm{cu}_-(R_*) \oplus \tilde{V}, \qquad
\tilde{E}^\mathrm{s}_+(R_*) := E^\mathrm{s}_+(R_*)
\]
and note that $\tilde{E}^\mathrm{cu}_+(R_*)\oplus\tilde{E}^\mathrm{s}_+(R_*)=X$. We can use these two subspaces to define an exponential dichotomy on $[R_*,R]$ with rates and constants that are uniform in $R$. In particular, we have that $\tilde{E}^\mathrm{cu}_+(R)$ and $\tilde{E}^\mathrm{s}_+(R)$ are $\rmO(\rme^{-(R-R_*)})$ close to $E^\mathrm{cu}_\infty$ and $E^\mathrm{s}_\infty$, respectively.

Proceeding in the same way at $r=R$, and using our assumption that there are no resonances at $\lambda_*$ caused by the boundary conditions, we see that 
\[
\dim(E^\mathrm{bc}\cap E^\mathrm{cu}_+(R)) = 1, \qquad
E^\mathrm{bc}\cap E^\mathrm{u}_+(R) = \{0\}.
\]
We conclude that there is a one-dimensional subspace $\hat{V}\subset E^\mathrm{c}_+(R)$ so that
\[
(E^\mathrm{bc}\cap E^\mathrm{cu}_+(R)) \oplus E^\mathrm{u}_+(R) \oplus \hat{V} = E^\mathrm{cu}_+(R).
\]
This allows us to define the complementary spaces
\[
\hat{E}^\mathrm{cs}_+(R) := E^\mathrm{bc} \oplus \hat{V}, \qquad
\hat{E}^\mathrm{u}_+(R) := E^\mathrm{u}_+(R)
\]
and use them to define an exponential dichotomy on $[R_*,R]$ with rates and constants that are uniform in $R$. The resulting subspaces $\hat{E}^\mathrm{cs}_+(R_*)$ and $\hat{E}^\mathrm{u}_+(R_*)$ are $\rmO(\rme^{-(R-R_*)})$ close to $E^\mathrm{cs}_+(R_*)$ and $E^\mathrm{u}_+(R_*)$, respectively.

It follows that $E^\mathrm{c}_\mathrm{new}(r):=\tilde{E}^\mathrm{cu}_+(r)\cap\hat{E}^\mathrm{cs}_+(r)$ is two-dimensional with
\[
\dim(E^\mathrm{c}_\mathrm{new}(R_*)\cap E^\mathrm{cu}_-(R_*)) = 1, \qquad
\dim(E^\mathrm{c}_\mathrm{new}(R)\cap E^\mathrm{bc}) = 1.
\]
We conclude that there is a solution $\mathbf{w}(r)$ of (\ref{e:1043}) with $\mathbf{w}(R_*)\in E^\mathrm{cu}_-(R_*)$ and $\mathbf{w}(R)\in E^\mathrm{bc}$ if and only if $\mathbf{w}(R_*)\in E^\mathrm{cu}_-(R_*)\cap E^\mathrm{c}_\mathrm{new}(R_*)$ and $\mathbf{w}(R)\in E^\mathrm{bc}\cap E^\mathrm{c}_\mathrm{new}(R)$.

To find such intersections, we now turn to the diagonal coordinates in the center subspace constructed in Proposition~\ref{p:absspecdiag}. Each intersection corresponds to a solution $z(r)$ of the boundary-value problem
\begin{equation}\label{e108p}
z'=(\lambda+h_\lambda(r))z,\qquad
z(R_*)=z_0(\lambda),\qquad
z(R)=z_+(\lambda),
\end{equation}
in projective space, where $0<|z_{0,+}(\lambda)|<\infty$ due to the absence of resonances. Defining $\eta:=\log z$ and $\eta_{0,+}(\lambda):=\log z_{0,+}(\lambda)$, we find that
\[
\eta' = \lambda+h_\lambda(r)  \quad\Longrightarrow\quad
\eta_+ = \eta_0 + \lambda (R-R_*) + \int_{R_*}^R h_\lambda(s)\rmd s,
\]
which we write in the form 
\[
\lambda = \frac{\eta_+-\eta_0}{R-R_*} \rev{\,-\,} \frac{1}{R-R_*} \int_{R_*}^R h_\lambda(s)\rmd s. 
\]
Exploiting the nonuniqueness of the logarithm, we set $\eta_+(\lambda)=\eta_+^0(\lambda)+2\pi\rev{\rmi} j$ (with $0\leq j<m$ and $j\in\Z$) and write $\zeta(\lambda)=\eta_+^0(\lambda)-\eta_0(\lambda)$ to arrive at the equation 
\begin{equation}\label{e:1042}
\lambda = \frac{2\pi\rmi j}{R-R_*} + \frac{\zeta(\lambda)}{R-R_*} \rev{\,-\,} \frac{1}{R-R_*}\int_{R_*}^R h_\lambda(s)\rmd s.
\end{equation}
Since $h$ has an expansion in $r^{-1}$, we conclude that the last term and its derivative with respect to $\lambda$ are bounded by $\frac{\log(R-R_*)}{R-R_*}$. In particular,  the right-hand side of (\ref{e:1042}) defines a contraction in $\lambda$ for all sufficiently large $R$, and we find eigenvalues $\lambda$ within a ball of radius $R^{-1}$ for each $0\leq j<m$, which proves the proposition. 
\end{Proof}

%%%%%%%%%%%%%%%%%%%%%%%%%%%%%%%%%%%%%%%%%%%%%%%%%%%%%%%%%%%%%%%%%%%%%%%%%

\section{Spectra of truncated spiral waves}\label{s:absglue}

This section extends the results from \S\ref{s:abs} to include corrections to the nonlinear spiral wave profile, constructed in \S\ref{s:trunc}. The solutions constructed can be thought of as spiral waves glued to a boundary sink that corrects for the influence of the boundary conditions. In comparison with the situation in \S\ref{s:abs},  the additional difficulty  due to this gluing procedure is to account for the boundary sink, effectively replacing the boundary spectrum $\Sigma_\mathrm{bdy}$ in the results of \S\ref{s:abs} with the extended point spectrum of the boundary sink.

Many of the proofs are analogous to the proofs in \S\ref{s:abs} and we will mainly point out the key differences. \rev{We start in \S\ref{s:11.0} by collecting some geometric information on the linearization at the boundary sink, depending on the eigenvalues $\lambda$. In \S\ref{s:11.1}, we characterize the resolvent set and point eigenvalues away from the absolute spectrum, before we consider accumulation of eigenvalues onto the absolute spectrum in \S\ref{s:11.2}}. 

\subsection{Geometry of the linearization at boundary sinks}\label{s:11.0}

The eigenvalue problem near a boundary sink can be written in spatial dynamics on $x<0$,
\begin{align*}
u_x = &  v \\ \nonumber
v_x = &  - D^{-1}[-\omega\partial_\tau u + f'(u_\mathrm{bs})u-\lambda u]
\end{align*}
with boundary subspace as in \eqref{e:tybc},
\[
\mathbf{u}(0) \in E^\mathrm{bc}_1 =
\{ (u,v)\in Y;\; (u(\tau),v(\tau))\in E^\mathrm{bc}_0 \;\forall\tau \}.
\]
More conveniently, we consider the equation in the corotating coordinates
\begin{align}\label{e:tyl2}
u_x = &  -k_* \partial_\sigma u + v \\ \nonumber
v_x = &  -k_* \partial_\sigma v - D^{-1}[-\omega\partial_\tau u + f'(u_\mathrm{bs})u-\lambda u],
\end{align}
where the coefficients $u_\mathrm{bs}(x,\sigma)\to u_\infty(\sigma)$ converge to an $x$-independent limit. 
We collect geometric information on this equation that results from our spectral assumptions.

\paragraph{The case $\lambda_*\notin \Sigma_\mathrm{abs}$.} In this case, we can conjugate the equation with an exponential weight $\eta\in J_0(\lambda_*)$, considering $(\tilde{u},\tilde{v})=\rme^{\eta x} (u,v)$, and find an exponential dichotomy on $x<0$ with projections  $P^\mathrm{s/u}(x)$. Moreover, the $P^\mathrm{s/u}$ converge exponentially to the corresponding projections $P_\mathrm{wt}^\mathrm{s/u}$ of the asymptotic wave train. If in addition $\lambda_*$ does not belong to the extended point spectrum, we may choose $\Rg P^\mathrm{s}(0)=E^\mathrm{bc}_0$. 
 
\paragraph{The case $\lambda_*\in \Sigma_\mathrm{abs}$.} In this case, we again conjugate with an exponential weight $\eta=-\Re\nu_{0}(\lambda_*)$ and we can then define a trichotomy for the shifted equation with projections $P^\mathrm{s/c/u}(x)$, $\,\mathrm{dim}\,\Rg(P^\mathrm{c})=2$. Moreover, each solutions in $\Rg(P^\mathrm{c})$ satisfies
\[
|(u,v)(x)-\alpha_0 e_0\rme^{\nu_0(\lambda_*)}-\alpha_{-1}e_{-1}\rme^{\nu_{-1}(\lambda_*)}|\longrightarrow 0
\]
at an exponential rate for some $\alpha_0,\alpha_{-1}\in\C$. 

\begin{Definition}[Boundary resonance in absolute spectrum]\label{d:resonance2}
For a point $\lambda_*\in\Sigma_\mathrm{abs}$ where the absolute spectrum is simple, we say that the linearization at the boundary sink possesses a resonance if there exists an solution to \eqref{e:tyl2} with $(u,v)(x,\sigma)\in L^2_{\eta-\varepsilon}$, for some $\varepsilon>0$, arbitrarily small,  such that the component in $\Rg P^\mathrm{c}$ satisfies $\alpha_0=0$ or $\alpha_{-1}=0$.
\end{Definition}

\paragraph{Continuity in $\omega$.}
Non-degenerate boundary sinks come in one-parameter families parametrized by $\omega$; see Lemma~\ref{l:tr}. In corotating coordinates $\sigma$, the boundary sinks depend smoothly on $\omega$ as functions in $L^\infty$.  We may then consider the spectral properties described above for nearby values of $\omega$.  Continuity of exponential dichotomies with respect to the parameter $\omega$, through explicit dependence and implicit dependence in $k$ and the profile, gives continuity of the exponential dichotomies in $\omega$ and thereby continuity of absolute and extended point spectra of the boundary sink. Similarly, absence of resonances is robust with respect to changes in $\omega$.

\subsection{Eigenvalues and resolvent outside of the absolute spectrum}\label{s:11.1}

We exclude eigenvalues in the complement of the limiting spectrum, that is, for $\lambda_*$ in the complement of extended point spectra of spiral wave and boundary sink, and not in the absolute spectrum. 

\begin{Lemma}[Resolvent continuity under truncation]\label{l:res1}
    Suppose that $\lambda_*$ does not belong to the absolute spectrum, the extended point spectrum, or the boundary spectrum; then there exists $\delta>0$ and $\bar{R}>0$ such that $B_\delta(\lambda_*)$ belongs to the resolvent set of $\mathcal{L}_{\mathrm{s},R}$ for all $R>\bar{R}$. Moreover, $\bar{R}(\lambda_*)$ can be chosen uniformly in compact subsets of the complement of absolute, extended, and boundary spectrum.
\end{Lemma}
\begin{Proof}
    Convergence estimates in the construction of the boundary sink give us that the truncated spiral is uniformly close to the profile of a boundary sink on $r\geq R-\kappa^{-1}\log r$ with nearby frequency and uniformly close to the spiral wave on $r\leq 
    R-\kappa^{-1}\log r$. Since stable and unstable subspaces of spiral waves converge to the stable and unstable subspaces of the wave trains for $r\to\infty$, and stable and unstable subspaces of the boundary sink similarly similarly converge for $x\to -\infty$, we find can conclude transversality of the unstable subspace for the spiral wave and the stable subspace for the \rev{boundary sink} at the gluing point $r=R-\kappa^{-1}\log R$, which implies existence of an exponential dichotomy near the glued profile and absence of spectrum for all $R$ sufficiently large and nearby values of the parameter $\lambda$. 
\end{Proof}

Establishing persistence of eigenvalues in the extended point spectrum of the spiral wave or the boundary sink is equivalent to the constructions in \S\ref{s:10.2} since the equation near the boundary sink possesses exponential dichotomies. 

\subsection{Accumulation of eigenvalues \rev{onto} the absolute spectrum}\label{s:11.2}

We now show how to adapt the techniques from \S\ref{s:10.4} to establish accumulation of eigenvalues near simple points of the absolute spectrum, assuming absence of resonances. We consider the linearized equation
\[
\mathbf{w}'=\mathcal{A}_{R,\lambda}(r)\mathbf{w}
\]

with parameters $\lambda$ and $R$ on $R_*<r<R$ with boundary conditions
\[
\mathbf{w}\in E_-^\mathrm{u} \mbox{ at } r=R_*\quad\mbox{and}\quad \mathbf{w}\in E_1^\mathrm{bc} \mbox{ at } r=R.
\]
We choose $\lambda\sim \lambda_*$, with $\lambda_*$ in the simple absolute spectrum, not a resonance for boundary sink or spiral wave. 
\paragraph{Step 1: Relaxing the boundary conditions.} The linearization at the boundary sink possesses an exponential trichotomy with subspaces $E^\mathrm{s,c,u}_\mathrm{bs,\infty}(r)$, $r\leq R$, \rev{where we shifted} the boundary sink profile and the associated linearized equation as in the construction in \S\ref{s:trunc} such that the boundary condition is situated at $r=R$. Absence of resonances implies that  $E_1^\mathrm{bc}\cap E^\mathrm{cu}_\mathrm{bs,\infty}(R)$ and we can choose a one-dimensional complement $V_\mathrm{bs}\subset  E^\mathrm{c}_\mathrm{bs,\infty}(R)$ such that $E_1^\mathrm{bc}\oplus V_\mathrm{bs} \oplus E^\mathrm{u}_\mathrm{bs,\infty}(R)=X$. We can then assume that $E^\mathrm{cs}_\mathrm{bs,\infty}(R)=E_1^\mathrm{bc}\oplus V_\mathrm{bs}$. 

Similarly, \rev{the linearization about the primary spiral wave has an exponential trichotomy with subspaces $E^\mathrm{s,c,u}_\mathrm{sp,\infty}(r)$ for $r\geq R_*$. The} intersection $ E_-^\mathrm{u}\cap E^\mathrm{cs}_\mathrm{sp,\infty}$ is one-dimensional and we can  choose a one-dimensional complement $V_\mathrm{sp}\subset  E^\mathrm{c}_\mathrm{sp,\infty}(R_*)$ such that $ E_-^\mathrm{u}\oplus V_\mathrm{sp} \oplus E^\mathrm{s}_\mathrm{sp,\infty}=X$. We can now assume that $E^\mathrm{cu}_\mathrm{sp,\infty}(R_*)=E_-^\mathrm{u}\oplus V_\mathrm{sp}$. 

\paragraph{Step 2: Robustness and transversality.}
The dichotomies constructed in the first step are robust and yield  exponential trichotomies with subspaces $E^\mathrm{s,c,u}_\mathrm{sp,bs}$ on $R_*\leq r<R_0$ and on  $R_0<r<R$, where $R_0=R-\kappa^{-1}\log(R)$. In particular, exponential dichotomies converge to the corresponding trichotomy of the wave train. As a consequence, at $r=R_0$, we have transversality
\[
E^\mathrm{cs}_\mathrm{bs}\oplus E^\mathrm{cu}_\mathrm{sp}=X,\qquad E^\mathrm{cs}_\mathrm{bs}\cap  E^\mathrm{cu}_\mathrm{sp}=:E^\mathrm{c}_R
\]
with 
\[
E^\mathrm{c}_R\cap E^\mathrm{u}_\mathrm{sp}=\{0\},\qquad E^\mathrm{c}_R\cap E^\mathrm{s}_\mathrm{bs}=\{0\}. 
\]
We can now continue this two-dimensional intersection  $E^\mathrm{c}_R$ along $r$ to find $E^\mathrm{c}_R(r)$, $R_*\leq r<R$. By construction, $E^\mathrm{c}(r)$ and the flow in these subspaces converge exponentially to the flow on $E^c_\mathrm{sp}(r)$. 

\paragraph{Step 3: Un-relaxing the boundary conditions.}
Eigenfunctions for finite $R$ are in one-to-one correspondence to solutions of the ODE in the subspace $E^\mathrm{c}_R(r)$ that also satisfy the boundary condition, that is, whose component in $V_\mathrm{bs}$ and $V_\mathrm{sp}$ vanish at $r=R$ and $r=R_*$, respectively. 

\paragraph{Step 4: Scattering and reduction to the pure spiral.}
The resulting equation on $E^\mathrm{c}_R(r)$ can be identified with an equation on $E^\mathrm{c}_\infty$ as a convenient trivialization of the two-dimensional bundle. Exponential convergence implies that this equation is, in suitable coordinates,  of the form given in Proposition~\ref{p:absspecdiag}, with an exponentially decaying correction
\[
 \mathbf{w}_\mathrm{c}'=\left(A_\mathrm{c}(r,\lambda)+ B(r;R,\lambda)\right)\mathbf{w}_\mathrm{c}, \qquad |B(r;R,\lambda)|\leq C\rme^{-\delta (R-r)}
\]
for some constants $C,\delta>0$ that are independent of $R,r,\lambda$. Exponential decay gives a continuous foliation over the asymptotic equation, that is, we have
\[
\mathbf{w}_\mathrm{c}(r)=\Psi(r;R,\lambda)\mathbf{w}_\mathrm{c}^\infty(r)
\]
where $\mathbf{w}_\mathrm{c}^\infty$ satisfies
\[
 (\mathbf{w}_\mathrm{c}^\infty)'= A_\mathrm{c}(r,\lambda)\mathbf{w}_\mathrm{c}^\infty,
\]
and $\Psi(r;R,\lambda)$ is continuous in $r$, decays exponentially so that $|\Psi(r;R,\lambda)|\leq C\rme^{-\delta (R-r)}$, and has the limits
\[
\Psi(r;R,\lambda)\to \Psi(r;\infty,\lambda) \mbox{ as }  R\to\infty,\qquad \Psi(r;\infty,\lambda)\to 0 \mbox{ as } \ r\to \infty.
\]
In summary, we reduced our eigenvalue problem to an boundary-value problem for the linearized equation for the primary spiral with boundary conditions pulled back from the foliation $\Psi$. Absence of resonances, as used in \S\ref{s:10.4}, follows from the assumptions on absence of resonances for boundary sink and spiral. 

\paragraph{Step 5: Conclusion.} As a consequence, we reduced the problem to precisely the problem studied in \S\ref{s:10.4}. We reduced to an equation of the form given in Proposition~\ref{p:absspecdiag} with boundary conditions at $r=R $ and at $r=R_*$. Proceeding as in Proposition~\ref{p:acc} now establishes accumulation of eigenvalues for the truncated problem near $\lambda_*$ as $R\to\infty$ and concludes the proof of Theorem~\ref{t:tl2}.

%%%%%%%%%%%%%%%%%%%%%%%%%%%%%%%%%%%%%%%%%%%%%%%%%%%%%%%%%%%%%%%%%%%%%%%%%

\section{Applications to spiral-wave dynamics and discussion}\label{s:appl}

The theory developed here can illuminate many experimental and numerical observations of spiral-wave dynamics. In order to illustrate the role of our results in the prediction and understanding of observations, we return to the phenomena alluded to in the introduction. In \S\ref{s:rigid}, we discuss the viewpoint that spiral waves are robust coherent structures that can be continued in parameter space, both analytically and numerically, on large but bounded domains. We then discuss possible instabilities and how they relate to the fine structure of spectra developed here in \S\ref{s:bif}. We conclude with a discussion of selected open problems in \S\ref{ss:op}. We focus here on the phenomena and relegate details of numerical algorithms, their implementation, and the PDE models used for our computations to the appendix. 
% \S\ref{s:sim}.

\subsection{Rigid rotation, truncation, and continuation.}\label{s:rigid}

\begin{figure}
\centering
\includegraphics{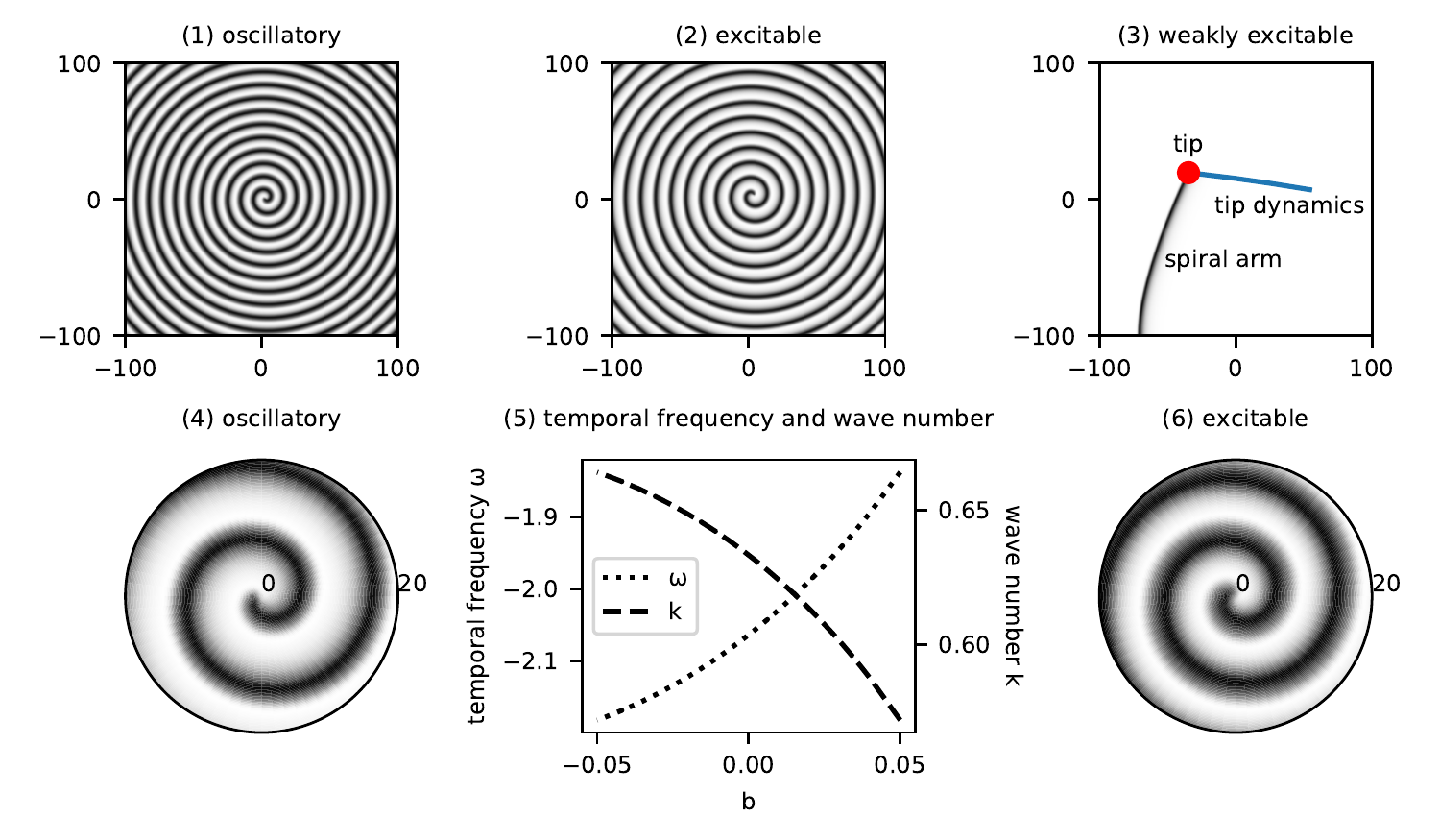}
\caption{Panels (1)-(3) show snapshots of spiral waves in the Barkley model in the oscillatory, excitable, and weakly excitable regimes. Panels~(4) and~(6) show snapshots of spiral waves in the oscillatory ($b=-0.05$) and excitable ($b=0.05$) regimes that were computed using a continuation algorithm together with the temporal frequencies and spatial wave numbers for spiral waves as functions of the system parameter $b$.}
\label{cf:4}
\end{figure}

\paragraph{Existence, continuation, and logarithmic phase.} Existence of spiral waves has been proved only in the special case of the complex Ginzburg--Landau equation and, extending from there by perturbative arguments, in the vicinity of a Hopf bifurcation in the reaction-diffusion kinetics \cite{s-siam}. In excitable media, good matched asymptotic approximations are available \cite{keener1,keener2,bernoff,barkleyrecipe}. From the point of view taken in this paper, spiral waves will exist in open classes of reaction diffusion systems, possibly containing a connected region that included both oscillatory and excitable media. We used numerical continuation to follow a spiral wave from the excitable to the oscillatory regime in Barkley's variant of the FitzHugh--Nagumo system; see Figure~\ref{cf:4}. Note that one typically thinks of excitable media as organized around excitation pulses and their periodic concatenation, so-called trigger waves, whereas periodic media are organized around spatially homogeneous oscillations and their spatial modulation, so-called phase waves. At the transition from oscillatory to excitable media, excitation pulses terminate in a saddle-node bifurcation \cite{bellay,carter1}, while homogeneous oscillations end in homoclinic or Hopf bifurcations. Phase waves can however be continued to trigger waves \cite{engel1,bellay}, and we show in Figure~\ref{cf:4} that spiral waves emitting those phase and trigger waves, respectively, are connected in parameter space. Spiral waves eventually terminate at a point where the temporal frequency $\omega_*$ approaches zero ($\omega_*\searrow 0$) in the regime of weak excitability. In this regime, the wavelength of wave trains selected by the spiral diverge. It is worth noticing however that during the crossover from excitable into oscillatory regimes, the spiral-wave profile changes very little. 

\begin{figure}
\centering
\includegraphics[scale=0.9]{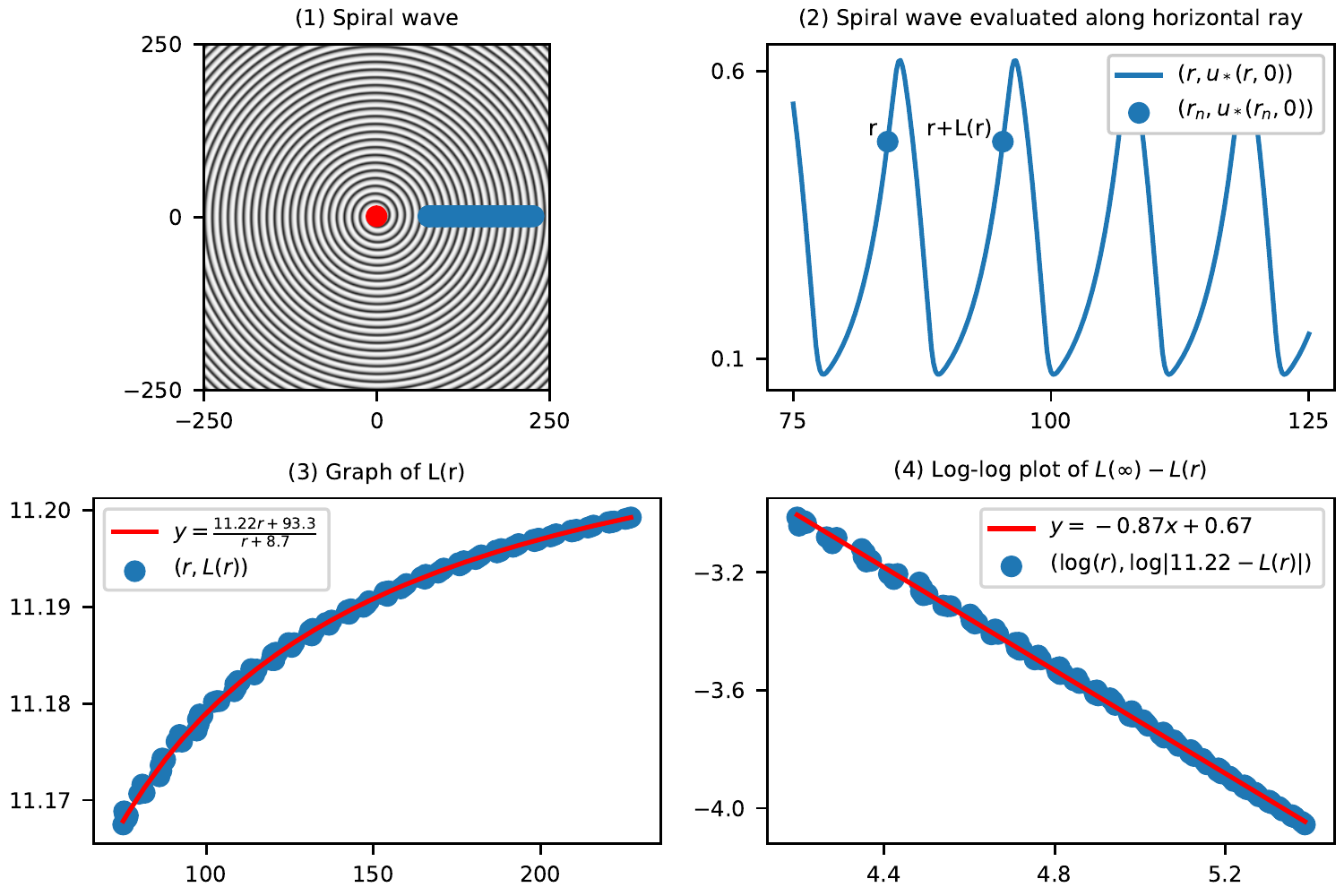}
\caption{Panel~(1) shows a rigidly-rotating spiral wave in the Barkley model. Panel~(2) shows the spiral wave evaluated along the horizontal ray starting at the center of the spiral: for each $r$, we denote by $r+L(r)$ the second-next value of the radius at which the spiral wave attains the same value. Panel~(3) shows the graph of the function $L(r)$ together with a curve fit of the quotient of two linear functions: the fit shows that the asymptotic period is $L(\infty)=11.22$. In panel~(4), we plot $\log(L(\infty)-L(r))$ against $\log(r)$: a fit with a linear function gives a slope of $-0.87$, which is close to the expected value of $-1$.}
\label{cf:7}
\end{figure}

Our main robustness result relies on the construction of a center manifold $\mathcal{M}^\mathrm{c}_+$, which continues wave-train solutions to finite radii $r$. Solutions in this center manifold can be expanded in the radius $r$ and the leading-order correction contains the effect of curvature on the speed and wavenumber of wave trains. Figure~\ref{cf:7} confirms the predicted logarithmic phase shift and corresponding algebraic $\frac{1}{r}$-convergence of the wavenumber. 

\begin{figure}
\centering
\includegraphics{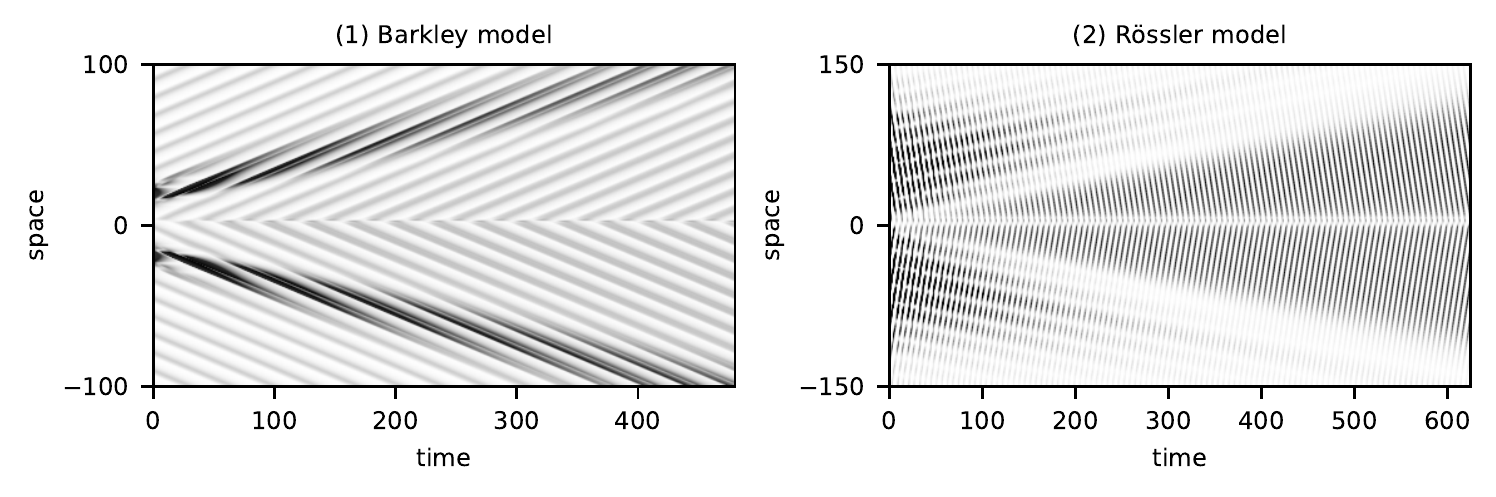}
\caption{To illustrate transport properties of spiral waves, we added a localized perturbation near the spiral core for spiral waves with positive group velocity, evaluate the difference of the original and the perturbed solution along a line through the spiral core, and plot the result overlaid with the original spiral waves as a space-time plot for spiral waves in the Barkley model (positive phase velocity) in panel~(1) and the R\"ossler system (negative phase velocity) in (2). These computations indicate that, in both cases, perturbations are transported away from the core}
\label{cf:3}
\end{figure}

\paragraph{Group velocities and transport.} A crucial property of spiral waves we assumed throughout this paper is that the group velocities of the asymptotic wave trains are directed outward in the far field. This basic property underlies the selection of wavenumber and frequency by the spiral core and determines growth and decay properties of eigenfunctions and adjoint eigenfunctions on the imaginary axis. Figure~\ref{cf:3} illustrates that positive group velocities imply outward transport via direct simulation in Barkley's model and in the R\"ossler system. Shown are temporal dynamics along a line section through the center of rotation, which clearly exhibit outward transport (diffusive decay and spreading) of perturbations. We emphasize that this transport is independent of the apparent phase velocity of wave trains, which indeed is directed towards the core of the spiral in the R\"ossler system. 

\begin{figure}
\centering
\includegraphics[width=\textwidth]{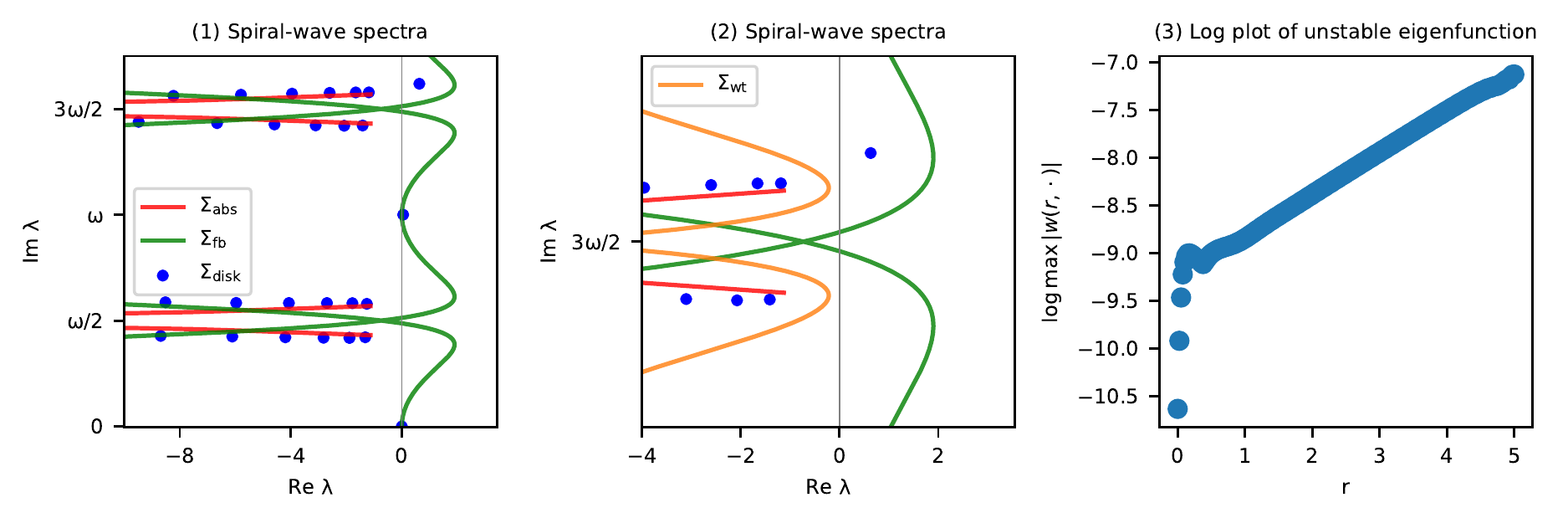}
\caption{Panel~(1) shows the spectrum \rev{$\Sigma_\mathrm{disk}$ (blue disks)} of an unstable spiral wave in the Karma model posed on a bounded disk with Neumann boundary conditions together with the stable absolute spectrum \rev{$\Sigma_\mathrm{abs}$ (red curves)} and the unstable Fredholm boundary \rev{$\Sigma_\mathrm{fb}$ (green curves)}. As shown in panel~(2), the Fredholm boundaries stabilize in exponentially weighted spaces (we used $\eta=-1$ \rev{to obtain $\Sigma_\mathrm{wt}$ (orange curves)}) and point spectrum may emerge as the spectral boundaries move. The instability is caused by a discrete point eigenvalue that belongs to the extended point spectrum: as indicated in panel~(3), the associated eigenfunction grows exponentially as $r$ increases.}
\label{cf:6}
\end{figure}

\paragraph{Spectra at linearization and shape of eigenfunctions.} 
Without using any information about the specific model, our results predict a number of structural and qualitative properties of the spectra\footnote{The different spectra we refer to in this paragraph are defined in \S\ref{s:mr.fredholm}} of the linearization about a spiral wave. Figure~\ref{cf:6} illustrates many of those basic properties in the Karma model. We computed \rev{Fredholm boundaries, Fredholm boundaries} in weighted spaces, and absolute spectra based on the wave train linearization using continuation \cite{ssr}. We compared those with spectra computed in a finite-size disk. As predicted, spectra stabilize when exponential weights are introduced that allow exponential growth of functions. Absolute spectra are stable and eigenvalues in finite-size disks cluster along the absolute spectrum. We also see an unstable isolated eigenvalue in the extended point spectrum. As predicted, the eigenfunction belonging to this unstable eigenvalue exhibits exponential growth in the radial variable and therefore contributes to the kernel only in the exponentially weighted space. 

\begin{figure}
\centering
\includegraphics{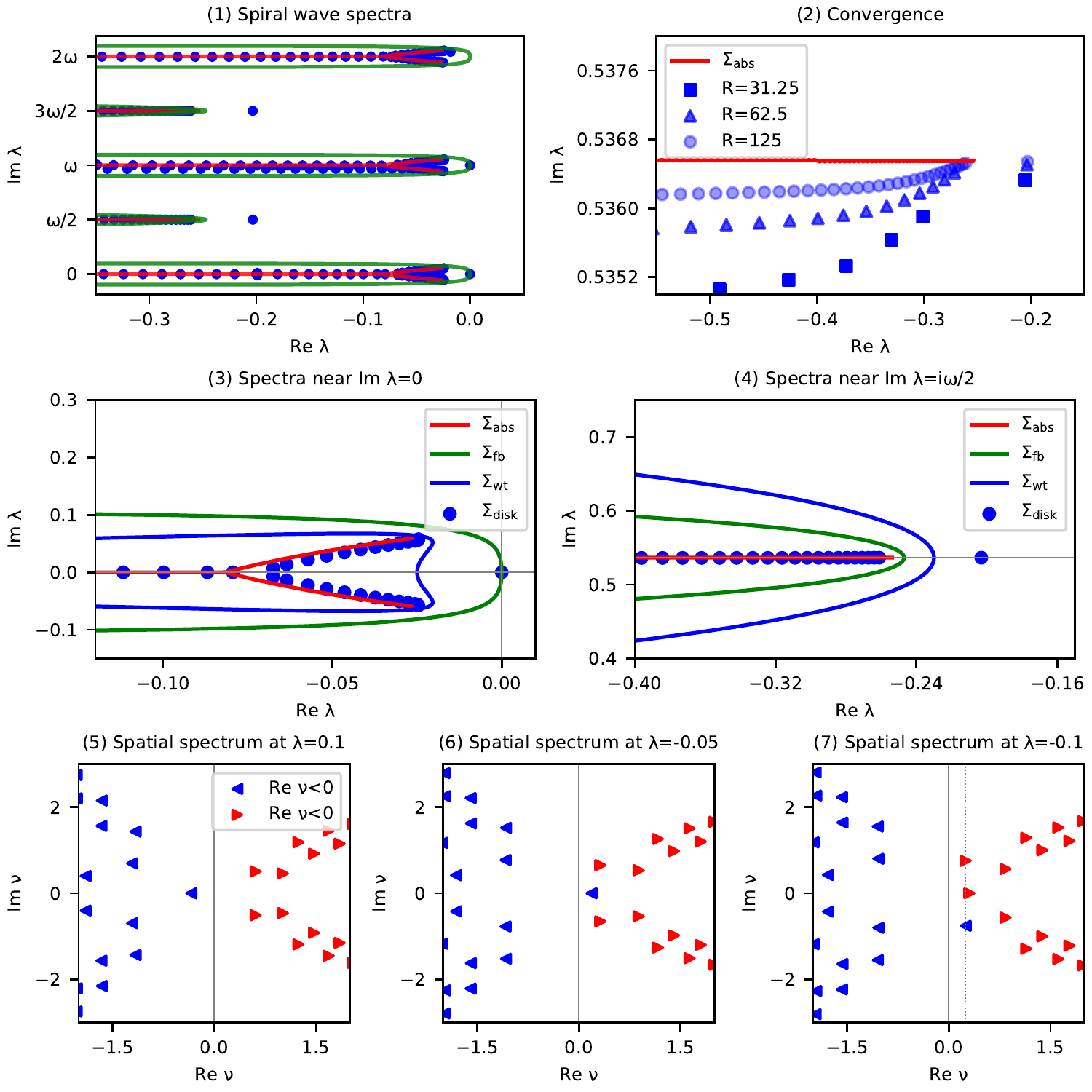}
\caption{For spiral waves in the R\"ossler model, panel~(1) contains the discrete spectrum of the spiral on a bounded disk of radius $R=125$ together with the absolute spectra (red curves) and Fredholm boundaries (green curves). Panel~(2) illustrates the predicted convergence towards the absolute spectrum, while panels~(3) and~(4) zoom in on the rightmost regions along $\Re\lambda=0$ and $\Re\lambda=\omega/2$, respectively, and also include Fredholm boundaries (blue curves) computed in an exponentially weighted norm with negative rate $\eta<0$. Panels~(5)-(7) contain the spatial spectra for $\lambda=0.1, -0.05, -0.1$, respectively.}
\label{cf:5}
\end{figure}

\begin{figure}
\centering
\includegraphics[scale=0.9]{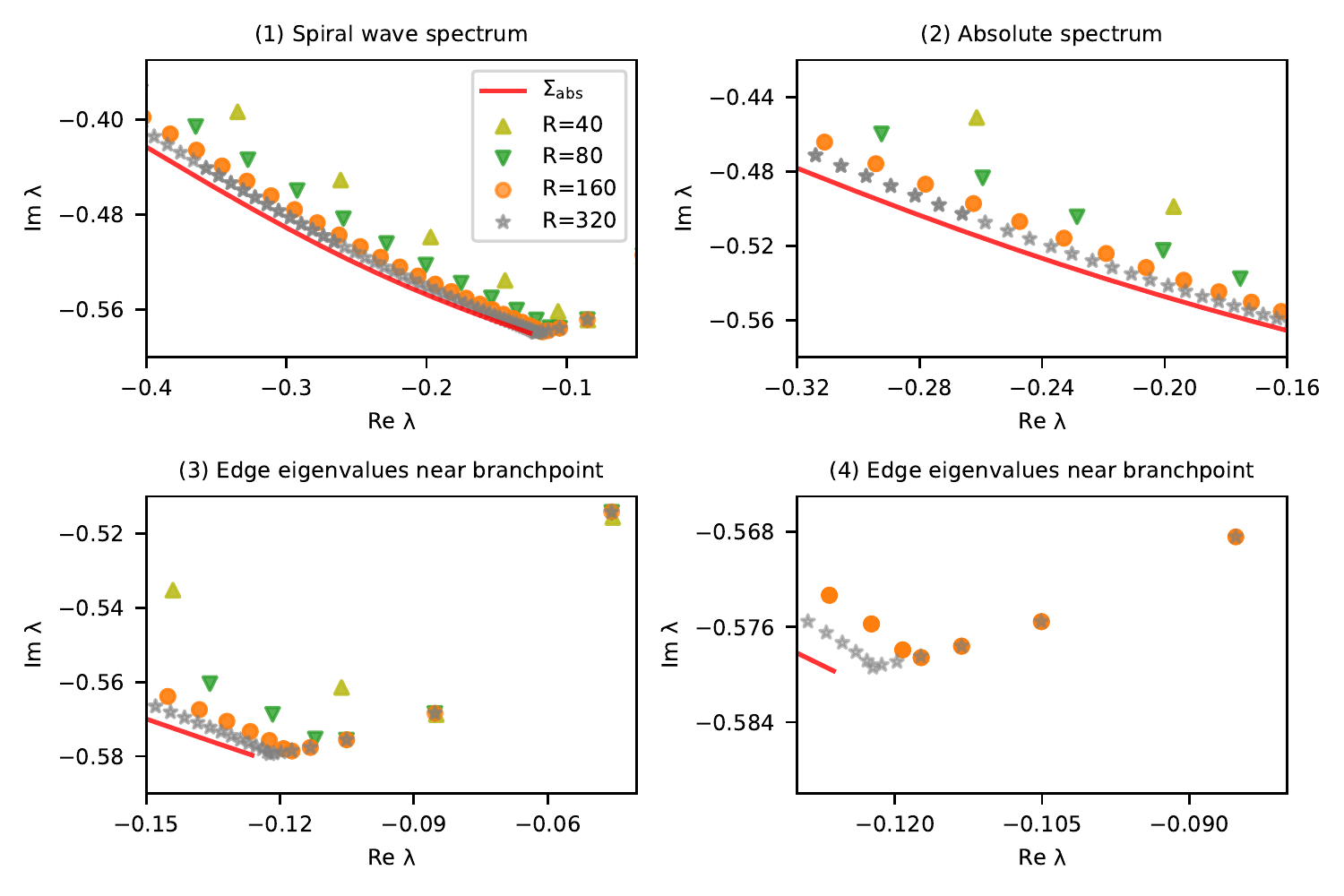}
\caption{Shown are the spectra of the spiral wave in the B\"ar--Eiswirth model for radii $R=40, 80, 160, 320$. Panel~(2) demonstrates convergence to the absolute spectrum. Panels~(3) and~(4) illustrate how eigenvalues in the extended point spectrum can emerge from the edge of the absolute spectrum.}
\label{cf:9}
\end{figure}

We note here that the location of the unstable eigenvalue near the edge of the absolute spectrum is not purely incidental (see Figure~\ref{cf:9} for another clearer example). In \cite{ss-edge}, we showed using formal asymptotics that eigenvalues in the extended point spectrum can accumulate on the edge of the absolute spectrum (or, conversely, that eigenvalues in the extended point spectrum can emerge from branch points of the absolute spectrum) and predicted the asymptotic locations for these eigenvalue clusters. Since these eigenvalues belong to the extended point spectrum, they converge exponentially in the radius $R$ of the domain, as opposed to the weak set-wise, algebraic convergence near the absolute spectrum. Their presence can be roughly attributed as follows to curvature corrections to the wave train linearization. Curvature effects can be thought of as slowly varying in space. In an adiabatic approximation, one can then consider the linearization at a curved wave train to predict possible eigenvalues. If, for instance, the curved wave train is more unstable than the planar wave train, this would then predict existence of eigenvalues to the right of the spectrum of the wave train. Though it appears to be difficult to analytically predict the rightmost of these eigenvalues, which would give rise to the first instability, complex conjugation $\lambda\mapsto \bar{\lambda}$ and Floquet-covering symmetry $\lambda\mapsto \lambda+\rmi\omega_*$ generally predict the robust presence of near-resonant eigenvalues at $\pm\rmi\ell\omega_*$ or $\pm\rmi(\ell+\frac{1}{2})\omega_*$ with $\ell\in\Z$, a fact that contributes to the rich phenomenology of spiral instabilities that we shall discuss briefly below. 

We use the R\"ossler model to illustrate our predictions for eigenvalue clusters near the absolute spectra and their relation to spatial spectra in more detail; see Figure~\ref{cf:5}. We observe, in particular, the predicted $\rmi\omega_*$-periodicity of eigenvalue clusters, algebraic $\frac{1}{r}$-convergence of eigenvalue clusters to the absolute spectrum together with increased density of clusters, and typical singularities of absolute spectra as triple junctions and branch-point termination. We also computed the spatial Floquet exponents $\nu_j$ and demonstrate how crossing of real parts on the imaginary axis induces essential spectrum and crossing real parts of separate eigenvalues corresponds to absolute spectra. We note that it was shown in \cite{dodson2019} that the discrete eigenvalue near $\rmi\omega_*/2$ arises as an eigenvalue of the boundary sink that accommodates Neumann boundary conditions.

Figure~\ref{cf:9} contains a refined numerical analysis near absolute spectra in the B\"ar--Eiswirth model, which shows the very rapid convergence of (extended) point spectrum versus algebraic convergence of clusters on the absolute spectrum and also demonstrated the emergence of point eigenvalues from the edge of the absolute spectrum.

\begin{figure}
\centering
\includegraphics[scale=0.8]{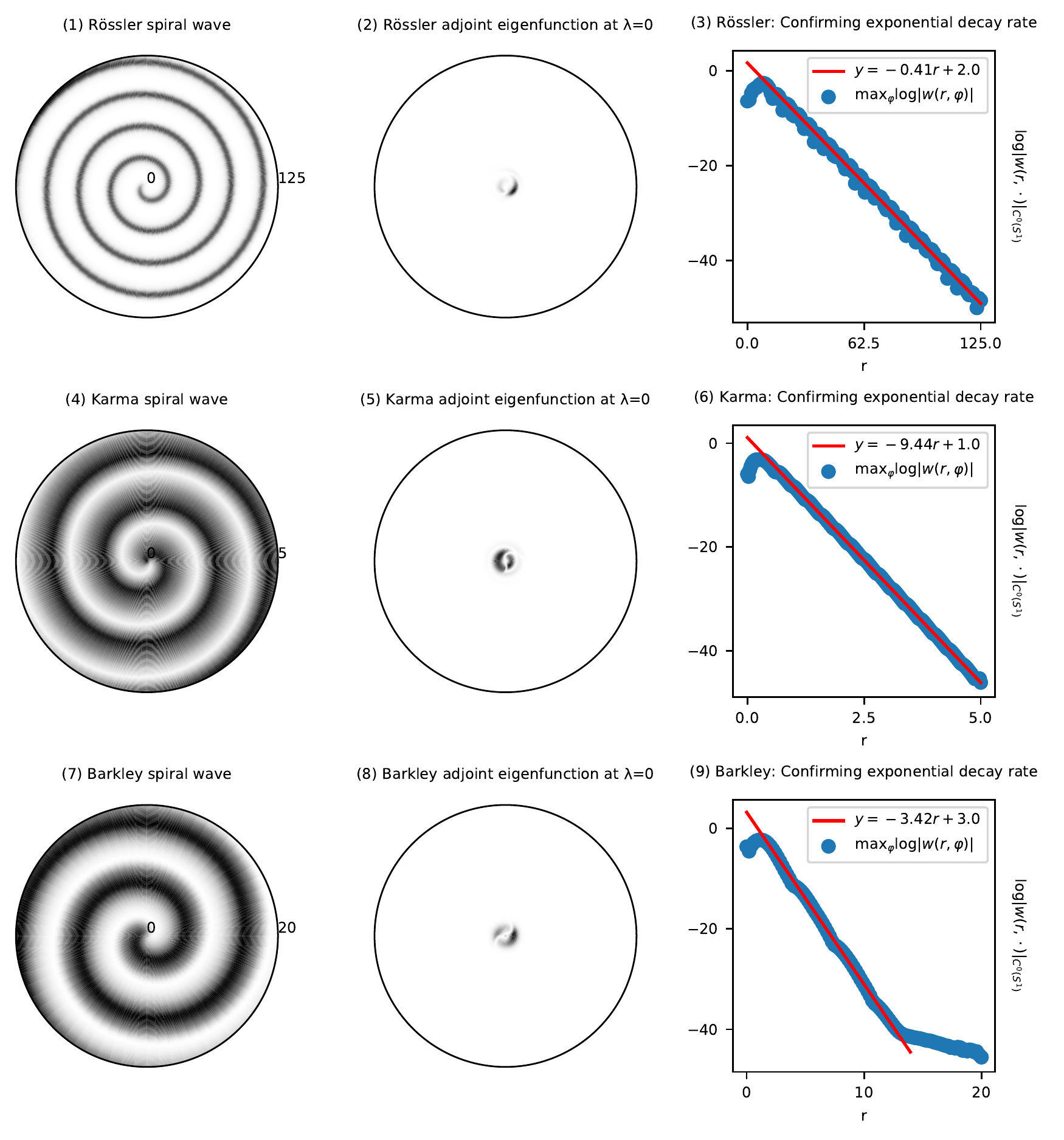}
\caption{Separately for the R\"ossler, Karma, and Barkley models, we show snapshots of spiral waves and the associated eigenfunctions $w(r)$ of the adjoint linearization belonging to the rotation eigenvalue $\lambda=0$ together with a log plot that indicates that the adjoint eigenfunctions decay exponentially as $r$ increases. The expected decay rates predicted by the associated spatial spectra are $-0.42$ for R\"ossler, $-9.47$ for Karma, and $-3.55$ for Barkley, thus indicating good agreement between theory and numerical computations. Note that the spiral wave for the Barkley model is in the oscillatory regime.}
\label{cf:2}
\end{figure}

\paragraph{Position and response.} Our results on spectral properties include characterizations of adjoint eigenfunctions. In particular, we proved that the adjoint eigenfunctions associated with the rotation eigenvalue $\lambda=0$ and the translation eigenvalues $\lambda=\pm\rmi\omega_*$ are exponentially localized \rev{as originally conjectured in \cite{Biktashev1994}}. Assuming that no other eigenvalues in the extended point spectrum are located on or to the right of $\rmi\R$, these three eigenvalues span the tangent space of a center manifold to a spiral wave in any large finite disk, consisting of rotated and translated spirals. Perturbations of the centered spiral will rapidly relax to this center manifold, with position on the center manifold computed to leading-order approximation by the spectral projection onto the center eigenspace. As a consequence, the effect of a spatially localized perturbation on a spiral wave is to leading order a phase shift in the rotation and a translation. The magnitude of the effect can be computed by evaluating the scalar product in $L^2$ of the perturbation and the adjoint eigenfunctions. As a consequence, the effect of perturbations decreases exponentially with distance from the center of rotation, making spiral waves extremely robust also against perturbations of initial conditions as long as those are centered away from the core. Adjoint eigenfunction were computed, for instance, in \rev{\cite{Biktasheva1998, Biktasheva2006, Biktasheva2009, Marcotte2}. We provide additional computations of adjoint eigenfunctions in Figure~\ref{cf:2}, where we also} compare the spatial exponential decay rates with the rates predicted by the spatial spectra.

%and comparisons of direct simulations with predictions of drift of spiral waves caused by external time-dependent forcing showed good agreement; see, for instance, \cite{Biktasheva2003, Biktasheva2010} and references therein) showed with direct simulations good agreement with direct simulations. 

\subsection{Instabilities of spiral waves}\label{s:bif}

\begin{figure}
\centering
\includegraphics{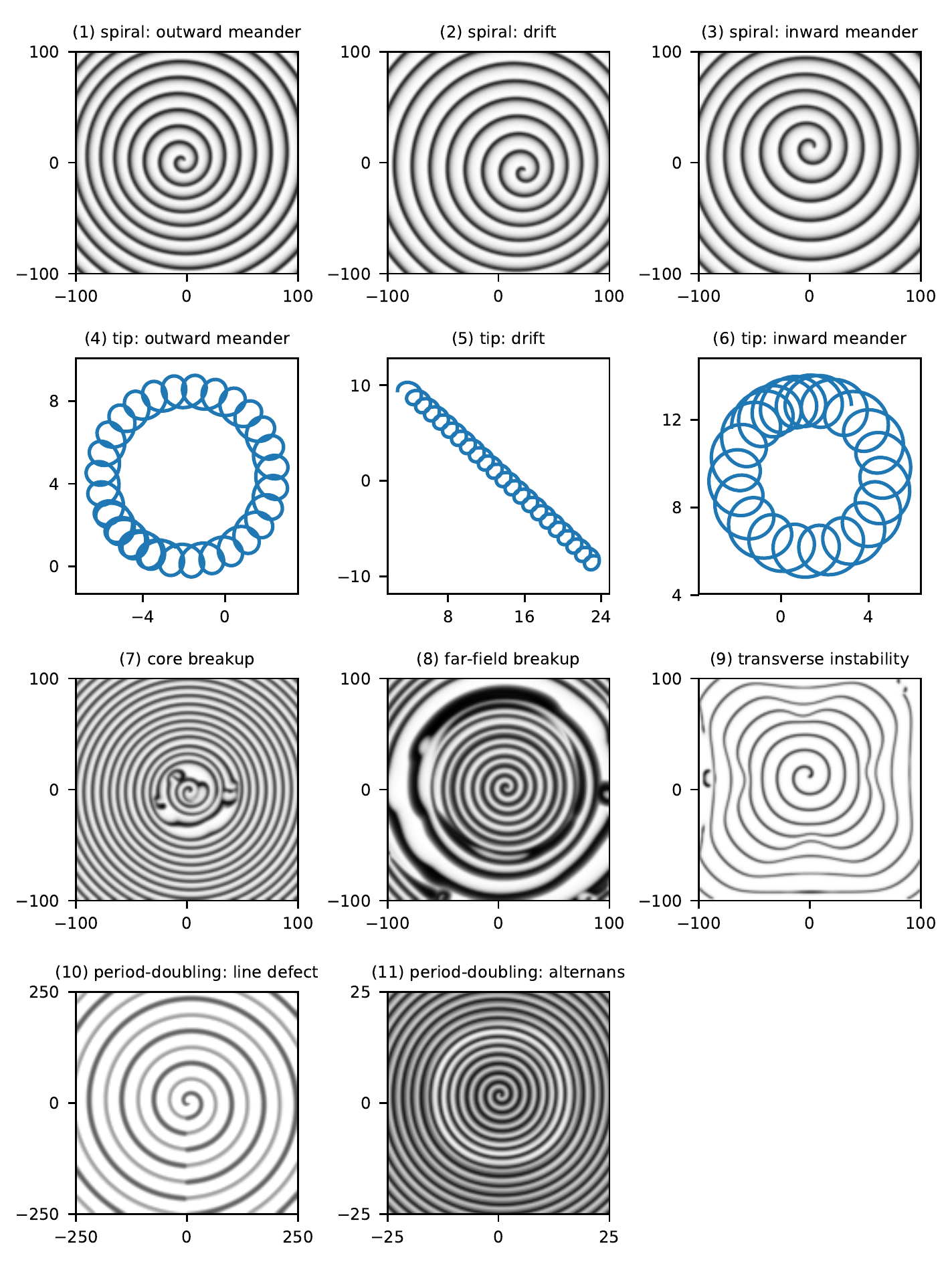}
\caption{We illustrate several typical instabilities of spiral waves. Panel~(1)-(3) show snapshots of outwardly meandering, drifting, and inwardly meandering spiral waves in the Barkley model; the curves traced out by the spiral tips are shown in panels~(4)-(6). Panels~(7)-(8) show core and far-field instabilities in the B\"ar--Eiswirth model, while panel (9) shows the snapshot of a spiral wave in the FitzHugh--Nagumo model that exhibits a transverse instability. Panels~(10)-(11) show period-doubling instabilities in the R\"ossler system and the Karma model, respectively.}
\label{cf:1}
\end{figure}

The structural description of essential, absolute, and point spectra allows us to classify instabilities of spiral waves. In the following we list common instabilities and explain the implications of our spectral analysis on the phenomenology. 
An overview  is shown in Figure~\ref{cf:1}.

\begin{figure}
\centering
\includegraphics{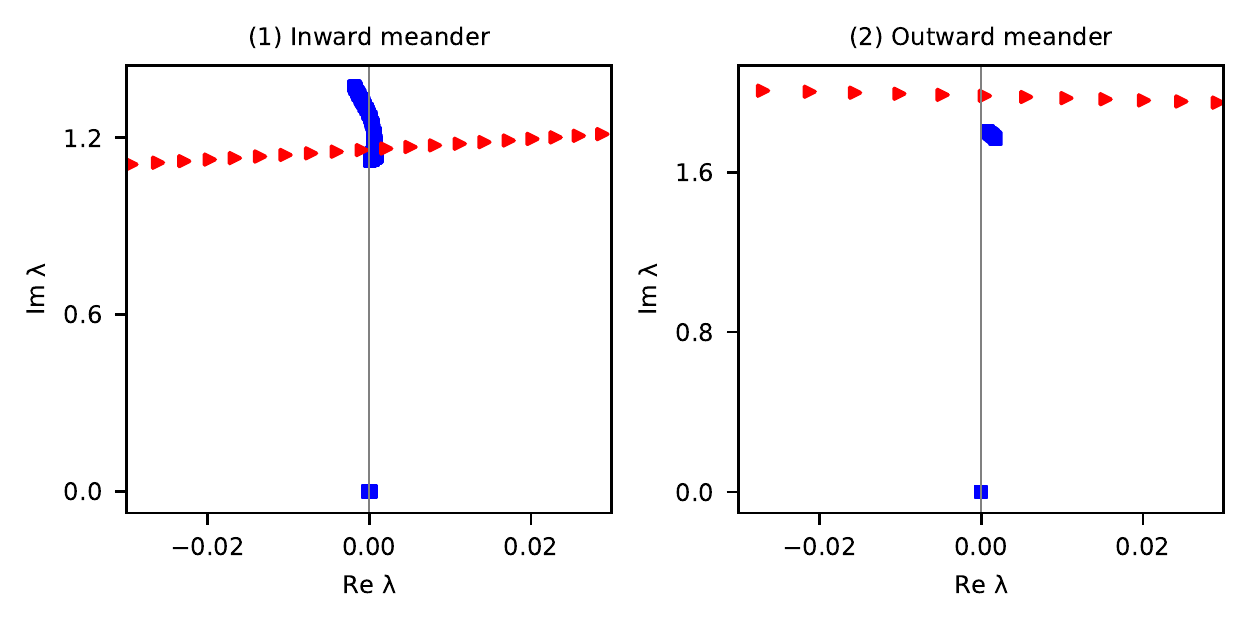}
\caption{Shown are the rotation and translation eigenvalues together with the Hopf eigenvalues that cause the transition to (1) inward and (2) outward meander in the Barkley model as the parameter $a$ is varied.}
\label{cf:8}
\end{figure}

\paragraph{Transition to meandering and drifting spiral waves}
The possibly most prominent spiral-wave instability, described in the introduction, is the transition from rigidly rotating to meandering and drifting spiral waves. Tracking the location of the spiral tip, defined for instance through the location $x(t)\in\R^2$ where $u(x(t),t)=\bar{u}$ for some fixed $\bar{u}\in\R^2$, one notices that, past a distinct threshold of a system parameter, the motion occurs on epicycloids rather than circles. In other words, small periodic circle motions are superimposed on the primary circular rotation. These superimposed rotations can occur with the same or the opposite orientation as the primary rotation, leading to outward and inward petals in the epicycloids, respectively; see Figure~\ref{cf:1}(4-6). At the codimension-one transition from outward to inward petals, the spiral wave moves along a straight line. An explanation of this striking motion was found by Barkley \cite{barkleyeuclid}, noticing the coupling of Hopf instability modes to the inherent neutral modes induced by translation and rotation. More formalized treatments, both in terms of center-manifold reductions and reduced dynamics followed in \cite{fssw,golubitskymeander,ssw1,ssw2}. We remark here that all of those rely on a \rev{spectral gap} which, for the Archimedean spirals considered here, is not present. In Figure~\ref{cf:8}, we illustrate the Hopf instability in the Barkley model by computing eigenvalues during the transition and showing that the instability is caused by point spectrum with frequency $\omega_\mathrm{H}>\omega_*$ for outward meander and $\omega_\mathrm{H}<\omega_*$ for inward meander. Since the Floquet spectrum of the wave trains touches the imaginary axis at $\rmi\omega_*$, eigenfunctions grow linearly in $r$ at resonance $\omega_\mathrm{H}=\omega_*$, and are localized only with small exponential rate for near-resonant Hopf bifurcations. Using the results on shape of eigenfunctions presented here, we were able to predict in \cite{ss-sup} striking superspiral patterns in the far field of meandering spirals. Meandering transitions and the associated super-spiral patterns were observed in \cite{Jahnke1989,Plesser1990,swinneymeander,Perez-Munuzuri1991,Ouyang1996}

\paragraph{Core and far-field breakup}
Instabilities caused by essential rather than point spectrum exhibit a more complex phenomenology. Often, the associated instabilities of wave trains are subcritical, leading to breakup of wave trains and spatio-temporal chaos. We investigated such instabilities from the point of view presented here in \cite{ss-spst}; see Figure~\ref{cf:1}, panels (7) and (8) for phenomenologies. Depending on parameter values, the absolute spectrum induces eigenfunctions corresponding to exponential growth or exponential decay in $r$. The resulting instability then is strongly localized in the far field or near the core, respectively. In the former case, the instability is convective at onset, with perturbations growing as they travel outwards, but decaying pointwise: the essential spectrum is unstable but the absolute spectrum is still stable. Only when the absolute spectrum destabilizes do perturbations grow pointwise and perturbations invariably lead to breakup of the primary spiral. The onset of convective and absolute instability matches well the prediction from computations of spectra of wave trains. In the regime of convective instability, the subcritical nature of the instability implies that basins of attraction of the spiral are exponentially small in the size of the domain. In the case of absolute spectrum with exponential decay, instabilities grow in the core region and the transition immediately leads to breakup and turbulence with small correlation length scales. Barkley and Wheeler \cite{BW} confirmed the predictions made in \cite{ss-spst} using numerical computation of spectra of spirals in bounded domains. In addition to the eigenvalue clusters along absolute spectra with the predicted exponential radial decay, they found an eigenvalue in the extended point spectrum near the edge of the absolute spectrum, thus showing that the core instability is in fact caused by subcritical Hopf bifurcation due to extended point spectrum. The location of the leading Hopf eigenvalue near the edge of the absolute spectrum can be attributed to curvature effects as analyzed in \cite{ss-edge}.

\paragraph{Period-doubling bifurcations and alternans}
As mentioned above, Floquet and complex conjugation symmetries of essential and absolute spectra can lead to robust resonances. One of those resonances is the robust period-doubling of a spiral wave, intrinsically linked to a period-doubling of a periodic orbit in the kinetics; see \cite{sspd,dodson2019} and references therein. The three-variable R\"ossler ODE exhibits periodic orbits that undergo a period-doubling cascade. When adding diffusion to all three components, the resulting system supports spiral waves that emit phase waves, which, in turn, can undergo a period-doubling instability. In the linearization of the spiral wave, this period-doubling instability corresponds to marginally stable spectrum at $\rmi\omega_*/2$ (half the spiral frequency). The instability causes the emergence of line defects in the far field which appear almost stationary; see Figure~\ref{cf:1}(10--11). In any finite-size domain, this resonant instability can be caused by resonant absolute spectrum, near-resonant extended point spectrum near the edge of the absolute spectrum, or by period-doubling instabilities through extended point spectrum of the boundary sink; see again \cite{sspd,dodson2019}. We emphasize that this robust period-doubling instability really can only be understood through an analysis of the far field and the limit of unbounded domains since a spiral, in any finite domain, is simply an equilibrium in a corotating frame, rendering the possibility of a period-doubling impossible. In particular, since time evolution of the spiral is simply given by rotation, the center manifold along the periodic orbit is a trivial bundle induced by symmetry rather than the non-orientable M\"obius strip typical in period-doubling instabilities. 

\subsection{Open problems}\label{ss:op}

We conclude with a discussion of open problems and possible extensions. 

\paragraph{Nonlinear perturbations: stability.}
The presence of essential spectra on the imaginary axis for the linearization at a spiral wave, induced by the wave trains in the far field, is essential to much of the implications of our results described above. In this regard, essential spectra, while inherently complicated, give us additional insight, while of course presenting many technical challenges. Our results exploit the linear theory to show robustness under parameter changes and domain truncation to large disks. In any finite-size, sufficiently large disk, our results give a rigidly rotating wave with a simple zero eigenvalue induced by the rotational symmetry. Standard semigroup methods \cite{hen} then yield nonlinear asymptotic stability of the spiral wave profile in any such large domain. A significant drawback of this argument is the fact that the size of the  basin of attraction established in such stability proofs depends on the norm of the resolvent, used to construct the spectral projections and the exponential decay estimates in the complement of the center subspace. Our results show that the norm of the resolvent grows exponentially in $R$, leading to exponentially small estimates in $R$ for the basin of attraction. The discussion of the dynamics in the case of far-field breakup show that such a conclusion is in fact optimal when only information from the bounded domain is used. 

To strengthen the result, one would need to incorporate the effect of the essential spectrum, tracking in particular  how perturbations are transported away from the spiral core and decay diffusively. While such results have been established in one spatial dimension \cite{becksource}, the two-dimensional radial geometry is likely to introduce difficulties due to the azimuthal stretching that prevents diffusive decay; see  \cite{roussier} for a related analysis. 

\paragraph{Nonlinear perturbations: boundaries and interaction.}
Due to the exponential localization of the adjoint, one expects more robust persistence results for truncation to bounded domains in the form of a slow manifold parametrized by translations and rotations of the spiral wave, with dynamics induced by the boundary, exponentially slow in the distance between core and boundary. In a similar way, one would expect to be able to describe the nonlinear interaction of multiple spirals on a reduced slow manifold parametrized by their respective position and rotational phase, with dynamics that are exponentially slow in the separation distance of the spiral cores. Approximate profiles for such multi-spiral solutions would be obtained by inserting sinks similar to the boundary sinks considered here in between the spiral domains. Similar to the boundary sinks we used in the domain truncation, we would not expect these domain boundaries to contribute neutral eigenvalues to the linearization, that is, their dynamics would be determined by phase matching of waves emanating from the spiral core; see \cite{ssdefect} for the one-dimensional analogue. 

\paragraph{Bifurcations: nonlinear aspects.}
Further extending the nonlinear analysis, beyond asymptotic stability and interaction dynamics, one would want to complement the linear predictions for spiral wave instabilities with nonlinear analysis. A simple example would be an existence proof for meandering spirals past a super-critical Hopf bifurcation. Again, one could resort to analysis in a large bounded disk. Assuming that the critical extended point spectrum consists of a simple pair of imaginary eigenvalues crossing the imaginary axis in addition to the neutral eigenvalues induced by translation and rotation, one finds a 5-dimensional center manifold in any bounded domain. We conjecture that the Taylor jet of the vector field on this manifold converges exponentially as the size of the disk radius $R$ increases and that the limiting equation is given by the skew-product equations from \cite{barkleyeuclid}. 

Again, this analysis is somewhat unsatisfactory since it only captures exponentially small neighborhoods of the primary spiral wave profile. Also, the interesting resonant case with a drifting spiral is not accessible by this approach, as generally the truncation to a bounded domain destroys the underlying Euclidean symmetry. Lastly, the approach fails to clearly describe nonlinear effects such as possible frequency locking on super patterns in the far field. 

More ambitious reductions would, especially in the case of bifurcations involving instabilities of wave trains, derive in a rigorous fashion equations that couple the localized core dynamics to far-field modulation equations such as the viscous eikonal equation or, in the case of instabilities, the Kuramoto--Sivashinsky equation for breakup or even coupled mode equations in period-doubling instabilities. 

% \begin{compactitem}
% \item are their spiral wave saddle-nodes? should be,...
% \item justification of [Ashwin et al.] and [LeBlanc,Wulff]
% \end{compactitem}

\paragraph{Other instabilities and bifurcations.}
Some instability mechanisms do not fit well into the formalism developed here. One prominent example are instabilities of wave trains against perturbations perpendicular to their direction of propagation; see Figure~\ref{cf:1}(9). In the simplest case, such instabilities arise when $d_\perp$ becomes negative. Inspecting our results, one readily notices that the two-dimensional stability of wave trains simply does not affect the spectra of spiral waves. In fact, as shown in the proof of Lemma~\ref{l:spiralperp}, the temporal frequency of transverse perturbations of a wave train in the corotating frame of a spiral wave converges to infinity as their distance from the spiral core grows. These effects are relevant when trying to establish even linear stability from the spectral-stability assumptions made in the present paper. In general, spectral stability for generators of strongly continuous semigroups may not imply exponential growth bounds without further assumptions on resolvent bounds; see, for instance, \cite{engelnagel} for such additional assumptions. We have shown that the semigroup associated with the linearization at a spectrally stable spiral waves whose asymptotic wave trains are transversely unstable exhibits exponentially growth with a strictly positive exponential rate. Note that this observation does not lead to contradictions as our results on convergence of spectra under truncation to finite disks hold only in compact subsets of the complex plane and therefore do not exclude unstable eigenvalues created near $\pm\rmi\infty$.

The scenario where the temporal frequency $\omega_*$ of the spiral tends to zero (so that $\omega_*\searrow0$) is not within the scope of the analysis presented here since the loss of the rotational term $\omega_*\partial\varphi$ changes the spatial dynamics at $r=\infty$ at leading order, rendering the equation completely degenerate. In this case of the so-called retracting-wave bifurcation, the spiral core grows and the branch of spiral waves in parameter space terminates on an asymptotically straight spiral arm that, while propagating in the normal direction, also retracts in the tangential direction. Symmetry considerations \cite{ashwin1} predict the growth of the spiral core as $\omega_*\searrow 0$ and a drifting wave at $\omega_*=0$; they also predict that the branch of spirals continues into the regime $\omega_*<0$, a phenomenon that has not been observed in experiments or simulations. 

\paragraph{Beyond spirals.}
We suspect that our results can also serve as a basis for the study of a variety of related and more complex phenomena in excitable and oscillatory media. Changing the winding number $\ell$ in the far field, we find target patterns $\ell=0$ and multi-armed spirals $\ell>1$. It appears that, in most simple models, neither structure exists as a stable periodic solution, although many of our analytical tools would apply to those structures with minor modifications. 

In three-dimensional physical space, one can ``stack'' spirals along filaments while possibly rotating the spiral. Straight filaments yield scroll waves and twisted scroll waves, circular filaments yield scroll waves; see \cite{winfree}. It seems that scroll waves and twisted scroll waves would be amenable to an analysis similar to the one presented here. More interestingly, the analysis here predicts that the filament dynamics as generalizations of tip dynamics should be described by a PDE for the three independent variables of local normal displacement of the filament and phase, as function of the arc length along the filament. Continuing the extended point spectrum of the spiral in a Fourier variable along the filament would then yield bending and torsion stiffness of the filament. 

%%%%%%%%%%%%%%%%%%%%%%%%%%%%%%%%%%%%%%%%%%%%%%%%%%%%%%%%%%%%%%%%%%%%%%%%%

\appendix
\section{Numerical computation of spiral waves in model systems}\label{s:sim}

\rev{In \S\ref{s:sim.models},} we outline the models we used to produce the computations and simulations summarized in \S\ref{s:appl} and provide a brief summary of the numerical algorithms used for these computations \rev{in \S\ref{s:sim.methods}}.

\subsection{Model systems}\label{s:sim.models}

\paragraph{Barkley model.}
The model
\begin{eqnarray*}
u_t & = & \Delta u + \frac{1}{\epsilon} u (1-u) \left( u - \frac{v+b}{a} \right) \\
v_t & = & \delta \Delta v + u - v
\end{eqnarray*}
was introduced in \cite{BKT} as a system that exhibits meander and drift of spiral waves. In \cite{Barkleyspectra}, spectral computations were used to demonstrate that these instabilities arise due to Hopf instabilities. This model also exhibits retracting spiral waves in the weakly excitable regime (see \cite{HG} and references therein). In all computations, we set $\delta=0.01$ and $\epsilon=0.02$. We used the following parameter values:
\begin{center}
\begin{tabular}{|llll|}\hline
\textbf{Description} & \textbf{Figure} & $\mathbf{a}$ & $\mathbf{b}$ \\ \hline
rigid (excitable) & \ref{cf:4}(2,6), \ref{cf:7}, \ref{cf:3}(1) & 0.8 & 0.05 \\
rigid (oscillatory) & \ref{cf:4}(1,4), \ref{cf:2}(7)-(9)  & 0.8 & -0.05 \\
retracting (weakly excitable) & \ref{cf:4}(3) & 0.44 & 0.05 \\
outward meander & \ref{cf:1}(1,4) & 0.67 & 0.05 \\
drift & \ref{cf:1}(2,5) & 0.63 & 0.05 \\
inward meander & ~\ref{cf:1}(3,6) & 0.59 & 0.05 \\ \hline
\end{tabular}
\end{center}
In Figure~\ref{cf:8}, $b=0.05$ is fixed and $a$ varies.

\paragraph{B\"ar--Eiswirth model.}
The model
\begin{eqnarray*}
u_t & = & \Delta u - \frac{1}{\epsilon} u (u-1) \left( u - \frac{v+b}{a} \right)\\
v_t & = & \delta \Delta v + g(u) - v
\end{eqnarray*}
with
\[
g(u) = \left\{ \begin{array}{lcl}
0                  & \quad & 0 \leq u < 1/3 \\
1 - 6.75 u (u-1)^2 &       & 1/3 \leq u \leq 1 \\
1                  &       & 1 < u
\end{array} \right.
\]
was introduced in \cite{PhysRevE.48.R1635}. As shown in \cite{PhysRevE.48.R1635}, it exhibits core and far-field instabilities of spiral waves. These instabilities were further studied using absolute and convective instabilities \cite{PhysRevLett.82.1160}, absolute spectra \cite{ss-spst, ss-edge}, and spectral computations \cite{BW}. Figure~\ref{cf:9} uses the parameter values $a=0.84$, $b=-0.045$, $\delta=0.1$, and $\epsilon=0.0751$. We used $a=0.75$, $b=0.0006$, $\delta=0.01$, and $1/\epsilon=13.15$ for core break-up in Figure~\ref{cf:1}(7) and $a=0.84$, $b=-0.045$, $\delta=0.01$, and $1/\epsilon=13.1$ for far-field break-up in Figure~\ref{cf:1}(8).

\begin{figure}
\centering
\includegraphics{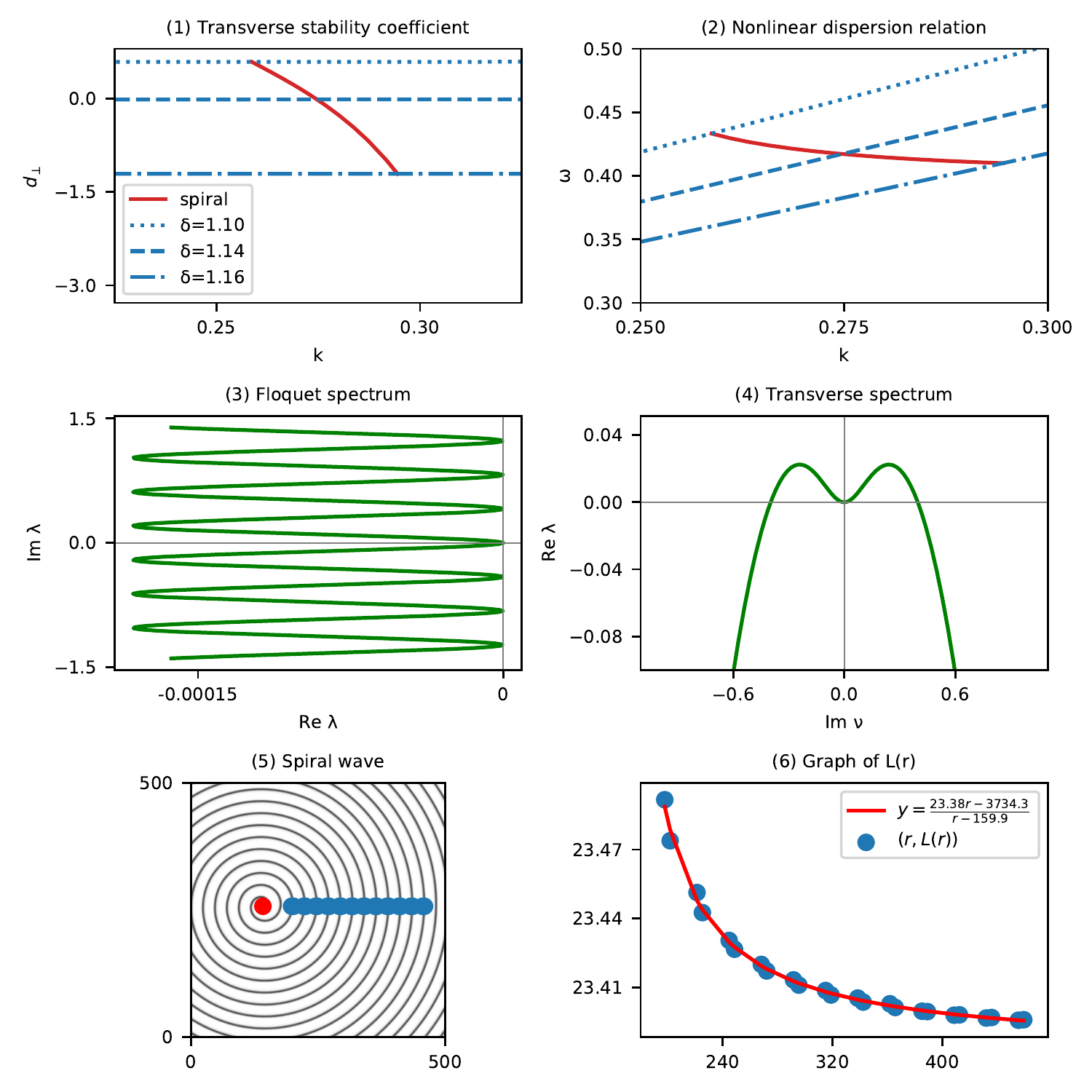}
\caption{For the FitzHugh--Nagumo system (\ref{m:fhn}), we show the transverse instability coefficient $d_\perp(k)$ and the nonlinear dispersion relation $\omega(k)$ in panels~(1) and~(2), respectively, as functions of the wavenumber $k$ of the wave trains for different values of $\delta$; also shown are the curves $(k_*,d_\perp)(\delta)$ and $(k_*,\omega_*)(\delta)$, respectively, evaluated along the spiral wave of (\ref{m:fhn}). Recall that $d_\perp<0$ corresponds to a transverse instability. Panels~(3)-(4) contain the Floquet and transverse spectra of the wave train with wavenumber $k=0.29$ and frequency $\omega=0.41$ selected by the spiral wave at $\delta=1.16$. Panels~(5)-(6) contain the spiral wave and the graph of successive wavelengths $L(r)$ at $\delta=1.16$, showing that wave trains are now compressed as $r$ increases (see Figure~\ref{cf:7} for additional details for the case $d_\perp>0$).}\label{mf:fhn}
\end{figure}

\paragraph{FitzHugh--Nagumo model.}
Transverse instabilities of spiral waves in the FitzHugh--Nagumo model
\begin{eqnarray}
u_t & = & \Delta u + \frac{1}{\epsilon} \left(u\left(u-0.5\right)(1-u) - \frac{v+b}{a}\right)
\label{m:fhn} \\ \nonumber
v_t & = & \delta \Delta v + u - v
\end{eqnarray}
were observed in \cite[Figure~9]{hagmer} (the model in \cite{hagmer} is written in a different form, which can be transformed into (\ref{m:fhn}) using a linear change of the dependent and independent variables). For Figure~\ref{cf:1}(9), we set $a=8$, $b=-0.45$, $\epsilon=1/57$, and $\delta=1.215$. Figure~\ref{mf:fhn}, which uses the same values for $(a,b,\epsilon)$, provides numerical evidence that the instability visible in Figure~\ref{cf:1}(9) is indeed caused by a transverse instability of the asymptotic wave train.

\paragraph{Karma model.}
The Karma model
\begin{eqnarray*}
u_t & = & 1.1\, \Delta u + 400 \left( -u + \left(1.5414 - v^4 \right) \left(1 - \tanh(u - 3)\right) \frac{u^2}{2} \right) \\
v_t & = & 0.1\, \Delta v + 4 \left( \frac{\vartheta(u-1)}{1 - \rme^{-Re}} - v \right)
\end{eqnarray*}
with $\vartheta(u)=(1+\tanh(4u))/2$ admits spiral waves that undergo period-doubling bifurcations to alternans. We refer to \cite{Karma1, Karma2} for the model and direct simulations, to \cite{Marcotte1} for spectral computations, and to \cite{dodson2019} for recent computations and further references. For our computations, we used the parameter value $Re=1.2$ in Figures~\ref{cf:6} and~\ref{cf:2}(4)-(6) and $Re=0.95$ in Figure~\ref{cf:1}(11).

\paragraph{R\"ossler model.}
The R\"ossler model
\begin{eqnarray*}
u_t & = & 0.4\, \Delta u - v - w \\
v_t & = & 0.4\, \Delta v + u + 0.2\, v \\
w_t & = & 0.4\, \Delta w + u w - c w + 0.2
\end{eqnarray*}
admits spiral waves with positive group velocity and negative phase velocity. Spiral waves exhibit spatio-temporal period-doubling bifurcations as $c$ is increased. We refer to \cite{kapralpd, davidsen, sspd, dodson2019} and the references therein for detailed studies of spiral waves in this model. In our computations, we used $c=2$ in Figures~\ref{cf:5} and~\ref{cf:2}(1)-(3), $c=3$ in Figure~\ref{cf:3}(2), and $c=4.2$ in Figure~\ref{cf:1}(10).

\subsection{Methods}\label{s:sim.methods}

Since our codes and data are publicly available \cite{code_dodson, code_location}, we discuss our numerical algorithms and the computational parameters only briefly.

\paragraph{Direct numerical simulations.}
We used the package \textsc{ez-spiral} written by Dwight Barkley \cite{ezspiral} for the direct numerical simulations shown in Figure~\ref{cf:4}(1)-(3) and Figures~\ref{cf:7}, \ref{cf:3}, \ref{cf:1}, and~\ref{mf:fhn}(5)-(6). In each case, we used a square domain of length $L$ with Neumann boundary conditions. The package \textsc{ez-spiral} uses a finite-difference scheme in space and provides both explicit and implicit Euler schemes for time integration. Details about the choices for $L$, grid sizes, and time steps are given in the repository \cite{code_location}.

\paragraph{Continuation and computations of spectra.}
To continue spiral waves in parameters and to compute their spectra, we used the \textsc{matlab} scripts developed in \cite{dodson2019, code_dodson}. The one-dimensional wave trains and two-dimensional spiral waves shown or used in Figure~\ref{cf:4}(4)-(6) and Figures~\ref{cf:6}-\ref{cf:2} and~\ref{cf:8}-\ref{mf:fhn} were computed using (\ref{e:wtode}) posed on a circle and (\ref{e:rds}) posed on bounded disks with Neumann boundary conditions, respectively. These equations were discretized in polar coordinates using a spectral Fourier scheme in the angular variable and finite differences in the radial direction, and the resulting systems were then solved using \textsc{matlab}'s \textsc{fsolve} routine. The point spectra of the linearization (\ref{e:lrds}) about spiral waves on bounded disks in Figures~\ref{cf:6}-\ref{cf:9} and~\ref{cf:8} were computed using \textsc{matlab}'s \textsc{eig} and \textsc{eigs} routines applied to the discretization of (\ref{e:lrds}). We also used these routines to compute the adjoint eigenfunctions shown in Figure~\ref{cf:2}. We computed the absolute and essential spectra (including transverse spectra) shown in Figures~\ref{cf:6}-\ref{cf:9} and~\ref{mf:fhn} using the algorithms developed in \cite{ssr}, which were implemented in \textsc{matlab}. The spatial spectra used in Figure~\ref{cf:2}(3,6,9) to predict the exponential decay rates of adjoint eigenfunctions and shown in Figure~\ref{cf:5}(5)-(7) to illustrate absolute and essential spectra were computed using \textsc{matlab}'s \textsc{eig} routine applied to the discretization of (\ref{e:spmtw}).

%%%%%%%%%%%%%%%%%%%%%%%%%%%%%%%%%%%%%%%%%%%%%%%%%%%%%%%%%%%%%%%%%%%%%%%%%

\begin{Acknowledgment}
Sandstede expresses his gratitude to Stephanie Dodson for very helpful conversations about the numerical computations of spiral waves and their spectra.
Sandstede was partially supported by the Alfred~P~Sloan Foundation and by the NSF through grants DMS-9971703, DMS-0203854, and DMS-1714429.
Scheel was partially supported by the NSF through grant DMS-0203301, DMS-0504271, DMS-1612441, and DMS-1907391.
\end{Acknowledgment}

\end{document}